\documentclass[11pt,A4paper]{book}
\newcommand{\e}{\varepsilon}
\newcommand{\n}{\noindent}
\newcommand{\x}{\mathbf x}
\newcommand{\uu}{\mathbf u}
\newcommand{\N}{\mathbb N}
\newcommand{\R}{\mathbb R}

\usepackage{epsfig,amssymb,latexsym,color,amsmath,pifont}
\begin{document}

\huge
\title{{\bf  MODULAR GENE DYNAMICS AND NETWORK THEORY AT MESOSCOPIC SCALE}  \\[0.8cm] {\it PhD Thesis} }

\author{\huge Zoran Levnaji\'c \\[.4cm] 
{\it \Large  Department of Theoretical Physics, Jo\v{z}ef Stefan Institute} \\ 
{\Large \it Jamova 39, SI-1000 Ljubljana, Slovenia} \\[4.0cm]
{\Large Adviser: Prof. Bosiljka Tadi\'c} \\[6.0cm]
{\Large Jo\v{z}ef Stefan International Postgraduate School}}
\date{}
\maketitle

\normalsize
\pagenumbering{Roman}

\addcontentsline{toc}{section}{Table of Contents}
\tableofcontents

\newpage
\addcontentsline{toc}{section}{Abstract}
\section*{Abstract}

Complex dynamical systems are often modeled as networks, with nodes representing dynamical units which interact through the network's links. Gene regulatory networks, responsible for the production of proteins inside a cell, are an example of system that can be described as a network of interacting genes. The behavior of a complex dynamical system is characterized by cooperativity of its units at various scales, leading to emergent dynamics which is related to the system's function. Among the key problems concerning complex systems is the issue of stability of their functioning, in relation to different internal and external interaction parameters. 

In this Thesis we study two-dimensional chaotic maps coupled through non-directed networks with different topologies. We use a non-symplectic coupling which involves a time delay in the interaction among the maps. We test the stability of network topologies through investigation of their collective motion, done by analyzing the departures from chaotic nature of the isolated units. The study is done on two network scales: (a) full-size networks (a computer generated scalefree tree and a tree with addition of cliques); (b) tree's characteristic sub-graph 4-star, as a tree's typical dynamical motif which captures its topology in smallest possible number of nodes and is suitable for time-delayed interaction. We study the dynamical relationship between these two network structures, examining the emergence of cooperativity on a large scale (trees) as a consequence of mesoscale dynamical patterns exhibited by the 4-star.

We find a variety of coherent dynamical effects on the networks, which include: regular motion (emergent periodicity), weakly chaotic behavior (different from the uncoupled case), and self-organized motion characterized by close to zero Lyapunov exponents and anomalous diffusion in the phase space. Dynamical regions given as the intervals of coupling strength with distinctive motion and stability patterns are identified, suggesting a mesoscale interpretation of collective tree's dynamics in terms of 4-star's behavior. The system shows dynamical clustering in form of the structured phase space organization of orbits for all coupling strengths. Furthermore, various manifestations of the non-symplectic coupling are explored, including quasi-periodic orbits and strange attractors with weakly positive Lyapunov exponents. In our extended 4-star system whose dynamical units are driving each other, for certain coupling strengths we find the evidence of strange nonchaotic attractors displaying quantitative features which are known to appear in non-periodically driven maps.

We employ the same two-dimensional chaotic maps for studying the stability of a real directed gene regulatory network of bacterium Escherichia Coli, the data on which are empirically known. The main cooperative effects including stability, clustering and long-range correlations are still present in the network's emergent dynamics. However, with increase of coupling strength the motion destabilizes on a sub-network of specific genes, although still maintaining some coherent properties. For comparison, a two-dimensional Hill model of gene interaction is implemented on the same Escherichia Coli network. We find the system to exhibit stable attractors and the flexibility of response to external stimuli, along with the robustness to fluctuations of the environmental inputs.\\[1.2cm]

\n \textbf{Keywords}: complex dynamical systems, gene regulatory networks, scalefree topology, modular networks, coupled chaotic maps, time delay, strange nonchaotic attractors, Escherichia Coli

\newpage
\addcontentsline{toc}{section}{Povzetek}
\section*{Povzetek}
Kompleksne dinami\v cne sisteme pogosto modeliramo v obliki omre\v zij, kjer vozli predstavljajo dinami\v cne enote sistema, ki so sklopljene prek mre\v znih povezav. Primer tak\v snega sistema je omre\v zje genske regulacije, ki uravnava tvorbo proteinov znotraj celice. Tu vozli sistema ustrezajo posameznim genom. Zna\v cilno za kompleksne dinami\v cne sisteme je sodelavanje gradnikov na ve\v cih ravneh kompleksnosti, kar se odra\v za v dinamiki, ki je dolo\v cena z ustreznimi funkcijami (oz. operacijami) sistema. Eden od glavnih problemov na podro\v cju kompleksnih sistemov je vpra\v sanje stabilnosti delovanja sistema ob raznih notranjih in zunanjih motnjah.

V disertaciji obravnavamo dvodimenzionalne kaoti\v cne preslikave, skopljene prek neusmerjenih omre\v zij z razli\v cnimi topologijami. V sistemu uporabljamo nesimplekti\v cno sklopitev, ki vklju\v cuje \v casovni zamik v sklopitvi med preslikavami. Stabilnost razli\v cnih mre\v znih topologij ocenimo z opazovanjem kolektivnega gibanja, ki ga dolo\v cimo z analizo odmika od kaoti\v cnega vedenja posameznih enot. Sistem podrobno obravnavamo na dveh ravneh: (a) na ravni celotnega omre\v zja (z numeri\v cno generiranim samopodobnim drevesom, ter z drevesom ki mu dodamo clique skupine); (b) na ravni drevesne mezostrukture, kjer obravnavamo podgraf '4-zvezda', ki je tipi\v cen dinami\v cni vzorec za topologijo drevesa z najmanj\v sim mo\v znim \v stevilom vozlov in je primeren za interakcije s \v casovnim zamikom. Prou\v cujemo dinami\v cne odnose med tema strukturama, zlasti pojav kooperativnosti na ravni celotnega drevesa, kot posledico dinami\v cnih vzorcev ki ih 4-zvezda ka\v ze na vmesni mezoravni.

V raziskavi najdemo vpliv ve\v c koherentnih dinami\v cnih pojavov: regularno gibanje (emergentna periodi\v cnost), \v sibko kaoti\v cno obna\v sanje (druga\v cno kot za nesklopljen primer), samo-organizirani razvoj s komaj pozitivnimi Lyapnovimi eksponenti ter anomalno difuzijo v faznem prostoru. V nekaterih obmo\v cjih sklopitve se pojavijo zna\v cilni vzorci razvoja in stabilizacije, ki sugerirajo na mo\v znost mezoskopske razlage kolektivnega razvoja drevesa preko razvoja na nivoju 4-zvezde. Sistem ka\v ze dinami\v cno zdru\v zevanje v obliki strukturirane organizacije orbit v faznem prostoru za vse vrednosti sklopitve. Poleg tega smo raziskali vpliv nesimplekti\v cne sklopitve, ki se poka\v ze zlasti s pojavom kvaziperiodi\v cnih orbit in \v cudnih atraktorjev s komaj pozitivnimi Lyapnovimi eksponenti. V sistemu raz\v sirjene 4-zvezde, kjer so dinami\v cne enote sklopljene med sabo, smo za dolo\v cene vrednosti sklopitve na\v sli \v cudne nekaoti\v cne atraktorje, ki nakazujejo na kvantitativne lastnosti ki so znane za neperiodi\v cno pogojene preslikave.

Enake dvodimenzionalne kaoti\v cne preslikave smo uporabili tudi za studij stabilnosti realnega omre\v zja genske regulacije na primeru bakterije Escherichia Coli, za katere so znani empiri\v cni podatki. Dinamika omre\v zja v tem primeru \v se vedno ka\v ze glavne kooperativne vplive, kot so stabilnost, zdru\v zevanje in korelacije dolgega dosega. Z nara\v s\v cajo\v co sklopitvijo se razvoj destabilizira enem podomre\v zju posameznih genov, kljub temu pa se dolo\v cene koherentne lastnosti ohranijo. Za primerjavo smo implementirali dvodimenzionalni Hillov model genske interakcije na enakem omre\v zju Escherichie Coli. Pri sistemu opazimo stabilne atraktorje in fleksibilen oddziv na zunanje dra\v zljaje, kot tudi neob\v cutljivost na motnje vplivov iz okolja.\\[1.2cm]

\n \textbf{Klju\v cne besede}: kompleksni dinami\v cni sistemi, omre\v zja genske regulacije, samopodobna (scalefree) topologija, modularna omre\v zja, sklopljene kaoti\v cne preslikave, \v casovni zamik, \v cudni nekaoti\v cni atraktoriji, Escherichia Coli


\chapter{Introduction}  \label{Introduction}

\begin{flushright}
\begin{minipage}{4.6in}
    This Thesis is a study of collective dynamics of two-dimensional units interacting through the links of various network topologies. In this Chapter 
    we introduce the basics necessary for understanding this work, which include the concept of  complex systems realized through networks and the idea 
    of network-coupling between the dynamical maps attached to the network nodes. We describe the process of gene regulation as the motivation
    behind the work presented here. In the last Section we precisely define the problem to be studied and set the directions of investigation.
\\[0.1cm]
\end{minipage}
\end{flushright}

Nature presents us with an immense variety of \textit{complex systems} which are composed of many simple units that through mutual interactions can achieve various patterns of coherent \textit{collective behavior} \cite{mikhailovsocieties,nicolisnicolis}. Their physically most intriguing aspect is  that their behavior can often be very complicated and structured, despite the simplicity of their elementary building units. The self-organization effects are typically the key mechanism behind the cooperativity of the system: due to mutual interactions, system's units "communicate" and often manage to organize their functioning, with each single unit playing a particular role. Complex systems are characterized by their \textit{emergent behavior}, which refers to the fact that the system's collective properties emerge as a consequence of interactions among the units, and are in general not exhibited by the isolated units without interaction. The nature of emergent behavior, along with the roles played by the system's units and small clusters of units are related to the origin of the system's complexity and its function.

\pagenumbering{arabic}

Examples of complex systems include climate, stock markets, telecommunication systems, World Wide Web etc \cite{nicolisnicolis}. All of these systems are inherently composed of many individual units (that can be atmospheric air masses, financial agents or web-pages), which develop cooperative behavior due to mutual communication/interaction, and/or due to being subjected to some external influence. Most of the exploration in this field is focused on life and biological sciences, where perhaps the largest variety of complex systems can be found in \cite{camazine}. Social insects like ant colonies, exhibit a remarkably complex community structure with division and organization of labor that includes optimization procedures in searching for food \cite{mikhailovsocieties}. Neural systems, widely investigated over the last few decades, explain the neural activity of a living organism based on the patterns of interaction and self-organization among the neurons, that are themselves very simple \cite{baryam}. Ecological systems related to various environments, include very structured predator-prey relationships that are sensitive and adaptive to different environmental conditions \cite{camazine}. Gene expression regulatory system within living cells allows for efficient production of the proteins required for functioning and development of the cell, based on the chemical inputs coming from cell's environment that signal the current needs of the cell \cite{uribook}. Fish schools are able to 
quickly adapt to changes in environmental conditions like predator attacks, without any hierarchical structure or leaders \cite{mikhailovsocieties}.

The emergent function of a complex system is generally a non-equilibrium process, which depends on the external inputs and is sensitive to various parameters which are typically showing some sort of fluctuations. The functioning of gene regulatory system is, for instance, very sensitive to cellular inputs and characterized by the ability to quickly respond to them. Stock market quickly adapts to constantly changing actions of agents, as well as to different external circumstances. Functioning of complex systems with very different origins often share certain features that are more related to the needs of the optimal operation of the system, rather than to its particular purpose \cite{nicolisnicolis}. The operation performed by a biological system can differ according to the environmental conditions that come in form of external inputs; gene regulatory system produces different concentrations of proteins depending on the current needs of the cell \cite{uribook}.

There are various levels of interplay between the behavioral patterns displayed by the single units and the emergent behavior of the corresponding  system involving many units. The nature of a complex system is typically either examined locally (on the scale of single units and their behavior), or globally (on the scale of the whole system). However, a complex system can also be studied at a \textit{mesoscopic} scale, i.e. on the level of small clusters of units \cite{hartwell}, which form \textit{functional modules}. Modularity of biological systems is recently attracting a lot of attention, specifically in the context of gene regulatory systems where many functional modules have been revealed \cite{uribook,milo}. Moduls are composed of few units which develop a certain level of cooperativity through particular interaction patterns. They play given roles in the emergent operation of the system, which is more involved than the behavior of a single unit, but still far simpler than the global system's functioning. The mesoscale approach to the complex systems bridges the gap between the single-unit level and the global level.

Structural and organizational diversity of the complex systems are investigated through a variety of disciplines ranging from experimental methods to mathematical and theoretical approaches. Besides studying natural systems, one is often interested in designing artificial complex systems that can serve as models for real (e.g. biological) systems. Computational modeling has a major role in this regard, as modern numerical techniques allow large and efficient simulations. By constructing the appropriate interactions among given isolated units, one can design a complex system with a specific technological or biological purpose. Due to numerical limitations, the isolated unit model is typically oversimplified; this however allows for big  simulations involving very large number of units with possibly very complicated interactions \cite{baryam}. The models are usually constructed with many parameters in order to allow for different types of collective behaviors to occur in relation to parameter variations. Last few decades saw a rapid development in this direction, particularly emphasizing the interdisciplinary nature of complex systems investigation \cite{nicolisnicolis,uribook}.

Computer generated complex systems can be designed in various ways. Typically, they can be represented as a \textit{network} where a single unit is attached to every node. The interactions between the units are then assumed to follow the network links \cite{arenasguilera,boccaletti}. In the context of real systems, the links between units are constructed following the  directions of interaction among them, which results in a network of interacting units. In the case of artificially designed complex system, one considers a pre-specified (large) network and attaches a simple (e.g. dynamical) unit with known behavior to every node. The cooperativity of the interacting units in then examined in relation to the network's architecture, nature of isolated units and external influences. Alternatively, a complex system can be constructed through a process of \textit{self-assembly}: units are added one after another and linked according to pre-given rules producing a stable structure \cite{camazine,nicolisnicolis}. The system's emergent behavior evolves with addition of new units, and is not dependent only on the nature of isolated units and their interactions, but also on the properties of self-assembly process. This "bottom-up" growth is ofter behind complex biological systems that arise through self-assembly of simple bio-units like cells or neurons \cite{camazine}.

In this Thesis we will study a complex system designed by attaching dynamical units to the nodes of a pre-specified network.

\section{Physics of Networks}

Networks are a typically used paradigm of representing complex systems, with nodes picturing individual units and network links specifying the directions and intensities of interaction \cite{arenasguilera,boccaletti}. In a directed network links go exclusively from one node to another, whereas in a non-directed network each link symbolizes mutual interactions between the two nodes. Links can be weighted in case interactions between different pairs of nodes have different intensities. During the system's time-evolution, links and nodes can appear and/or disappear.

Neural networks are ensembles of neurons interacting through synaptic contacts that can be represented as complex networks \cite{camazine}. Predator-prey networks (food webs) are networks with species as nodes, and predator relations as directed links \cite{albert}. Telecommunication systems can be viewed as a network of users communicating through phone/Internet connections \cite{boccaletti}. Regulation of gene expression involves networks of interacting genes that produce desired amounts of the appropriate proteins in response to the cell's needs, by directed activatory or inhibitory connections among the genes \cite{uribook}. Social networks consider persons as nodes and friendship connections as links \cite{danon,guimera}.

Complex system realized through complex network topologies are often \textit{adaptive} to certain inputs -- by adjusting their interaction structure and/or topology according to set of environmental variables, they are able to modify their emergent operation and adapt to the assigned tasks. Study of adaptive networks is a fastly expanding field \cite{thilo}. Adaptivity is recognized to be among the most prominent features of natural complex systems, which is relevant for their optimal design.

The data behind the complex networks studied in physics usually come from experimental results found in certain measurement: World Wide Web (number of links per web-site), gene regulatory networks (directions of gene interactions), social networks (friendship relations) etc. The network's architecture in reconstructed upon these empirical data, and studied using various statistical and topological methods. Ultimately, many features of natural complex systems have been explored through studying architecture and statistical properties of the underlying networks \cite{albert,dorogovtsev}. Alternatively, as already mentioned, networks can be computer generated by given algorithms that produce connection patterns among a specified number of nodes.\\[0.1cm]

\textbf{Elements of Graph Theory.}  The topological properties of networks are from mathematical viewpoint studied through graph theory \cite{bollobas,diestel}. We distinguish between the \textit{connected} and \textit{unconnected} graphs -- the former is made of only one component (there is a path between any two nodes), whereas the latter is not. As in the case of networks, we distinguish between \textit{directed} and \textit{non-directed} graphs. Graphs are represented as ensembles of $N$ nodes indexed by the index $i=1,\hdots , N$ with each node having $k_i$ links (to other nodes) which defines its \textit{degree}. They are characterized through their topological properties, which besides the size (number of nodes) $N$, also include: \\[0.06cm]
- \textit{average degree} $<k>$ -- total number of links divided by the total number of nodes $<k> = \sum_i k_i/N$.  \\
- \textit{degree distribution} $P(k_i)$ -- the distribution of node's degrees, i.e. the probability that a given node has a certain number of links. It reports about global graph structure and is used to distinguish between types of graphs/networks (see below) \cite{dorogovtsev}. \\
- \textit{diameter} -- the maximal shortest path between two nodes belonging to the graph. Gives a rough idea of how compact is a graph: chains have biggest, and cliques (all nodes mutually linked) smallest diameters \cite{diestel}. \\
- \textit{clustering coefficient} -- characterizes how connected are a node's neighbors among them, which is often relevant for social (friendship) networks \cite{danon}. \\
- \textit{modularity} -- expresses the structure of topological scales and presence of moduls in a network \cite{arenas,milo,marija}

In computational simulations, a network/graph of size $N$ is usually defined through its $N\times N$ \textit{adjacency matrix} $C_{ij}$ that specifies whether there exists a connection from node $[i]$ to node $[j]$. The adjacency matrix is symmetric in the case of non-directed networks, while this is not necessarily true in the case of directed networks.

In the work presented here we will deal with connected graphs/networks only, that will vary in size and degree distribution which shall be specified accordingly. We will be concerned with investigations on local, global and mesoscopic scale, using both directed and non-directed networks.\\[0.1cm]

\textbf{Networks Relevant for our Study.} Different types of networks/graphs are referring to their architecture and/or their construction algorithms, which is particularly important for the applications in physics of complex systems. In the context of computer generated networks which are relevant for our study of complex dynamical systems, the considered types include: \\
- \textit{Lattices and Chains} -- simple arrays of nodes connected deterministically in a form of a regular grid (e.g. linear chain with first-neighbor links) \cite{diestel}. \\ 
- \textit{Erdos-R\'enyi} random graph -- constructed by connecting a given ensemble of nodes by linking any two nodes with uniform probability \cite{bollobas}. \\ 
- \textit{Small world} networks (Watts-Strogatz model) -- created from linear chains with random re-wiring of links between pairs of pre-selected nodes with a prescribed small probability \cite{watts}. \\
- \textit{Scalefree} networks -- characterized by a power-law degree distribution (the probability of node having a certain degree is proportional to a power of that degree: $P(k) \sim k^{-\gamma}$), and grown by preferential attachment of the new nodes to the ones on the existing structures depending on their current degrees \cite{dorogovtsev,samukhin}. \\
- \textit{Modular} networks -- characterized by the presence of specific sub-networks occurring far more frequently than expected (with respect to the random networks), hence exhibiting a modular mesoscopic structure \cite{uribook,milo,marija}. Modules (in dynamical terms also called \textit{motifs}) of different sorts can be found in natural modular networks, having particular roles in network functioning.

Natural and technological networks are widely studied in the recent years \cite{arenasguilera,boccaletti}. Much have been understood regarding the topology and architecture of the networks that underline many complex systems, specifically in the context of relationship between the emergent function of a complex network and its topology \cite{dorogovtsev,albert}. In particular, many networks exhibit a scalefree structure, with a power-law degree distributions having the power exponent between $\gamma=2$ and $\gamma=4$. Scalefree structure was found in an ample and diverse variety of networks -- World Wide Web, transport connections, social networks, gene regulatory networks etc. \cite{barabasi}. Some of these networks are natural while some other are technological, i.e. artificially designed: nevertheless, power-laws with very similar power exponents are found. It appears that hierarchical organization of links within these networks increases their efficiency in performing certain task \cite{crucitti}; in a transport network for instance,  hierarchy of connections characterized by existence of few "hubs" (well connected central nodes) reduces the path between any two nodes, thus allowing a more efficient transport \cite{boccaletti}. Evolutionary advantages of scalefree topology were also emphasized and its role in the flexibility and operational robustness of many biological networks was underlined \cite{panos}. In addition, these networks often exhibit a modular structure, with identifiable families of motifs with different network functions \cite{oltvai,jeong}. In particular, gene regulatory networks are known to posses an architecture which is a combination of scalefree and modular topology \cite{uribook} (to be described in detail later).\\[0.1cm]

\textbf{Dynamics on Networks}. Of particular interest for modeling complex systems is the study of dynamics on networks: isolated units (nodes) are described as simple dynamical systems interacting through the network links. Their interaction leads to emergence of \textit{collective dynamics}, which has a specific relationship with the network topology, interaction parameters and the nature of the isolated units. The emergent behavior of neural or gene regulatory networks is primarily related to the dynamical features of single neurons/genes, and the properties of their interaction \cite{camazine,oltvai}. The efficiency of transport on complex networks is having an intricate relationship with the network structure   \cite{bosatraffic}, with the peculiarity of scalefree structure playing a crucial role \cite{bosapacket,bosarev}. Specifically, the transport of information in World Wide Web and similar complex networks is of great interest for a variety of modern applications \cite{bosadirected,bosaadaptive}. The advantages of scalefree topology was moreover confirmed in a general context of network dynamics \cite{panos,ljupco,willeboordse}.

Furthermore, study of dynamics on networks can give insights into the topology and structure of the networks, by observing the collective patterns and its respective transients \cite{zhou,arenas}. The phenomena of \textit{synchronization} (simultaneous motion of separate dynamical units) is of particular interest as it represents the most simple dynamical self-organization pattern \cite{arenasguilera}. General aspects of synchronization were investigated in a large variety of systems, including complex networks of dynamical systems \cite{restrepo}, interacting chaotic systems \cite{pisarchik}, systems of coupled cells \cite{rajesh}, interacting strange non-chaotic attractors \cite{rama} and randomly coupled chaotic dynamical networks \cite{manrubiaemergence}. Interactions between both continuous-time and discrete-time systems can lead to synchronization which is very easy to detect; in particular, more chaotic systems are also able to synchronize while maintaining their chaotic behavior \cite{osipov}.

Dynamics on network is studied in diverse areas of physics related to complex dynamical systems, including heat conduction in chains of alternate masses \cite{prosenliwang}, and in dynamic models of relaxor ferroelectrics \cite{rasa}.

The interplay between the complexity of emergent behavior and the dynamical nature of the isolated units can generally be of two types: 
\begin{itemize} 
\item collective dynamics is very structured or possibly chaotic due to the \textit{nonlinear} nature of interactions among its elementary units, that are themselves exhibiting very simple (linear) dynamics \cite{korientrainment}.\\
\item isolated units behave chaotically/erratically, but due to specific type of interactions among them, the emergent collective dynamics exhibits regular/periodic patterns with a high degree of internal dynamical organization \cite{manrubiaemergence}.
\end{itemize}
In the present work we will examine a complex systems of the second type, with very chaotic individual units that exhibit a regular collective behavior due to neighbor-neighbor interactions through the links of the underlying network.

\section{Coupled Maps on Networks}

Discrete-time dynamical systems (\textit{maps}) are often used for modeling complex dynamics, as they can exhibit a vast range of dynamical behaviors \cite{kanekobook}. Physically, they typically represent discretizations (Poincar\'e maps) of realistic continuous-time systems, e.g. oscillators \cite{wiggy,strogatzbook}. A map is usually defined as iterative (recursive) sequence of values:
\begin{equation}  x_{n+1} = f (x_n), \; n \in \N \;\;\; \mbox{or simply} \;\;\; x' = f (x) , \; x \in \R^N \label{map} \end{equation}
with function $f$ determining the map's nature. Very simple maps often exhibit extremely rich dynamics that can be easily studied due to their computational simplicity. A famous example of one-dimensional map is the \textit{logistic map}:
\begin{equation}  f(x) = c x (1-x)  \label{logistic}  \end{equation}
which displays a variety of behaviors when the parameter $c$ is varied from 0 to 4. This includes periodic oscillations, period-doubling cascade, chaotic intermittency and strong chaos (ergodicity) \cite{strogatzbook}. Logistic map belongs to the class of \textit{measure-preserving maps}, whose dynamics conserves the phase space measure (linear length in this case), and is often used as a simple prototype of a chaotic system (displaying strong chaos for $c=4$).

A well-known example of two-dimensional chaotic system is the Chirikov-Taylor map, usually referred to as \textit{standard map} \cite{chirikov}, defined as:
\begin{equation} \left( \begin{array}{c}
\theta' \\
I'
\end{array} \right) = f(\theta,I)
= \left(
\begin{array}{l}
  \theta + I + \bar{\e} \sin \theta \;\;\;\;\;\;\;  \mbox{mod} \; 2\pi  \\
  I + \bar{\e} \sin \theta 
\end{array} \right) \label{standardmap} \end{equation}
It arises as a Poincar\'e section of the kicked rotator (frictionless plane rotation with a periodic kicking), with variables $\theta$ and $I$ denoting the angle and the angular momentum of the rotator, respectively. The map gives the values after the kick in function of the values before the kick. It is also a measure-preserving (area-preserving) map, as it is derived from a Hamiltonian dynamical system \cite{chirikov,ll}. Parameter $\bar{\e}$ that denotes the kicking strength is also the chaotic parameter for this map, in function of which the system Eq.(\ref{standardmap}) displays an extremely wide spectrum of dynamical behaviors. They include periodic and quasi-periodic orbits, families of periodic islands, weakly and strongly chaotic behavior which has ergodic and mixing properties (demonstrated only numerically). As the standard map is a measure-preserving map that exhibits chaos, it represents a prototype of widely studied \textit{Hamiltonian chaos} \cite{ll,zaslavsky}. Standard map will serve as an example of isolated unit used for designing the complex system studied in this Thesis. 

Network of Coupled Chaotic Maps (CCM) is constructed by assigning a map to each node of a given network and allowing them to interact through the network links due to dynamical \textit{coupling} between them, giving a simple way to model dynamics on networks. The time-evolution of a map/node on a network of CCM is given by the map's original update described by the Eq.(\ref{map}), and an additional coupling contribution from its network neighbors at each iteration. The network is assumed to have $N$ nodes/maps labeled by $i=1,\hdots,N$. A coupled evolution equation for a CCM is usually written as:
\begin{equation}
 x_i ' =  (1-\mu) f(x_i) + \frac{\mu}{N_i} \sum_j g(x_j) \label{genericCCM}
\end{equation}
where the parameter $\mu$ controls the \textit{network coupling strength}, and the sum goes over all the neighbors of the node $i$ (where the normalization is done over the number of $[i]$'s neighbors denoted $N_i$) \cite{lind}. While the function $f$ defines the original map, the function $g$ represents the coupling (interaction) on the network that can be modeled in various ways.

Network of CCM represent a simple and computationally easy model of complex systems, with logistic map being very frequently used as the dynamical unit (function $f$ in Eq.(\ref{genericCCM})). Systems of this sort are extensively studied for various examples of underlying network topologies, interaction forms $g$ and network coupling strengths $\mu$, as they capture the essence of complex systems by exhibiting various cooperative phenomena \cite{kaneko1,kanekobook}. The essence of CCM network emergent behavior lies in dynamically correlated motion of maps at different network nodes, whose type of correlation generally depends on the network coupling strength. This primarily refers to the various \textit{synchronization effects} \cite{restrepo,lind,pineda}, including partial synchronization \cite{zhangcerdeira}, chaotic synchronization \cite{osipov,anteneodo} and examples of entrainment of oscillators (that can also be modeled in continuous-time versions) \cite{korientrainment,koriarchitecture}. Generalized synchronization phenomena include dynamical clustering \cite{manrubiaemergence,amritkar}, driven phase-synchronization \cite{jalan} and various other examples of dynamical self-organization \cite{zahera,ja-jsm,sinha}. CCM were investigated for the cases of local and global coupling including external driving \cite{jalan}, for regular lattices and complex networks \cite{kanekobook,gadecardeirarama}, with homogeneous and inhomogeneous coupling \cite{zanette} and synchronous and asynchronous updating \cite{abramson}. Recently, special emphasis was put on CCM collective effects on scalefree topologies \cite{zhou,lind,ja-jsm,ja-lncs-1,ja-lncs-2}, small-world networks \cite{gade} and modular networks \cite{arenas,kahng}, in the context of the mentioned developments in understanding architectures of these networks, including synchronization fittness of specific motifs \cite{vega}. Networks of CCM were also successful in modeling complex phenomena like dynamical phase transitions \cite{ahlers} and their classification \cite{baroni}. Furthermore, coupled maps on networks are being increasingly applied for modeling of biological systems like cell-cycle dynamics \cite{li} and regulation of gene expression \cite{tamayo}. In this context the collective dynamical effects are known to be relatively robust to interaction parameters and perturbations \cite{li,ja-pramana}.

In regard of biological and technological applications of networks of CCM, recent investigations involve coupling between the units/nodes with a \textit{time delay}. That is to say, a node receives input from the neighboring nodes "seeing" them in their dynamical states one (or few) time-steps in the past. Time delay in communication is ubiquitous in all natural systems. As no information travels instantaneously, real complex dynamical systems on networks ought to be modeled including a time delay in the node interactions. Synchronization and similar collective phenomena are actively investigated for  time-delayed systems of CCM \cite{masoller} for a vast range of topologies, interactions and lengths of time delay \cite{nunes}. Specifically, for the systems of coupled logistic maps mentioned above, time delays are known to enhance the synchronization properties \cite{atay,kurths}.

However, the case of coupling between two-dimensional maps is still poorly investigated. The works that examine 2D coupled maps typically study the  statistical system's properties only, like the anomalous diffusion \cite{ja-jsm,altmann}, without investigating various non-symplectic coupling forms and their dynamical manifestations. As two-dimensional dynamics can be far more complex than its one-dimensional counterpart, one expects that systems of CCM involving two-dimensional maps may have a larger spectrum of possible behaviors. Two-dimensional dynamics furthermore allows modeling of a wider range of complex systems, given that two degrees of freedom attached at each node give more coupling options, in addition to a richer local dynamics. Needless to mention, many real complex system posses elementary units whose internal isolated behavior cannot be well described by only a single degree of freedom (e.g. gene dynamics \cite{uribook,schuster}).

In this work we will be dealing with a time-delayed system of CCM involving two-dimensional maps, effectively showing the richness of its collective dynamical spectrum. As described in detail later, we examine CCM on large complex networks in comparison to the same dynamics on small sub-graphs (motifs).

\section{Models of Gene Regulatory Networks}

The regulation of gene expression is a process of fundamental importance for the functioning and growth of biological cells. Gene regulatory network represents a complex system of genes interacting by activating or repressing each other's \textit{expression level} that defines the production rate of proteins \cite{lewin}. Collective function of gene regulatory system responds to the cell's needs and provides the appropriate proteins at all times. Gene regulatory networks involve different scales of functions, and often have modular structure with specific groups of genes preforming particular tasks \cite{dobrin,uribook}.\\[0.1cm]

\begin{figure}[!hbt]
\begin{center}
\includegraphics[height=4.1in,width=5.0in]{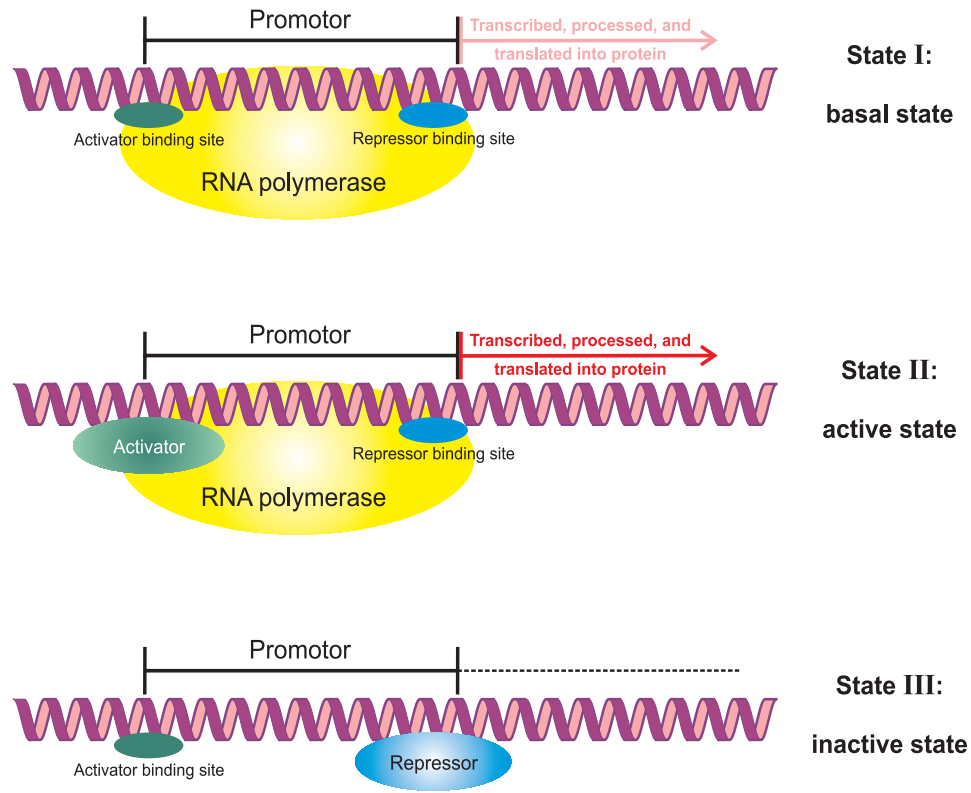}
\caption[Schematics of gene regulation process (with permission from \cite{schuster})]{Process of gene regulation (with permission from \cite{schuster} ) with its three stages of operation pictured schematically: basal state, active state and inactive state.} \label{fig-regulationscheme}
\end{center}
\end{figure}

\textbf{Process of Gene Regulation.} Genes are building blocks of DNA that are responsible for synthesizing mRNA molecules from RNA polymerase present in the cell, through the process called \textit{transcription}. The rate of transcription is determined by presence/absence of \textit{transcription factors}, activators and repressors that can bind to the promotor region of a gene, located upstream from each gene on DNA strand that carries them. As pictured in Fig.\,\ref{fig-regulationscheme}, gene's transcriptional activity has three states depending on presence/absence of activators/repressors. Absence of both puts the gene in \textit{basal state} (also termed leaky transcription) characterized by a faint transcription rate (much smaller than in active state). Presence of an activator in the absence of all repressors defines \textit{active state} of the gene when the transcription is maximal. 
When a repressor is bounded to its site, the gene is in \textit{inactive state} regardless of the activators, and the transcription is fully inhibited. After transcription, mRNA initiates the production of proteins by the \textit{RNA translation} process. Thus created proteins have various functions in the cell, including operating as activators/repressors for transcription of other genes \cite{uribook,lewin,schuster}. 

The operation of gene regulatory network is hence given by production of proteins by genes, that activate/repress other genes, creating collective functioning of the network that provides the products needed by the cell \cite{uribook}. Besides proteins themselves, transcription factors can also be various environmental inputs able to modify the global operation of the network according to the cell's current needs. They are referred to as \textit{external transcription factors} (ETF) \cite{uribook}.

Gene regulatory networks are directed networks: genes are attached to the nodes, while the regulatory interactions are pictured as the directed links. Moreover, they have two possible types of directed connections: activatory and repressory. Recent studies revealed their architecture in many details; in particular, for the cases of bacterium \textit{Escherichia Coli} \cite{thieffry,orr} and yeast \textit{Saccharomyces Cerevisiae} \cite{tong} where all of the data on gene regulation has been mapped. The network structure often shows scalefree properties, with few very connected genes (hubs) having central role in global network operation. In addition, these networks typically exhibit a modular structure \cite{milo,hartwell}, which has been extensively studied for the case of Escherichia Coli (E.Coli) \cite{orr,dobrin,osbaldo}. Modularity in the gene regulatory networks was found to be related to its global (biological) functioning \cite{dobrin,zaslaver} and evolutionary design \cite{mazurie,spontaneous}. Various computation algorithms have been designed for estimating the number and distribution of modules in a given network \cite{itzkovitz,segal}, and the software applications are already available, like \texttt{MFinder} program by U. Alon \cite{mfinder}. The central issue of modular networks regards the relationship between the structure/ frequency of different modules and their functional role in the the network operation \cite{vazquez}. In the context of gene regulatory networks where modularity plays a crucial role, the presence of \textit{functional motifs} has been revealed, whose topologies and network locations are directly related to the tasks they perform \cite{dobrin,orr}. Specifically, the \textit{Feed Forward Loop} (FFL) motif found in different regulatory networks is well investigated and its function fully understood, both theoretically and experimentally \cite{uribook,mangan}. As it appears, the coherent FFL always serves as a \textit{sign-sensitive delay element} in all the gene regulatory networks investigated so far \cite{zaslaver}. 

Appearance of modular networks is not limited to gene regulation systems only: functional motifs have been revealed in other real networks as well, notably in the case of brain networks \cite{sporns}, metabolitic networks \cite{jeong,ravasz} and some technological networks \cite{milo}. Various motifs can play different roles in networks of different origin, that may not be necessarily related to their topological properties, but to the needs and functioning of the network in question \cite{prill}. This was demonstrated on the example of \textit{Bi-Fan} motif, another recurrent gene regulatory sub-network (motif) \cite{ingram}. Modularity seems to have the central role in functioning of many natural networks \cite{kashtan}, indicating that a mesoscale approach might lead to applications in engineering of the systems that perform specifically assigned operations. Also, recent works are considering the possibility of engineering \textit{genetic circuits} involving few genes that would perform pre-defined tasks, with potentially vast range of applications \cite{elowitz,guido,collins}. \\[0.1cm]

\textbf{Mathematical Models of Gene Interaction.} As already described, genes interact by activating and repressing each others expression levels. They  define the production rates of mRNA, which translate into proteins that may act as a transcription factors regulating the behavior of other genes \cite{lewin}. Activation/repression processes are characterized by their respective thresholds: if the concentration of a given transcription factor is above a certain value, activation/repression starts \cite{uribook}. A gene can be activated/repressed by other specific genes with interaction thresholds depending on the gene -- its expression level will be determined by the interplay of these effects. The precise protein production rate is however not determined only by the quantity of activator/repressor present (as long as this quantity is above the threshold), but by other interaction parameters.

The gene interaction and the function of gene regulation network is known experimentally for some organisms, including those mentioned above \cite{thieffry,orr,osbaldo}. However, the mathematical modeling of gene interaction is not simple, as the details of this process can vary in time and from gene to gene in the same network. Still, a variety of models are currently in use, picturing the gene interaction with different levels of simplification \cite{jong,smolen}. Among them are the following:
\begin{itemize}
 \item \textit{Boolean systems} describe the interactions in truth-tables, where a gene at each time-step can be either "on" (activated and producing
 proteins) or "off" (repressed and not producing proteins), depending on the states of the genes influencing it \cite{kauffman}. While in terms of
 dynamics this model is strongly oversimplified, it still offers a computationally easy model that allows large-scale numerical investigations of gene 
 regulatory networks. Robustness and flexibility in a boolean dynamical model of the E.Coli's gene network was studied recently \cite{areejit}. 
 \item \textit{Step-function 1D coupled maps} are a more elaborate model, picturing threshold interactions as step-functions as follows
\cite{coutinho}:
\begin{equation}
  x[i]_{t+1} = a x[i]_{t} + (1-a) \sum_{j} K_{ij} \Theta ( s_{ij} (x[j]_{t} - T_{ij}) )  \label{coutinho}
\end{equation}
where $x[i]$ is the state of $i$-th gene/node, $T_{ij}$ is the interaction threshold of node $i$ influencing node $j$, $K_{ij}$ are the interaction strengths and the binary value $s_{ij}$ (being either -1 or +1) describes the type of interaction (activatory/repressory). The step-function nature of activation/repression is captured in Heaviside function defined by:
\[ \Theta (x) = \left\lbrace  \begin{array}{cc}  
0 & \;\;\; \mbox{if} \;\;\; x<0 \\
1 & \;\;\; \mbox{if} \;\;\; x\geq 0 
\end{array} \right. \]
This is a simple and elegant model allowing more detailed studies of gene regulation, including theoretical studies of dynamical complexity and stability \cite{lima,cros}. In particular, robustness of gene regulation was recently confirmed in the context of time-delayed interaction using this model \cite{klemm}. This model generalized the boolean approach, but also involves a simple summation of neighbor inputs on each gene similarly to the usual CCM model mentioned above. A further generalization would include a truth-table for each gene, describing how neighbor inputs are processed for each gene separately.
\item \textit{Piecewise-linear 1D Ordinary Differential Equations (ODEs)} are the simplest continuous-time model of gene interaction constructed from the above discrete model \cite{jong}: 
\[ \frac{dx[i]}{dt} = \sum_j g_{ji} (x[j]) - \gamma_i x[i] , \] 
where $x[i]=x[i](t)$ denotes the continuous-time state of the gene $[i]$ at the time $t$. The piecewise-linear function $g_{ji}$ describes the way gene $[j]$ influences the gene $[i]$, defining the input to the gene $[i]$ by summing all the neighbor contributions, equivalently to the discrete model Eq.(\ref{coutinho}) (again, instead of a sum, one could use truth-tables to specify relevance of different inputs for a given gene). The degradation constants $\gamma_i$ describe the rate of self-degradation of each protein/transcription factor corresponding to gene $[i]$. This approach has also been amply studied both analytically \cite{casey} and computationally \cite{aziza,kappler}, confirming a possibility of chaotic dynamics in gene interactions \cite{mestl}. It is also simple to implement numerically, although more directed towards concrete studies of dynamical phenomena, rather than collective  network behavior.
\item \textit{Hill-function 2D ODEs} include both mRNA concentration ($q$) and protein concentration ($p$) as system variables, rendering internal gene dynamics two-dimensional, with protein production being governed by the concentration of mRNA. The equations are given by \cite{uribook,schuster}:
\begin{equation}
\begin{array}{ccc}
 \dfrac{dq[i]}{dt}  & = & \sum_j F^\pm_{ji} p[j] - \gamma^{Q}_{i} q[i]  \\
     & \; & \\ 
 \dfrac{dp[i]}{dt}  & = & \alpha_{i} q[i] - \gamma^{P}_{i} p[i] 
\end{array}  \label{schuster}
\end{equation}  
where the function $F^{\pm}$ (similarly to the function $g$ above) models the gene to gene interaction, which is in this case either 
 activation ($+$) or repression ($-$), given by Hill's functions:
\begin{equation}
\begin{array}{ccc}
 F^{+} (p) & = \xi + \dfrac{\beta p^n}{p^n + T^n}, & \;\;\; \mbox{activation}  \\
  & \; & \\
 F^{-} (p) & = \xi + \dfrac{\beta T^n}{p^n + T^n}, & \;\;\; \mbox{repression}
\end{array}  \label{hill}
\end{equation} 
This model has also been extensively studied theoretically \cite{muller}, and in particular, the two-gene system was analytically examined in detail for all combinations of interactions \cite{schuster}. Note that this model besides being two-dimensional, also incorporates the largest variety of parameters describing the interaction: the Hill coefficient $n$ determines the steep of activation/repression curve when the threshold $T$ is crossed (at the limit of $n \rightarrow \infty$ function $F^{\pm}$ becomes a step-function). We distinguish between non-cooperative binding ($n=1$) and cooperative binding ($n \geq 2$) among the genes \cite{schuster}. The factor $\beta$ defines the maximal expression level that a gene interaction can produce. Both activation and repression expressions include leaky transcription state due to the presence of $\xi > 0$, which makes sure both $F^{+}$ and $F^{-}$ are always bigger than zero. Note again that instead of summing over neighbor inputs, one can design more elaborate versions of input organization, as we shall describe in what follows.
\end{itemize}
This 2D continuous-time approach to the internal dynamics of a gene (termed Hill model from now on) will be used in this Thesis to design a model for gene regulatory network of E.Coli, which will be investigated along the lines of 2D network of coupled chaotic standard maps. It represents a two-dimensional model for gene interactions that locally exhibit regular dynamics, as opposed to the coupled maps system that locally displays only chaotic behavior. We will be confronting these two dynamical models through the same directed topology of gene regulatory network of E.Coli. 
\begin{figure}[!hbt]
\begin{center}
\includegraphics[height=3.4in,width=3.4in]{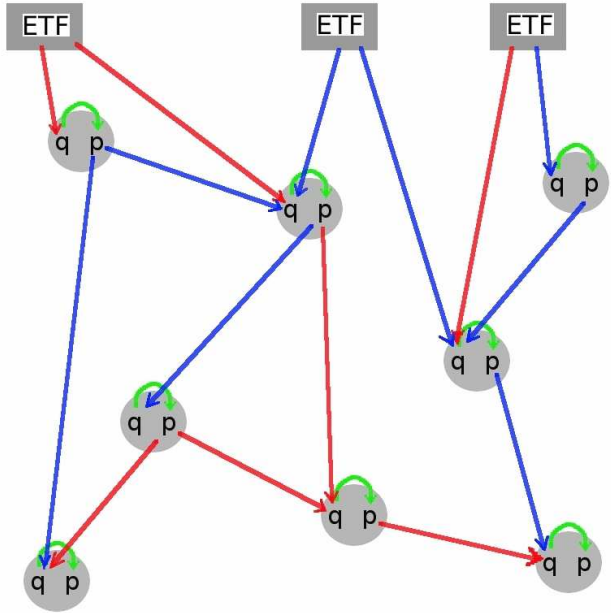}
\caption[The coupling structure of continuous-time 2D Hill model of gene interactions \cite{uribook,schuster}]{The coupling structure of continuous-time 2D model of gene interactions \cite{uribook,schuster} showing local $(q,p)$ variables at each node. They interact mutually and are influenced by external transcription factors (ETFs), with directed interactions being either activatory (red) or repressory (blue).} \label{fig-widderscheme}
\end{center}
\end{figure}
The schematics of interactions in Hill model are shown in Fig.\,\ref{fig-widderscheme}. \\[0.1cm]

\textbf{Logic Gates.} The remaining detail needed to complete the continuous-time model defined above regards the case when a gene is regulated by more than one transcription factor (among which can be $p$-variable of that gene itself, i.e. \textit{self-loop}). The situation of more transcription factors regulating one gene is frequent, and the corresponding empirical data is usually provided gene by gene. As already mentioned, the simplest version of modeling this situation involves summation over all the neighbor inputs, while more detailed models include truth-tables or logic gates. Namely, some genes may need simultaneous presence of more activators to be maximally expressed, while some other genes enforce competition among many activators, and the final expression level is determined by the activator present in the highest concentration. The former case can be viewed as logic \textit{AND-gate}, while the latter corresponds to the case of logic \textit{OR-gate}. In naturally occurring regulatory networks logic gates in general have very complex operation patterns, and never operate as simple AND-gates or OR-gates. However, for the purposes of mathematical modeling we shall simplify the genes' logic gates, and determine the final expression level of some gene regulated by $N$ transcription factors $p_i$ through Hill functions $F^{\pm}_i$ in one of the following ways:
\begin{itemize}
 \item OR-gate:  ${\mathcal G}_i^O = \max_{j=1,\hdots,N} F^\pm_{ji} (p[j])$ 
 \item AND-gate: ${\mathcal G}_i^A = \min_{j=1,\hdots,N} F^\pm_{ji} (p[j])$
 \item SUM-gate: ${\mathcal G}_i^S = \frac{1}{N} \sum_{j=1}^{j=N} F^\pm_{ji} (p[j])$
\end{itemize}
Although with lesser biological motivation, the SUM-gate has certain relevance, as it represents an analogy with the usual input-sum approach behind the typical design of coupled maps systems. In Chapter \ref{Dynamics on E.Coli Network}. we will be considering a dynamical model of E.Coli network, comparing two versions of the system with SUM-gates and AND-gates, as the logic gates involved in the E.Coli gene regulatory network appear to have properties of these two logic gate types \cite{uribook}.

\section{The Subject of This Thesis}

The central subject of this Thesis is the computational investigation of collective dynamics of chaotic two-dimensional maps coupled with time delay on non-directed complex networks and their dynamical motifs. The Thesis summarizes the research results obtained in these directions and describes the computational techniques employed.\\[0.1cm]

\textbf{Motivation.} We study a network of coupled chaotic dynamical units, seeking the appearance of regular behavior due to the network interactions. Our aim is to test the ability of complex networks to induce stability into the dynamics of coupled chaotic units. In that context we consider two-dimensional standard map given by Eq.(\ref{standardmap}) which exhibits strong chaos when isolated, and is characterized by phase space mixing and chaotic diffusion \cite{ll}. We develop a suitable model for testing the network stability by attaching a standard map to every node of a scalefree tree, which is taken as a representative of complex networks. In opposition to classical models involving one-dimensional units, we introduce two degrees of freedom at each node; many natural complex systems include units that cannot be appropriately described with a single degree of freedom.
Gene interactions underlying gene regulatory networks are one of the examples, as they involve concentrations of proteins and mRNA \cite{uribook}. Our system also includes time delay in modeling communication among the dynamical units.

We seek to recognize and study the departures from chaotic behavior of the isolated standard maps generated by the network-coupling. Our hypothesis is that two-dimensional network-interacting systems generally posses equal propensity for self-organization as their one-dimensional counterparts, whose tendency towards cooperative behavior is well known \cite{zahera,kurths,lind}. 

As two-dimensional maps are far more dynamically rich, specially in the context of chaotic dynamics, our system provides insight into general nature of two-dimensional discrete-time oscillators with non-symplectic coupling. We study the behavior of separate nodes attached to the tree, and precisely examine departures from their nature as isolated nonlinear maps, as a further manifestation of the system's collectivity.

Furthermore, we address the dynamical manifestations of network's modularity, by developing a mesoscale approach to understanding the emergent behavior of the system. We emphasize the role of the dynamical motifs, by examining the motion of the large scalefree tree in relation to the motion of its smallest typical sub-graph termed 4-star. The 4-star motif captures the tree topology in only four nodes and is suitable for time-delayed interaction. We expect that the topological relationship between these two structures might be reflected into a relationship of their dynamical properties. This envisages the \textit{bottom-up} approach to complex networks, where one seeks to reveal the properties of the large system by examining the properties of its local and mesoscale components -- functional motifs.

For comparison, we examine the standard maps coupled through the directed gene regulatory network of bacterium E.Coli. We claim that even a directed topology possesses equal tendency towards the dynamical regularity of coupled chaotic units. By investigating the stability of the emergent motion of standard maps on a real biological network, we are establishing a result that regards the general stability of the systems realized through the examined network. The result applies to other (less chaotic) nonlinear maps, demonstrating functional stability of the studied real network. As an extension, we examine the E.Coli's gene regulatory network model with two-dimensional (regular) units, showing its robustness and flexibility, which is expected by the biological origin of this network. We are showing E.Coli's topology to be able to generate self-organization that reflects the nature of the particular model, with both chaotic maps and regular Hill gene interaction model. \\[0.1cm]

\textbf{The Statement of the Problem.} We construct the system of Coupled Chaotic Maps (CCM) by considering a scalefree network (tree) and attaching a two-dimensional standard map given by Eq.(\ref{standardmap}) to every network's node. The maps are allowed to interact through time-delayed coupling in the angle coordinate that goes along the network's (tree's) links. We comparatively examine the collective dynamics on the scalefree tree and a 4-node graph connected in a form of a star, representing the smallest dynamical structure capturing the tree topology. We study the  emergent dynamical properties of this systems in relation to the variations of coupling strengths regulating the intensity of interactions among the network nodes/maps. We furthermore examine the dynamics of the same CCM realized through the directed graph of the gene regulatory network of E.Coli, and a two-dimensional Hill model of E.Coli's gene interactions given by Eqs.(\ref{schuster})\&(\ref{hill}).

Various dynamical regions of our systems of CCM are explored in relation to the coupling strength. The statistical properties of the emergent dynamics appearing on all networks are investigated using various methods, describing and quantifying the presence of collective effects in the systems. The stability of motion is addressed, demonstrating the ability of network structures to inhibit the chaoticity of isolated node's dynamics due to inter-node interactions. The dynamical relationship between the motion exhibited by the scalefree tree and its dynamical 4-star motif is investigated and discussed. We study the orbits of individual nodes attached to the network and investigate different phase space manifestations of non-symplectic interaction among the maps, like strange attractors and quasi-periodic orbits. A similar analysis is performed for the same CCM realized using the directed topology of E.Coli (which is downloaded from a database -- see later). Relationships between the nature of dynamics on directed and non-directed topologies are discussed. In regard of two-dimensional gene regulation Hill model, we study the robustness and flexibility of the emergent behavior in relation to the environmental inputs and fluctuations (noise). 

The directions of our study and exposition of results will be the following: 
\begin{itemize}
\item appearance of regular behavior in network dynamics of coupled chaotic standard maps and the types of single-node orbits in relation to the coupling strength; dynamical regions with different types of collective dynamics for various ranges of coupling strengths for different network topologies
\item general characterization of emergent dynamics, statistical properties of global motion depending on the dynamical region; transient times related to occurrence of regular behavior
\item stability of emergent dynamics and its relationship with dynamical regions; effects of non-symplectic coupling (strange attractors) and their  characterization using non-linear dynamical systems approach
\item dynamical relationship between large structures (scalefree tree) and small structures (4-star motif) in function of the coupling strength; relationship between collective effects inducing emergent properties on large-scale vs. small-scale structure
\item dependence of emergent properties on the time delay, coupling form and standard map's chaotic parameter $\tilde{\e}$
\item examination of the same CCM on a directed network of E.Coli: regularization properties, single-node orbits, dynamical regions, statistical characterization of the emergent motion
\item investigation of E.Coli's gene regulatory network designed via Hill model; system's robustness and flexibility of response to environmental inputs and noise \\[0.1cm]
\end{itemize}

\textbf{Details and Organization of this Thesis.} This Thesis is a summary of research results obtained in the directions mentioned above, performed as a requirement for Candidate's PhD degree. The research methods are purely computational, involving various programming codes written in \texttt{C++} programming language, that compute the quantities of interest according to the investigation directions listed above. The obtained data are analyzed using smaller \texttt{C++} codes, along with other software packages like \texttt{MatLab} and \texttt{Python}. 

As for the pictures representing the results, color plots were done with \texttt{MatLab}, while all other pictures were done using \texttt{Gnuplot}. The graphical representations of networks were produced in \texttt{Pajek} \cite{pajek}. The research work was performed at the Department of Theoretical Physics at the Jo\v{z}ef Stefan Institute over a period of two and half years. The numerical simulations and data analysis were done using the Department's computing resources, specifically \texttt{minos.ijs.si}, \texttt{eurus.ijs.si} and \texttt{zephyrus.ijs.si}.

The exposition of results in this Thesis follows the named directions, in form of Chapters dedicated to specific subjects. The structure of programming algorithms and the discussion of results is given along with the presentation. The Thesis is organized as follows: in Chapter \ref{Coupled Maps System on Networks with Time delay}. we introduce the structure of our network of CCM, along with the origin of different topologies to be examined. In Chapter \ref{Collective Dynamical Effects in CCM on Networks}. the general properties of CCM system are investigated in function of the coupling strength. This includes main types of emergent orbits, dynamical regions and clustering, and the process of dynamical regularization. We also examine various statistical properties of the collective motion using network-averaged orbit approach. This Chapter finishes with a brief consideration of CCM system realized with other coupling forms. In Chapter \ref{Stability of Network Dynamics}. we analyze in detail the stability of the emergent dynamics in relation to different dynamical regions, using three techniques: Finite-time Maximal Lyapunov Exponents, Standard Maximal Lyapunov Exponents and Parametric Instability Analysis. Occurrences of strange (nonchaotic) attractors and weak chaos, along with quasi-periodic orbits are found and examined. In Chapter \ref{Dynamics on E.Coli Network}. our CCM is investigated using the directed graph of the largest connected component of E.Coli gene regulatory network. A similar analysis is done in this case, discussing the differences with the case of non-directed topology. The same network is used for implementing a continuous-time Hill model of gene regulatory network for E.Coli based on Eqs.(\ref{schuster})\&(\ref{hill}). The expected stable behavior is found, and the flexibility of network's response to environmental changes and the robustness to noise is analyzed. The Chapter \ref{Conclusions}. is devoted to conclusions, including a systematic outline of the main results and the discussion of open questions. The extensive list of relevant references is provided in the end.\\[0.1cm]

\textbf{Our Publications of Relevant for this Thesis.} Most of the results presented in this Thesis have already been published. In particular:
\begin{itemize}
\item Z. Levnaji\'c and B. Tadi\'c: "Dynamical Patterns in Scalefree Trees of Coupled 2D Chaotic Maps", ICCS 2007, Lecture Notes on 
Computer Science 4488, p.633-640, 2007 (\cite{ja-lncs-1}), contains preliminary results on scalefree tree collective dynamics from Chapter \ref{Collective Dynamical Effects in CCM on Networks}. 
\item Z. Levnaji\'c and B. Tadi\'c: "Self-organization in Trees and Motifs of Two-Dimensional Chaotic Maps with Time Delay", Journal of 
Statistical Mechanics: Theory and Experiment, P03003, 2008 (\cite{ja-jsm}) contains the core material on statistical and stability properties of CCM on scalefree tree and 4-star reported in Chapters \ref{Collective Dynamical Effects in CCM on Networks}. and \ref{Stability of Network Dynamics}.
\item Z. Levnaji\'c: "Dynamical Regularization in Scalefree-trees of Coupled 2D Chaotic Maps", ICCS 2008, Lecture Notes on 
Computer Science 5102, p.584-592, 2008 (\cite{ja-lncs-2}) presents investigation of regularization of the collective dynamics on scalefree tree, covered in Chapter \ref{Collective Dynamical Effects in CCM on Networks}.
\item B. Tadi\'c and Z. Levnaji\'c: "Robust dynamical effects in traffic and chaotic maps on trees", Pramana Journal of Physics, 70, 6, p.1099-1108, 2008 (\cite{ja-pramana}) includes results on clustering of periodic orbits and dispersion of time series of CCM on scalefree tree, from Chapter \ref{Collective Dynamical Effects in CCM on Networks}.
\item Z. Levnaji\'c and B. Tadi\'c: "Collective Phenomena in Time-delayed Coupled Chaotic Maps on Directed E.Coli Gene Regulatory Network" is a work in preparation, aimed to include results from Chapter \ref{Dynamics on E.Coli Network}. regarding the CCM on directed networks
\end{itemize}
The first two references from the list are provided at the end of the Thesis.


\chapter{Networks of Coupled Chaotic Maps with Time Delay}  \label{Coupled Maps System on Networks with Time delay}

\begin{flushright}
\begin{minipage}{4.6in}
    The chaotic properties of Chirikov standard map are discussed and the construction of our network of CCM is explained in detail. 
    The coupling form follows the oscillatory nature of the standard map and models the interaction of real two-dimensional units. 
    The network structures to be employed are exposed both graphically and through their topological properties.\\[0.1cm]
\end{minipage}
\end{flushright}

The Network of Coupled Chaotic Maps (CCM) that we will study in this Thesis is designed by placing a Chirikov standard map Eq.(\ref{standardmap}) on every node of a given network structure, and assuming the nodes/maps to interact via network links through a pre-defined coupling form.\\[0.1cm]

\textbf{The Properties of Standard Map.} For computation purposes we rewrite the standard map Eq.(\ref{standardmap}) by introducing a new angular variable $x=\theta/2\pi$ and a new angular momentum variable $y=I/2\pi$ reducing the map's phase space to $[0,1] \times \R$. The standard map now reads:
\begin{equation} \left( \begin{array}{c}
x' \\
y'
\end{array} \right) = \left(
\begin{array}{l} 
  x + y + \e \sin (2 \pi x)  \;\;\;  \mbox{mod} \; 1  \\
  y + \e \sin (2 \pi x)
\end{array} \right) \label{oursm} \end{equation}
with parameter $\e$ replacing the chaotic parameter $\bar{\e}=2\pi\e$. We will use the standard map in this form to design our CCM, with prime (') denoting the map's update according to Eq.(\ref{oursm}). 

\begin{figure}[!hbt]
\begin{center}
$\begin{array}{cc}
\includegraphics[height=3.in,width=3.1in]{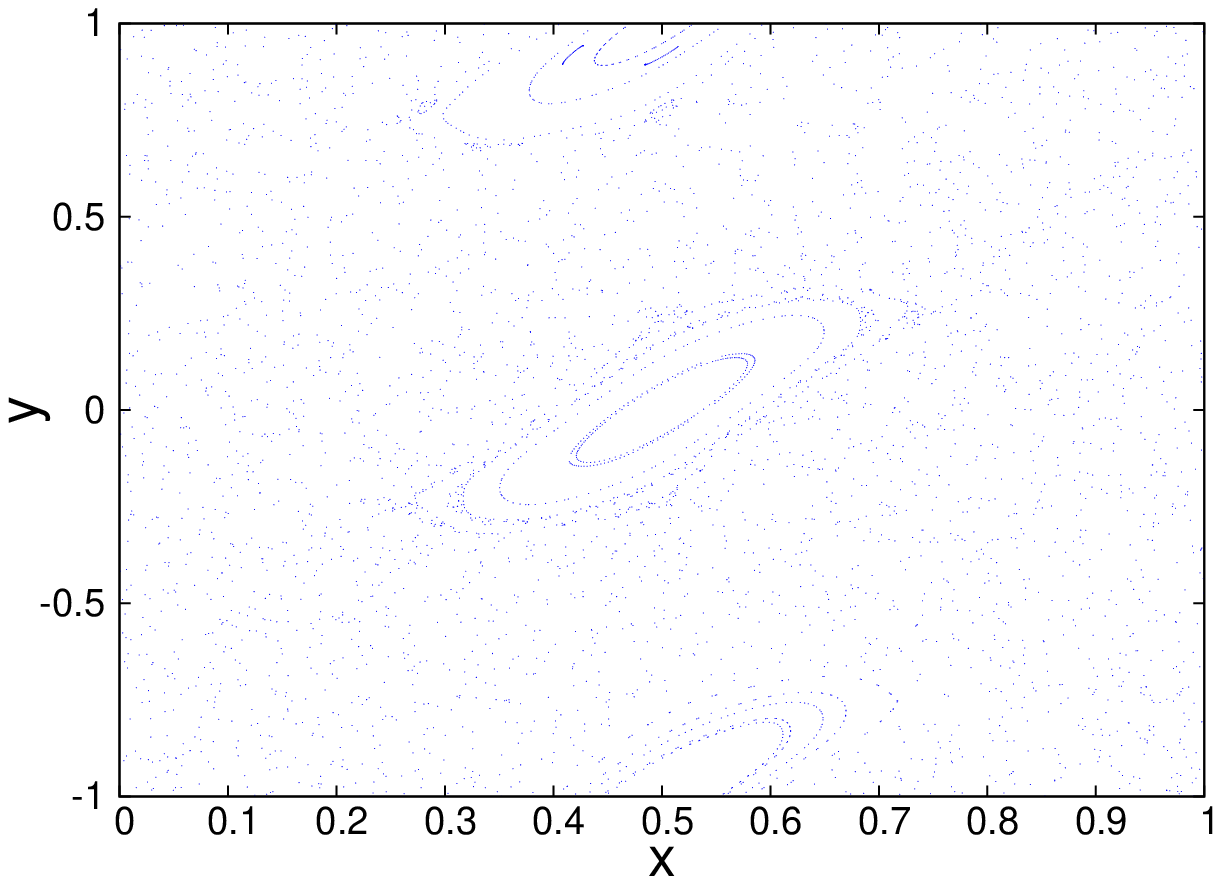}&
\includegraphics[height=3.in,width=3.1in]{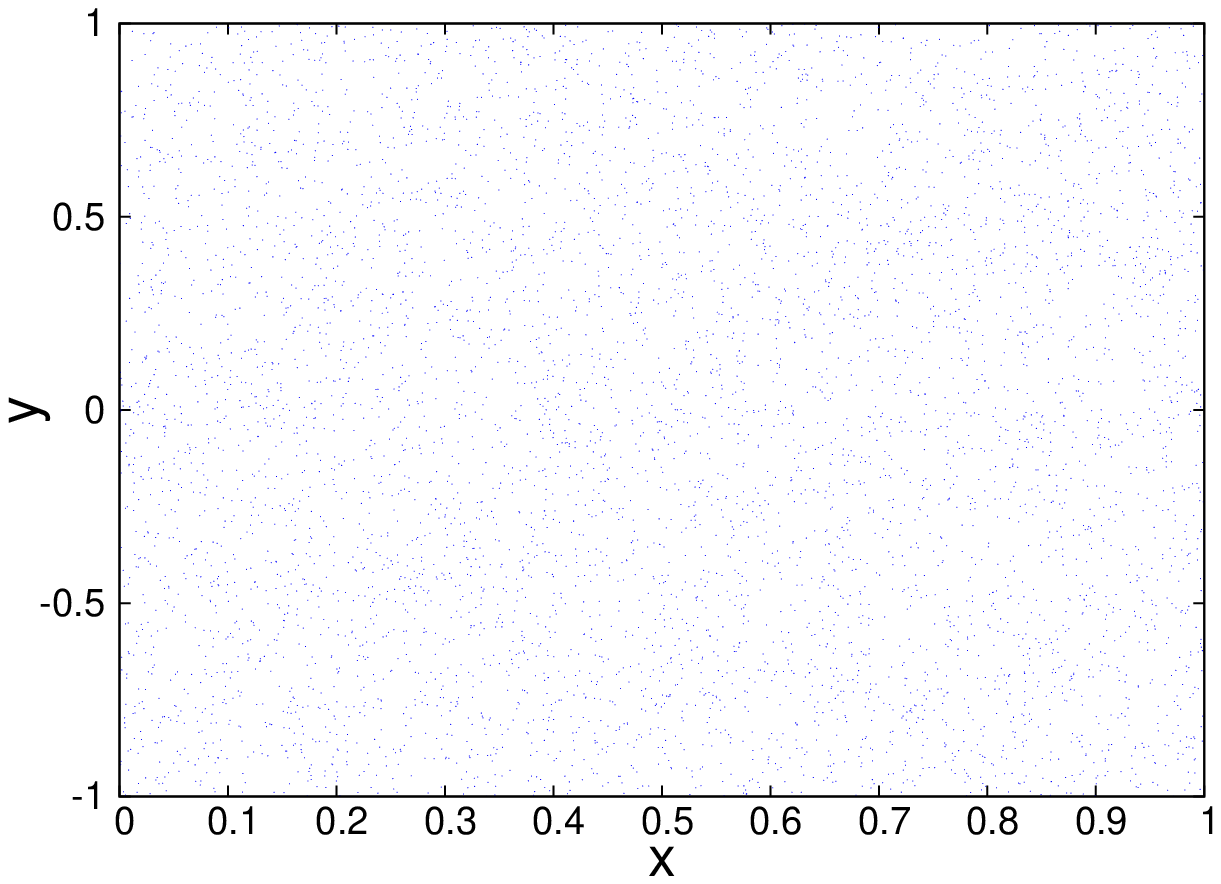} \\
\mbox{(a)} & \mbox{(b)} 
\end{array}$ 
\caption[Phase space portraits for the standard map]{Phase space portraits for the standard map Eq.(\ref{oursm}) done with $10\times 10$ randomly chosen trajectories for $\e=0.5$ in (a) and for $\e=0.9$ in (b).} \label{fig-oursm}
\end{center}
\end{figure}

As mentioned, the standard map represents a prototype of two-dimensional system displaying Hamiltonian chaos \cite{ll,zaslavsky}. In particular, much attention has been devoted to its \textit{chaotic transition} as the paradigm of stochastic transition in dynamical systems \cite{greene,GH,zaslavsky}: at $\bar{\e} \cong 0.971$ ($\e \cong 0.155$) the phase space regions with locally chaotic behavior merge into a single zone of global chaos allowing transport through phase space. In Fig.\,\ref{fig-oursm} we show the phase space portrait of this map for two values of $\e$ displaying strong chaos. As standard map is an area-preserving system, remnants of the KAM surfaces persist at any $\e$ value, along with the
portion of the phase space dominated by strongly chaotic dynamics (Fig.\,\ref{fig-oursm}a). As in this work we are primarily interested in the effects that inter-node coupling has on regularization of chaotic dynamics, we shall focus on the standard map with the chaotic parameter fixed to $\e=0.9$ ($\bar{\e}=5.65$). At this parameter $\e$-value, the motion is strongly chaotic and mixing over the whole phase space without any KAM surfaces remaining, as illustrated in Fig.\,\ref{fig-oursm}b. Strong chaos at this parameter value is characterized by \cite{wiggy,chirikov,ll,zaslavsky,GH}: 
\begin{itemize}
 \item normal diffusion in angular momentum coordinate $y$ governed by $< y^2 > = D_0 t$ with constant $D_0$ for the regime of strong chaos given by 
$D_0=\bar{\e}^2/2=\pi^2 \e^2/2$ 
 \item positive Kolmogorov-Sinai entropy $h_{KS}=\ln (\bar{\e}/2)=\ln (\pi \e)$ measuring the rate of information loss due to strong phase space mixing  
 \item positive Maximal Lyapunov Exponent (2D area-preserving system has at most one positive Lyapunov Exponent) which is equal to Kolmogorov-Sinai 
entropy by the Pesin Theorem 
\end{itemize}
The remarkable dynamical properties of standard map are a continuous source of past and recent explorations  \cite{shevchenko}.

\section{Construction of Our CCM on Networks}

We design our Coupled Map System by assigning a standard map Eq.(\ref{oursm}) with $\e=0.9$ to every node of a given network, which is assumed to have $N$ nodes indexed by $i=1,\hdots,N$. Each node $[i]$ becomes a 2D phase space, whose dynamical state at a discrete time $t$ is denoted by $(x[i]_t,y[i]_t)$. We construct the coupling by following the usual framework of diffusive phase-coupling in $x$-variable (angle) of the oscillatory standard map Eq.(\ref{oursm}), as done in \cite{lind,amritkar,zahera,masoller}. A one-step time delay in the difference of the angle coordinate values of the network neighbors is added. We also adopt the usual schematic approach to diffusive coupling: 
\begin{equation}
 (1-\mu) \times (\mbox{isolated map-update}) \;\; + \;\; \mu \times (\mbox{coupling}) \label{couplingscheme}
\end{equation}
as in \cite{lind} and other works. The full system is defined recursively, as follows:
\begin{equation} 
\left(\begin{array}{l}
x[i]_{t+1} \\
y[i]_{t+1}
\end{array}\right)
=(1- \mu) 
\left(\begin{array}{l}
x[i]_t' \\
y[i]_t'
\end{array}\right)
+
\frac{\mu}{k_i}
\left(\begin{array}{c}
\sum_{j \in {\mathcal K_i}} (x[j]_t - x[i]_t') \\ 
0
\end{array}\right),
\label{main-equation} \end{equation}
in the sense that each node's update is a function of its own current state and the previous states of the neighboring nodes. Prime ($'$) denotes the isolated map's next iterate as in Eq.(\ref{oursm}), while $t$ stays for the global discrete time. The parameter $\mu$ measures the network coupling strength, $k_i$ is node's $[i]$ degree and ${\mathcal K_i}$ denotes the network neighborhood of the node $[i]$. The update of each node is therefore the sum of a contribution given by its isolated standard map-update (the $'$ part) plus a coupling contribution given by the sum of differences between the node's phase-value and the phase-values of the neighboring nodes in the previous iteration, normalized by the node's degree. The sum of these two contributions is then implemented as pictured by Eq.(\ref{couplingscheme}).

Note that time delay is realized here by comparing the updates of the isolated nodes (the considered node $[i]$ is an update ahead of its neighboring nodes in Eq.(\ref{main-equation})). Some authors however define time delay by comparing the current iterate of the full system (time $t$) with its previous iterate (time $t-1$) \cite{masoller,atay,kurths}. For the purposes of our study, we will remain with the present definition (the CCM with  alternative definitions of time delay were examined as well, without observing qualitative differences). \\[0.1cm]

\textbf{Discussion of the Coupling Form.} In designing our system, we followed the oscillatory origin of standard map to construct an appropriate coupling form that models the interactions among these 2D discrete oscillators. We couple the maps in the phase (action) variables only, in a way to make each node/oscillator "attempt" to adjust its phase to all of its neighboring oscillators' phases, but with receiving a time-delayed information about their dynamical phase states. The CCM system can therefore be seen as a system of one-dimensional phase oscillators (in action $x$-variables), which are moreover internally coupled to a second variable (angular momentum $y$). Our system represents a generalization of well-studied examples of one-dimensional oscillators (in both discrete-time and continuous-time) on diverse types of lattices/networks \cite{lind,koriarchitecture}. It also a generalization of typical diffusive systems (which are usually considered as chains and lattices of 1D units \cite{zahera,ahlers}), in the sense of employing two-dimensional units.

As mentioned, time delay comes from the natural context of realistic physical interaction/ communication between the units of a complex system. Given that we are dealing with maps, time delay is easy to implement in the form of iteration difference. The imposed time delay is kept constant for all nodes (following \cite{atay} and in contrast to \cite{masoller}). For simplicity, we consider only one time-step delay applied to the first network neighbors; however, there is no specific reason why the number of delay-iterations should match the network distance between the nodes (\cite{masoller} considers random time delay values). For comparison, we also show the results for the equivalent non-delayed system (see end of Chapter \ref{Collective Dynamical Effects in CCM on Networks}.). 

We are using the coupling scheme given by Eq.(\ref{couplingscheme}) as this way the parameter $\mu$ measures the ratio between the isolated-map and the  coupling contribution. Also, $(1-\mu)$ factor in front of the $y$-variable inhibits the standard map's diffusion in the angular momentum \cite{GH}, therefore allowing more interaction among the nodes. Moreover, we normalize the coupling contribution by the respective node's degree, as it is typically done \cite{lind,kahng,atay}, in order to make each node receive equal total amount of signal (and hence leaving the signal's two-dimensional direction to define the effect that the network neighborhood is having on the node at each time-step).

This coupling form is non-symplectic, i.e., it does not maintain the area-preserving nature of the original isolated map, as opposed to the coupling form used in the context of coupled standard maps in \cite{altmann,moyano}. This comes from both time delay and $(1-\mu)$ factor through which the coupling is realized. Non-symplectic coupling is also typical in the context of CCM models of complex systems, actually it is rather difficult to construct a naturally motivated symplectic coupling, specially in the context of 2D coupled systems. Non-symplectic (dissipative) nature of coupling will actually enhance the synchronization and other collective proprieties of our system.

Finally, as shown in \cite{ja-jsm}, the $x$-coordinate part of our system can be seen from the continuous-time point of view. By introducing a space-continuous variable $x'(i,t) \equiv x[i]'(t)$ (defined over the discrete network node-space and continuous time), the $x$-part of our CCM equation Eq.(\ref{main-equation}) takes the form:
\begin{equation} \begin{array}{l}
x'(i,t+1) = 
(1- \mu)x[i]' + \frac{\mu}{k_i} \left[ \sum_{j \in {\mathcal K_i}} (x[j]' - x[i]')  + \sum_{j \in {\mathcal K_i}} (x[j] - x[j]') \right] = \\
(1- \mu)x[i]' + \frac{\mu}{k_i} \left[ \nabla^2 x[i]'  - \sum_{j \in {\mathcal K_i}} \partial_t x[j]' \right]
\cong  \left[(1- \mu) + \frac{\mu}{k_i} (\nabla^2 - k_i \partial_t) \right] x'(i,t)
\end{array}   \label{continuous-version}  
\end{equation}
which emphasizes the phase-diffusive nature of our network coupling (last expression in Eq.(\ref{continuous-version})), and relates it to the usual expression for phase-coupling in CCM.

\section{The Network Structures Employed}

We study the collective dynamics of CCM Eq.(\ref{main-equation}) on networks with various topologies and $N$ nodes. Computationally, we identify each network by its $N\times N$ \textit{adjacency matrix} $C_{ij}$. The results for different topologies are systematically compared. In the rest of this Section we expose topological details for each considered network and show their graphical representations. \\[0.1cm]

\begin{figure}[!hbt]
\begin{center}
\includegraphics[height=5.8in,width=6.2in]{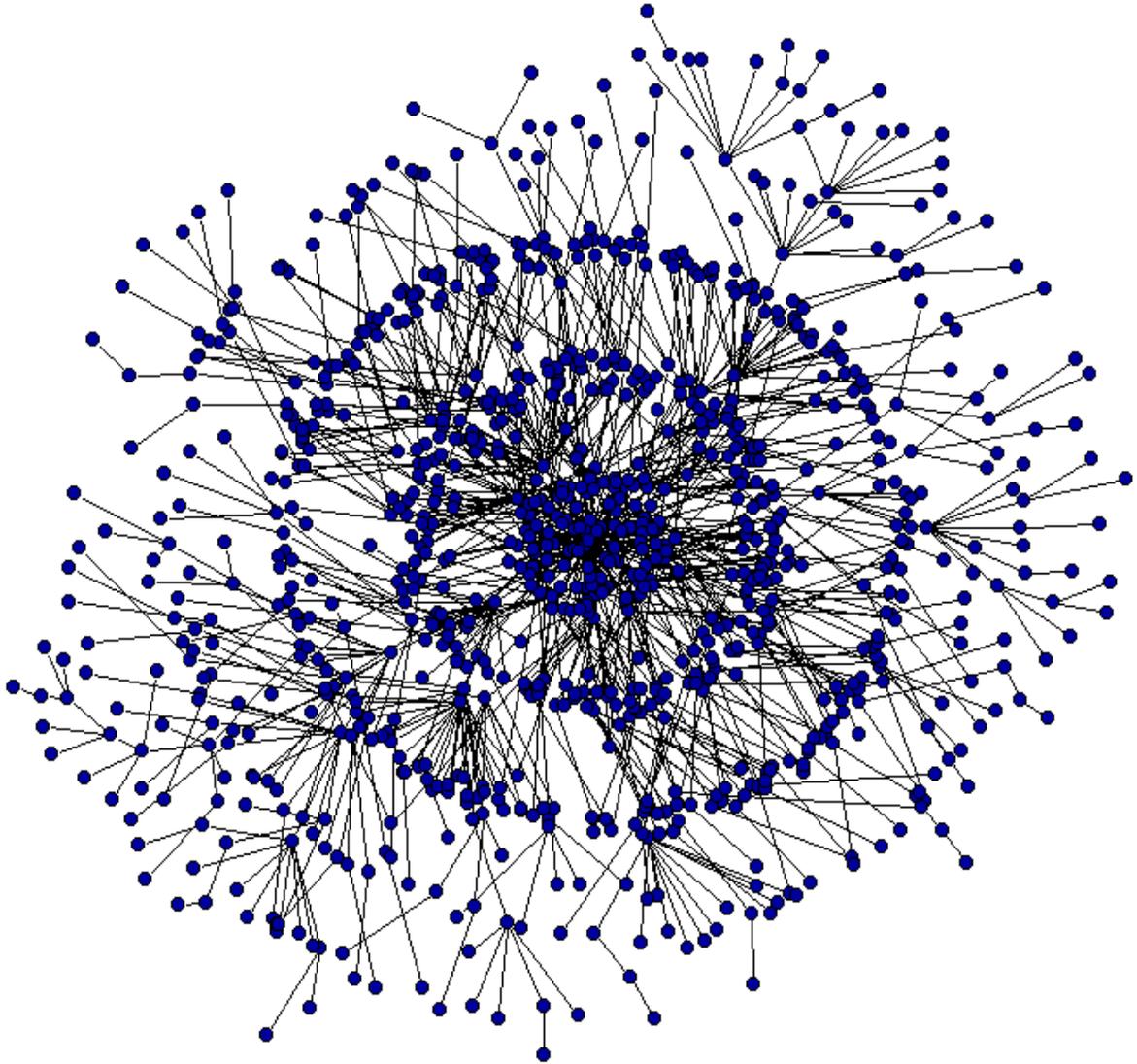}
\caption[Non-directed computer generated scalefree tree]{Non-directed computer generated scalefree tree with $N=1000$ nodes.} \label{fig-sftree}
\end{center}
\end{figure}

\n 1. \textbf{Scalefree Tree} with $N=1000$ nodes and symmetric links shown in Fig.\,\ref{fig-sftree}. This is the simplest example of the scalefree topology. The network is computer generated using the procedure of preferential attachment by 1 link/node \cite{dorogovtsev}, which is run until the size of $N=1000$ nodes is reached. We grow the network by adding a new node $[i]$ at each step which is symmetrically attached to one of the existing nodes $[j]$ with the probability
\begin{equation}
  P([j]) = \frac{k_j+\alpha}{\sum_n (k_n+\alpha)} . \label{sc-attachment}
\end{equation} 
$P([j])$ evolves with growth of the network and is re-calculated at each attachment-step. For the purposes of our study we fix $\alpha=1$. As it a tree structure, this network has no loops and there is a unique path between any pair of nodes. The scalefree tree's degree distribution is shown in Fig.\,\ref{fig-networkproperties}a in blue (averaged over 100 such trees), and has a clear power-law behavior:
\begin{equation}
  P(k) = \mbox{const} \times k^{-\gamma} . \label{sc-degreedistribution}
\end{equation} 
The distribution's slope is $\gamma=3$ as expected, given the analytically known results $\gamma=2+\alpha$ which holds for the case of scalefree trees \cite{dorogovtsev}. In our investigations we always consider the same scalefree tree with $N=1000$ nodes, grown as described above.

The system of CCM is constructed from this network by assuming each node to represent a standard map which is coupled to its network neighbors in accordance with Eq.(\ref{main-equation}). 
Due to symmetric (non-directed) tree's topology, every pair of connected nodes/maps influence each other. As mentioned above, our system of 2D CCM can be seen as a network of 1D CCM with each map/node influencing itself (cf. Eq.(\ref{oursm})), which represents a dynamical self-loop (recall that tree does not posses topological self-loops). Throughout this Thesis we will be primarily concerned with the emergent properties of CCM on this network. \\[0.1cm]

\begin{figure}[!hbt]
\begin{center}
\includegraphics[height=2.05in,width=2.05in]{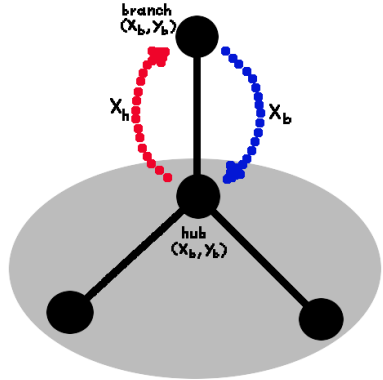}
\caption[4-star motif with a schematic view of the coupling interactions]{The 4-star motif with a schematic view of the coupling interactions between the central node (hub) and a branch node, as defined by Eq.\,(\ref{main-equation}).} \label{fig-4star}
\end{center}
\end{figure}

\n 2. \textbf{4-star Motif} which is a typical tree's sub-network with four nodes connected in a form of a star, shown in Fig.\,\ref{fig-4star} along with the schematics of coupling structure between the maps. 
This is the essential dynamical motif of the tree topology, capturing its structure in only four nodes and having symmetric (non-directed) coupling. It can also be thought of as the simplest non-trivial graph (smallest graph that is neither a chain or a clique). It displays graph diameter of 2 (links), necessary for introduction of one time-step delay into system of CCM (in a sense on time-delayed interplay between the central node and the branches). We will systematically examine 4-star's dynamics in comparison to the tree's dynamics. We will also refer to the 4-node motif in the form of clique (all nodes mutually connected), called \textit{4-clique}, in order to emphasize the peculiarity of the tree topology with respect to other topologies. The dynamics of CCM on 4-star represents an eight-dimensional discrete-time nonlinear system. The manifestations of non-symplectic coupling will be explored in detail from the prospective of nonlinear dynamics in Chapter \ref{Stability of Network Dynamics}.\\[0.1cm]

\begin{figure}[!hbt]
\begin{center}
\includegraphics[height=5.7in,width=6.3in]{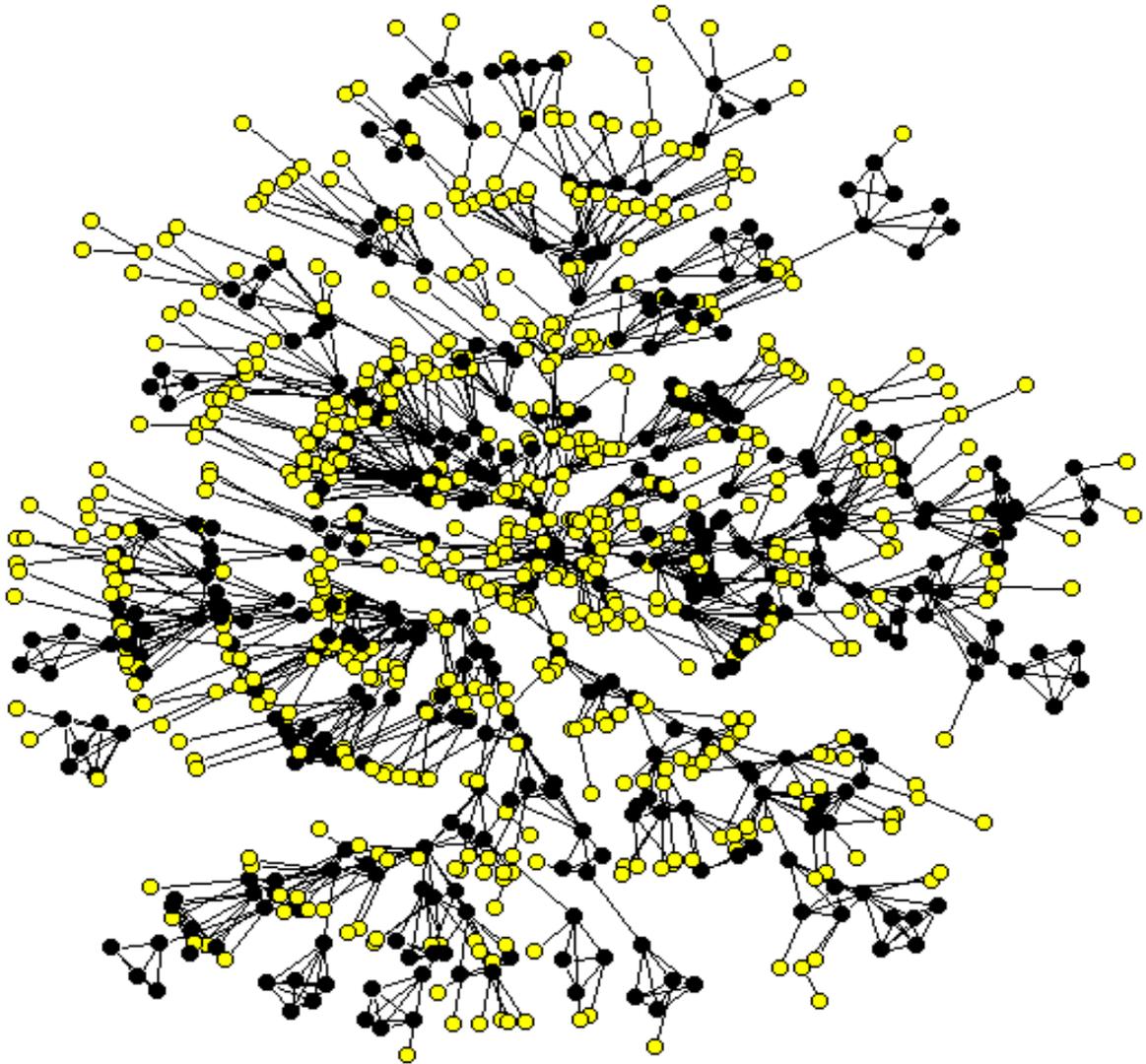}
\caption[Non-directed computer generated modular network constructed as a tree with addition of cliques]{Non-directed computer generated modular network with modules indicated in black, constructed by repeatedly attaching a clique of size 4-6, combined with attachment of 5-10 tree nodes, together totaling $N=1000$ nodes.} \label{fig-modularnetwork}
\end{center}
\end{figure}

\n 3. \textbf{Modular Network} generated from the scalefree tree by adding clique motifs, shown in Fig.\,\ref{fig-modularnetwork}. We use a modular network whose modules are explicitly known in order to explore its relationship with tree and 4-star, as well as with its own modules (cliques). This network is grown by a preferential attachment of tree nodes combined with a non-preferential attachment of cliques, following the algorithm:
\begin{itemize}
\item a certain number of tree nodes $v_{tree}$ is randomly selected with uniform probability from the interval $[v_{min}, v_{max}]$. New nodes are then attached to the existing structure one after another. The attachments are done preferentially (as for the tree above): each new node $[i]$ links to a selected node $[j]$ on the existing structure with the probability $P([j]) = \frac{k_j+\alpha}{\sum_n (k_n+\alpha)}$. The probabilities $P([j])$ are updated with each attachment of a tree node. As before, we fix $\alpha=1$ for all the networks
\item a clique of random size $w_{clique} \in [w_{min}, w_{max}]$ is selected and attached to the existing structure. The attachment is done non-preferentially: one node belonging to the clique is attached to a randomly selected node on the existing structure with a uniform probability. After the attachment of the clique, all $k_i$-s are updated accordingly with clique's size
\item the growth proceeds with a new selection of certain number of tree nodes $v_{tree}$. In all further attachments, both tree nodes and clique nodes are regarded as nodes belonging to the existing structure
\end{itemize}
The network growth starts from the a tree with 10 nodes, after which the named two steps are repeated until the network size exceeds $N=1000$ nodes. As previously, all the links are symmetric (non-directed). Note that in our  modular network the moduls are integrated in the network, and not appended as on a "Christmas tree". 

For the purposes of our system of CCM we took $v_{min}=5$, $v_{max}=10$, $w_{min}=4$ and $w_{max}=6$, to obtain a network from Fig.\,\ref{fig-modularnetwork}. Since the average number of tree nodes attached is 7.5 and the average size of attached cliques is 5, the ratio of tree nodes vs. clique nodes is 3:2. Our modular network on average has 600 tree nodes and 400 clique nodes, implying it contains on average 80 clique motifs.
The asymptotic degree distribution for this modular structure (averaged over many equivalent realizations of this network) is shown in Fig.\,\ref{fig-networkproperties}a in red. It also exhibits a power-law degree distribution $P(k) = \mbox{const} \times k^{-\gamma}$ with somewhat steeper slope than in the case of the tree.
As opposed to the scalefree tree exposed above, modular network contains topological loops, as cliques include loops (in addition to dynamical self-loops on each node due to two-dimensional nature of dynamical units). The system of CCM was implemented on this network assigning a standard map to each node and allowing them to interact according to Eq.(\ref{main-equation}), equally treating tree node maps and clique node maps. \\[0.1cm]

\begin{figure}[!hbt]
\begin{center}
\includegraphics[height=5.8in,width=6.3in]{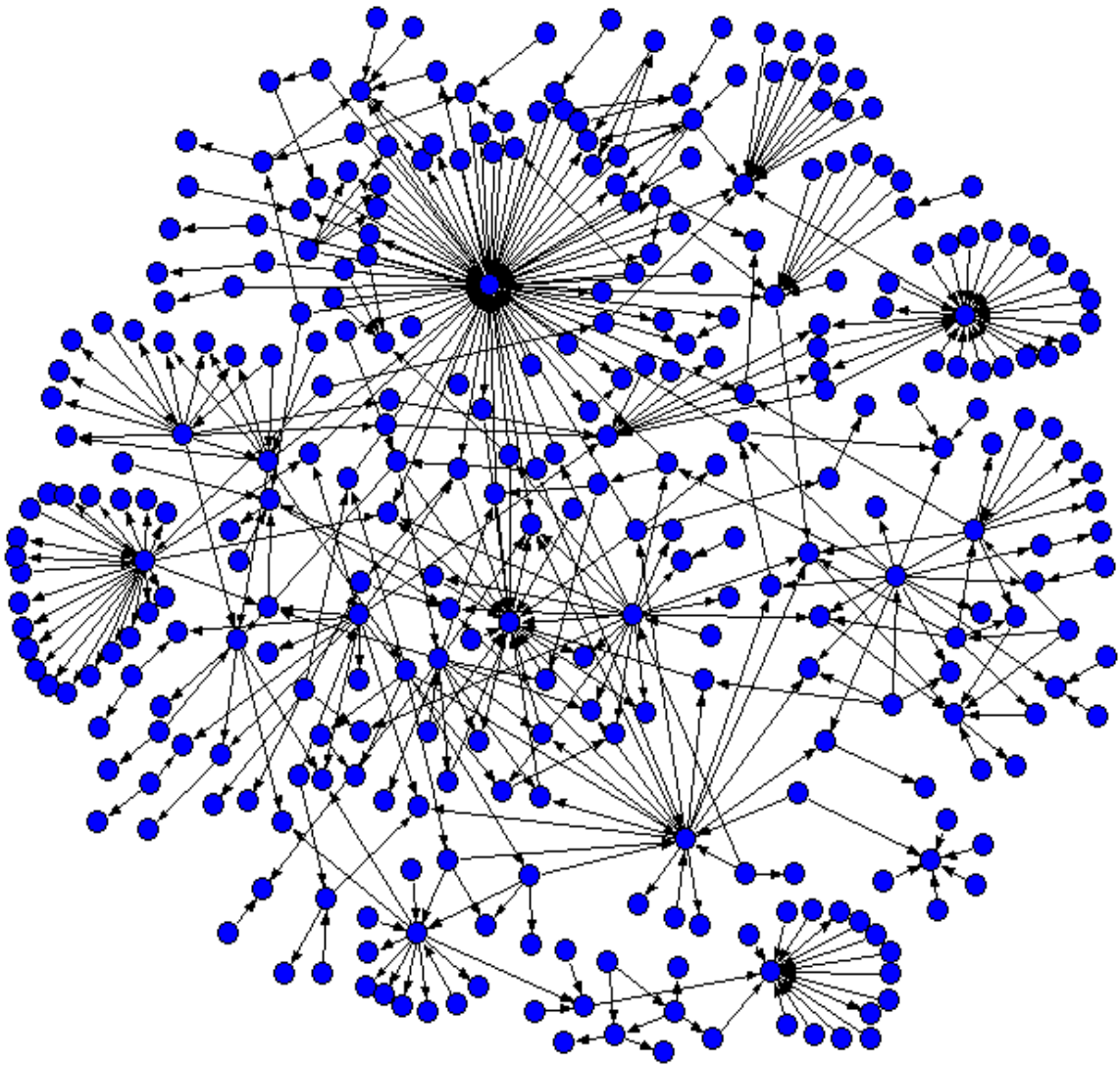}
\caption[The largest connected component of the directed gene regulatory network of bacterium E. Coli (taken from \cite{orr,mangan})]{The largest connected component of the directed gene regulatory network of bacterium E. Coli (interaction types, self-loops and names of the genes are not shown).} \label{fig-connectedcomponent}
\end{center}
\end{figure}

\n 4. \textbf{Gene Regulatory Network of Bacterium E.Coli}, which was experimentally found in 2003 \cite{orr,mangan} and whose adjacency matrix we downloaded from: \\ \texttt{http://www.weizmann.ac.il/mcb/UriAlon/} \\ 
We consider only the largest connected component of this directed network having $N=328$ nodes, shown in Fig.\,\ref{fig-connectedcomponent}. This is our main applicational example, where we examine the collective dynamics on a real biological network. Both CCM Eq.(\ref{main-equation}) and the 2D gene regulation Hill model Eqs.(\ref{schuster})\&(\ref{hill}) will be investigated using this network. 
For the system of CCM we will perform a study similar to the study of non-directed network, in order to observe the differences introduced through a directed topology of a real network. For the 2D Hill model of gene regulation we will see how the same network can display properties of a biological complex system, like flexibility to environmental inputs.

\begin{figure}[!hbt]
\begin{center}
\includegraphics[height=5.in,width=6.3in]{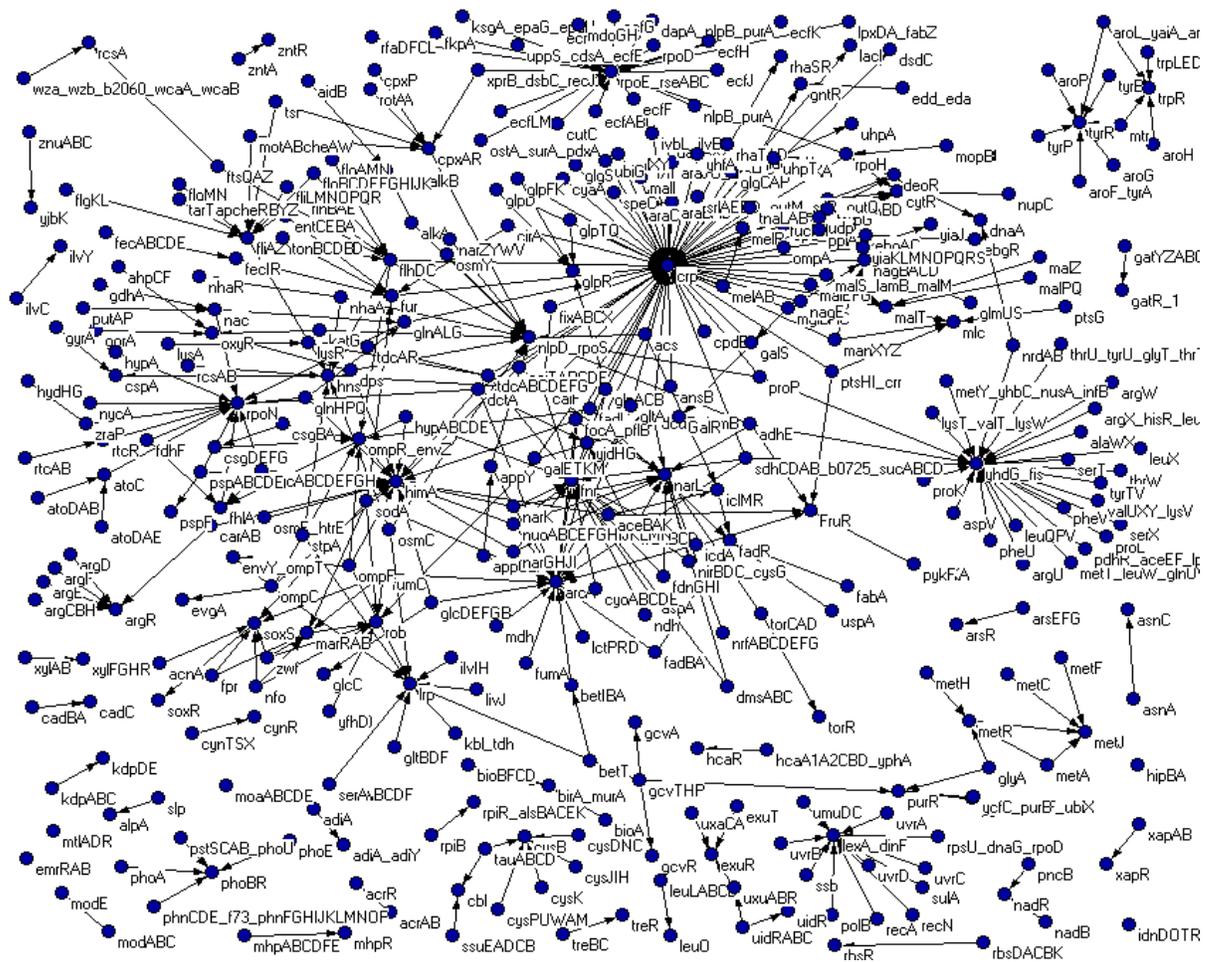}
\caption[Full E.Coli gene regulatory network including genes' biological names (taken from \cite{orr,mangan})]{Full E.Coli gene regulatory network including genes' biological names (without self-loops and the types of interactions), according to the data from \cite{orr,mangan}.} \label{fig-ecolifullnetwork}
\end{center}
\end{figure}

\begin{figure}[!hbt]
\begin{center}
$\begin{array}{cc}
\includegraphics[height=2.6in,width=3.3in]{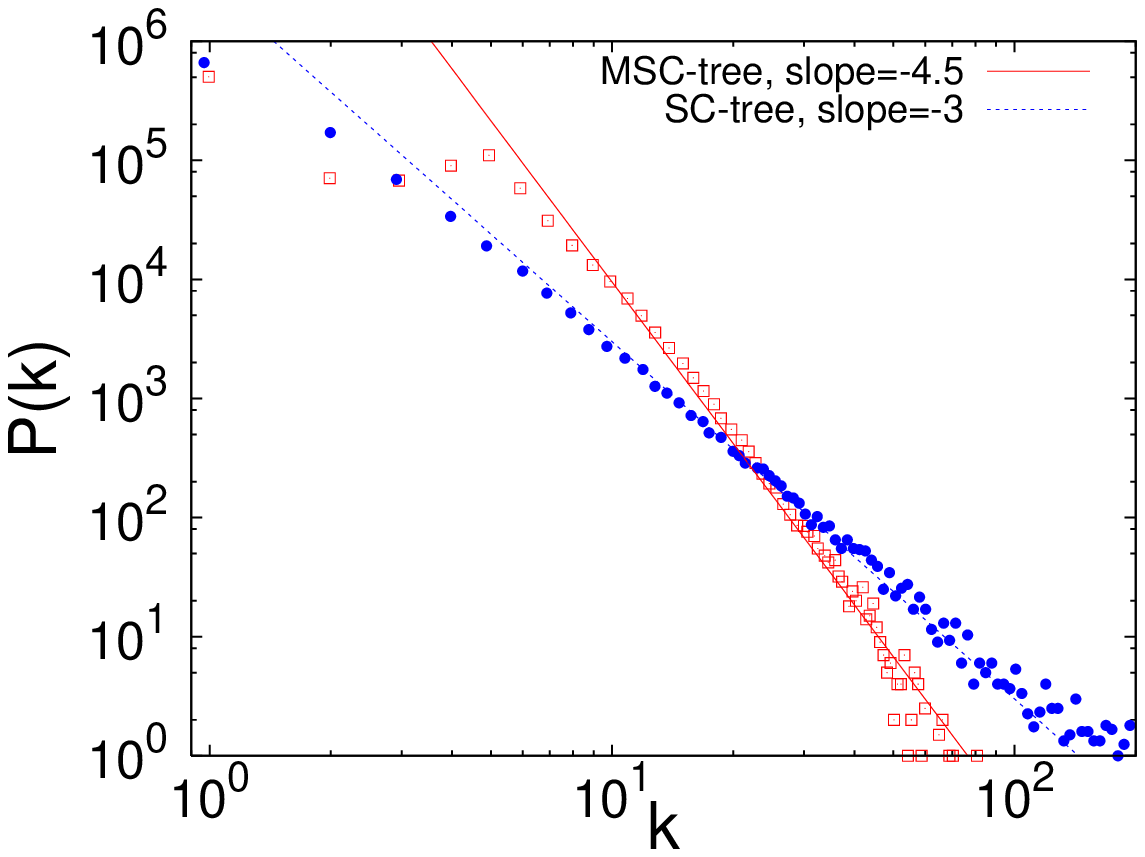}&
\includegraphics[height=2.6in,width=3.1in]{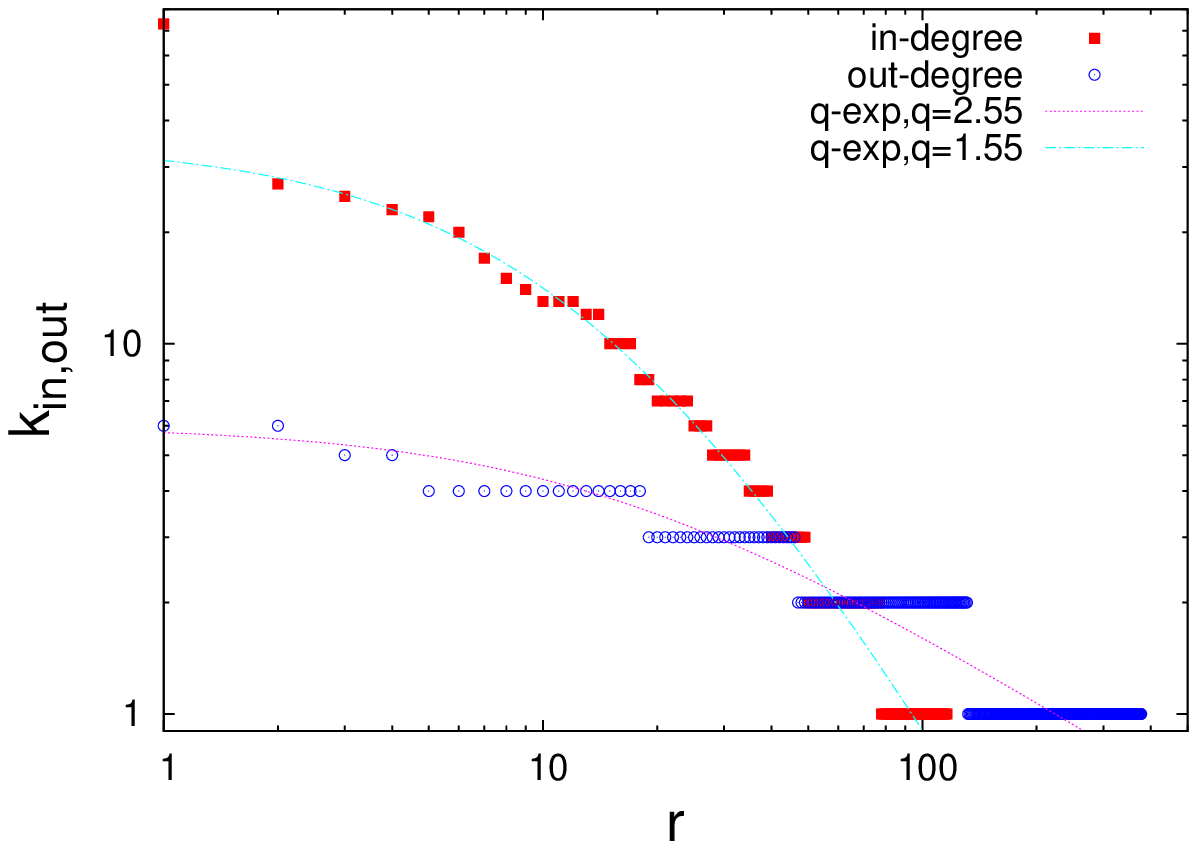} \\
\mbox{(a)} & \mbox{(b)} 
\end{array}$ 
\caption[Degree distributions for computer generated networks in use, and comparison of E.Coli's rankings for in-degrees and out-degrees]{Degree distributions averaged over many samples of the networks used for our systems of CCM with $N=1000$ nodes; scalefree tree like the one from Fig.\,\ref{fig-sftree} (blue), and tree with addition of $\sim 80$ cliques such as the one in Fig.\,\ref{fig-modularnetwork} (red), fitted with respective slopes in (a). Ranking distribution of in-degrees and out-degrees for all nodes of full gene regulatory network of E.Coli's from Fig.\,\ref{fig-ecolifullnetwork}, fitted with q-exponential distribution (which is to be introduced in Eq.(\ref{qexp}) in (b).} \label{fig-networkproperties}
\end{center}
\end{figure}

The full E.Coli gene regulatory network as available in the mentioned web-site, originally includes $N=423$ nodes which have their biological names as shown in Fig.\,\ref{fig-ecolifullnetwork}. For this network the rankings of nodes according to their in-degrees and out-degrees was computed and is shown in Fig.\,\ref{fig-networkproperties}b. Some newer E.Coli databases include more genes, but are oriented towards biological applications rather than towards physical (network) studies.

For the purposes of our investigations, we will focus only on the largest connected component of the E.Coli network from Fig.\,\ref{fig-connectedcomponent}, as it represent the dynamically most interesting part. Smaller connected components have their own dynamical behaviors which are unrelated to the dynamics of the largest component. In our exploration of the dynamics of CCM we will use the directed topology of E.Coli network only, while in the investigation of the 2D Hill gene regulation model the nature of gene-to-gene interactions (activatory or repressory) will also be included \cite{orr,mangan} (see Fig.\,\ref{fig-ecoli-colorpajek} in Chapter \ref{Dynamics on E.Coli Network}. for representation of the same network with types of regulatory interactions). 


\chapter{Collective Dynamical Effects in CCM on Networks} \label{Collective Dynamical Effects in CCM on Networks}

\begin{flushright}
\begin{minipage}{4.6in}
    Numerical implementation of our CCM on networks is described, and the main types of orbits exhibited by the system are shown. 
    The regularization process leading to periodic orbits is explored. The dynamical regions related to various types of emergent dynamics 
    are identified as intervals of coupling strength. A variety of collective effects arising due 
    to inter-node interactions are recognized and statistically investigated, in particular the clustering of periodic orbits and 
    return times with respect to phase space partitions. The Chapter ends with a brief discussion of networks of maps 
    with alternative coupling forms.\\[0.1cm]
\end{minipage}
\end{flushright}

In this Chapter we identify and investigate examples of collective effects in dynamics of networks of CCM defined in the previous Chapter. The study of our system is build upon observing its time-evolution for various values of coupling strength, using standard techniques of statistical characterization of motion, like return times distribution with respect to a phase space partition. We will focus on analyzing the departures from strongly chaotic properties of the isolated (uncoupled) standard map, which testify the presence of cooperativity among the system units. In particular, we will examine if the system's emergent behavior for small coupling strength can be characterized as regular, which would be a key argument towards the stability of real complex systems described through networks of this type.

\section{Numerical Implementation and Studied Types of Orbits}

We investigate our system computationally, by running numerical simulations in \texttt{C++} programming language that implement the dynamics of Eq.(\ref{main-equation}) with a pre-selected network topology. Each run consists in the following steps:
\begin{enumerate}
 \item set the network's size $N$, total number of iterations and the coupling strength $\mu$ 
 \item introduce the topology by uploading the computer generated network adjacency matrix $C_{ij}$, which has non-zero values $C_{ij}=1$ if a link 
between the nodes $[i]$ and $[j]$ exist 
 \item select the initial conditions for all the nodes by taking them randomly from $(x,y) \in [0,1] \times [-1,1]$ with uniform probability 
 \item iterate the network of CCM by updating all the $x$ and $y$ values for all the nodes simultaneously, for the chosen total number of iterations
 \item analyze the final network state by extracting the information on the desired dynamical properties, which possibly includes application of 
additional program subroutines
\end{enumerate}
We examine the coupling strength interval of $0 < \mu < 0.08$, focusing on certain regions more specifically as discussed below. As mentioned, standard map's chaotic parameter $\e$ is always fixed at $\e=0.9$ (regime of very strong chaos for the isolated standard map). The dynamics is typically run for $10^5$ iterations in order for transients to settle. The \textit{emergent dynamics} describes the system's motion after the transients, which is also the final dynamical state of the system: once the transients are over, the system's overall motion pattern does not qualitatively evolve further.

In our study, depending on the dynamical aspect under investigation, we examine system's orbits following three different approaches ($t_0$ denotes the length of transients), as in \cite{ja-jsm}:
\begin{itemize}
\item orbit of an individual node $[i]$ given simply as: 
 \begin{equation} 
\left( x[i]_t,y[i]_t \right)_{t>t_0}, \;\;\; i=1,\hdots, N   \label{eo} 
\end{equation}
which allows the usual study of nonlinear dynamics exhibited by a 2D discrete-time map (a node) coupled to other nodes, using standard dynamical system's tools.
\item network-averaged orbit (n.a.o.) defined as:
\begin{equation} 
\left( \hat{x}_t,\hat{y}_t \right)_{t>t_0} = \frac{1}{N}\sum_{i=1}^{N} (x[i]_t,y[i]_t)_{t>t_0}.
\label{naeo}  \end{equation}
reduces the network's 2N-dimensional dynamics to the evolution of a single 2D phase space point. Useful in qualitatively illustrating global network's behavior, particularly in the context of statistics and stability of the dynamics
\item time-averaged orbit (t.a.o.), defined for each network node as:
\begin{equation} 
\left( \bar{x}[i],\bar{y}[i] \right) = \lim_{t \rightarrow \infty} \frac{1}{t-t_0}\sum_{k=t_0}^{t} (x[i]_k,y[i]_k),
\label{taeo}  \end{equation}
reduces the whole orbit of a node to a single phase space point that qualitatively captures the emergent motion. Good for examining the differences among the nodes in the network's final dynamical state. 
\end{itemize}

The employed programming codes are structured as follows: the main program starts by setting the simulation parameters according to steps 1., 2. and 3. from the algorithm above. A separate subroutine is called, which iterates the coupled dynamics for the desired number of iterations until final state is reached. A new subroutine is then called from the main program which performs the relevant analysis of the final dynamical state or simply prints one of the system's orbits as described above.\\[0.1cm]

\begin{figure}[!hbt]
\begin{center}
$\begin{array}{ccc}
\includegraphics[height=1.9in,width=2.in]{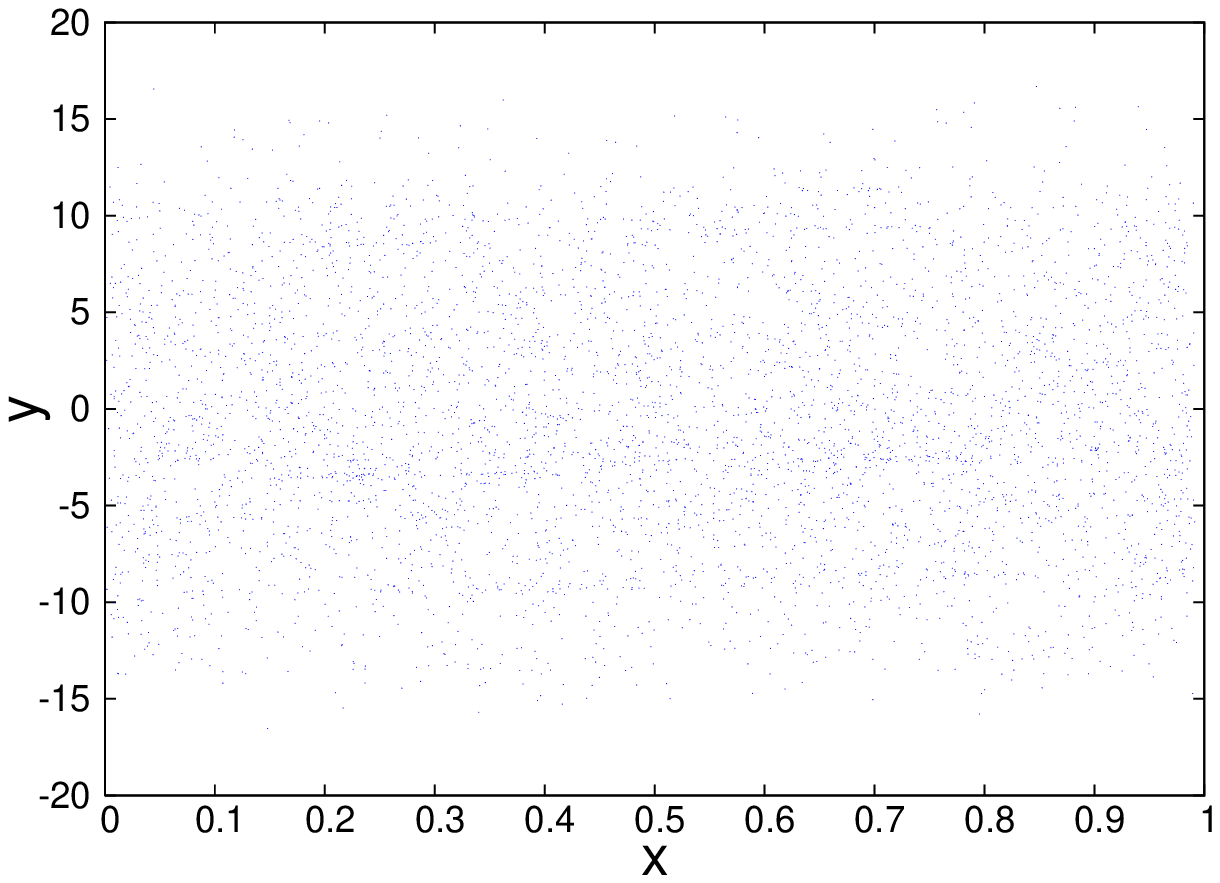} & 
\includegraphics[height=1.9in,width=2.in]{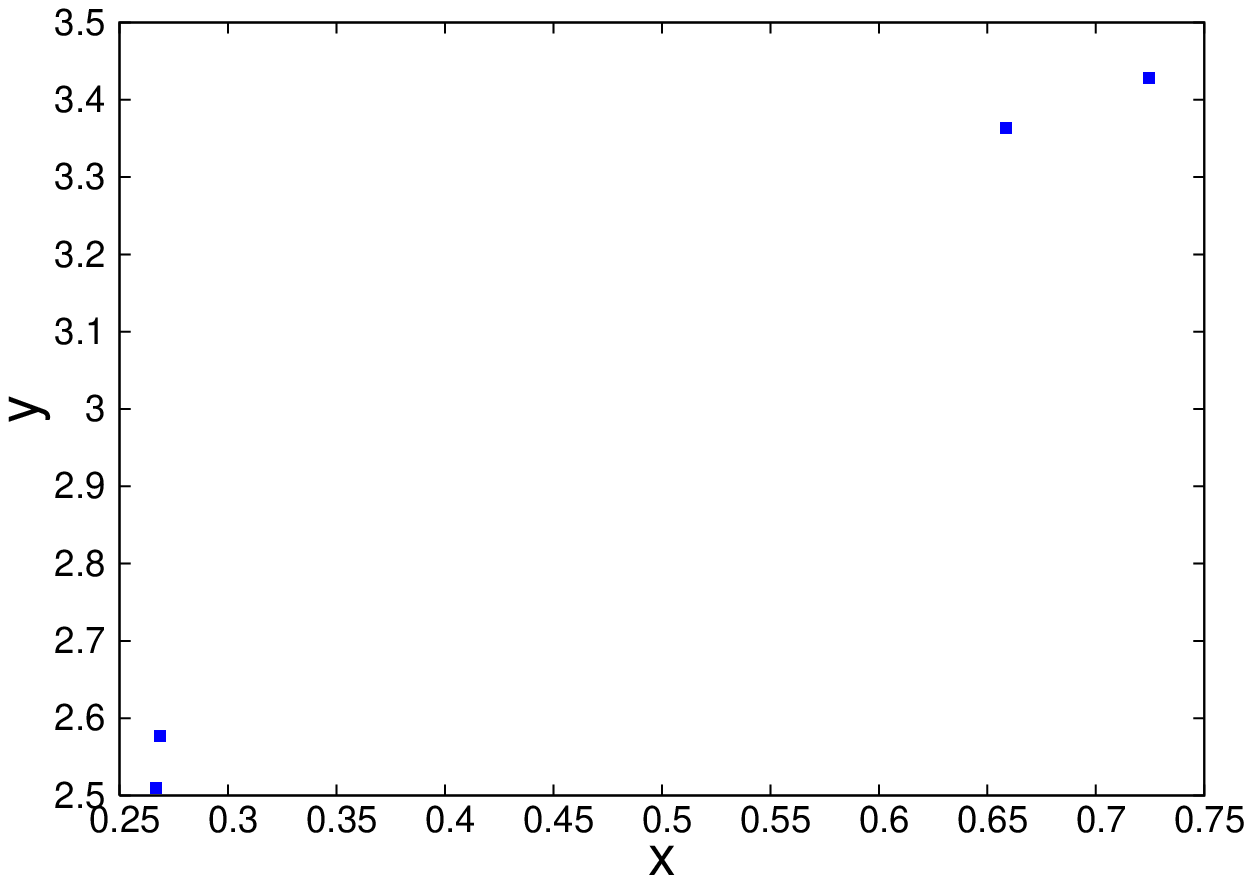} & 
\includegraphics[height=1.9in,width=2.in]{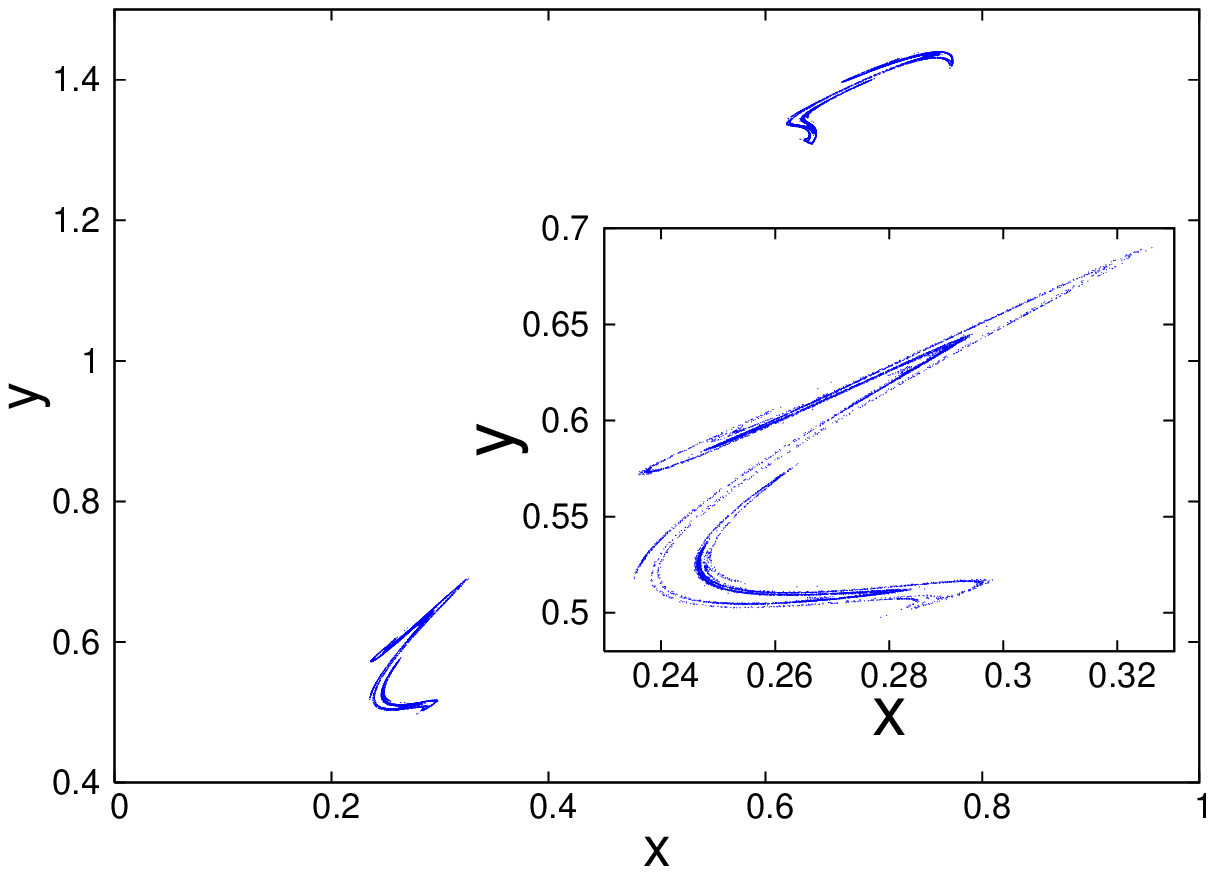} \\
\mbox{(a)} & \mbox{(b)} & \mbox{(c)} \\[0.2cm]
\includegraphics[height=1.9in,width=2.in]{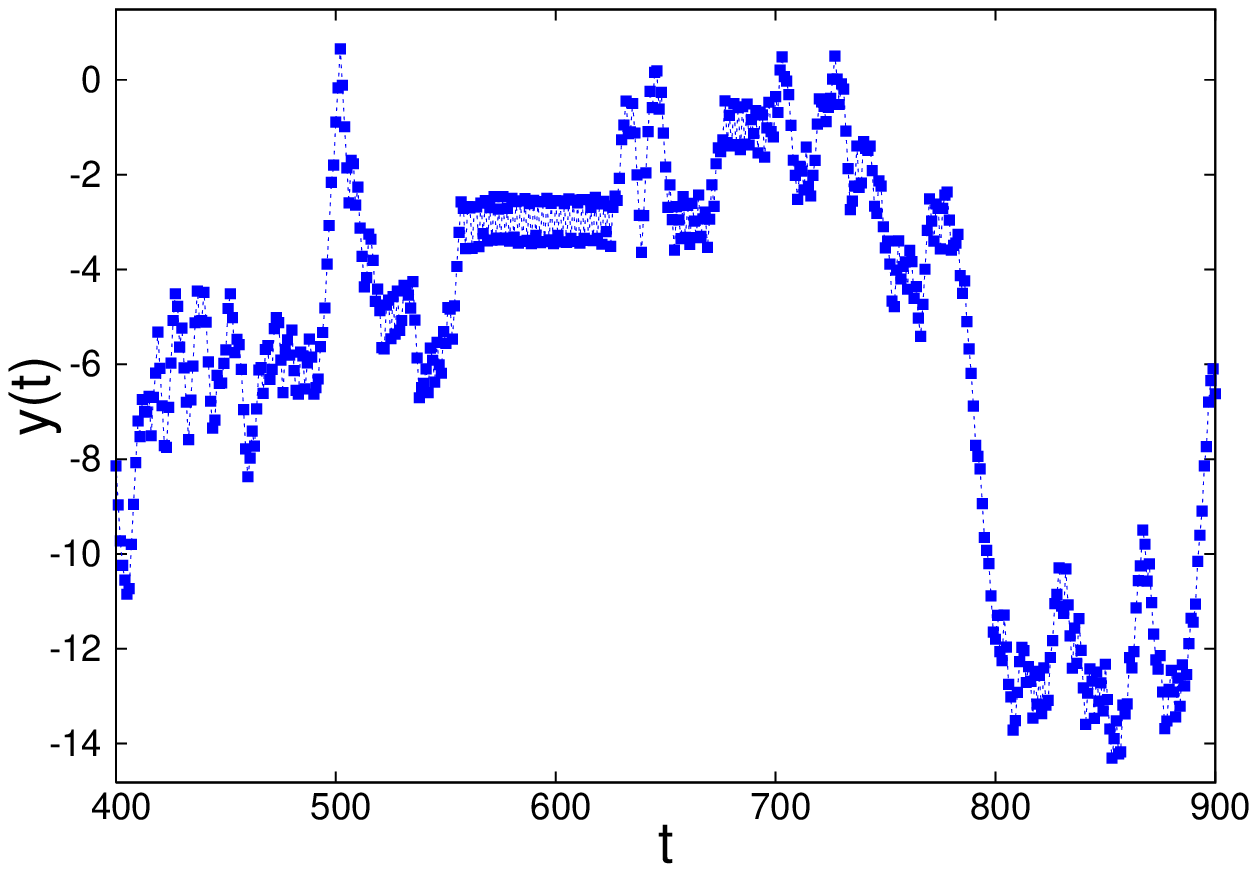} & 
\includegraphics[height=1.9in,width=2.in]{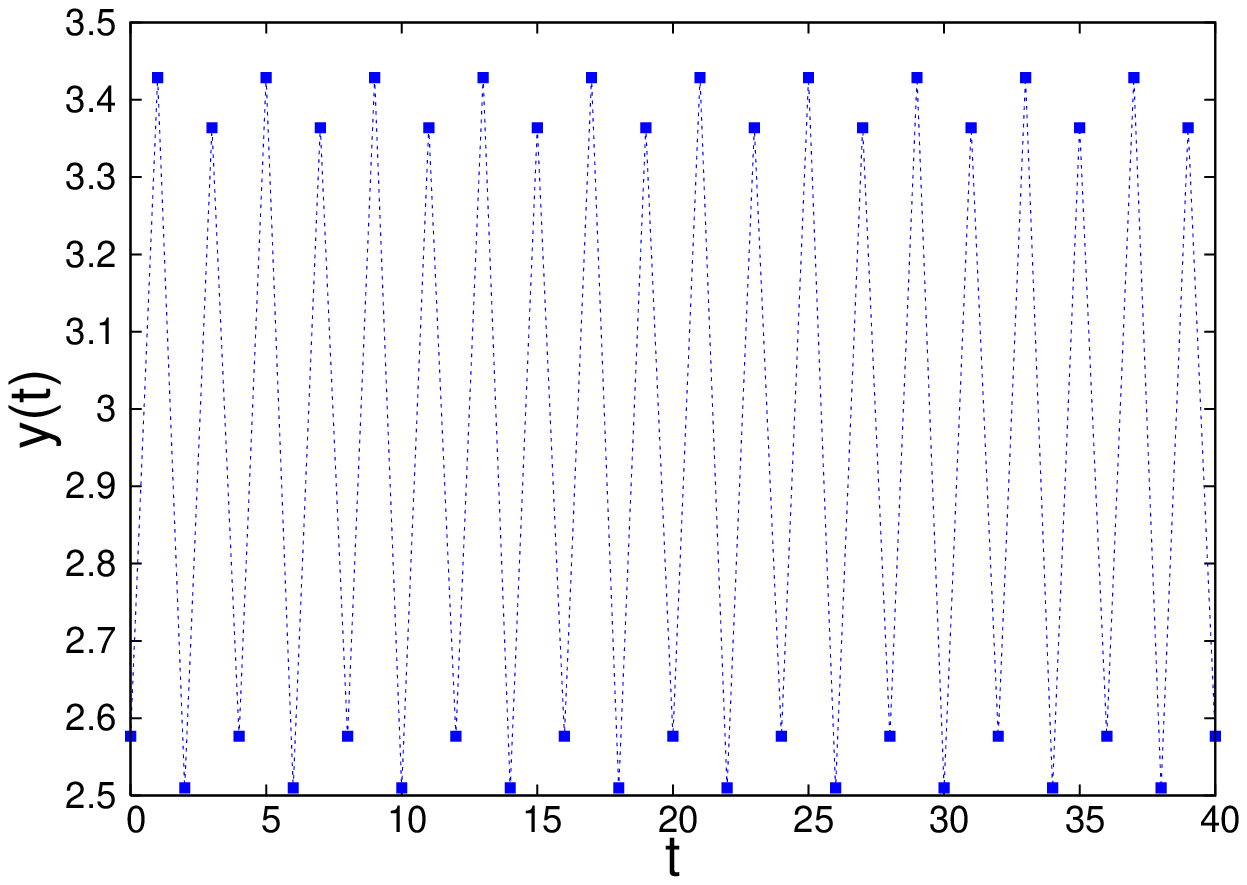} & 
\includegraphics[height=1.9in,width=2.in]{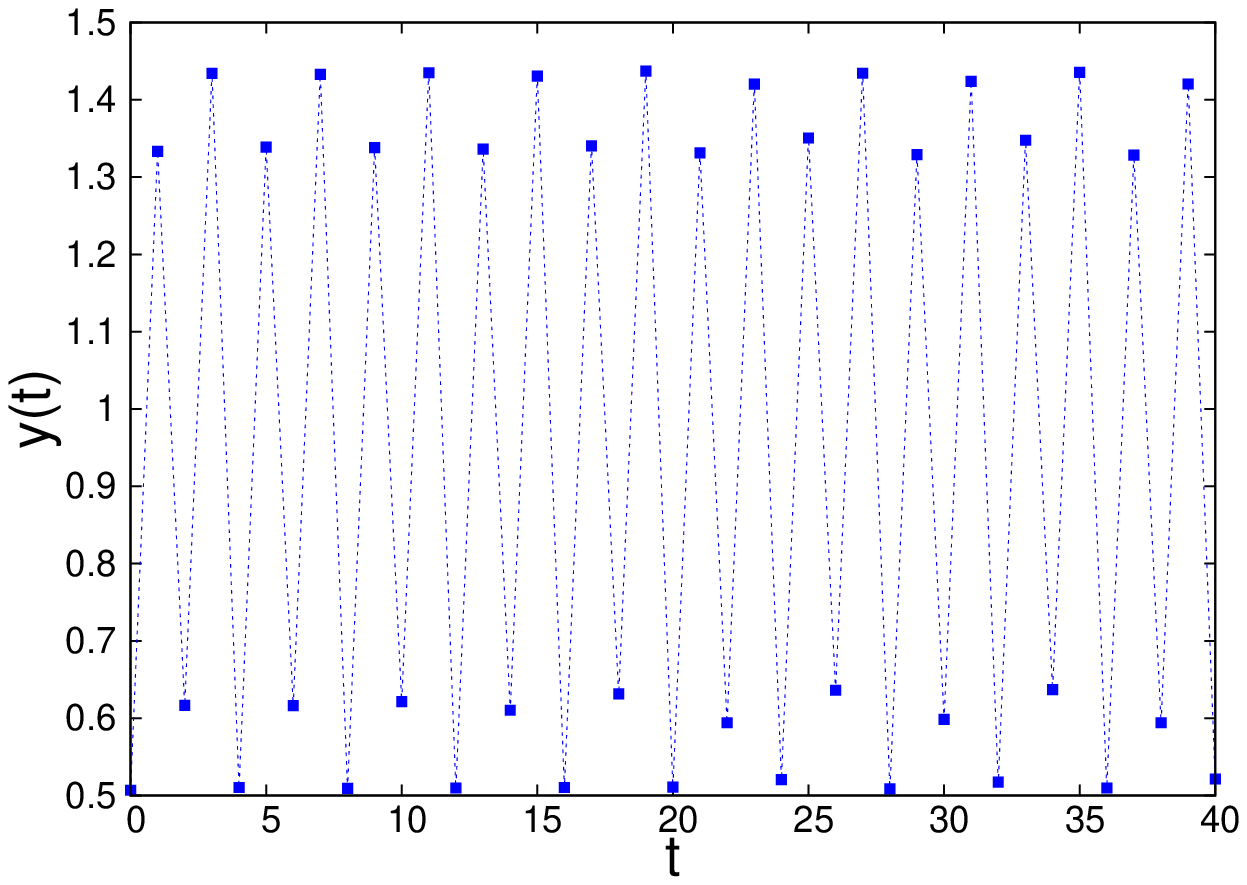} \\
\mbox{(d)} & \mbox{(e)} & \mbox{(f)} 
\end{array}$
\caption[Three typical emergent orbits appearing on the branch-node of the 4-star, along with their $y$-coordinate time-evolutions]{Three typical emergent orbits appearing on the branch-node of the 4-star (a--c), with their $y$-coordinate time-evolutions (d--f). A chaotic orbit at $\mu=0.007$ in (a\&d), a periodic orbit at $\mu = 0.012$ in (b\&e) and a strange attractor at $\mu=0.049$ in (c\&f). Inset in (c) 
zooms the left part of the orbit for a better visibility (we will often show only a half of an orbit).} \label{fig-orbitsexamples}
\end{center}
\end{figure}

\textbf{Orbits of a Single Node attached to Network.} To illustrate the behavior of our network of CCM, we show some orbits exhibited by single nodes attached to a network (tree or 4-star), for various coupling strengths. In Fig.\,\ref{fig-orbitsexamples} we show typical (emergent) orbits exhibited by the branch-node (leaf-node) of the 4-star, which are also typical for many nodes on the scalefree tree for the same $\mu$-values. They are characteristic orbits corresponding to three main dynamical regions that we will examine in detail throughout this Thesis (the regions will be defined precisely as intervals of $\mu$-values in the next Section). 
The difference between dynamical regions is given by the general motion properties which are captured by these three orbits: chaotic motion persists at small $\mu$-values (Figs.\,\ref{fig-orbitsexamples}a\,\&\,d), becoming regular/periodic with increase of $\mu$ (Figs.\,\ref{fig-orbitsexamples}b\,\&\,e), and exhibiting interesting dynamical phenomena for even larger $\mu$, including strange attractors (Figs.\,\ref{fig-orbitsexamples}c\,\&\,f). While standard map itself does not exhibit strange attractors (as it is an area-preserving system) \cite{wiggy}, our network of CCM with non-symplectic (dissipative) coupling allows this possibility.

\begin{figure}[!hbt]
\begin{center}
$\begin{array}{ccc}
\includegraphics[height=1.9in,width=2.in]{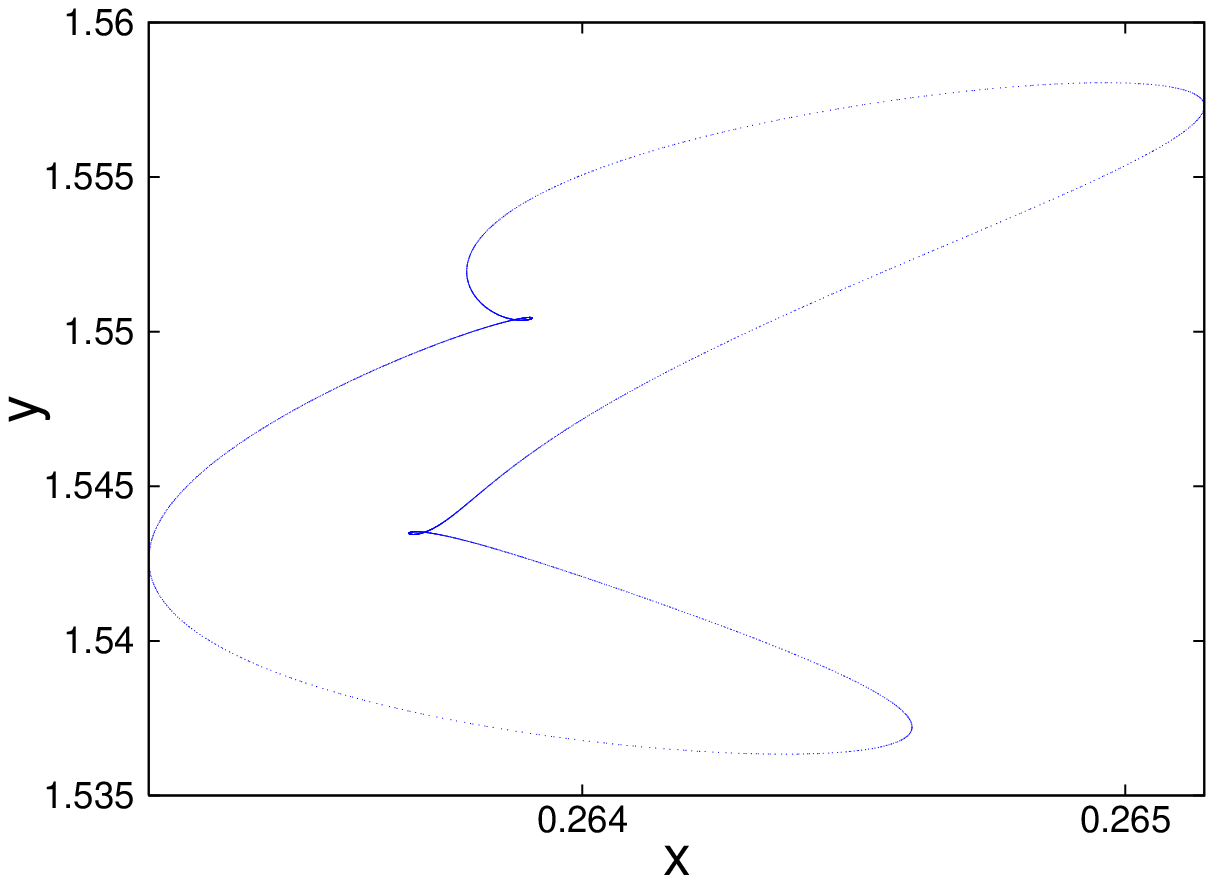} & 
\includegraphics[height=1.9in,width=2.in]{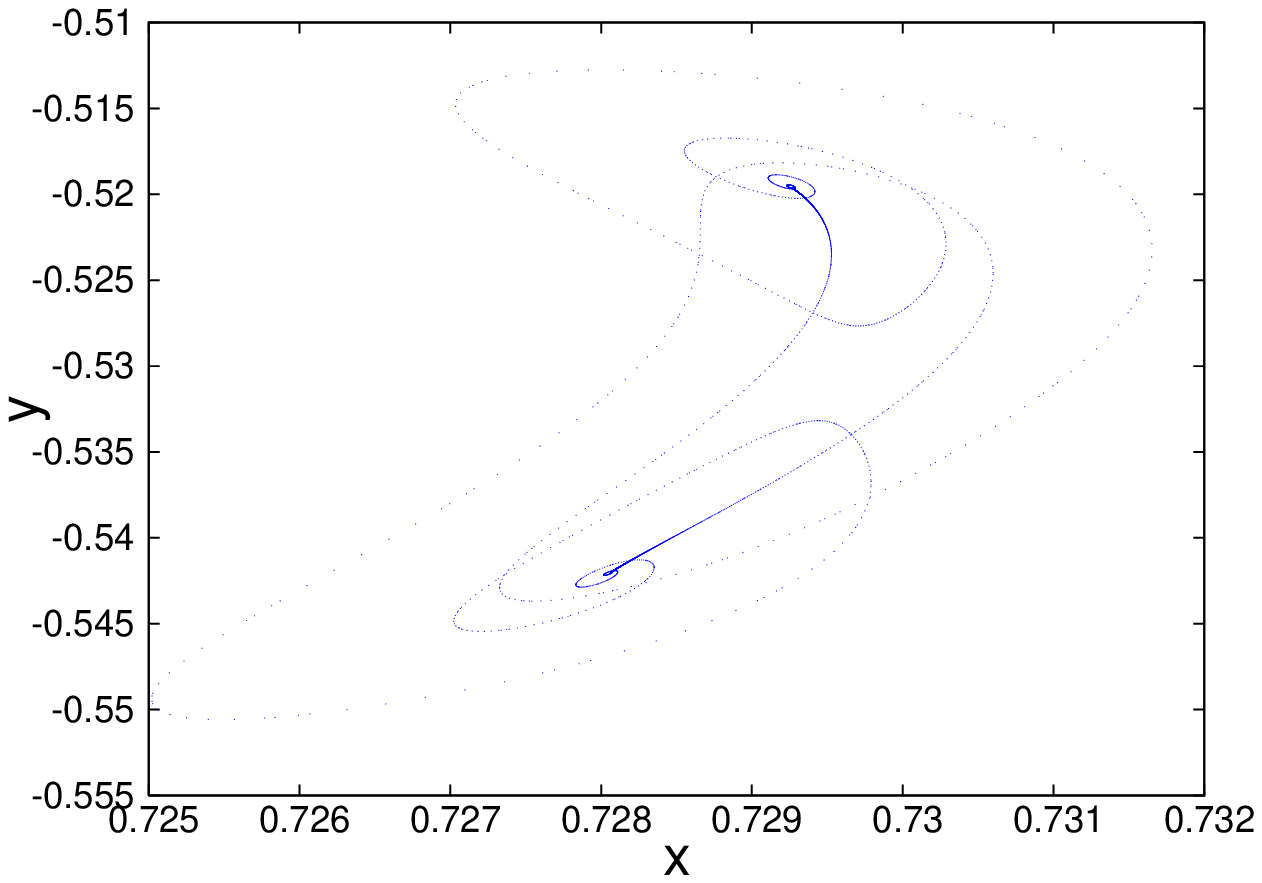} & 
\includegraphics[height=1.9in,width=2.in]{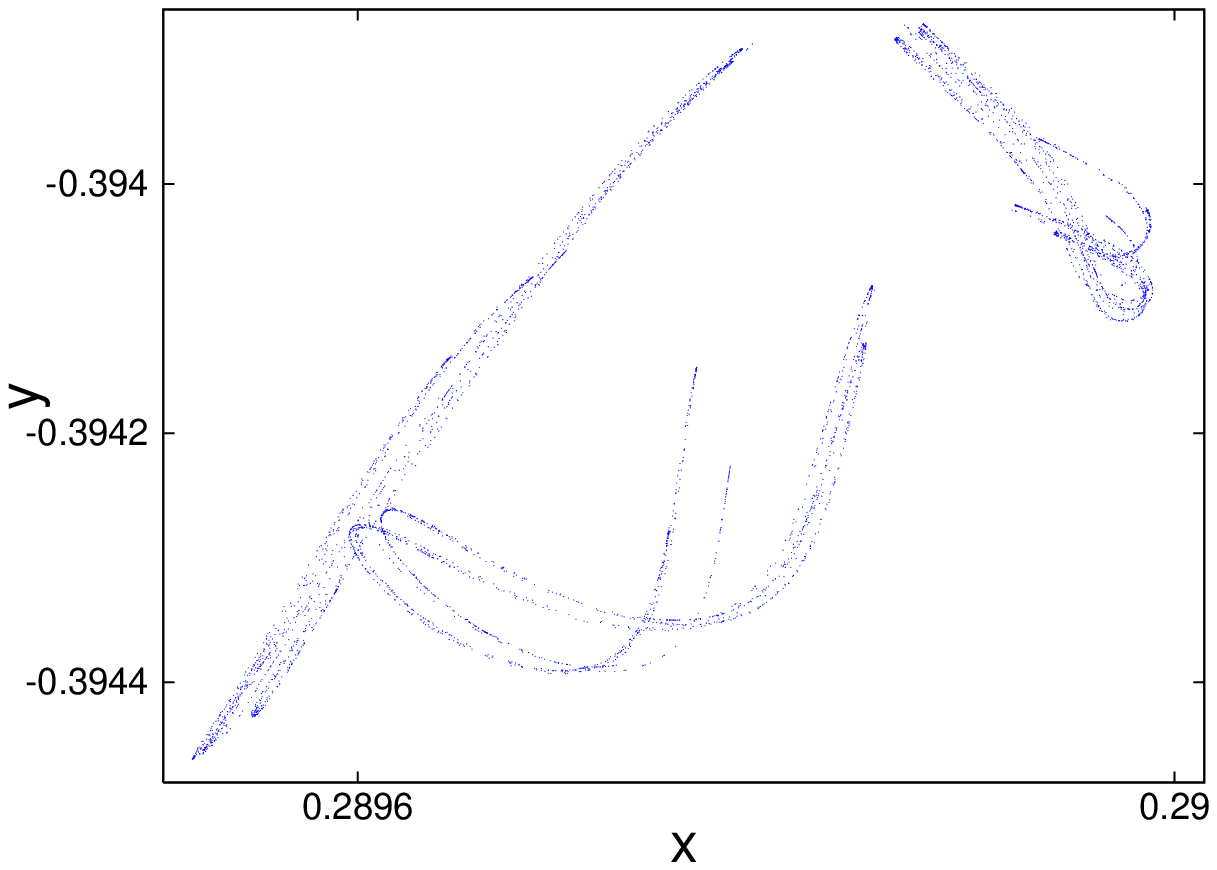} \\
\mbox{(a)} & \mbox{(b)} & \mbox{(c)} \\[0.2cm]
\includegraphics[height=1.9in,width=2.in]{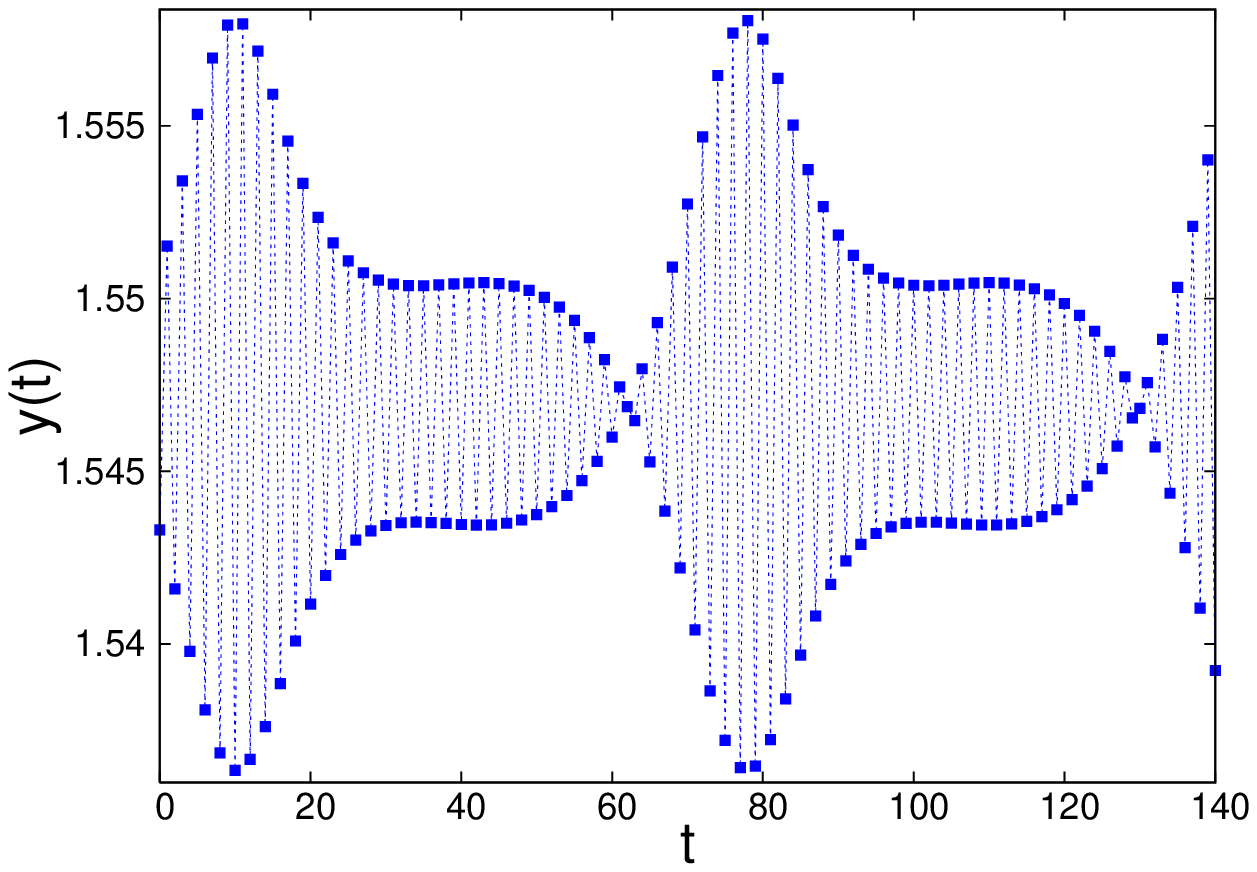} & 
\includegraphics[height=1.9in,width=2.in]{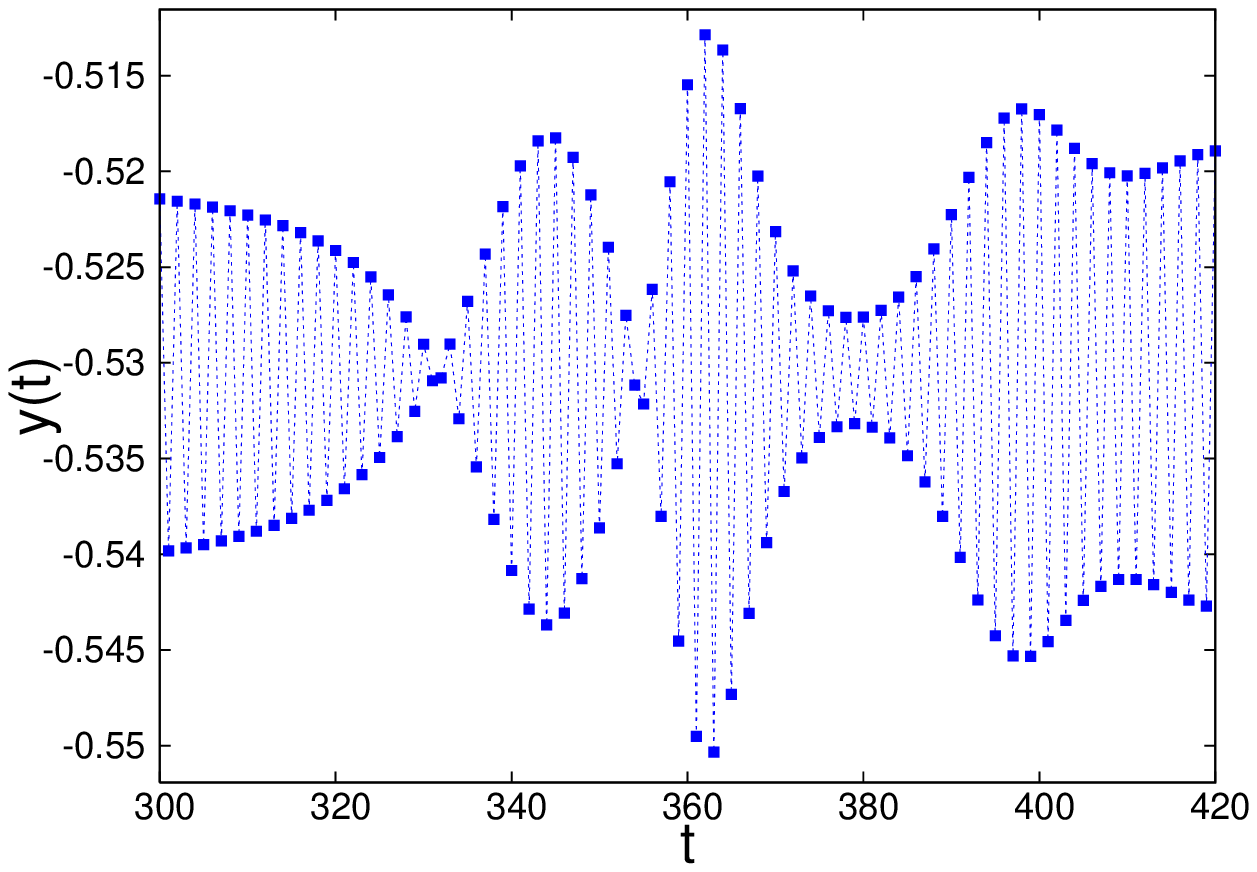} & 
\includegraphics[height=1.9in,width=2.in]{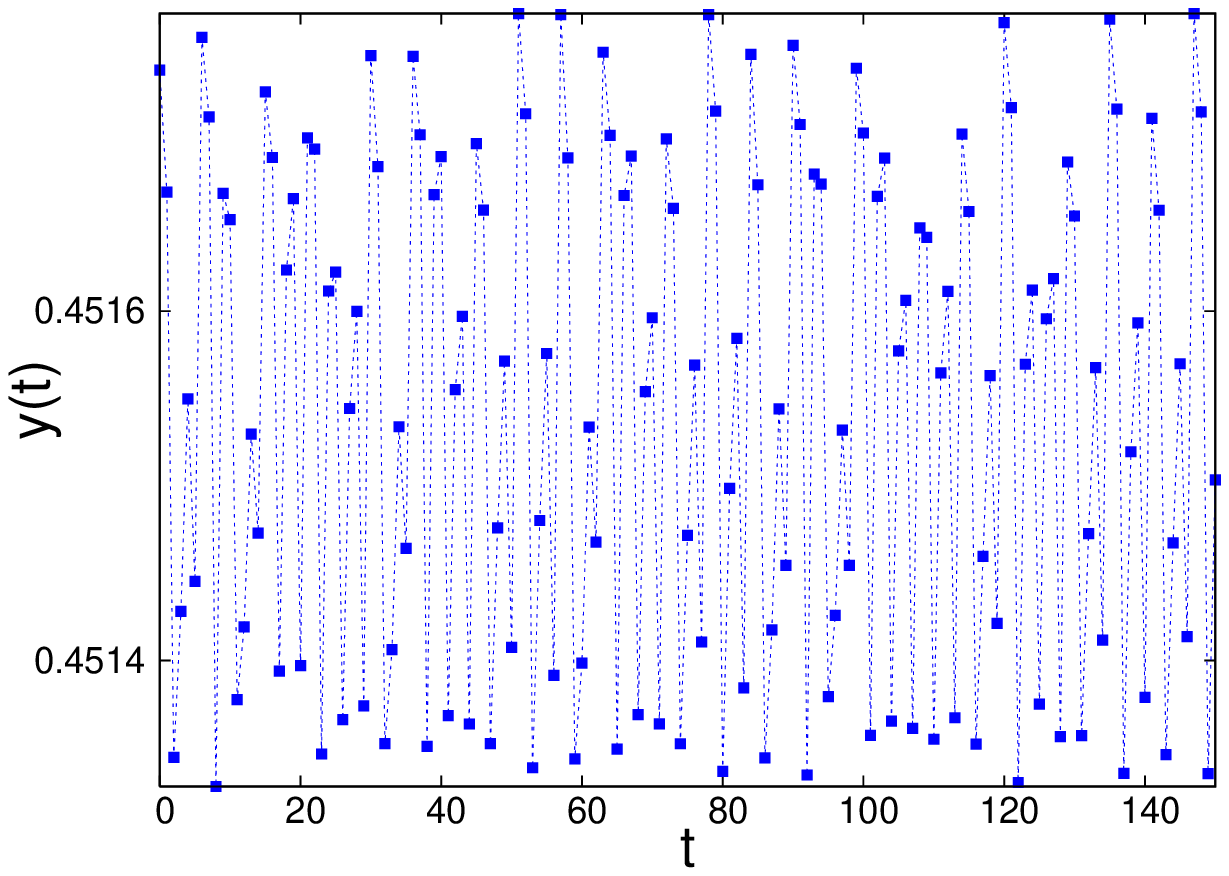} \\
\mbox{(d)} & \mbox{(e)} & \mbox{(f)} 
\end{array}$
\caption[Some additional orbits of our network of CCM with their $y(t)$ time-evolution]{Additional orbits of our CCM system. For simplicity, only a half of each orbit is shown (as in the inset of Fig.\,\ref{fig-orbitsexamples}c), together with its $y(t)$ time-evolution. Quasi-periodic orbits on 4-star's branch node at $\mu = 0.014$ (a\,\&\,d) and at $\mu = 0.027$ (b\,\&\,e). Strange attractor on 4-clique's node (all clique's nodes are topologically equivalent which is not the case with 4-star) at $\mu = 0.049$ (c\,\&\,f).} \label{fig-additionalorbits}
\end{center}
\end{figure}

Additional types of orbits are also displayed within the studied coupling strength range of $0 < \mu < 0.08$ as shown in Fig.\,\ref{fig-additionalorbits}. Quasi-periodic orbits Figs.\,\ref{fig-additionalorbits}a\,\&\,d,b\,\&\,e construct another dynamical region, although the main attention shall be devoted to three types of motion mentioned above. 
Strange attractors also appear on the 4-clique motif for the same coupling strength, as illustrated in Figs.\,\ref{fig-additionalorbits}c\,\&\,f. Clearly, our network of CCM posses a rich dynamics which is sensitive to both topology and the coupling strength. The (emergent) orbits on single-nodes of the scalefree tree are qualitatively similar to the orbits mentioned here (except for strange attractors), for the same $\mu$-values.

\section{The Process of Dynamical Regularization} 

In this Section we investigate the process of creation of regular orbits as a consequence of inter-node interactions during the time-evolution of the coupled dynamics, termed \textit{regularization process}. We will quantitatively show how the chaotic dynamics of isolated standard maps becomes self-organized, by examining the increase of number of nodes exhibiting periodic orbits. \\[0.1cm]

\textbf{Dynamical Localization and Periodic Orbits.} The dynamics for very small coupling strengths is characterized by chaotic irregular behavior of all nodes, irrespectively of the underlying topology (cf. Figs.\,\ref{fig-orbitsexamples}a\,\&\,d). However, this chaotic motion does not exhibit normal diffusive behavior in angular momentum as in the case of isolated standard map \cite{ll,GH}; instead, the motion remains localized in $y$-coordinate to a band of certain width, despite still being chaotic within the band. The network of CCM inhibits standard map's chaotic diffusion due to the interactions between the nodes, localizing the motion to a band in $y$-coordinate (Fig.\,\ref{fig-orbitsexamples}a illustrates what we intend as a band). The band's width furthermore shrinks with time-evolution, eventually becoming a periodic orbit after a transient which depends on the coupling strength value.

\begin{figure}[!hbt]
\begin{center}
$\begin{array}{cc}
\includegraphics[height=2.55in,width=3.15in]{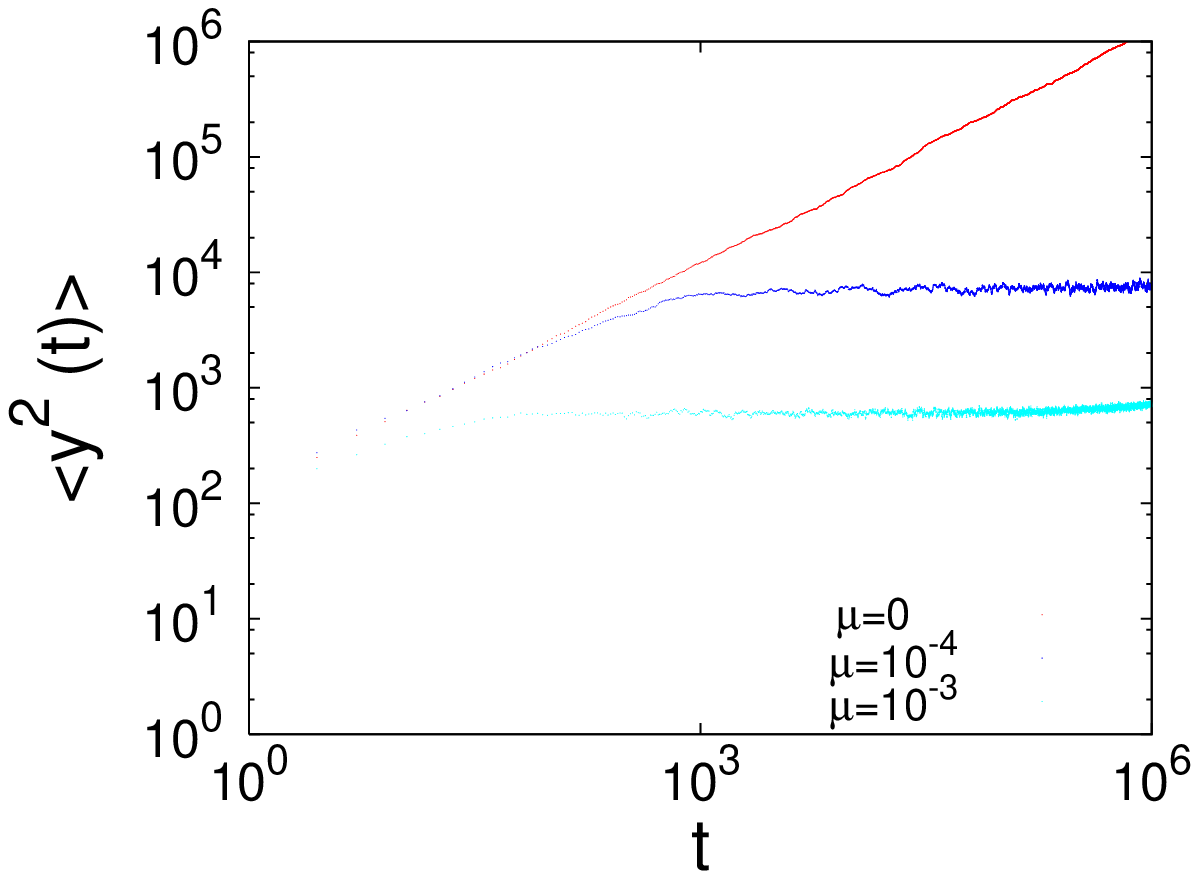} & 
\includegraphics[height=2.55in,width=3.15in]{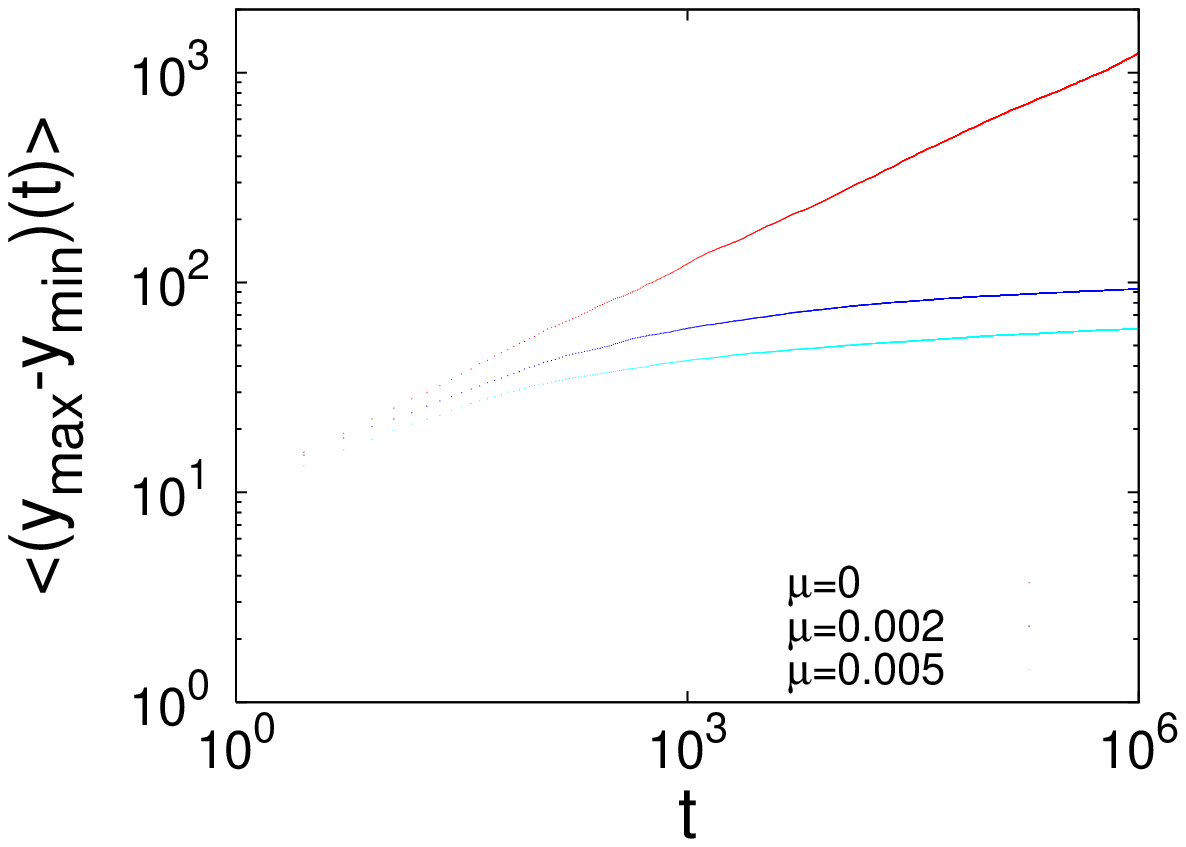} \\
\mbox{(a)} & \mbox{(b)}  
\end{array}$ 
\caption[Dynamical localization of emergent orbits and inhibition of chaotic diffusion]{Dynamical localization of orbits for various coupling strengths: angular momentum diffusion $<y^2(t)>$ in (a) and average band width stretching rate $<(y_{\mbox{max}}-y_{\mbox{min}})(t)>$ in (b). Plots are obtained by averaging over an ensemble of orbits of 4-star's branch node.}  \label{fig-diffusion}
\end{center}
\end{figure}

We expose this quantitatively in Fig.\,\ref{fig-diffusion}a where we examine the time-evolution of the mean-square distance $<y^2(t)>$ for the 4-star's branch node, averaged over many initial conditions. Observe that even for very small coupling vales the diffusion is inhibited after a certain time, and a  sub-diffusive behavior of $<y^2(t)>$ emerges with slope $\gamma \sim 0$. 
We moreover examine the time-evolution of the maximal band widths $<(y_{\mbox{max}}-y_{\mbox{min}})(t)>$, measuring it for the same node averaged over many initial conditions, with results shown in Fig.\,\ref{fig-diffusion}b. Again, for arbitrarily small coupling strengths the stretching rate is suppressed after a certain time (which is to be identified as the transient time). Once a band width reaches very small values ($\sim O(1)$), the orbit's band does not shrink further, but instead transforms into a regular (periodic) orbit as in  Fig.\,\ref{fig-orbitsexamples}b. As we shall clarify shortly, the chaotic orbits as the one in Fig.\,\ref{fig-orbitsexamples}a are of transitory nature, as they gradually disappear with the system's time-evolution. As it appears from Fig.\,\ref{fig-diffusion}, the transient times are independent from the initial conditions.

\begin{figure}[!hbt]
\begin{center}
\includegraphics[height=2.9in,width=4.3in]{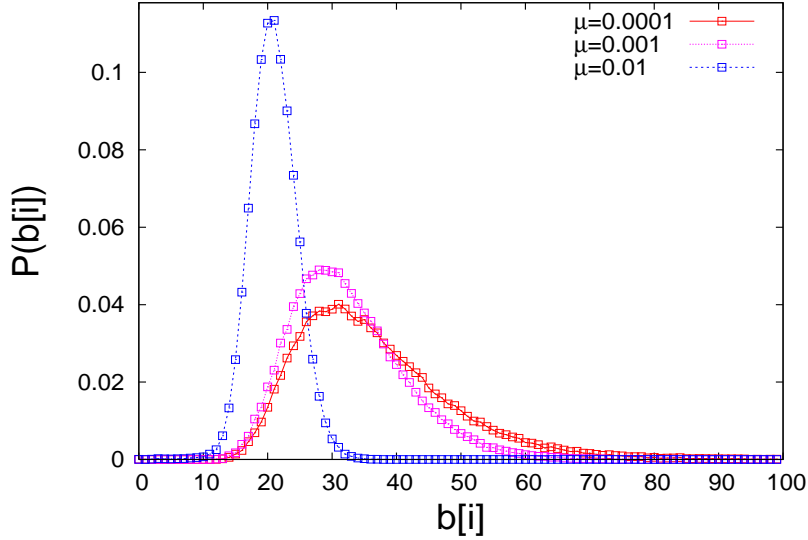} 
\caption[Distributions of band withs of all scalefree tree nodes in chaotic regime]{Distribution of band withs of all scalefree tree nodes averaged over many initial conditions, for three values of coupling strength. A transient of $10^5$ iterations was not included.}  \label{fig-bandwiths-histogram}
\end{center}
\end{figure}

We also examine the distribution of the band widths for the case of scalefree tree, considering the band widths of all tree's nodes. In Fig.\,\ref{fig-bandwiths-histogram} we show the results for three values of coupling strength, averaged over initial conditions. The increase of coupling reduces the band widths towards smaller and more uniform values, eventually shrinking the motion of nodes to a narrow region in phase space with a band width $\sim O(1)$.

As the 2D phase space domain available for nodes becomes smaller, the interaction among them becomes more efficient in terms of motion's collectivity. The accessible region for $x$-coordinate is always $[0,1]$, whereas a $(1-\mu)$ term is present in front of the momentum variable, which tends to reduce the accessible region in $y$-coordinate. Moreover, the dynamics in  $x$-coordinate is coupled, with nodes receiving each other's inputs and adjusting to them. This interplay defines the system's time-evolution, which for small coupling strengths results in shrinking of the accessible band in $y$-coordinate. Finally, the shrinking of bands results in the creation of regularity in the coupled motion appearing in the form of \textit{periodic orbits} of various periodicities, which are the central collective phenomena of our system of CCM Eq.(\ref{main-equation}). 

Periodic orbits are oscillatory and characterized by a period value; orbit shown in Fig.\,\ref{fig-orbitsexamples}b has period value of four. The chaotic region is essentially a transient region, as the inter-node interaction eventually creates periodic orbits for any non-zero (but small) coupling. However, the transient times before the emergence of periodic orbits significantly increases for very small coupling strengths (for $\mu \lesssim 10^{-4}$ it is typically $O(10^7)$ iterations). This regards both tree and 4-star: after a transient that depends on the coupling strength, all the nodes exhibit periodic behavior similar to the one shown in Fig.\,\ref{fig-orbitsexamples}b. The phase space locations and other characterization of periodic orbits will be extensively examined in the Sections to follow, along with the case of larger coupling strengths where the emergent orbits may not be periodic, but exhibit other self-organizational properties. \\[0.1cm]

\begin{figure}[!hbt]
\begin{center}
$\begin{array}{cc}
\includegraphics[height=2.55in,width=3.15in]{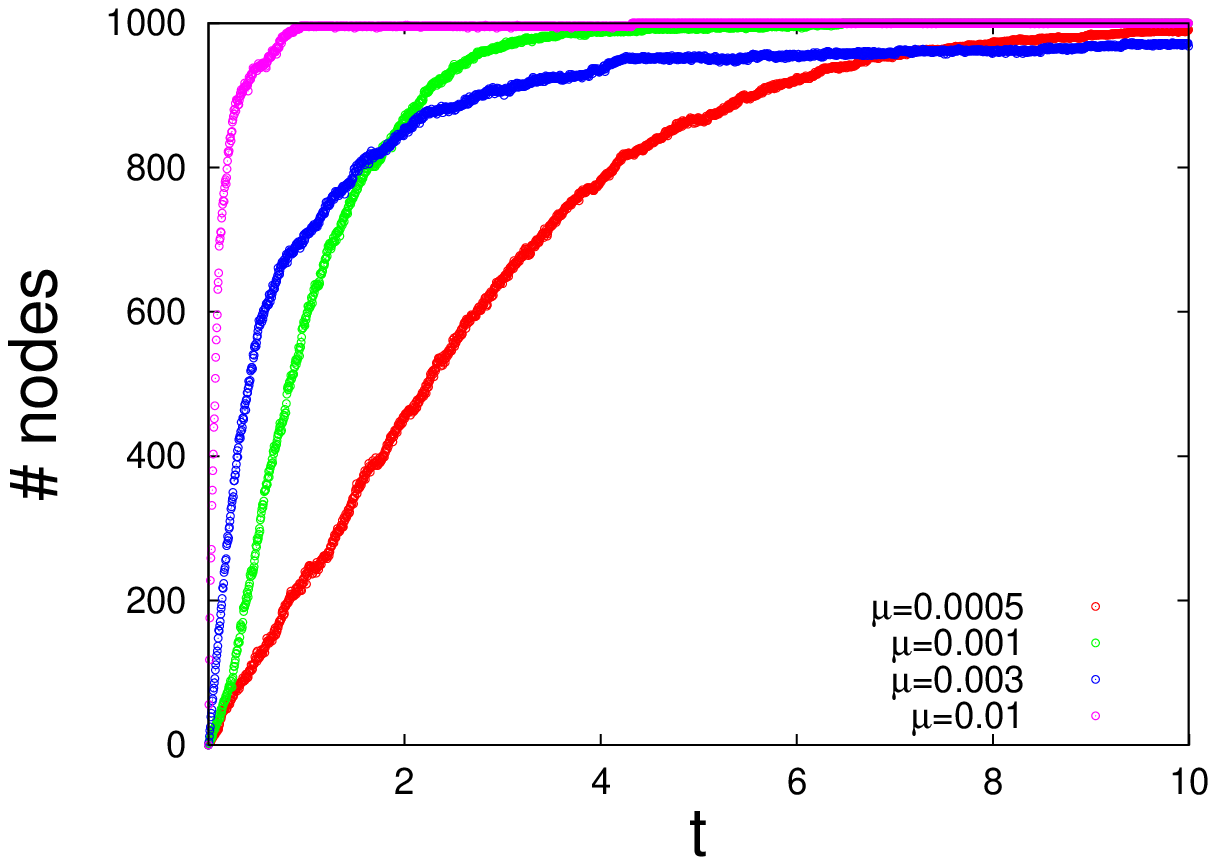} & 
\includegraphics[height=2.55in,width=3.15in]{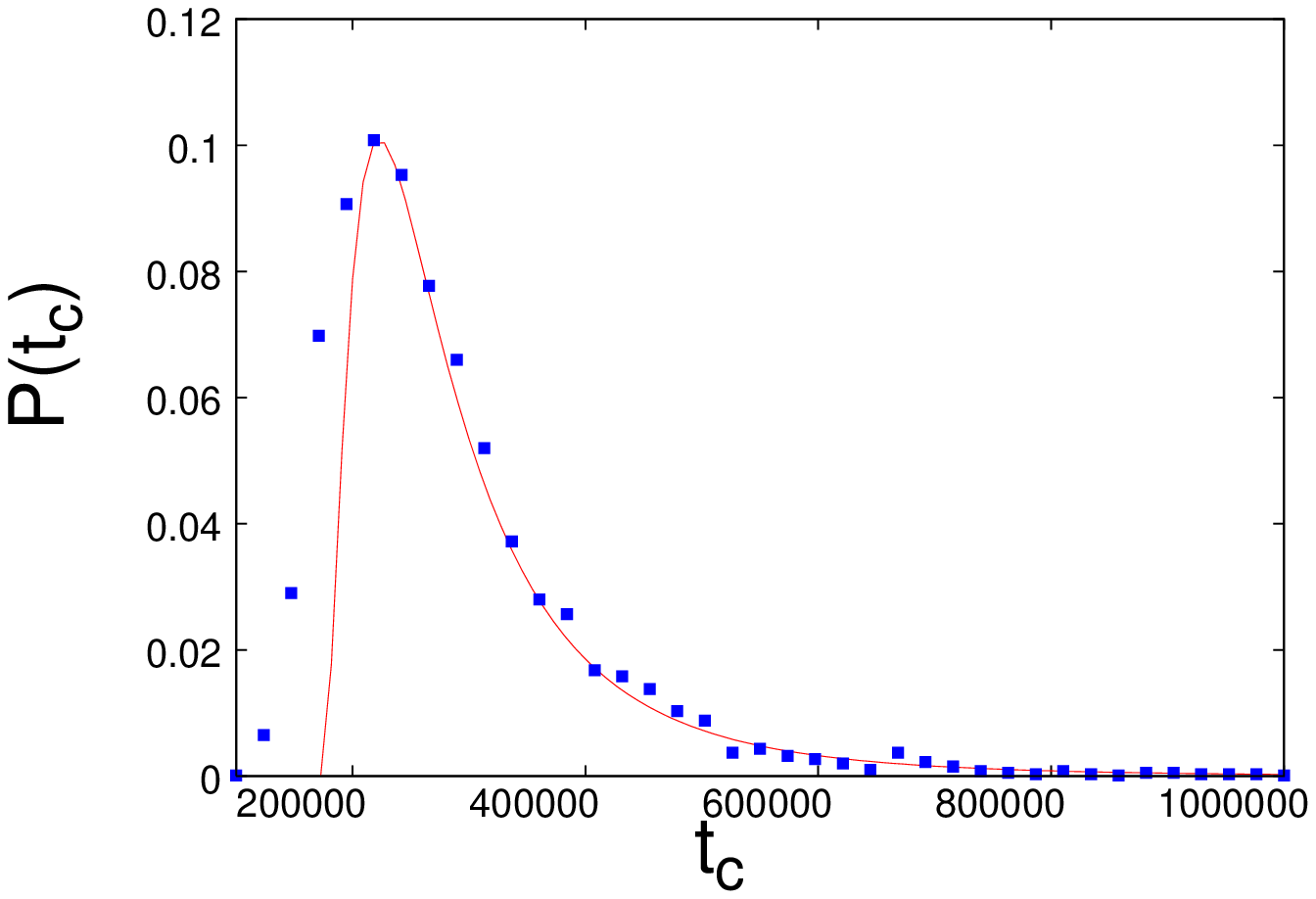} \\
\mbox{(a)} & \mbox{(b)}  
\end{array}$ 
\caption[Regularization process for the scalefree tree and transient times]{Regularization process for the scalefree tree. Number of periodic nodes in function of time for a single initial condition and various $\mu$-values in (a), distribution of transient times $t_c$ averaged over many initial conditions, with log-normal fit for $\mu=0.012$.}  \label{fig-regularization}
\end{center}
\end{figure}

\textbf{Dynamical Regularization.} For the case of 4-star all the nodes achieve periodic orbits simultaneously. However, in the case of tree with 1000 nodes, periodic motion firstly arises on certain nodes and then successively spreads with time-evolution, eventually reaching every network node.  The increase of fraction of tree nodes having periodic orbits allows a direct examination of the regularization process. We illustrate this in Fig.\,\ref{fig-regularization}a 
where we show the increase of number of tree nodes exhibiting periodic orbits in function of time-steps, for few values of coupling strengths (following \cite{ja-lncs-2}). As observed already, for stronger inter-node interaction, the regularization of the whole tree proceeds much faster, and quickly reaches the final steady state with all the nodes periodic. Interestingly, for some coupling strengths the process of dynamical regularization never (or extremely slowly) reaches the entire tree (cf. $\mu=0.003$ in Fig.\,\ref{fig-regularization}a). We also report the distribution (over many initial conditions) of transient times needed for the network to settle in the steady state for a fixed coupling strength of $\mu=0.012$ in Fig.\,\ref{fig-regularization}b.

\begin{figure}[!hbt]
\begin{center}
$\begin{array}{cc}
\includegraphics[height=2.5in,width=2.8in]{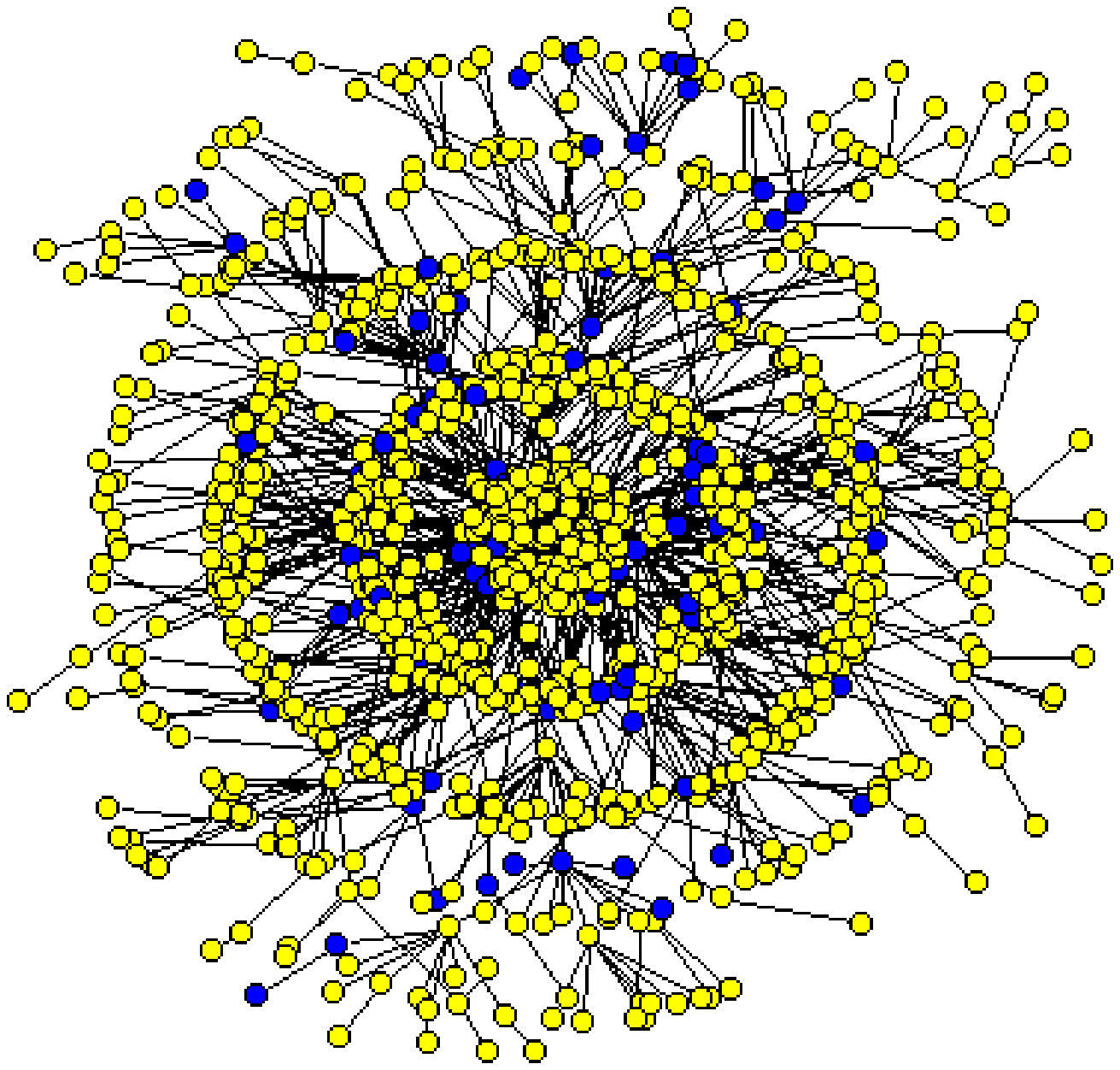} & 
\includegraphics[height=2.5in,width=2.8in]{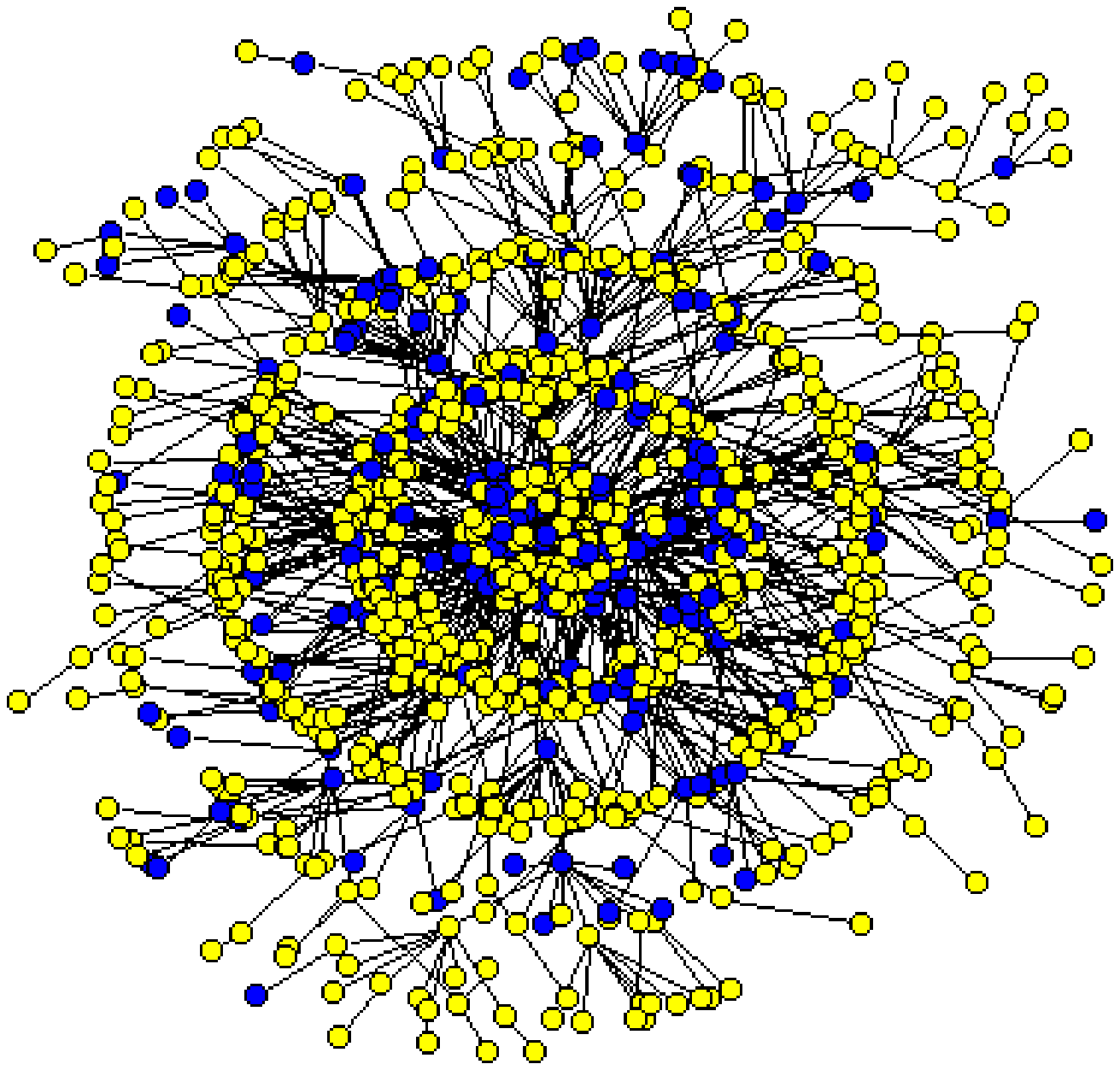} \\
\includegraphics[height=2.5in,width=2.8in]{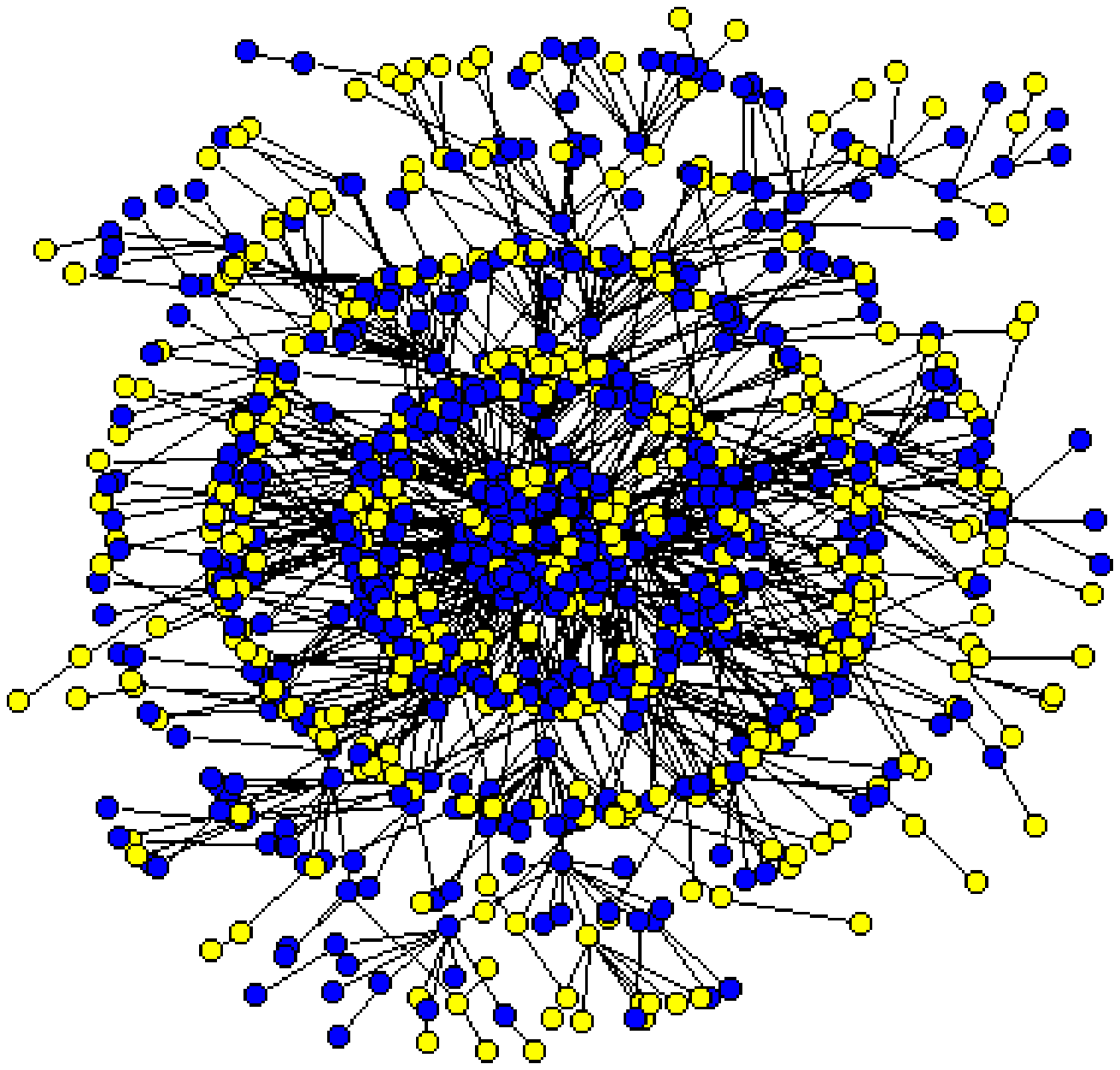} & 
\includegraphics[height=2.5in,width=2.8in]{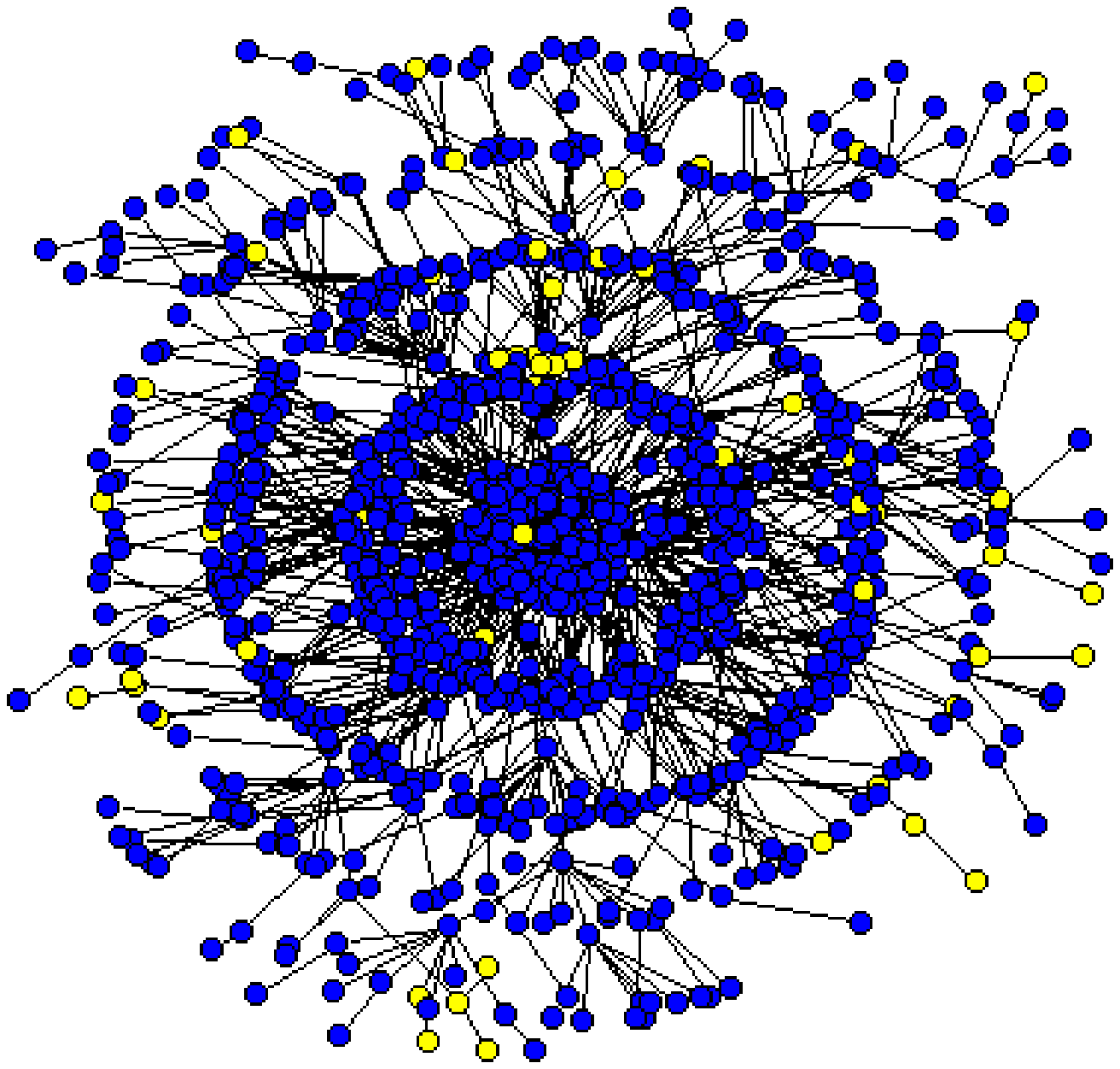}   
\end{array}$ 
\caption[Dynamical regularization process of the tree through graphical network representation]{Dynamical regularization process of the tree for $\mu=0.001$: the periodic nodes (blue) and non-periodic nodes (yellow) at different stages of CCM on networks time-evolution for a single initial condition -- after $2 \times 10^5$, $4 \times 10^5$,  $10^6$ and $3 \times 10^6$ iterations.}  \label{fig-pajek-regularization}
\end{center}
\end{figure}

It is also instructive to consider the regularization process through the network structure: in Fig.\,\ref{fig-pajek-regularization} we show four stages in tree's time-evolution by coloring periodic and non-periodic nodes differently at each stage. Outer less connected nodes seem to be somewhat faster in acquiring periodic orbits; the regularization process also seems to follow the tree's structure by spreading from one regularized node to another. Nodes neighboring periodic nodes do seem to reach regularity somewhat more easily (as they appear to be inter-connected among them in all the pictures).
However, other outer nodes seem to remain chaotic despite neighboring a regular node for a long time, and finally reach regularity together with some highly connected central nodes.

\section{Non-periodic Orbits and Identification of Dynamical Regions}

Periodic orbits are present in the collective dynamics of CCM system Eq.(\ref{main-equation}) at all non-zero coupling values. The entire network either develops periodic orbits on all  nodes, or on none of them: depending on the $\mu$-value, there is a fraction of initial conditions yielding periodic orbits and the remaining fraction that never leads to periodic dynamics. This pattern is universal for all the studied network structures and it will be employed to define the dynamical regions for the emergent dynamics of our system of CCM. We will define the dynamical regions by considering what kind of emergent motion corresponds to those initial conditions that does not lead to periodic orbits.

We define a periodic orbit of the node $[i]$ by looking at its individual emergent orbit, as defined in Eq.(\ref{eo}), for all the network nodes. We search for $t_1>t_0$ such that 
\begin{equation}
\dfrac{|x[i]_{t_0} - x[i]_{t_1}|}{|x[i]_{t_0}|} < \delta \;\;\;\; \mbox{and} \;\;\;\; 
\dfrac{|y[i]_{t_0} - y[i]_{t_1}|}{|y[i]_{t_0}|} < \delta, \label{periodicorbit}
\end{equation}
where $t_0$ denotes the end of transients and $\delta$ accounts for numerical errors. In the simulations we generally use $\delta=10^{-8}$. If the value $t_1$ is found within some considered range of iterations the orbit is defined to be periodic, otherwise the orbit is assumed non-periodic (of course, the results may depend on $\delta$-value and the range of iterations $t_1>t_0$). The difference 
\begin{equation}
 \pi[i] = t_1 - t_0 \label{period}
\end{equation}
is defined to be the \textit{period} of the periodic orbit for the node $[i]$.

The parameter we use to distinguish between the dynamical regions is the \textit{fraction of non-periodic orbits} $fr(np)$, that we examine in function of the coupling strength $\mu$. We obtain the value of this fraction by averaging over initial conditions and the nodes of the network. The value of $fr(np)$ reports how many nodes/initial conditions for a given $\mu$-value have not developed periodic orbits after transients. Due to the finite length of transients, some initial conditions for the tree might only lead to periodic orbits on a part of the tree (cf. Fig.\,\ref{fig-pajek-regularization}). Fully periodic dynamics on network, given by $fr(np)=0$ can be understood as an analog of synchronization, occurring in other systems of CCM as described in the Chapter \ref{Introduction}.

In Fig.\,\ref{fig-nonperiodic} we investigate the fraction of non-periodic orbits $fr(np)$ for the [0,0.08] range of $\mu$-values, comparing the tree with the 4-star. After initial stabilization to $fr(np)=0$ occurring at $\mu \cong 0.012$, both systems destabilize again, but with the portions of non-periodic orbits that vary with ranges of $\mu$-values: 
\begin{figure}[!hbt]
\begin{center}
\includegraphics[height=3.8in,width=5.15in]{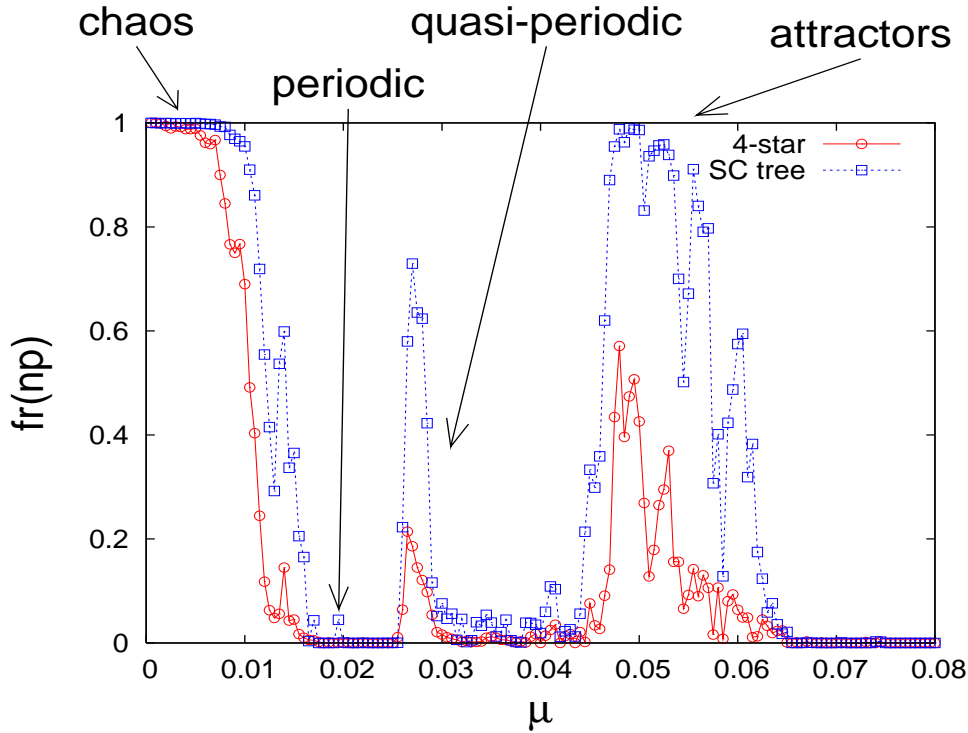}
\caption[The fraction of non-periodic orbits in function of the coupling strength for tree and 4-star, with identification of dynamical regions]{The fraction of non-periodic orbits in function of the coupling strength $\mu$ for tree and 4-star, averaged over 10 and 1000 initial conditions, respectively, with periods up to $\pi=10000$ considered. Four main dynamical regions are roughly indicated through their respective peaks and ranges of $\mu$-values. We will refer to this plot as for definition of dynamical regions.} \label{fig-nonperiodic}
\end{center}
\end{figure}
we identify these ranges as \textit{dynamical regions}, according to the scheme indicated in the picture. Note that these $\mu$-regions of destabilization seem to be (at least qualitatively) common to all the investigated structures, which already suggests a dynamical relationship between them. In particular, we focus on four main dynamical regions, defined by their ranges of $\mu$-values and characterized by specific collective behaviors emerging for the case of initial conditions leading to non-periodic orbits. 
\begin{itemize}
 \item \textit{Chaotic Orbits Region} at small coupling values $0 < \mu \lesssim 0.012$ is characterized by chaotic motion of all nodes 
(cf. Figs.\,\ref{fig-orbitsexamples}a\,\&\,d) that eventually evolve into periodic orbits after a given transient time of diffusive behavior (the region is somewhat shorter for the case of 4-star). This definition is given in accordance with the transient of $10^5$ iterations.
 \item \textit{Periodic Orbits Region} between $0.012 \lesssim \mu \lesssim 0.026$ exhibits periodic orbits of all the nodes at all topologies 
(cf. Figs.\,\ref{fig-orbitsexamples}b\,\&\,e), with no other types of emergent motion (after transients that are long enough). The details regarding structure and phase space properties of periodic orbits will be provided in the Sections to follow. 
 \item \textit{Quasi-periodic Orbits Region} between $0.026 \lesssim \mu \lesssim 0.04$ is identified by quasi-periodic orbits 
(cf. Figs.\,\ref{fig-additionalorbits}b\,\&\,e) emerging for some initial conditions for both structures (the remaining part of 
initial conditions leads to periodic orbits). Quasi-periodic orbits are displayed by all or none of the nodes, depending on the initial conditions, and represent final steady state of the coupled dynamics. They will be addressed in more detail in the Chapter \ref{Stability of Network Dynamics}.
 \item \textit{Strange Attractors Region}  for large coupling strengths $0.04 \lesssim \mu \lesssim 0.065$ is characterized by a variety of strange 
attractors appearing in the dynamics of 4-star for some fraction of initial conditions (cf. Figs.\,\ref{fig-orbitsexamples}c\,\&\,f). For the tree, this region is characterized by localized noisy emergent orbits, that despite not being periodic, display a high degree of self-organization. This region will
 be studied in much more detail later.
\end{itemize}
The dynamics with coupling strengths above $\mu=0.065$ is again characterized by periodic orbits for all initial conditions, and will be not investigated further here. Despite that precise $\mu$-range of dynamical regions depends on the topology in question, the qualitative analogy between 4-star and tree is to be emphasized, suggesting the expected dynamical relationship between these two topologies that derives from their network structures. This relationship will be further examined in terms of statistical properties of emergent motion.

\begin{figure}[!hbt]
\begin{center}
$\begin{array}{cc}
\includegraphics[height=2.55in,width=3.2in]{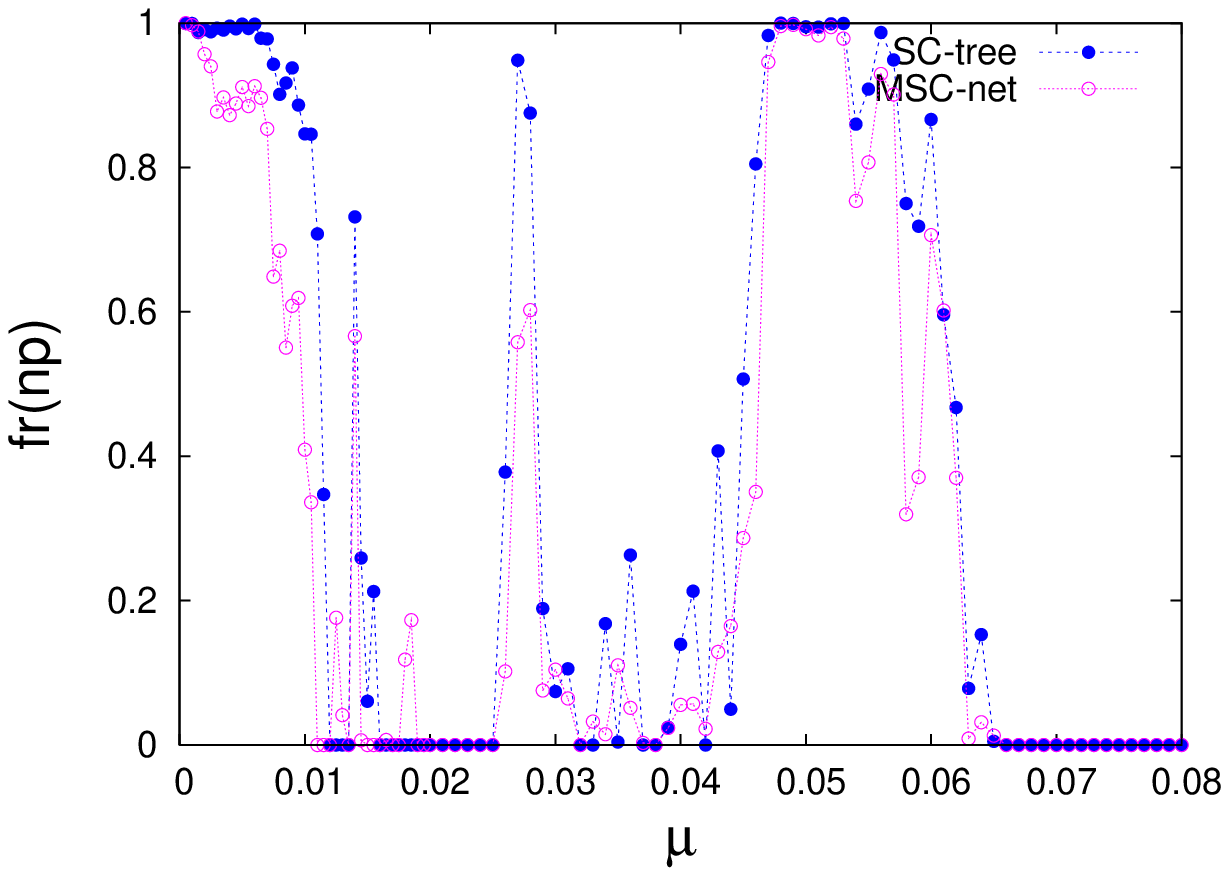} & 
\includegraphics[height=2.55in,width=3.2in]{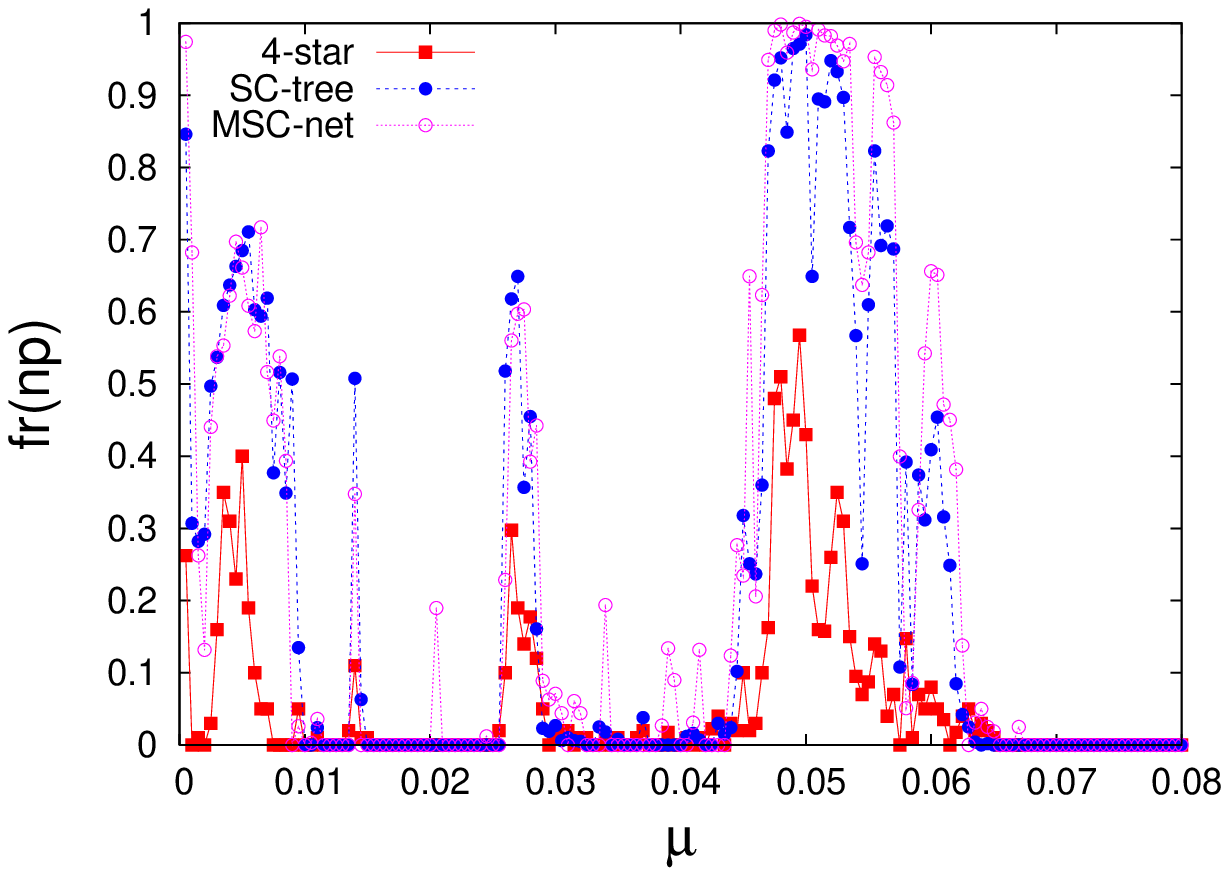} \\
\mbox{(a)} & \mbox{(b)}
\end{array}$ 
\caption[Fractions of non-periodic orbits in function of the coupling strength for much longer transient]{Fraction of non-periodic orbits in function of coupling strength for a single initial condition with periods up to $\pi=10000$ considered. Comparison of scalefree tree and modular scalefree network in (a); comparison of all structures for the transient of $2 \times 10^6$ iterations in (b).} \label{fig-nonperiodic-comparisons}
\end{center}
\end{figure}

The robustness of dynamical regions with respect to topology is further studied in Fig.\,\ref{fig-nonperiodic-comparisons}a where we compare the tree's profile from Fig.\,\ref{fig-nonperiodic} with the analogous profile for the modular scalefree network shown in Fig.\,\ref{fig-modularnetwork}. 
Note that the same dynamical regions can be equivalently defined here as well, despite modular network being structured differently. Furthermore, in Fig.\,\ref{fig-nonperiodic-comparisons}b we compare the profile for all the mentioned networks, computed for a much longer transient of $t_0=2 \times 10^6$ iterations. The profiles seem to maintain the same dynamical regions, except for the chaotic region which appears shorter due to more initial conditions reaching final periodic steady states due to longer transient.

As opposed to classical studies of 1D CCM mentioned earlier (\cite{lind,amritkar,zahera}) our network of CCM presents more intricate dynamical behaviors with respect to the various values of coupling strength. This is primarily given by the dynamical regions, which are not typical for 1D CCM, and so far studied examples of 2D CCM. In the remainder of this Chapter we will statistically explore the nature of the emergent motion in order to gain better characterization of all dynamical regions. Since periodic orbits can be easily quantified, we will persist in differing among the dynamical regions by referring to the properties of the non-periodic motion in function of the coupling strength $\mu$.

\section{Dynamical Clustering of Emergent Orbits} 

The nature of periodic orbits of system of CCM Eq.(\ref{main-equation}) is always given by oscillations between two groups of phase space locations, one on the left and the other on the right half of $x$-axis, just as in Fig.\,\ref{fig-orbitsexamples}b. This is to say the period of each periodic orbit is (at least) a multiple of two (four, in the case of Fig.\,\ref{fig-orbitsexamples}b). However, periodic orbits of more distinct nodes can sometimes approximately share the same phase space locations (both left and right), without necessarily synchronizing or having the same period. This occurs for many initial conditions, $\mu$-values and all topologies: some nodes group their final periodic motion by oscillating between same two phase space locations. All the nodes (orbits) that share the mentioned phase space locations for some initial conditions will be called a \textit{cluster of nodes (orbits)}. 

\begin{figure}[!hbt]
\begin{center}
$\begin{array}{cc}
\includegraphics[height=2.5in,width=2.75in]{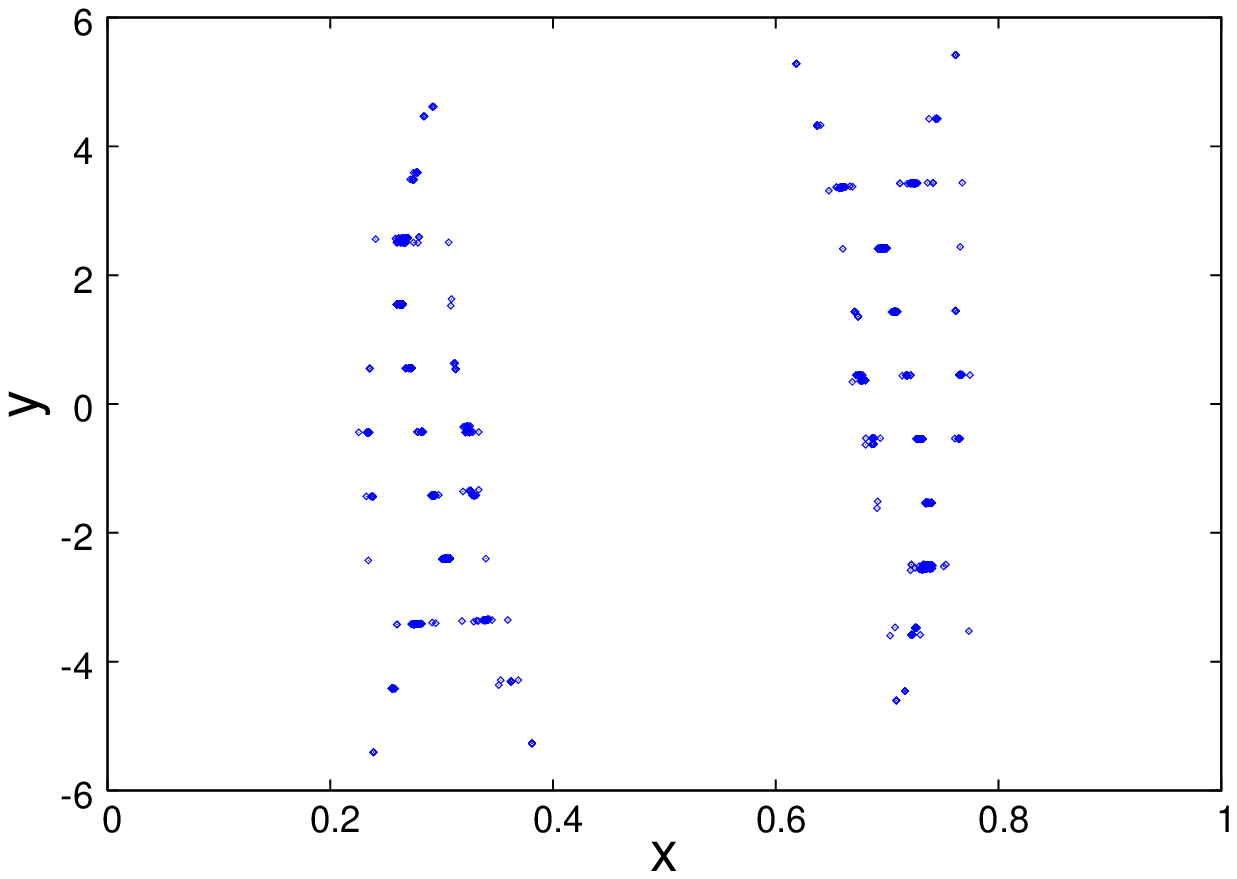} & 
\includegraphics[height=2.5in,width=3.5in]{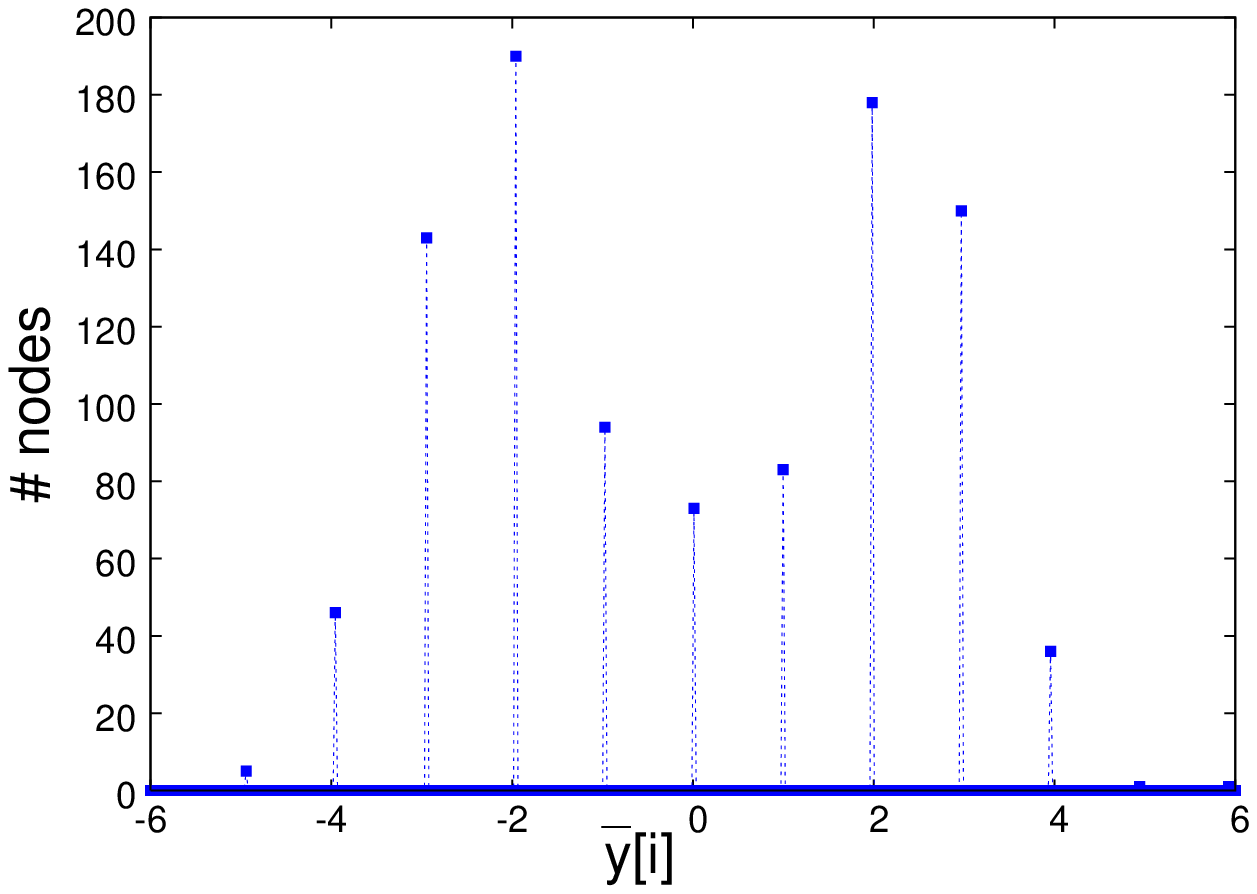} \\[0.1cm]
\mbox{(a)} & \mbox{(b)} 
\end{array}$ 
\caption[Clustering of the final dynamical tree state at $\mu=0.012$]{Final steady state of the tree for $\mu=0.012$ after a transient of $5 \times 10^6$ iterations. Ten iterations on the same phase space for all nodes in (a), and distribution of $\bar{y}[i]$-values for all nodes computed over 1000 iterations and averaged over many initial conditions in (b).}  \label{fig-clustering}
\end{center}
\end{figure}

The global phase space organization of periodic orbits in our CCM on networks is characterized by each orbit/node belonging to one of the clusters (this holds for both  periodic dynamical region and other dynamical regions, in case initial conditions lead to periodic orbits). To illustrate this graphically, we show ten iterations of all the tree nodes put on the same 2D phase space in Fig.\,\ref{fig-clustering}a, for the case of tree with $\mu=0.012$.
Oscillations are organized into 11 clusters, with each cluster being composed of two phase space locations, similarly to the orbit in Fig.\,\ref{fig-orbitsexamples}b. Each of the 1000 tree nodes belongs to one of the clusters, as it oscillates between one location on the left and one on the right side of the phase space, separated by a distance in $y$-coordinate of about 1. While the $x$-coordinates of all the clusters' phase space locations nearly overlap, the $y$-coordinates are equally spaced on the $y$-axis.

This \textit{dynamical clustering} of nodes can be conveniently quantified by referring to each node's time-averaged orbit (which is basically a phase space point) defined in Eq.(\ref{taeo}) and denoted by $(\bar{x}[i],\bar{y}[i])$. It is easy to recognize that time averages of orbits/nodes will be clustered, as the nodes sharing a cluster will have similar time-averaged orbits. Specifically, we shall examine the values of $\bar{y}[i]$, as through them the clusters can be easily defined, given the nearly uniform separation of clusters in $y$-coordinate. We report in Fig.\,\ref{fig-clustering}b the distribution of $\bar{y}[i]$-values corresponding to the situation in Fig.\,\ref{fig-clustering}a (but averaged over many initial conditions): well-defined peaks are in a clear analogy with Fig.\,\ref{fig-clustering}a, suggesting a phase space symmetric organization of nodes into clusters with respect to $y$-axis.

\begin{figure}[!hbt]
\begin{center}
$\begin{array}{cc}
\includegraphics[height=2.5in,width=3.2in]{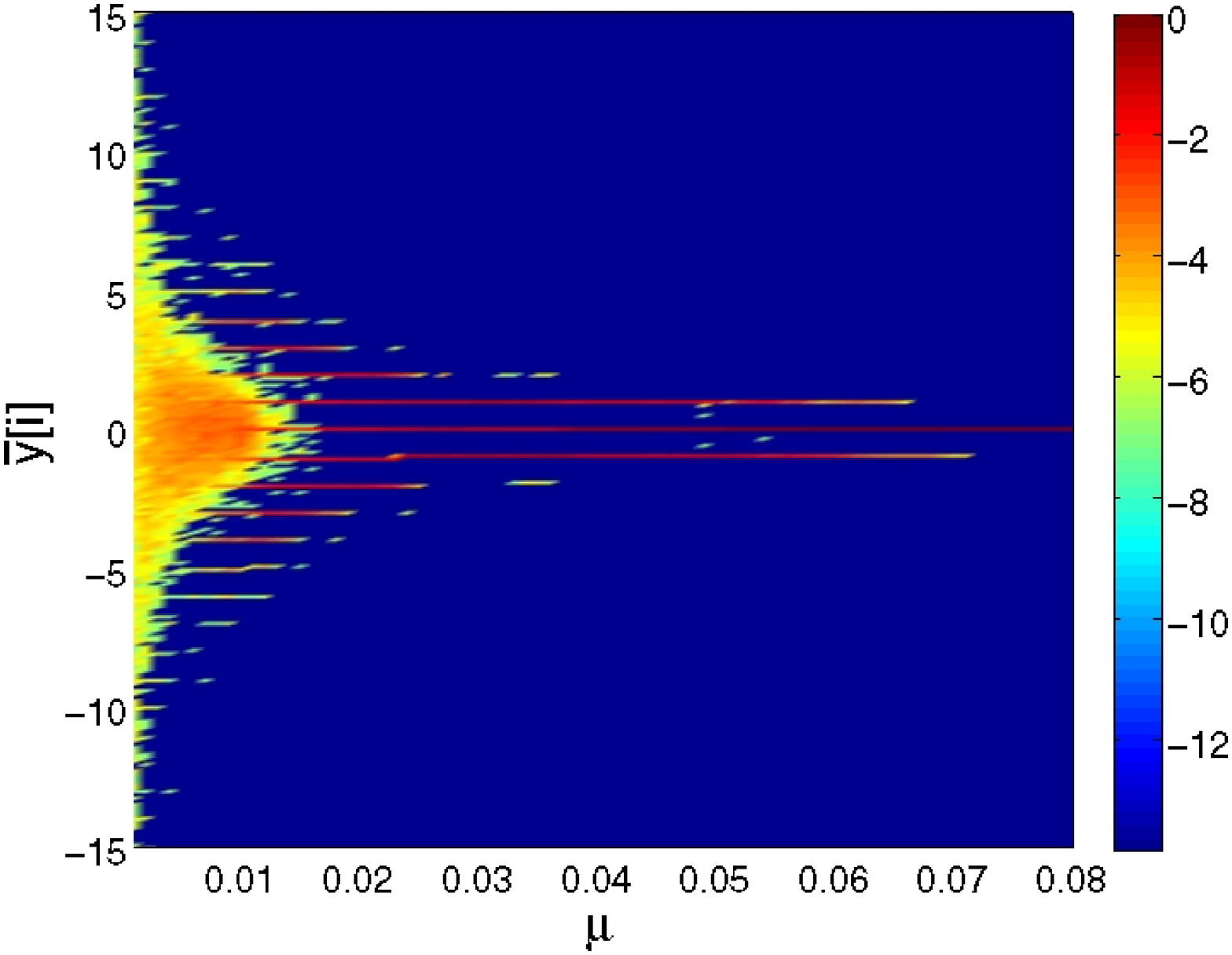} & 
\includegraphics[height=2.5in,width=3.2in]{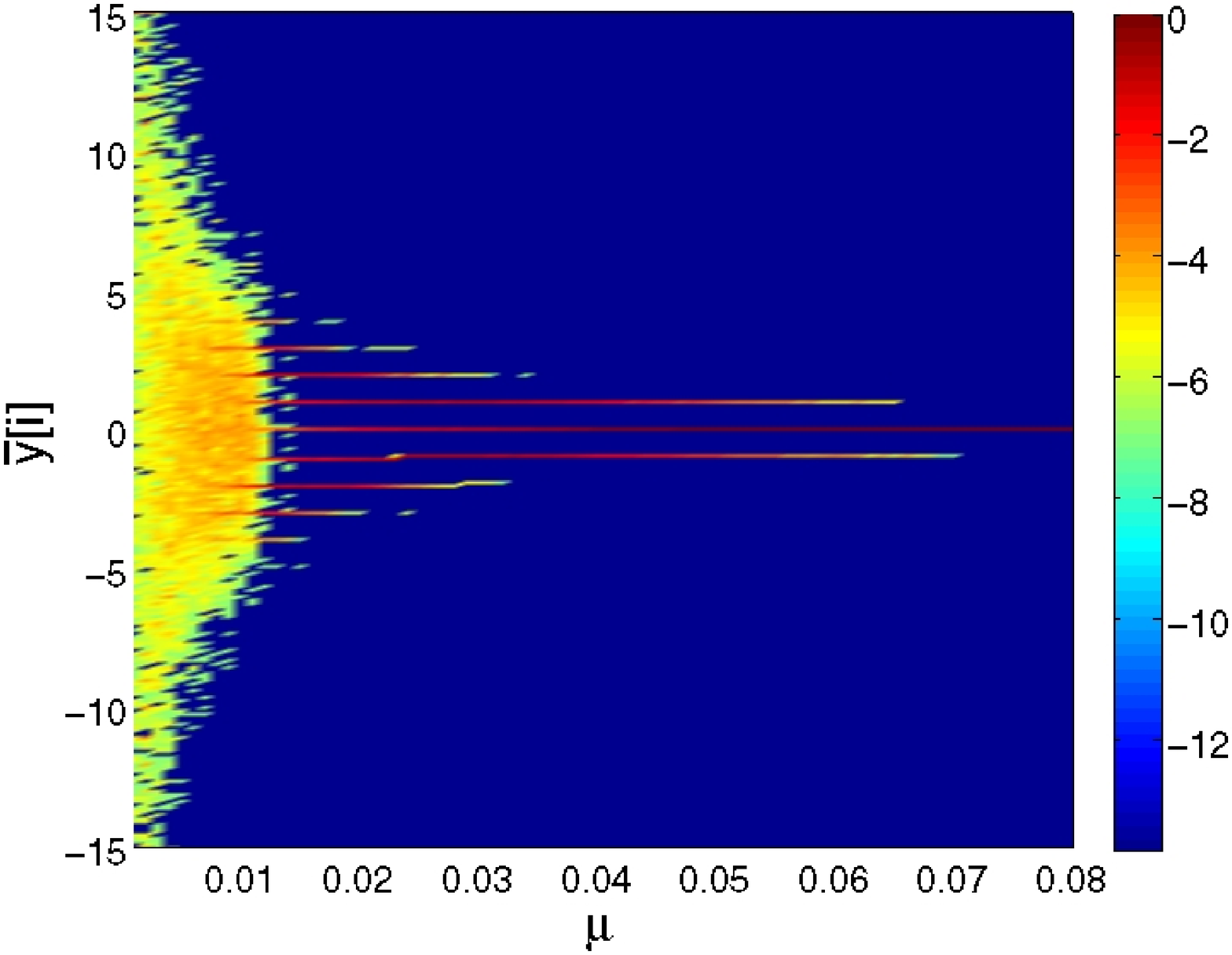} \\
\mbox{(a)} & \mbox{(b)} \\
\includegraphics[height=2.5in,width=3.2in]{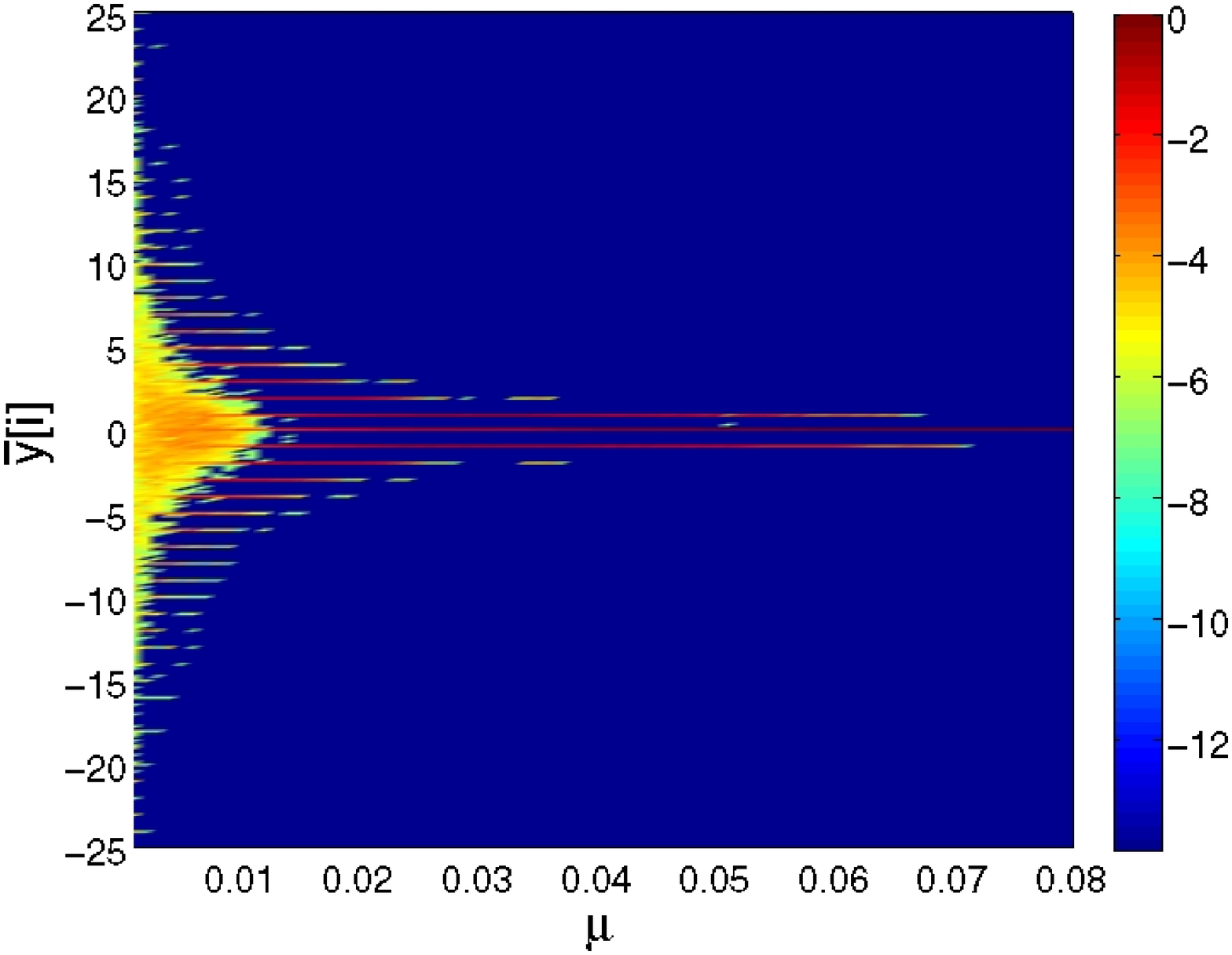} &
\includegraphics[height=2.5in,width=3.2in]{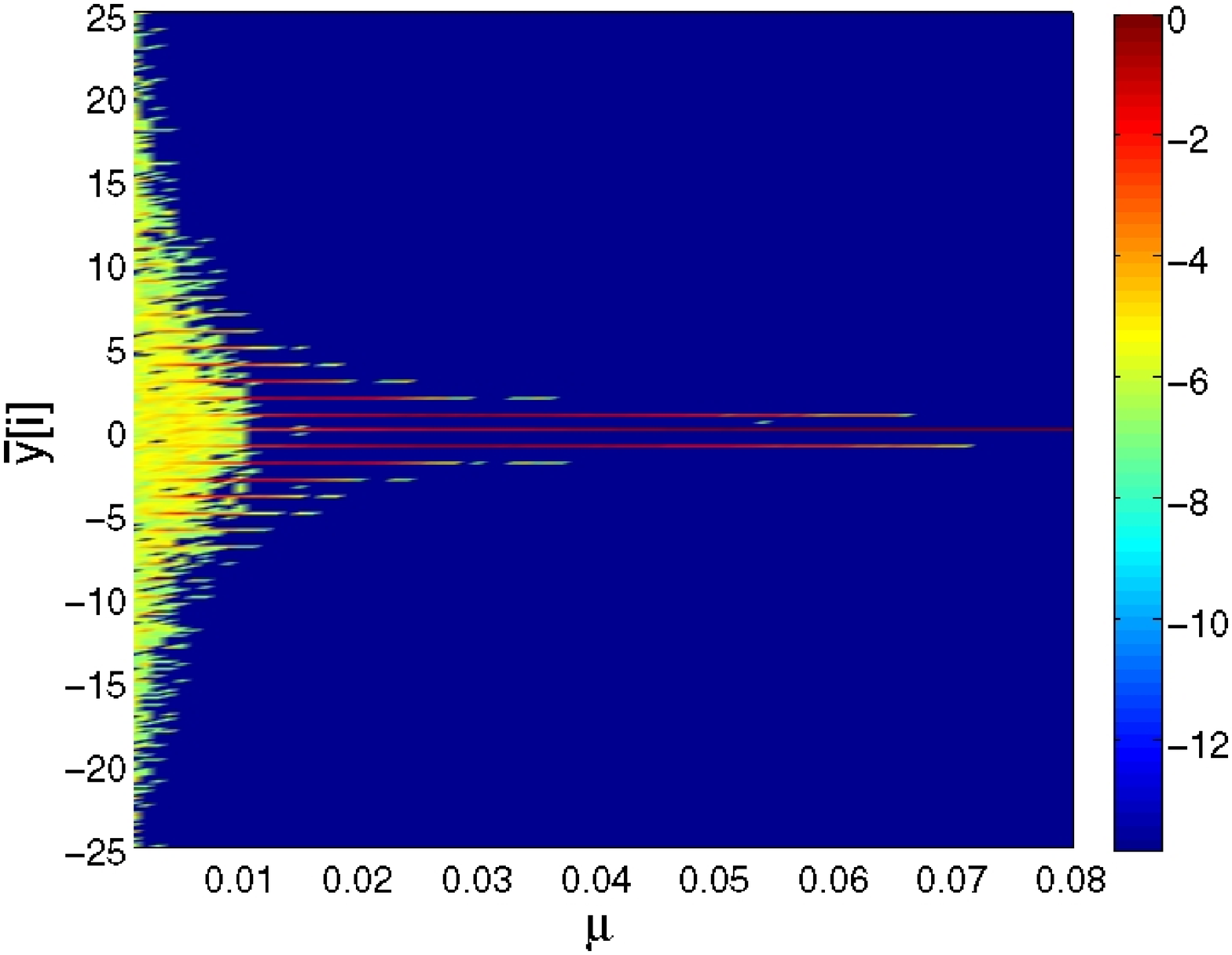} \\
\mbox{(c)} & \mbox{(d)} 
\end{array}$ 
\caption[2D color histogram showing the $\bar{y}$-values in function of the coupling $\mu$ for various topologies]{2D color histogram showing the $\bar{y}[i]$-values in function of the coupling $\mu$. 4-star's branch node in (a) and a clique's node in (b) averaged over many initial conditions. All the nodes for a single initial condition for tree in (c), and the modular network in (d).}  \label{fig-ybar-comparisons}
\end{center}
\end{figure}

We also examine the distribution of $\bar{y}[i]$-values for the whole range of coupling strengths, comparing all the examined structures. The results are shown in Fig.\,\ref{fig-ybar-comparisons} where for each $\mu$-value we report the histogram of $\bar{y}[i]$-values on the log-scale, thus making a 2D color histogram of values. 
The difference between the profiles for a 4-star's node (Fig.\,\ref{fig-ybar-comparisons}b) and the profile for a clique's node (Fig.\,\ref{fig-ybar-comparisons}a) indicates outer less connected nodes have more tendency to organize into clusters, even for smaller values of coupling strengths. This is also visible by comparing the profiles of scalefree tree (Fig.\,\ref{fig-ybar-comparisons}c) and modular network (Fig.\,\ref{fig-ybar-comparisons}d) -- tree's structure more rapidly develops clustered organization of nodes at smaller couplings. The clear structural similarity of profiles for 4-star's branch node and tree derives from tree having many branch nodes (nodes having only one link to the rest of the network), whereas modular network (despite its scalefree degree distribution) contains many cliques. The similarity between profiles of 4-star and tree, together with the similarity of profiles for 4-clique and modular network, clearly suggest a dynamics--topology relationship between a large network structure and its dynamical motifs.

The clustering appears around the same coupling $\mu$-value where the regularity transition occurs ($\mu=0.012$ with the transient of $10^{5}$ iterations), thus allowing a view of the regularization process from the clustering prospective. Furthermore, as it appears from the profiles in Fig.\,\ref{fig-ybar-comparisons}, the attractor and quasi-periodic dynamical regions seem to produce only faint departures from the very organized cluster profiles (visible for 4-star with $\mu \sim 0.05$). This is because all emergent motions (almost) entirely respect the observed dynamical clustering in terms of time-averaged orbits; nevertheless, their precise phase space locations may differ, as clear for the example of strange attractor in Fig.\,\ref{fig-orbitsexamples}c and quasi-periodic orbits in Figs.\,\ref{fig-additionalorbits}a\,\&\,b.

\begin{figure}[!hbt]
\begin{center}
\includegraphics[height=2.83in,width=4.1in]{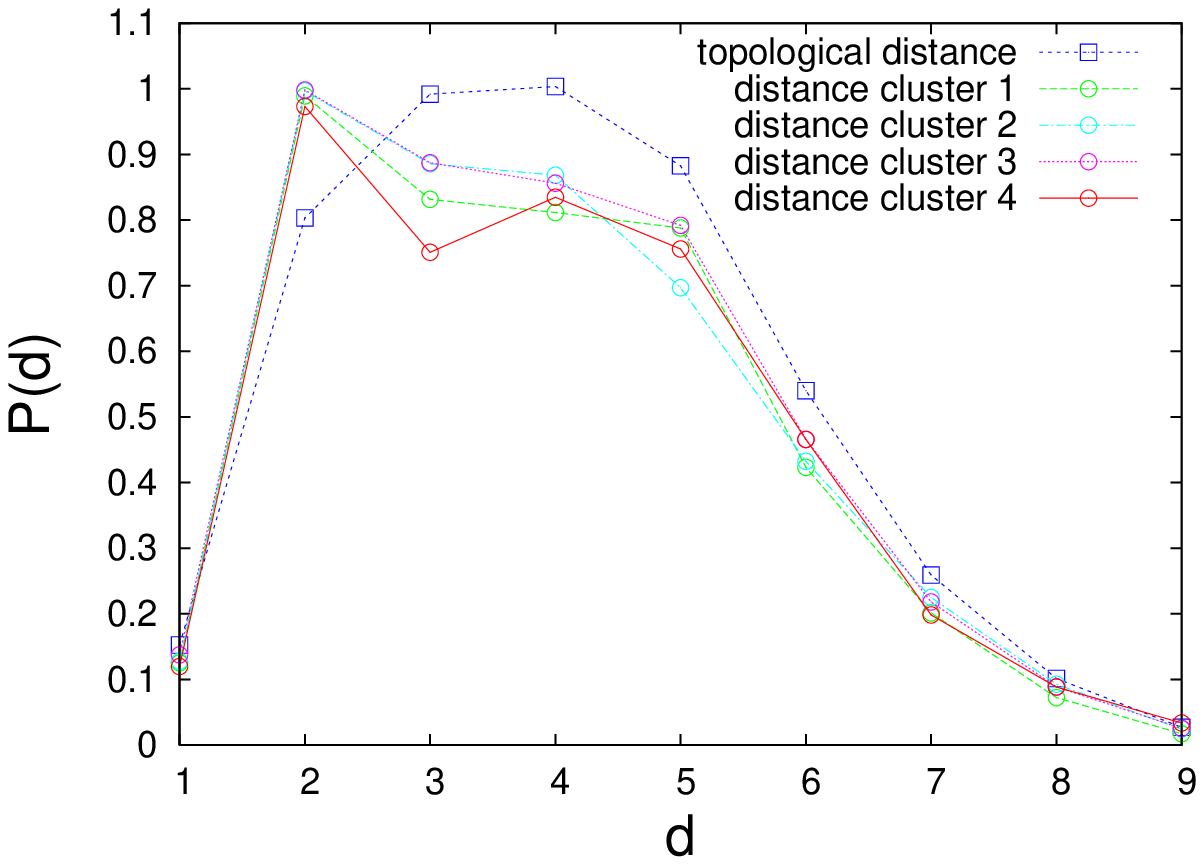}
\caption[Distributions of the topological network-distances for the tree vs. the network-distances between the nodes sharing a cluster at $\mu=0.017$]{Distributions of the topological network-distances for the tree, and the network-distances between the nodes sharing a cluster, for four different clusters. A single final state of tree with  $\mu=0.017$ was considered.} \label{fig-clusterdistances}
\end{center}
\end{figure}

\begin{figure}[!hbt]
\begin{center}
\includegraphics[height=3.87in,width=4.3in]{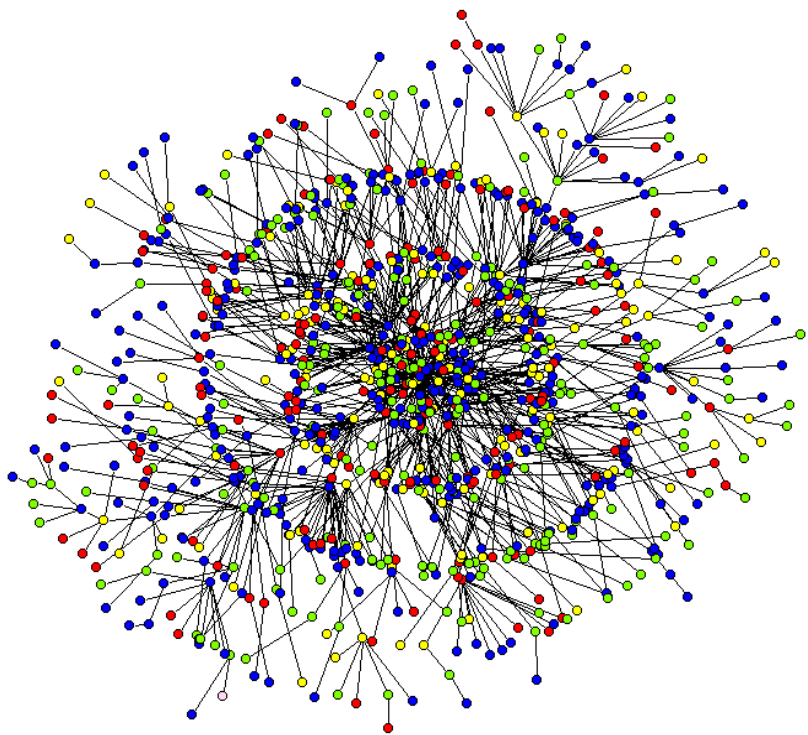}
\caption[Graphical representation of the tree with nodes sharing the same cluster marked with the same color, with $\mu=0.02$]{The network locations of nodes sharing the same cluster (colored with the same color) for a single initial condition on the tree with $\mu=0.02$.} \label{fig-clusternetwork}
\end{center}
\end{figure}

We furthermore study the distribution of network-distances of the nodes sharing a cluster for a single final state on the tree, which is reported in Fig.\,\ref{fig-clusterdistances} together with distribution of tree's topological distances \cite{ja-lncs-1}. The prevailing distance of 2-links between the same-cluster nodes seems to be common for all the clusters, regardless of their respective $\bar{y}[i]$-values. 
Interestingly, it is at smaller network-distances where the difference between same-cluster distances and topology-distances is the largest. The origin of this self-organization effect might lie in the time delay that induces local dynamical organization, that is however not present globally (as at larger network-distances the distribution follows the topological profile). Finally, in Fig.\,\ref{fig-clusternetwork} we examine the network locations of the clusters by showing the picture of the tree (cf. Fig.\,\ref{fig-sftree}) with the same-cluster nodes sharing a color. 
The network distance of 2-links is indeed often visible for all of the five considered clusters.

The dynamical clustering represents a clear collective effect present in the emergent dynamics of our CCM system, that was characterized using time-averaged orbit approach. Clustering of some sort is often found in 1D CCM, although typically in the context of synchronization \cite{amritkar}. The clustering found here refers to a "weaker" form of synchronization involving only shared phase space locations, but nevertheless indicating a high degree of dynamical order \cite{ja-lncs-1,ja-jsm}.

\section{Statistical Properties of Dynamics of CCM on Networks}

The emergent behavior of CCM Eq.(\ref{main-equation}) on networks is governed by collective effects arising as a consequence of inter-node interaction of isolated chaotic maps. In this Section we quantitatively characterize global collective effects using various statistical measures. 

\begin{figure}[!hbt]
\begin{center}
$\begin{array}{ccc}
       \includegraphics[height=1.9in,width=2.02in]{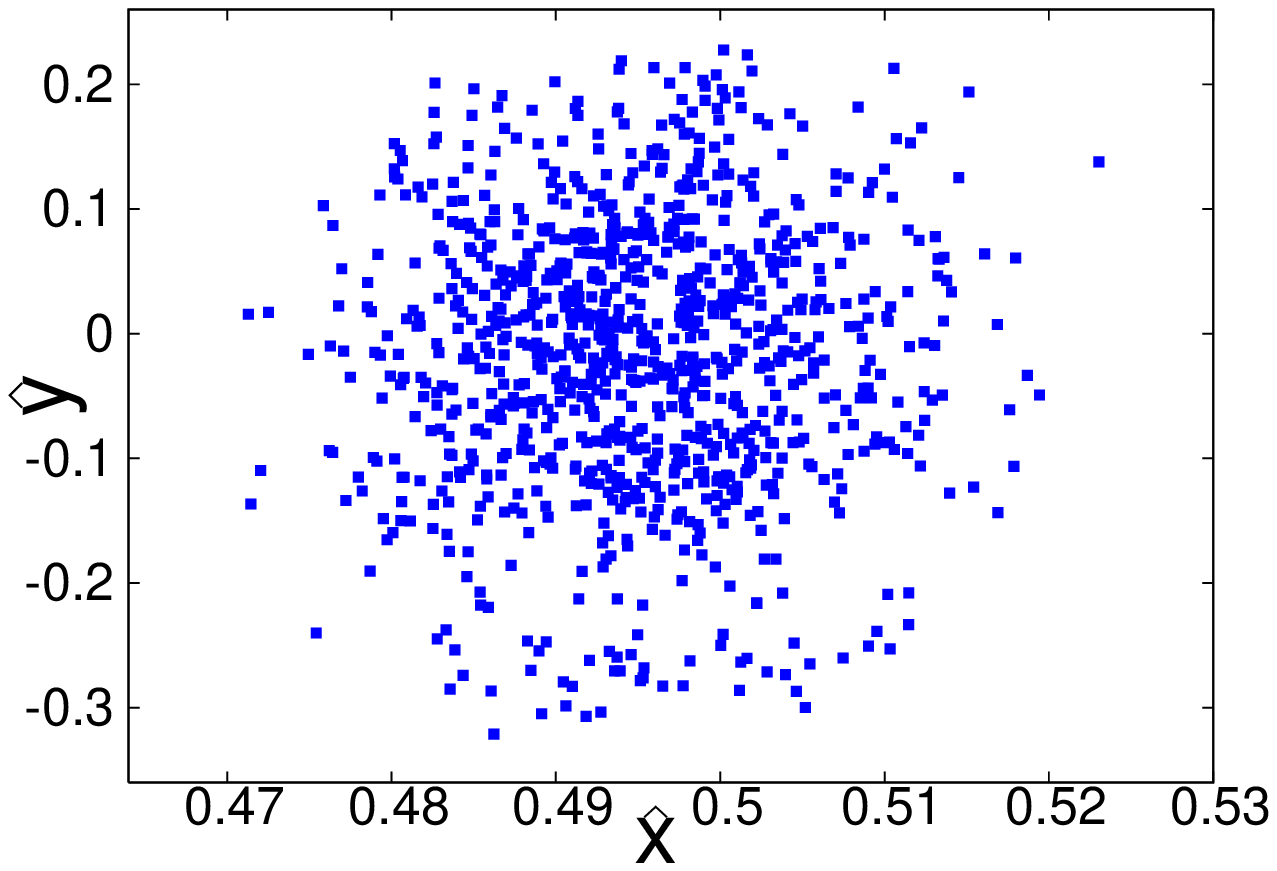} & 
       \includegraphics[height=1.9in,width=2.02in]{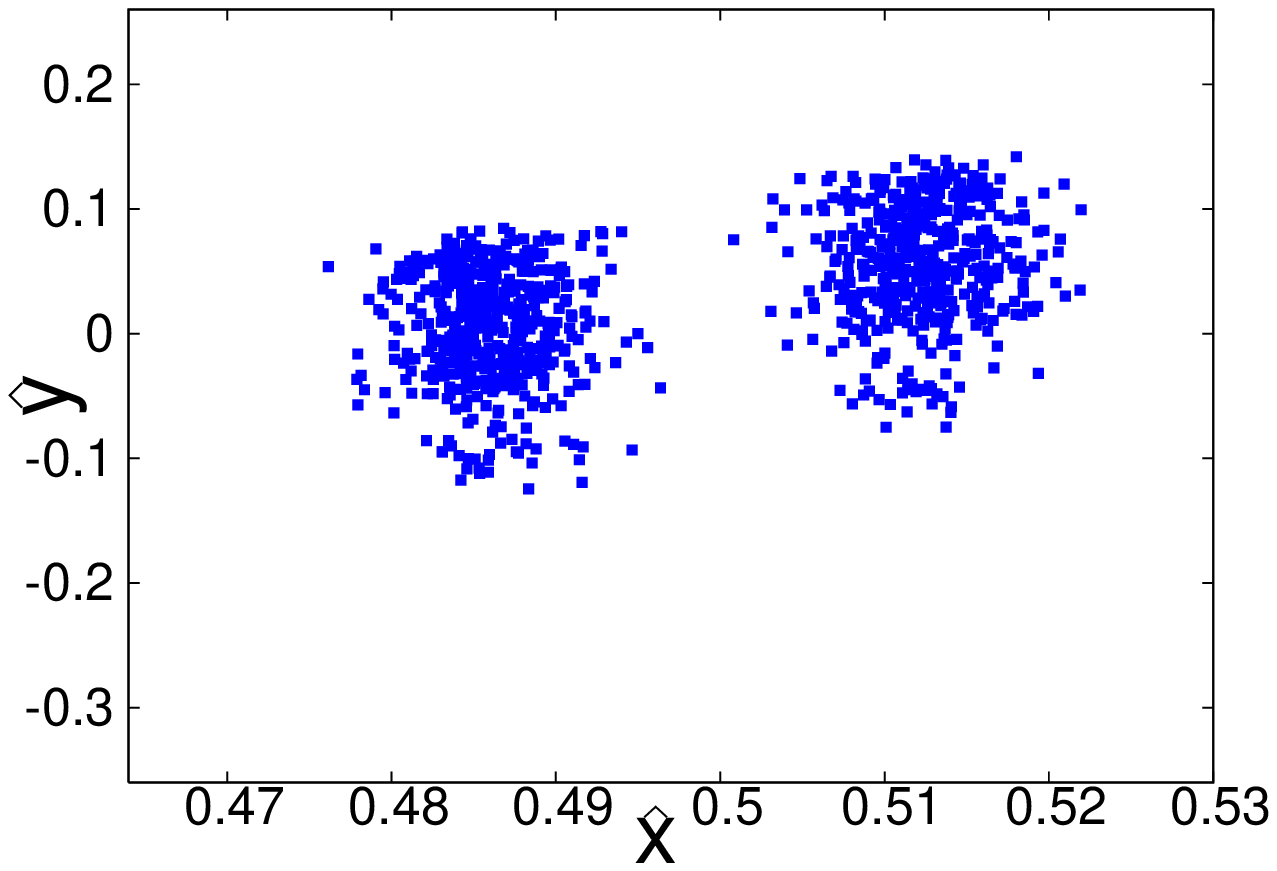} & 
       \includegraphics[height=1.9in,width=2.02in]{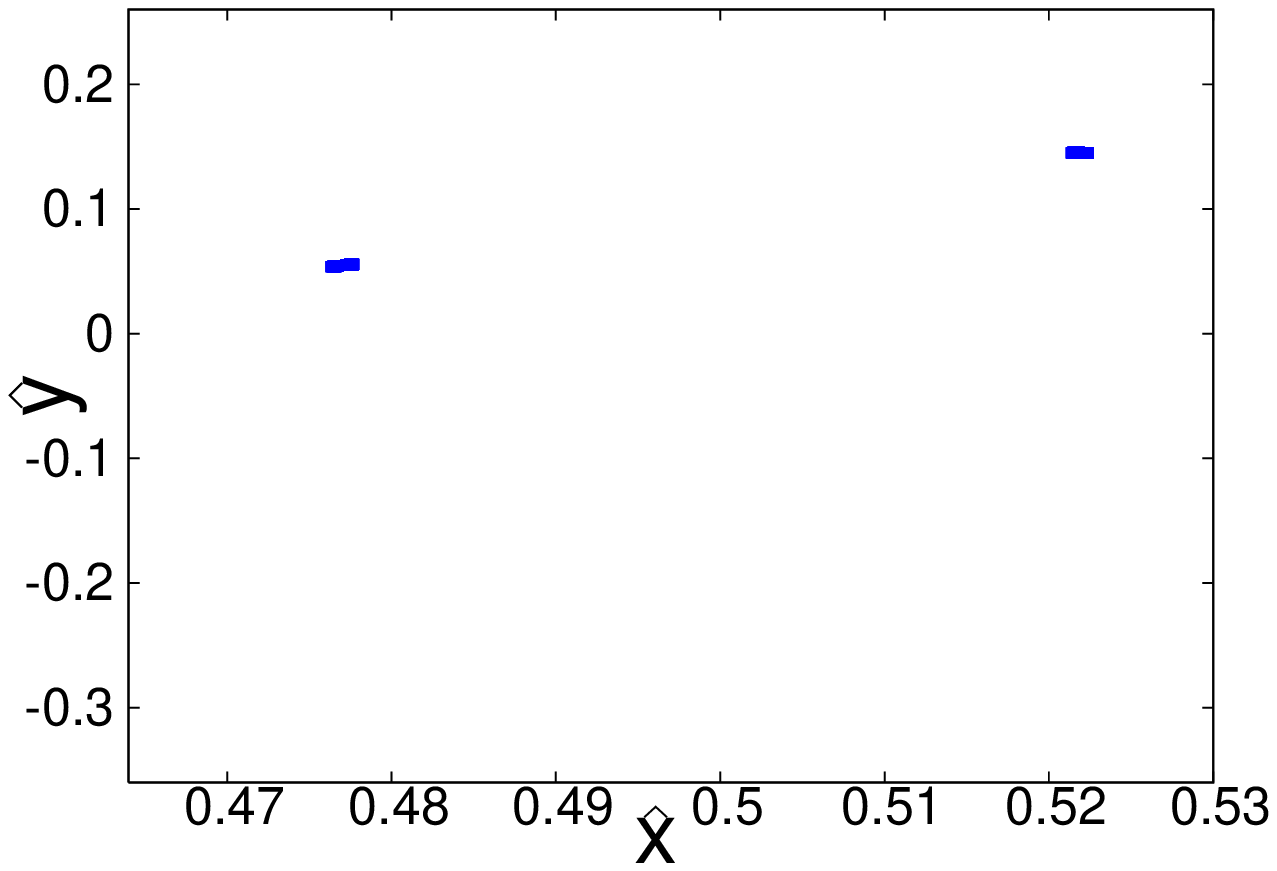} \\
       \mbox{(a)} & \mbox{(b)} & \mbox{(c)} 
\end{array}$ 
\caption[The time-evolution of the node-average orbit on the tree at $\mu=0.012$]{1000 iterations of the node-average orbit on the tree for a single initial condition at $\mu=0.012$. After 0 iterations in (a), $5 \times 10^4$ iterations in (b) and $5 \times 10^5$ iterations in (c).}
\label{fig-tree-av-mu0012}
\end{center} 
\end{figure}

In this context the node-averaged (or network-averaged) orbit defined by Eq.(\ref{naeo}) and denoted as $(\hat{x}_t,\hat{y}_t)$ can be conveniently used, as it captures the global time-evolution of the whole CCM system in a single trajectory that evolves in a single phase space. To illustrate this we show in Fig.\,\ref{fig-tree-av-mu0012} the time-evolution of our CCM system through node-averaged orbit starting from a single initial condition.
The formation of oscillatory behavior in a single cluster is visible, as expected from the known emergent cluster structure of the phase space at $\mu=0.012$ (cf. Fig.\,\ref{fig-clustering}a). The regular oscillatory behavior arises from an initial "cloud" after it splits and localizes, which leads to regularization, as captured here by the node-average orbit \cite{ja-lncs-2}. Due to large number of tree nodes, the final state of node-average orbit shown in  Fig.\,\ref{fig-tree-av-mu0012}c is very robust to the initial conditions. Because of the mentioned properties, we shall often refer to the node-average orbit in studying global aspects of our CCM on networks.\\[0.1cm]

\textbf{Return Times with respect to Phase Space Partitioning.} We examine the distributions of return times associated with phase space partitions  (sub-divisions), defined as follows: consider a partition of the phase space (e.g. a rectangular grid), and a system's specific orbit (node-average orbit for instance). Consider a single element of the phase space partition and observe the times $\Delta t$ (number of iterations) between the consecutive visits that orbit makes to this partition element. The distribution of the time intervals $\Delta t$ considered over all the partition elements for a time much longer than transients we call \textit{return times distribution with respect to a given phase space partition}. Alternatively, return times distribution can by created by considering separate node's trajectories and averaging over them (although this is numerically far more demanding), or by simply considering the distribution for an orbits of a single-node attached to the tree. The investigation of return times distributions reveals long-range correlations of the motion in question, which report the presence of self-organization effects.

\begin{figure}[!hbt]
\begin{center}
$\begin{array}{cc}
\includegraphics[height=2.55in,width=3.15in]{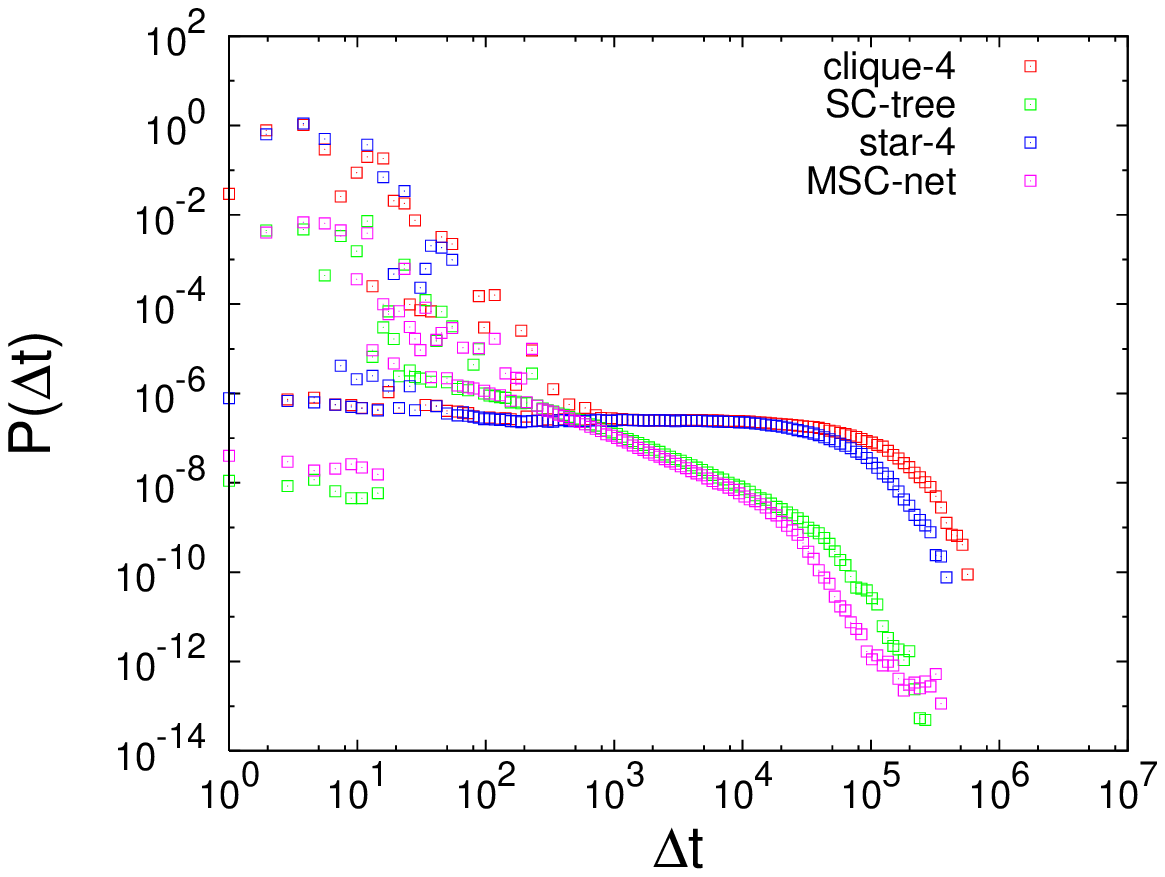} &  
\includegraphics[height=2.55in,width=3.15in]{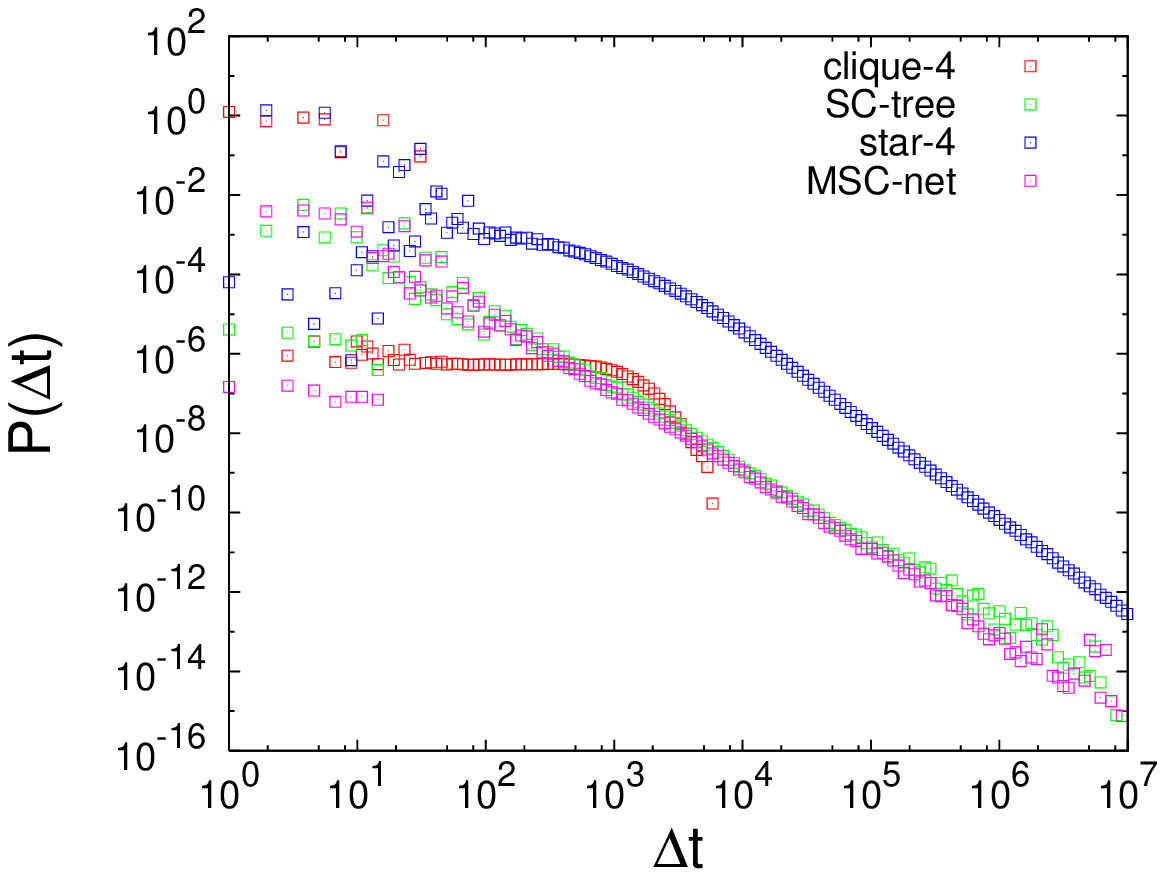}  \\
\mbox{(a)} & \mbox{(b)} 
\end{array}$ 
\caption[Return times distributions of node-averaged orbits for various topologies]{Return times distributions of node-averaged orbits for various topologies, averaged over many initial conditions. Partition in form of rectangular grid with $1000 \times 100000$ elements covering $(x,y) \in [0,1] \times [-50,50]$ was used. The case of $\mu=0.012$ in (a) and $\mu=0.051$ in (b).} \label{fig-rtcomparison}
\end{center}
\end{figure}

In Fig.\,\ref{fig-rtcomparison} we plot the return times distribution for node-averaged orbits for various topologies and for two specific coupling strength values, representing periodic motion ($\mu=0.012$) and self-organized motion in the attractor dynamical region ($\mu=0.051$). In the periodic dynamical region (Fig.\,\ref{fig-rtcomparison}a) the short return times $\Delta t \lesssim 100$ appear most frequently, which is expected as the CCM system motion exhibits periodic orbits with relatively small period values for all the structures. Bigger return times in this region for the cases of small structures show constant profiles followed by exponential decays, whereas large networks show some correlations in form of power-laws.
However, in the attractor region examined in Fig.\,\ref{fig-rtcomparison}b both tree and modular network display power-laws indicating a presence of long-range correlations in the dynamics of network of CCM at this coupling strength. Moreover, node-average orbit of 4-star also shows a similar power-law behavior, in opposition to 4-clique which exhibits fully periodic behavior similar to what seen Fig.\,\ref{fig-rtcomparison}a. Mentioned power-law distribution is a sign of anomalous diffusion in phase space, which presents a sharp contradiction with the chaotic uncoupled case. The relationship between these profiles again suggests the collective dynamics of large scalefree tree to be generated on a smaller scale of a motif that captures its topology in only four nodes: the 4-star. Also, the profiles for tree and modular network are almost overlapping for both $\mu$-values, which indicates 4-star to represent modular network's collective dynamics to some extent as well. This could be understood by observing our modular network to be constructed (cf. Chapter \ref{Coupled Maps System on Networks with Time delay}.) with many free links which become branch nodes, similar to the ones of the tree/4-star. 

\begin{figure}[!hbt]
\begin{center}
$\begin{array}{cc}
\includegraphics[height=2.55in,width=3.15in]{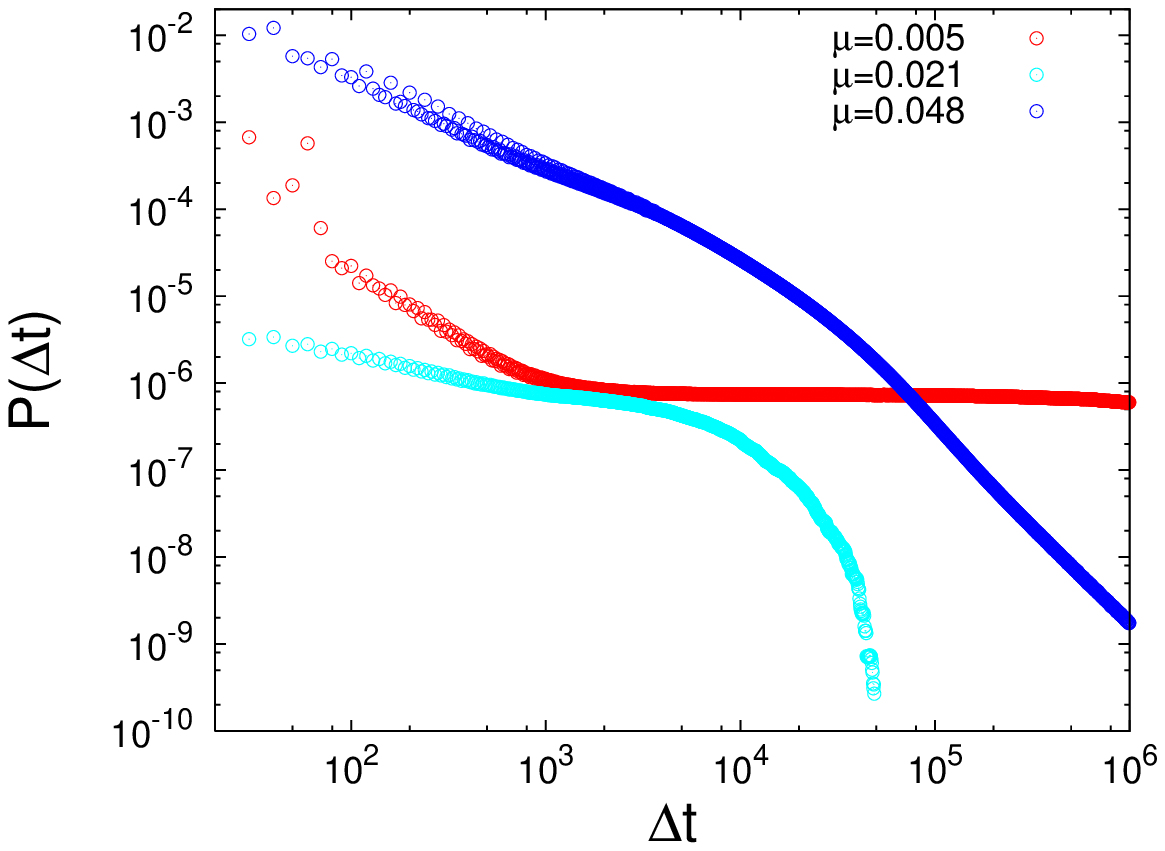} &
\includegraphics[height=2.55in,width=3.15in]{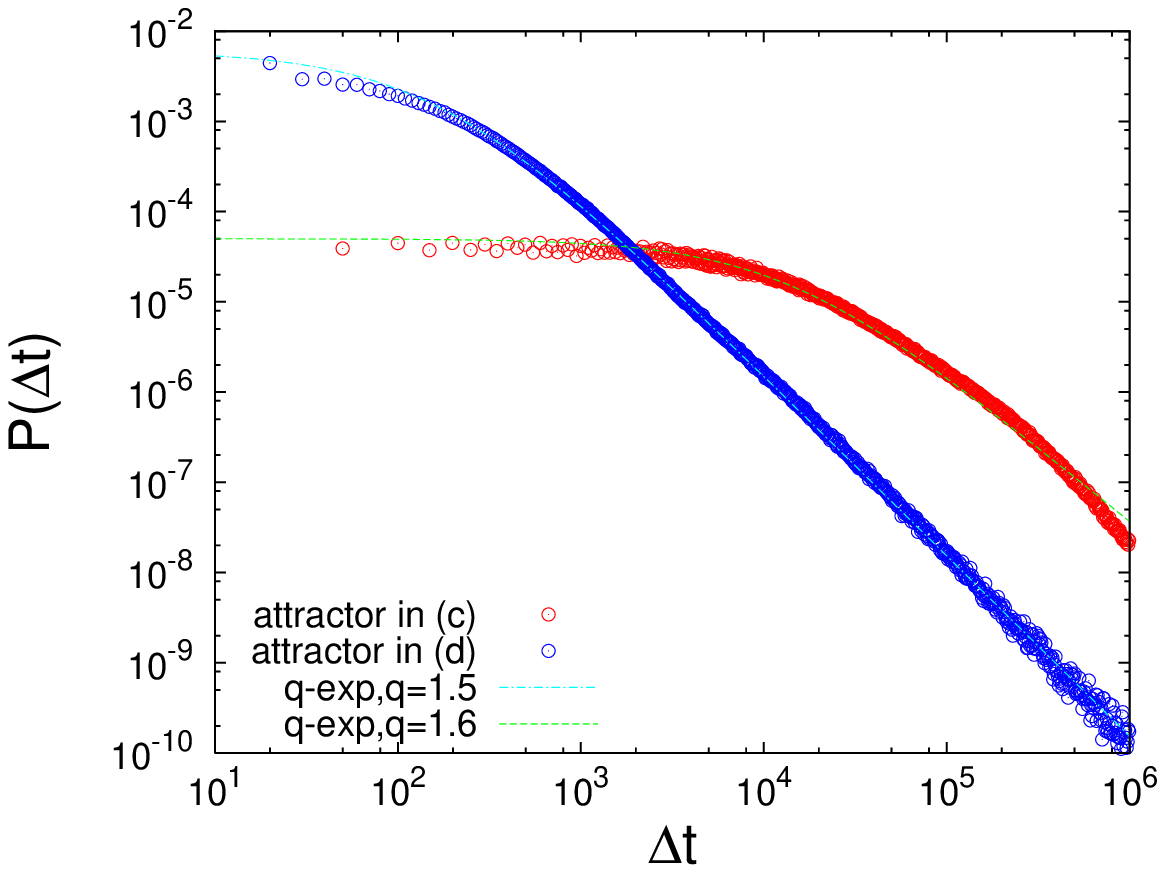} \\
\mbox{(a)} & \mbox{(b)}  \\
\includegraphics[height=2.55in,width=3.12in]{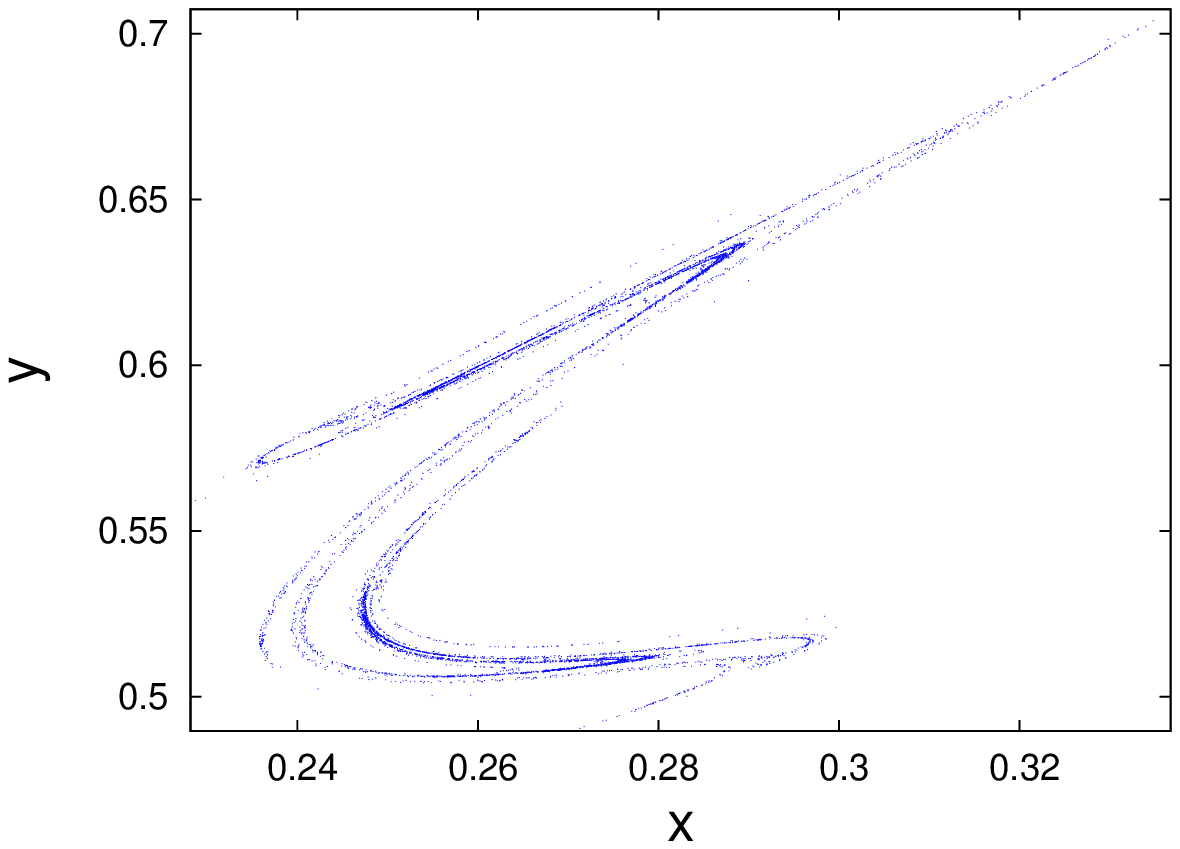} & 
\includegraphics[height=2.55in,width=3.12in]{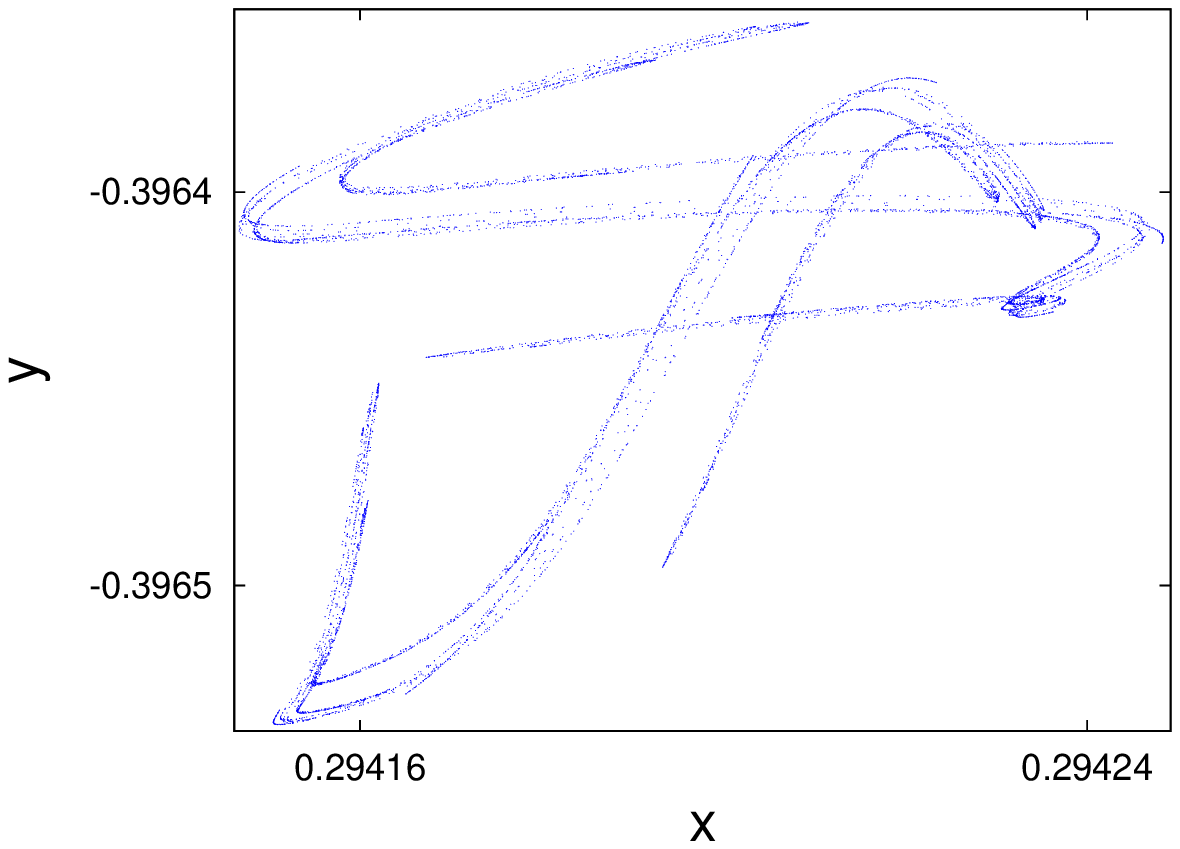} \\
\mbox{(c)} & \mbox{(d)}
\end{array}$ 
\caption[Return times distributions for 4-star, with node-averaged and single-node orbits]{Return times distributions and attractors for 4-star. Node-average orbit in three dynamical regions, averaged over many initial conditions in (a), two single-node attractor orbits occurring for $\mu=0.048$ at the 4-star's branch node, whose parts are shown in (c) \& (d). Same grid partition was used,  covering only the portion of phase space containing attractors.} \label{fig-rtstar}
\end{center}
\end{figure}

In Fig.\,\ref{fig-rtstar}a we show in more details the return times distributions for various dynamical regions for node-average orbit of 4-star, obtained by averaging over many initial conditions \cite{ja-jsm}. 
Chaotic behavior for small coupling strengths at the limit produces a constant distribution, as all return times are equally possible (this profile would however change due to regularization process, in the case when transients longer than $10^6$ are considered). In periodic region only small return times are prominent, while larger ones show similar behavior as in Fig.\,\ref{fig-rtcomparison}a. In the attractor region, the profile exhibits a structure similar to  Fig.\,\ref{fig-rtcomparison}b, but without a clear slope as it comprises many initial conditions leading to different phase space structures. We examine the attractor region at $\mu=0.048$ in Fig.\,\ref{fig-rtstar}b where we show the return times distributions for two single-node orbits (attractors) shown in Figs.\,\ref{fig-rtstar}c\,\&\,d. Both attractors are exhibited by 4-star's branch node at this $\mu$-value, but for different initial conditions. The return times distributions can be fitted with the q-exponential function, defined by \cite{tsallis88}
\begin{equation}
  e(x) = B_q \left(  1 - (1-q)\frac{x}{x_0} \right)^{\frac{1}{1-q}} ,  \label{qexp}
\end{equation} 
with $q=1.6$ for the attractor in Fig.\,\ref{fig-rtstar}c and $q=1.5$ for the attractor in Fig.\,\ref{fig-rtstar}d. We shall study these attractors further in the context of dynamical stability in Chapter \ref{Stability of Network Dynamics}.

\begin{figure}[!hbt]
\begin{center}
$\begin{array}{cc}
\includegraphics[height=2.55in,width=3.15in]{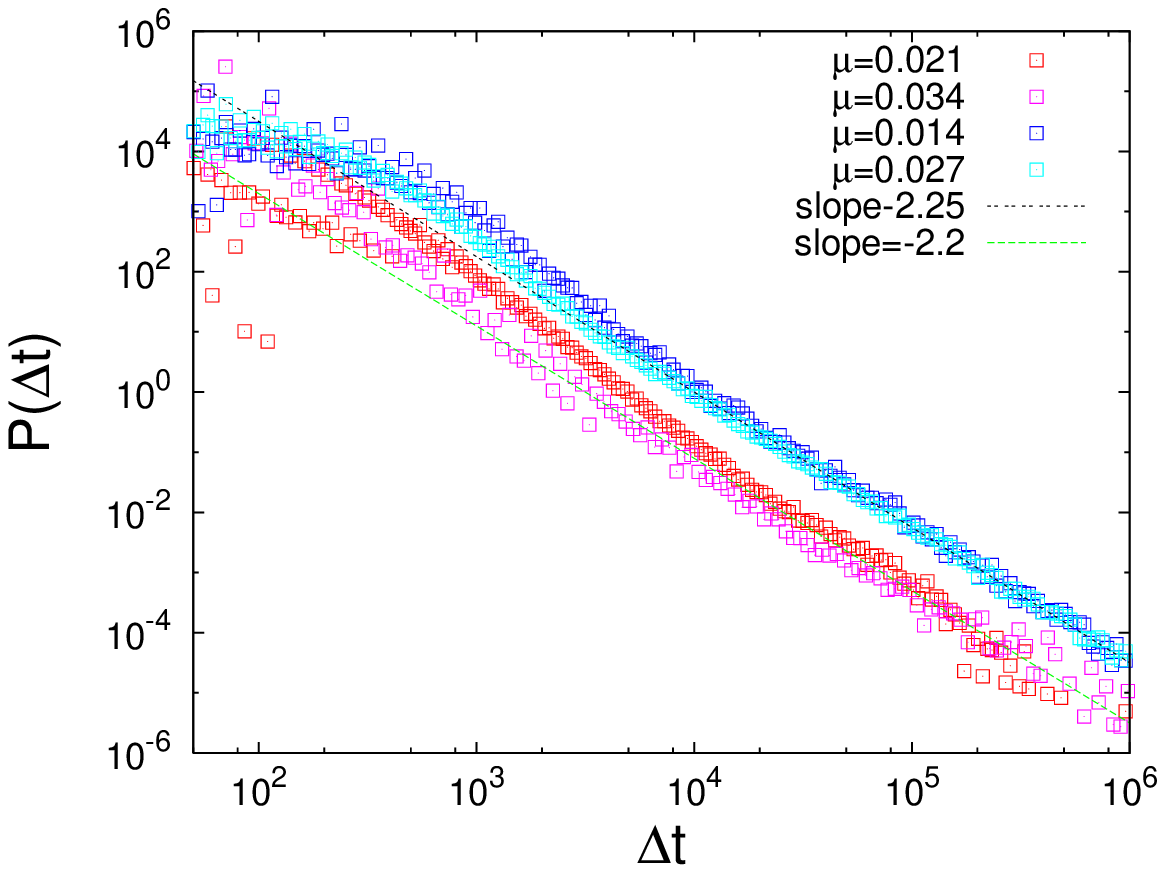} &  
\includegraphics[height=2.55in,width=3.15in]{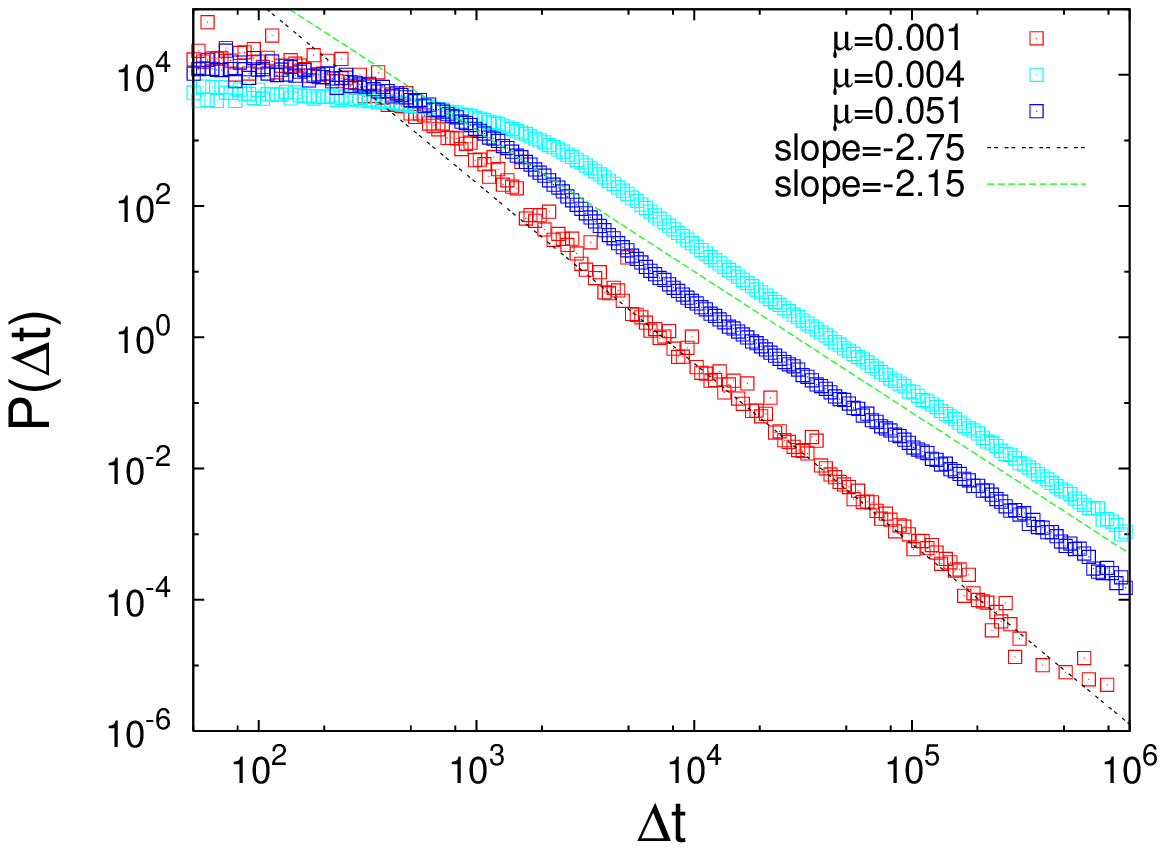}  \\
\mbox{(a)} & \mbox{(b)} 
\end{array}$ 
\caption[Return times distributions of tree's node-averaged orbit with respect to $x$-coordinate partition]{Return times distributions of tree's node-averaged orbit with respect to $x$-coordinate partition involving 100000 elements, averaged over many initial conditions, and for various $\mu$-values. Transients were not included.} \label{fig-rt-tree-x}
\end{center}
\end{figure}

Finally, we can distinguish between various dynamical regions by examining slopes of return times distributions. We consider for simplicity only the $x$-coordinate of the node-averaged orbit of the tree. The distributions for the periodic and quasi-periodic regions averaged over many initial conditions are reported in Fig.\,\ref{fig-rt-tree-x}a: note the similarity between two profiles referring to the periodic region at $\mu=0.021$ and $\mu=0.034$, and between other two profiles referring to the quasi-periodic region at $\mu=0.014$ and $\mu=0.027$. 
We also show in Fig.\,\ref{fig-rt-tree-x}b the distributions that regard the case of initial regularization at $\mu=0.001$, $\mu=0.004$ and the attractor region for $\mu=0.051$. Each region seems to have a unique power-law slope in the return times distribution, as the dynamics of the tree appears to be dependent on the $\mu$-value, with various types of collective effects present in all examined $\mu$-values. Although tree has many non-periodic orbits for all examined $\mu$-values, the presence of anomalous diffusion in the phase space motion indicates departure from the uncoupled dynamical regime.

The patterns of power-law and q-exponential fit observed in the return times distributions for various topologies and coupling strengths, testify clearly about the presence of network collectivity in the emergent dynamics of our system of CCM. We noted the analogies between 4-star's and tree's emergent statistical properties, pointing again towards a presence of a dynamical relationship between them induced by their known topological relationship. Similar statistical approaches, e.g. recurrence time statistics, are extensively in use in the context of coupled standard maps \cite{altmann}.\\[1.2cm]

\textbf{Period Values Statistics.} System's periodic orbits emerging in periodic and other dynamical regions, can be described by both their cluster-structure organization in the phase space, and by their period values. For the period value associated with a given periodic orbit of the node $[i]$ we intend the value $\pi[i]$ as defined by Eq.(\ref{period}). The period of an orbit clearly depends on the choice of $\delta$-value in Eq.(\ref{periodicorbit}); throughout this Section we will maintain $\delta=10^{-8}$ as previously. We checked our main results using different values of $\delta$ ($10^{-4},10^{-6},10^{-10}$), concluding their general properties to be qualitatively independent from this choice.

\begin{figure}[!hbt]
\begin{center}
$\begin{array}{cc}
\includegraphics[height=2.55in,width=3.15in]{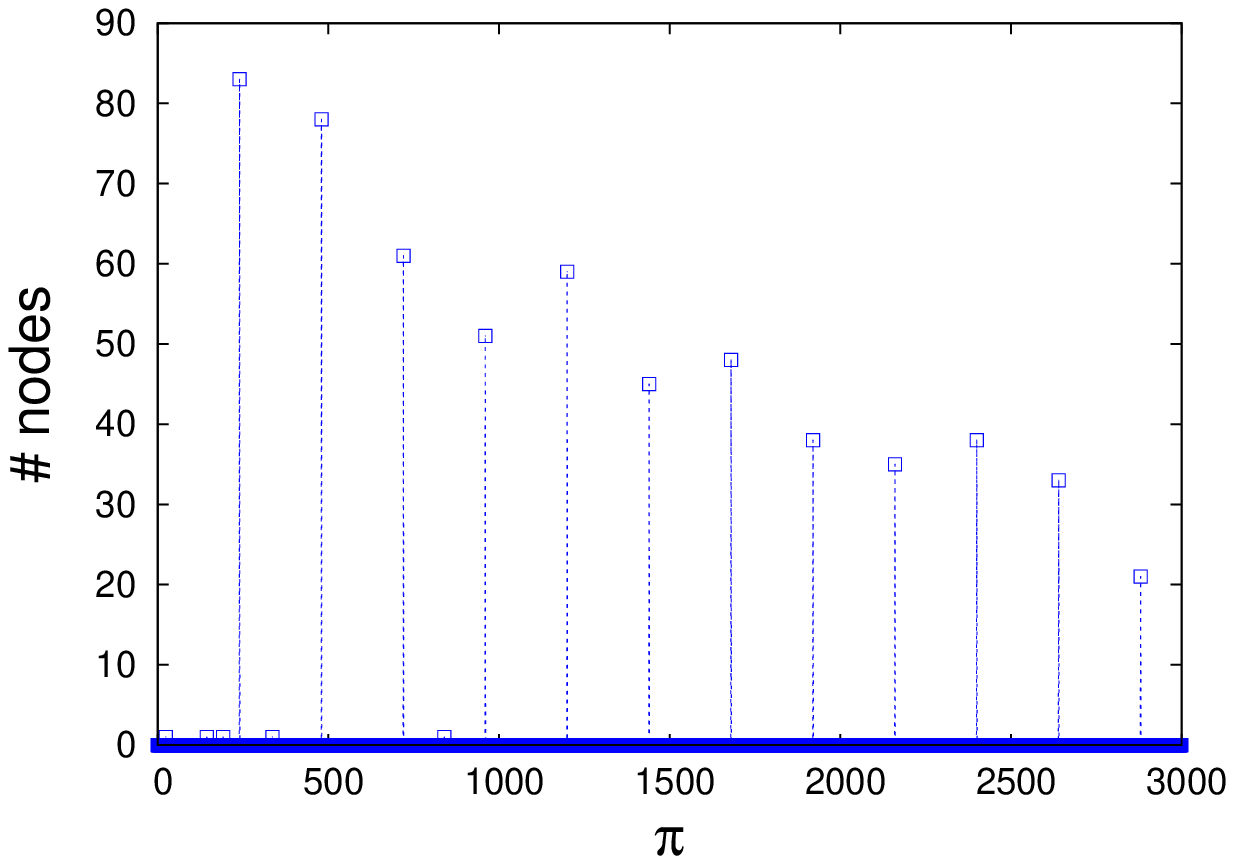} &
\includegraphics[height=2.55in,width=3.2in]{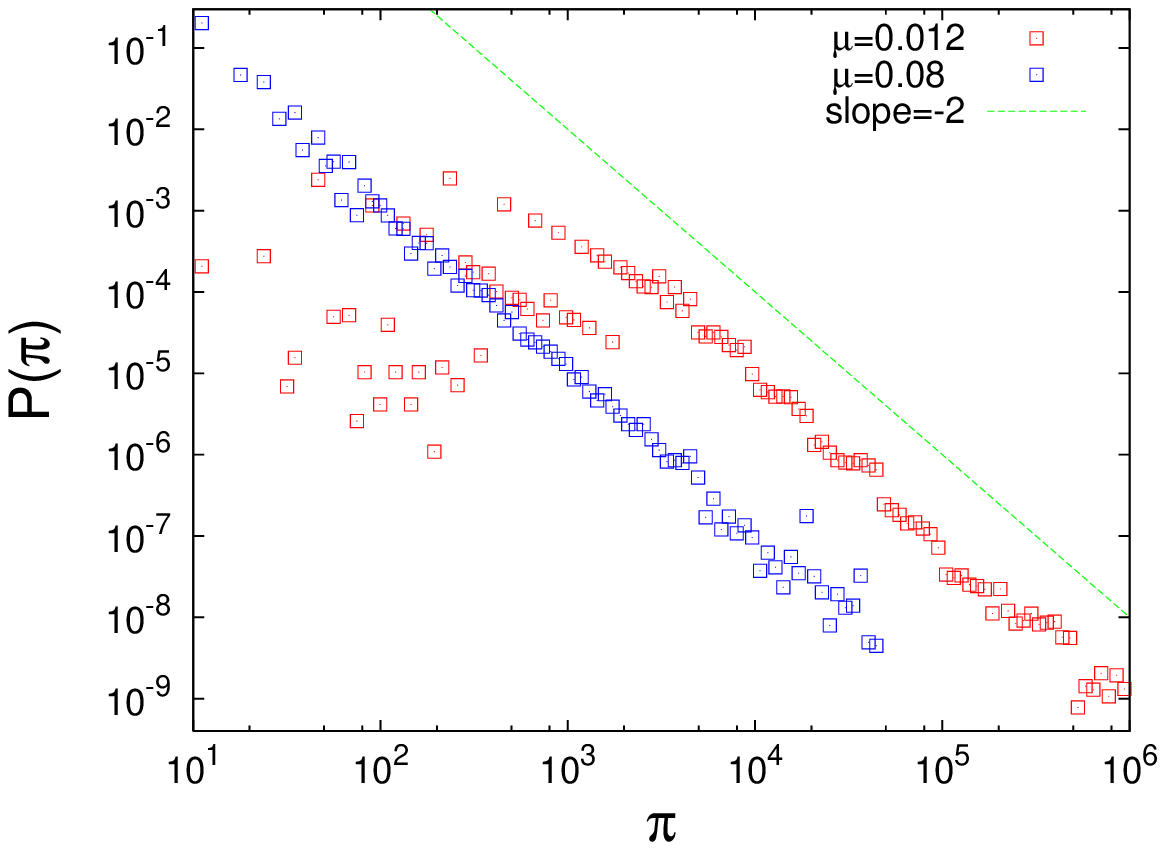} \\
\mbox{(a)} & \mbox{(b)} 
\end{array}$
\caption[Period values for the tree nodes at a final periodic state for $\mu=0.012$]{Period values of tree nodes at a final periodic state. Histogram of period values (up to $\pi=3000$) for a single initial condition at $\mu=0.012$ in (a), distribution of period values for all nodes averaged over many initial condition for two $\mu$-values fitted with a slope of -2 in (b).} \label{fig-periodstree}
\end{center}
\end{figure}

In Fig.\,\ref{fig-periodstree}a we show the histogram of period values obtained for a single final periodic state of the tree at $\mu=0.012$. In the picture we include the periods up to $\pi=3000$ (some nodes have bigger periods) showing the number of nodes (out of total $1000$) having certain $\pi$-value. The period values appear in a discrete sequence, with (almost) each node having $\pi$-value multiple of $240$, as it can be seen from the profile. This structure of values is robust to the initial conditions for this value of coupling strength.
The discretization of periods is also found for other $\mu$-values, but instead of 240 the basic period (basic multiplier) can be different (smaller for larger coupling values). The property is moreover independent of $\delta$-value in the definition of periodicity Eq.(\ref{periodicorbit}), although the basic period for the same coupling strength may vary. In Fig.\,\ref{fig-periodstree}b we report two distributions of period values obtained by averaging over many initial conditions, for $\mu=0.012$ (periodic region) and $\mu=0.08$ (also a periodic region, although not specifically studied here, cf. Fig.\,\ref{fig-nonperiodic}). Both distributions in the limit of large $\pi$-values exhibit power-laws with slopes close to -2, again indicating a presence of long-range correlations in the final dynamical states at those $\mu$-values. As expected, the case $\mu=0.08$ displays predominantly smaller periods up to $\pi=10^{4}$, whereas for $\mu=0.012$ the periods range up to $10^6$ (which was the maximal $\pi$-value we measured). This effect is systematically present in our system -- \textit{smaller} interaction strength leads to richer collective behavior. The statistics of period values is difficult to produce in the context of 4-star as the period values are much smaller.

The node-average orbit approach is also less convenient in the case of period values statistics, as it typically has extremely large periodicity (at least equal to the largest node period). However, it is instructive to examine the \textit{ranking statistics} of period values for specific nodes in relation to the coupling strength, which can be easily done for all the topologies. The ranking is obtained by ordering periods from the largest to the smallest observed value, and hence associating a rank $r$ with each period value $\pi$, without repeating the period values. 
\begin{figure}[!hbt]
\begin{center}
$\begin{array}{cc}
\includegraphics[height=2.55in,width=3.15in]{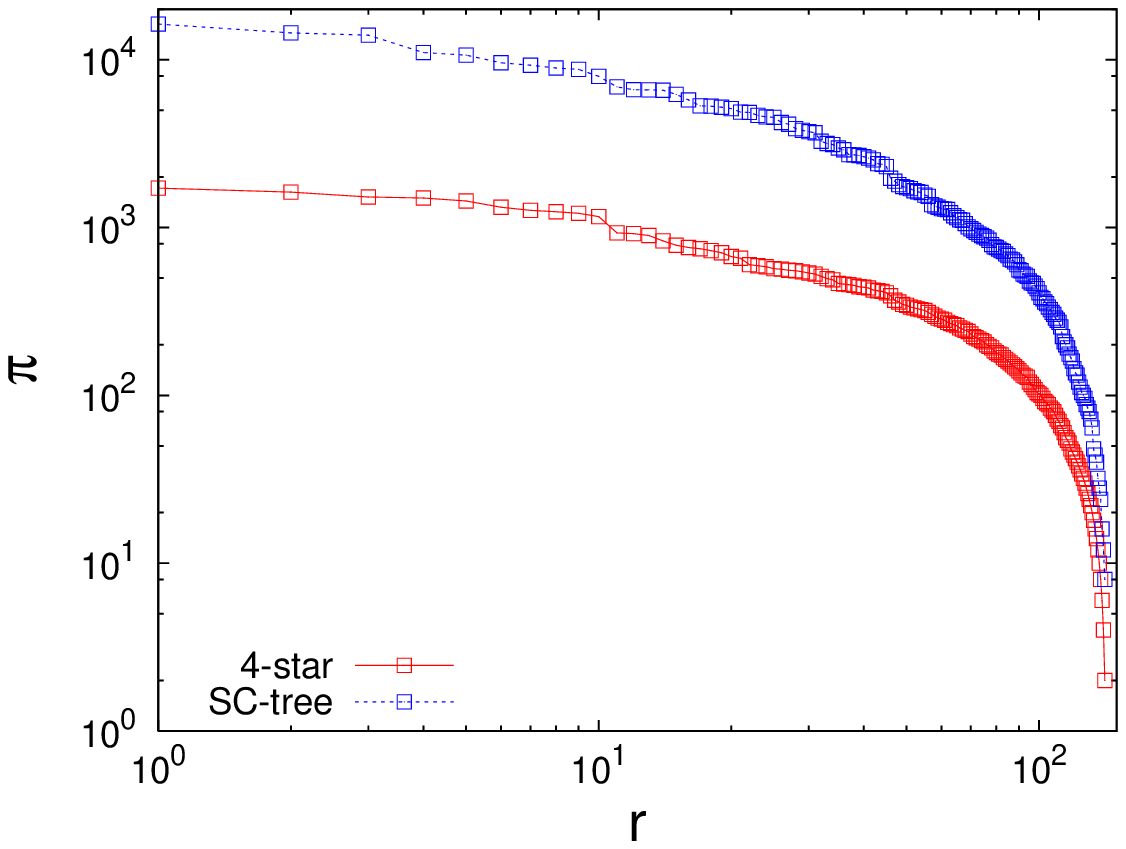} & 
\includegraphics[height=2.55in,width=3.15in]{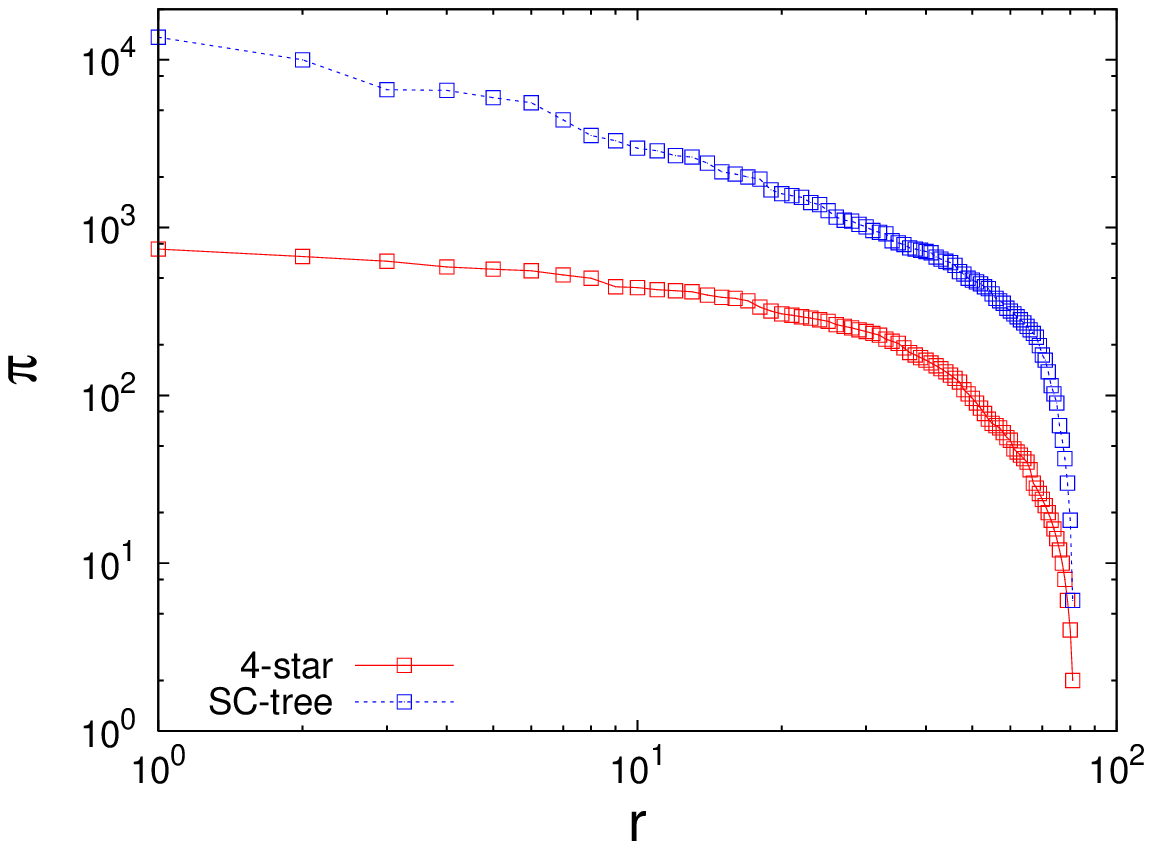} \\
\mbox{(a)} & \mbox{(b)}
\end{array}$ 
\caption[Ranking of period values for an outer tree node vs. 4-star's branch node]{Ranking of period values considered over many initial conditions for an outer tree node vs. 4-star's branch node. (a) $\mu=0.012$, (b) $\mu=0.08$.} \label{fig-periodranking}
\end{center}
\end{figure}
In Fig.\,\ref{fig-periodranking} we consider the rankings of period values of 4-star's branch node with an outer node on the tree (a node far away from the hub with degree 1), obtained for many initial conditions for two coupling values. Again, the network of CCM with $\mu=0.08$ (Fig.\,\ref{fig-periodranking}b) mainly exhibits smaller period values, while the profile is broader for $\mu=0.012$ (Fig.\,\ref{fig-periodranking}a). It is also clear the profiles resemble each other for both coupling values, in terms of shape, structure and the density of ranks with respect to the period value. This is expected from the network-location analogy between 4-star's branch node and the outer tree's node, and provides a further confirmation of the dynamical relationship between these two structures.

It is to be noted periodic orbits are not common in the context of 1D CCM. They are a distinctive feature of two- (and more) dimensional CCM which indicate collectivity of system's motion. Periodic orbits are very rare for the uncoupled standard map, which proves them to be a consequence of (non-symplectic) coupling in our CCM \cite{ja-lncs-2}. As opposed to other studies of CCM on lattices and networks (cf. Chapter \ref{Introduction}.), our system does not exhibit synchronized behavior. Nevertheless, given our context of 2D coupled units, the presence of dynamical clustering and emergent periodicity are a sign of self-organization in the system.\\[0.1cm]

\begin{figure}[!hbt]
\begin{center}
$\begin{array}{cc}
\includegraphics[height=2.6in,width=3.15in]{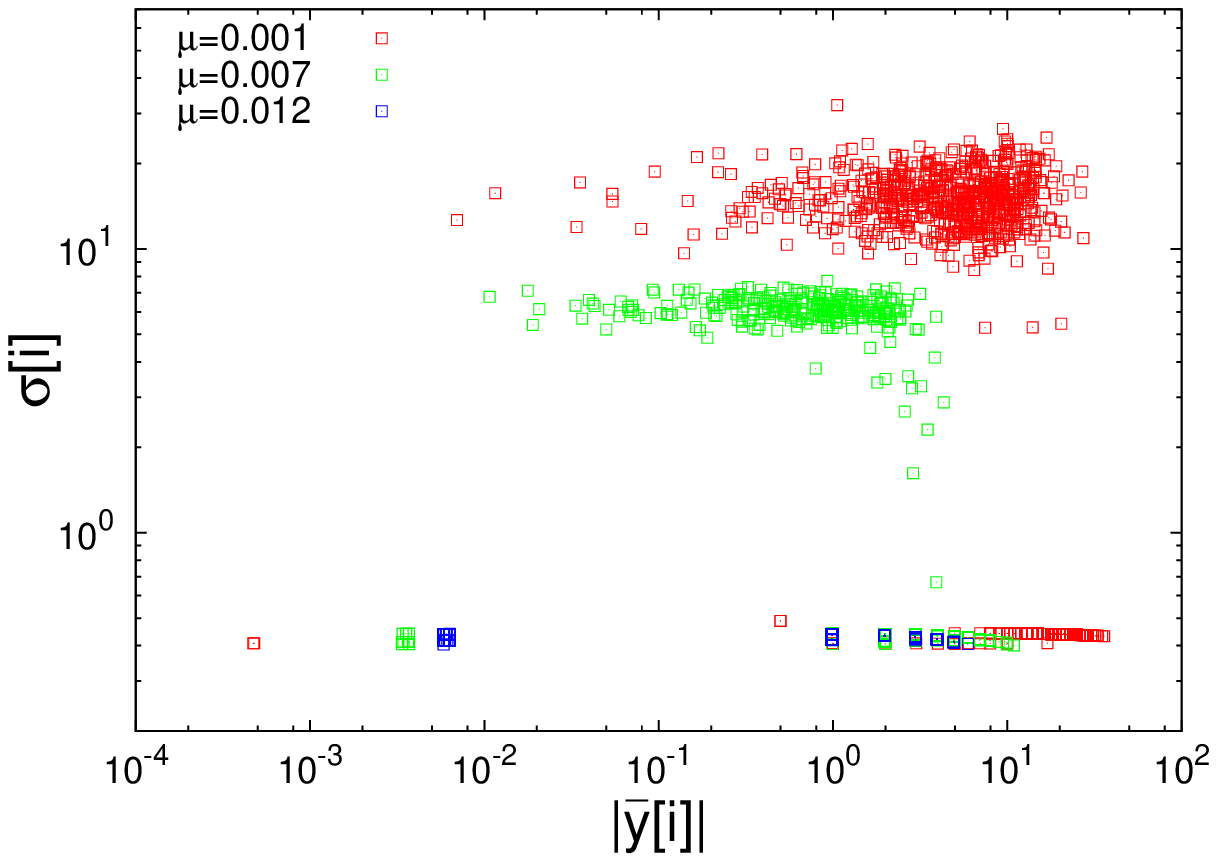} & 
\includegraphics[height=2.6in,width=3.15in]{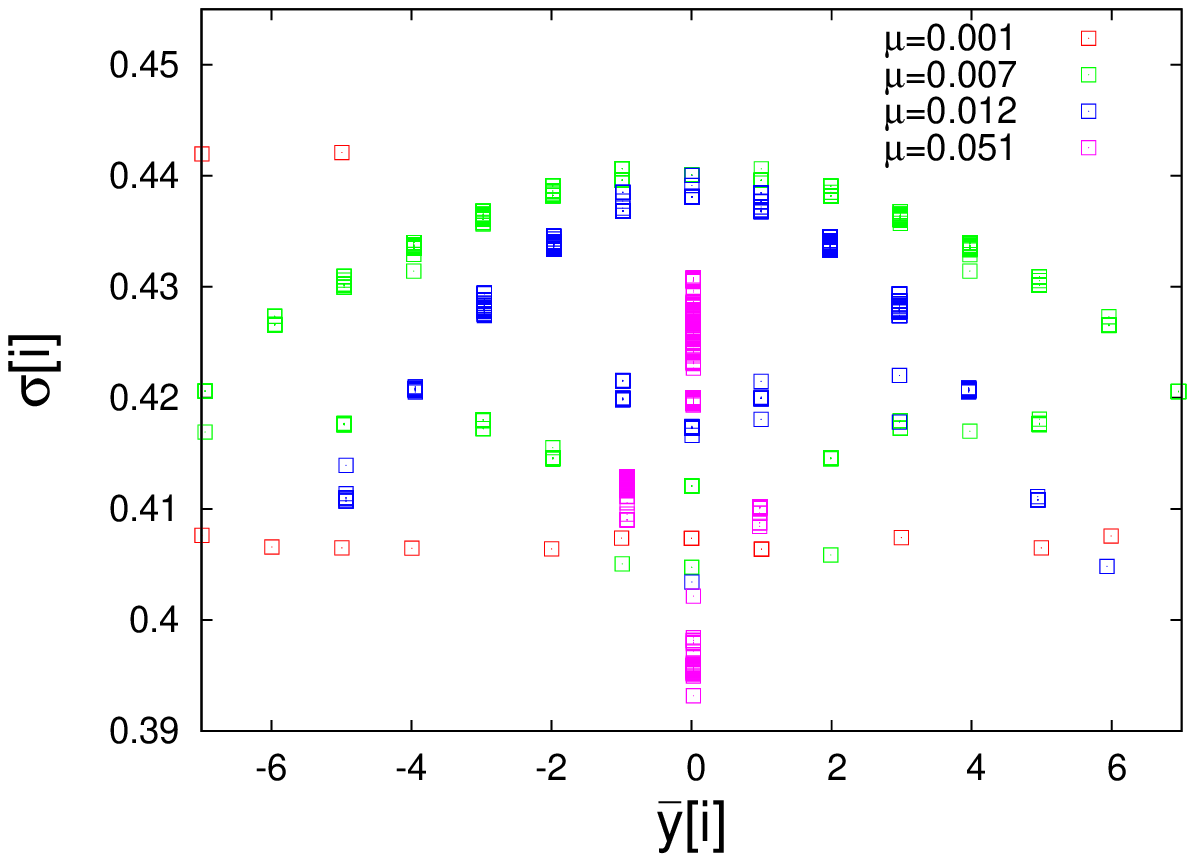} \\
\mbox{(a)} & \mbox{(b)}
\end{array}$ 
\caption[Dispersion of the network nodes time series $y_t$]{Dispersion of time series $y[i]_t$, obtained by computing $\bar{y}[i]$ and $\sigma [i]$ over 10000 iterations for each tree node at a single steady state for various coupling strengths. Log-log scale plot for all nodes in (a), and a zoom of the central part of the dispersion profile on linear scale in (b).}  \label{fig-dispersion}
\end{center}
\end{figure}

\textbf{The Dispersion of Time-series.} Another illustrative way to study the collective properties of network of CCM on the tree is to look at the dispersion of the system's time signals, i.e. to investigate the correlation between signal's time average and its standard deviation. For the the case of our system we consider time series of angular momentum $y[i]_t$ as the signal, and compare their time averages $\bar{y}[i]$ with their standard deviations $\sigma [i]$ taken over the averaging interval. The results are reported in Fig.\,\ref{fig-dispersion} where we examine $y[i]$ time series dispersion on two scales for various $\mu$-values. In Fig.\,\ref{fig-dispersion}a we show a log-log scale plot for all the tree nodes: for each coupling strength the variations of $\sigma [i]$ are very small in relation to the variations of $|\bar{y}[i]|$, in contrast to the general scaling law referring to the dispersion of time signals in (see \cite{ja-pramana} and references therein):
\begin{equation}
  \sigma [i] \propto \mbox{const} \times ( \bar{y}[i] )^\xi 
\end{equation} 
and typically displays exponents $\xi > \frac{1}{2}$. This profile also offers an alternative view on regular vs. chaotic dynamical region as the periodic clusters have $\sigma [i] < 1$, whereas chaotic nodes show $\sigma [i] \sim 10$. 
To illustrate this further we show in Fig.\,\ref{fig-dispersion}b the zoom of the cluster region of dispersion plot on a linear scale, with cluster organization of $\bar{y}[i]$ clearly visible for all coupling strengths. Observe the correlation between node's $\bar{y}[i]$-value and their standard deviation values $\sigma [i]$, which goes from no organization ($\mu=0.001$), via clusterization process ($\mu=0.007$ and $\mu=0.012$), to fully clustered phase space organization ($\mu=0.051$).\\[0.05cm]

In this Section we have shown various statistical feature our system's behavior, which uniformly suggest presence of cooperativity in the emergent motion. All the found properties are in a clear contradiction with the case of uncoupled standard map. Furthermore, we found more evidence of dynamical relationship between tree and 4-star, which was initially known to exist only in the sense of their topologies.

\section{Networks of Maps with Other Coupling Forms}

The coupled discrete-time equations defining our system are given in Eq.(\ref{main-equation}). They include a specifically selected time-delayed non-symplectic coupling form, and involve a fixed the standard map parameter $\e=0.9$. As described previously, through this particular coupling form we seeked to model a realistic coupled system of 2D oscillator units that would resemble a system of interacting genes. The particular system's set-up is also used to test the stability of networks of CCM realized through various network architectures. However, it is clear that Eq.(\ref{main-equation}) does not exhaust the possible versions of coupling forms, in terms of their structure and parameters. In this Section we will examine the CCM established through the same networks, but with different coupling forms and interaction parameters, showing the peculiarity of our specific choice.\\[0.1cm]

\textbf{The CCM without Time Delay.}  Our network of CCM is designed including the time delay of information traveling between the network's nodes that is ubiquitous for all natural systems. For the sake of completeness, we shall here present some results concerning the version of our system without the time delay. The coupled non-delayed equations equivalent to Eq.(\ref{main-equation}) read:
\begin{equation} 
\left(\begin{array}{l}
x[i]_{t+1} \\
y[i]_{t+1}
\end{array}\right)
=(1- \mu) 
\left(\begin{array}{l}
x[i]_t' \\
y[i]_t'
\end{array}\right)
+
\frac{\mu}{k_i}
\left(\begin{array}{c}
\sum_{j \in {\mathcal K_i}} (x[j]_t' - x[i]_t') \\ 
0
\end{array}\right),
\label{nodelayCCM} \end{equation}
and in contrast to Eq.(\ref{main-equation}) have prime ($'$) on both $x$ terms in the coupling expression, implying no time delay (an alternative version of non-delayed equations includes no primes at any of $x$ terms). We will examine the collective dynamics of system of CCM defined by Eq.(\ref{nodelayCCM}) in a way analogous to previously examined time-delayed CCM, using the same network topologies.

\begin{figure}[!hbt]
\begin{center}
$\begin{array}{cc}
\includegraphics[height=2.5in,width=3.15in]{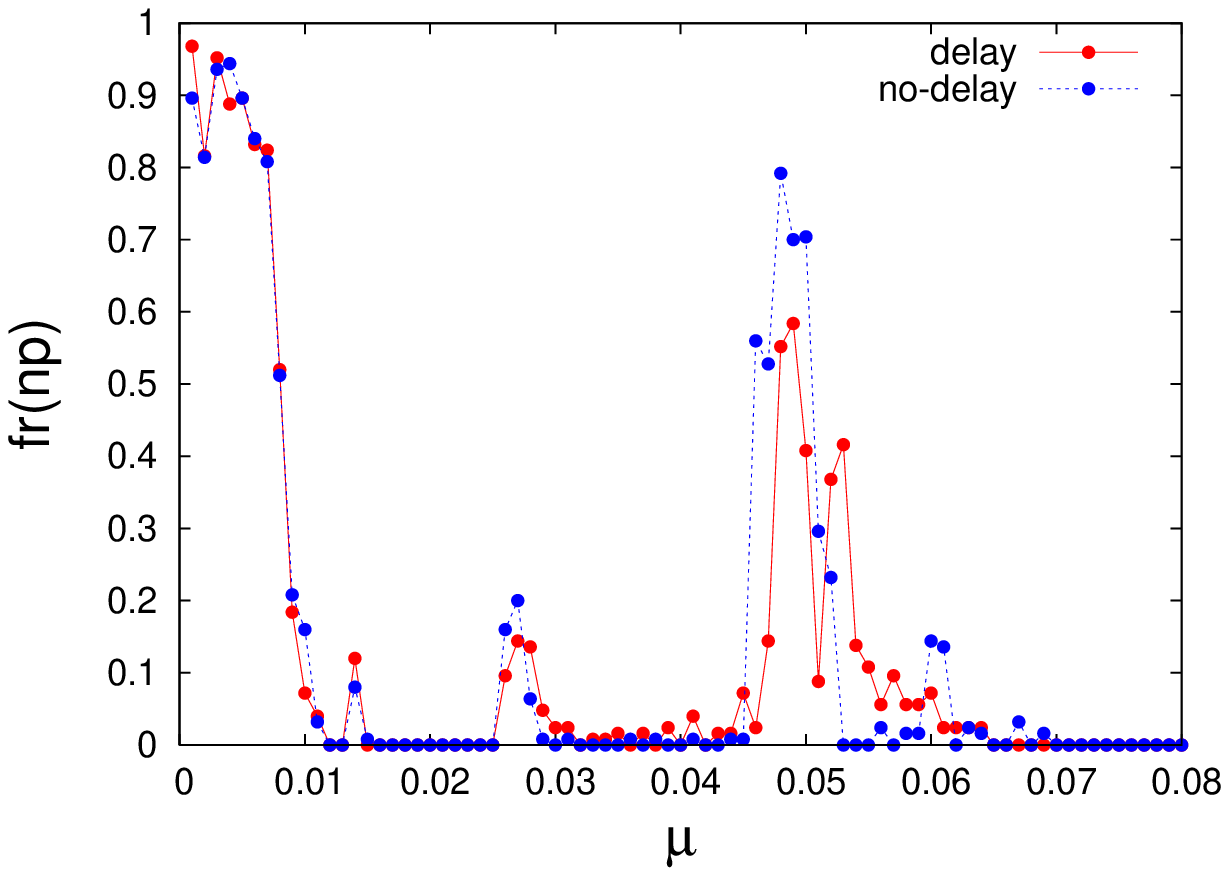} & 
\includegraphics[height=2.5in,width=3.15in]{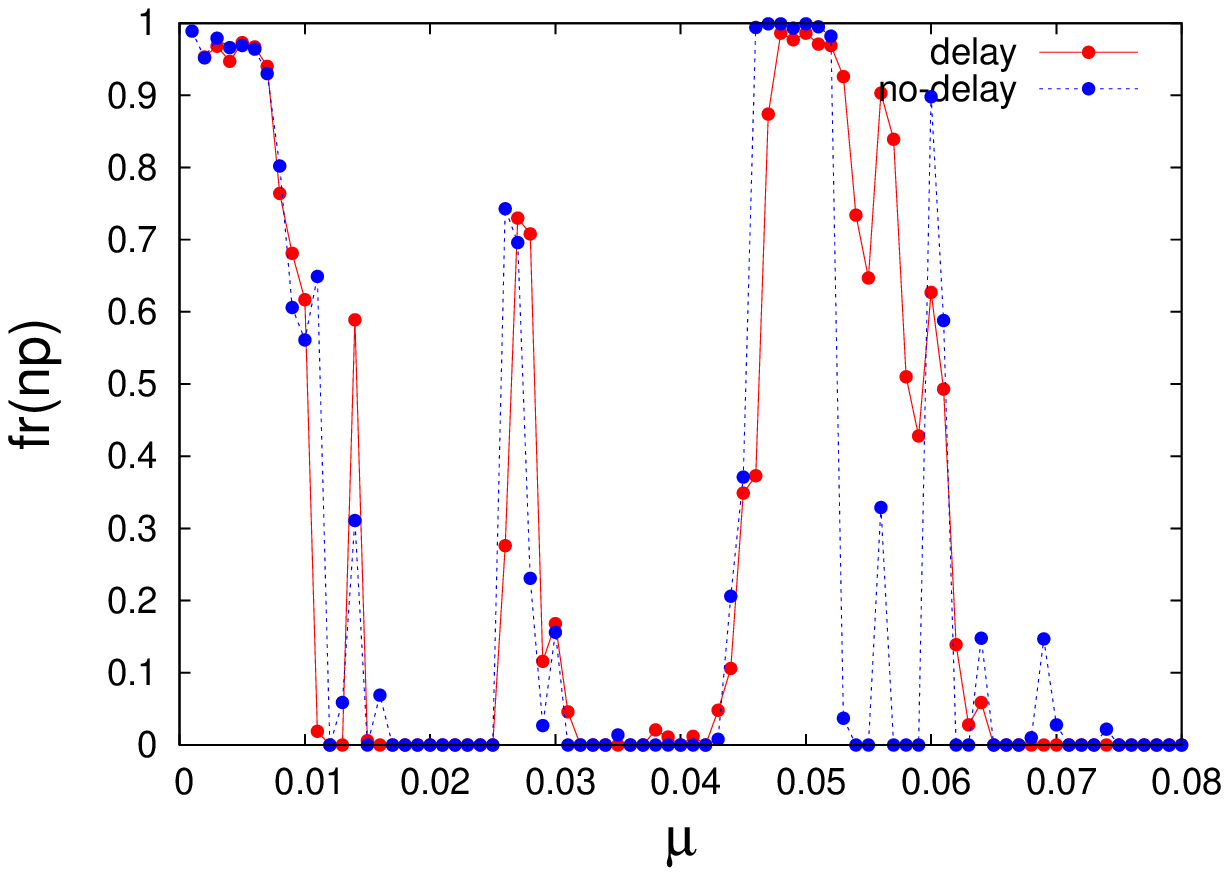} \\
\mbox{(a)} & \mbox{(b)}
\end{array}$ 
\caption[Fraction of non-periodic orbits in function of the coupling strength, comparing delayed and non-delayed versions of the system]{Fraction of non-periodic orbits in function of the coupling strength $\mu$, comparing delayed vs. non-delayed versions of network of CCM. Averaged over many initial conditions for 4-star in (a), and for a single initial condition for nodes of the tree in (b).}  \label{fig-nodelay-nonperiodic}
\end{center}
\end{figure}

\begin{figure}[!hbt]
\begin{center}
$\begin{array}{cc}
\includegraphics[height=2.6in,width=3.2in]{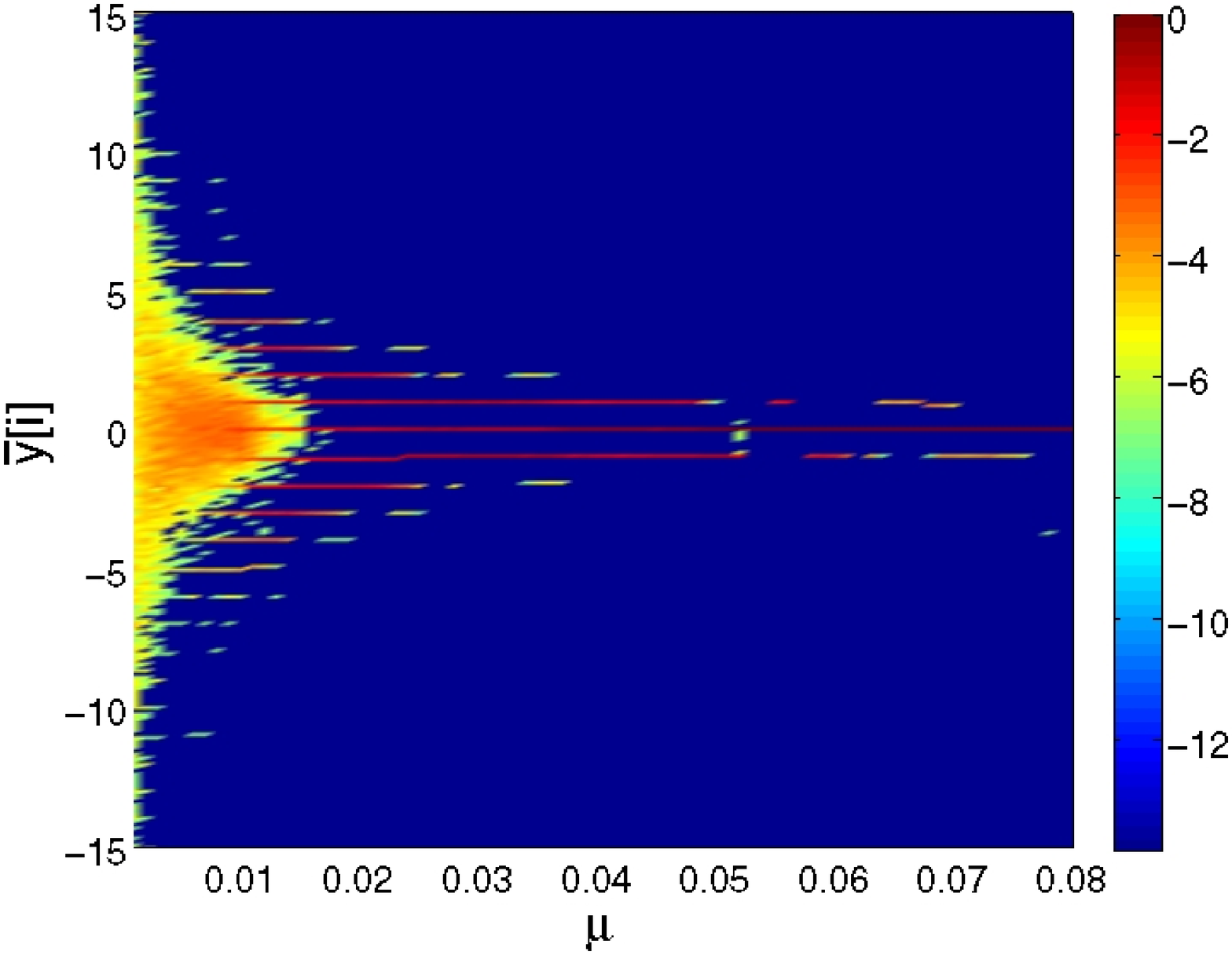} & 
\includegraphics[height=2.6in,width=3.2in]{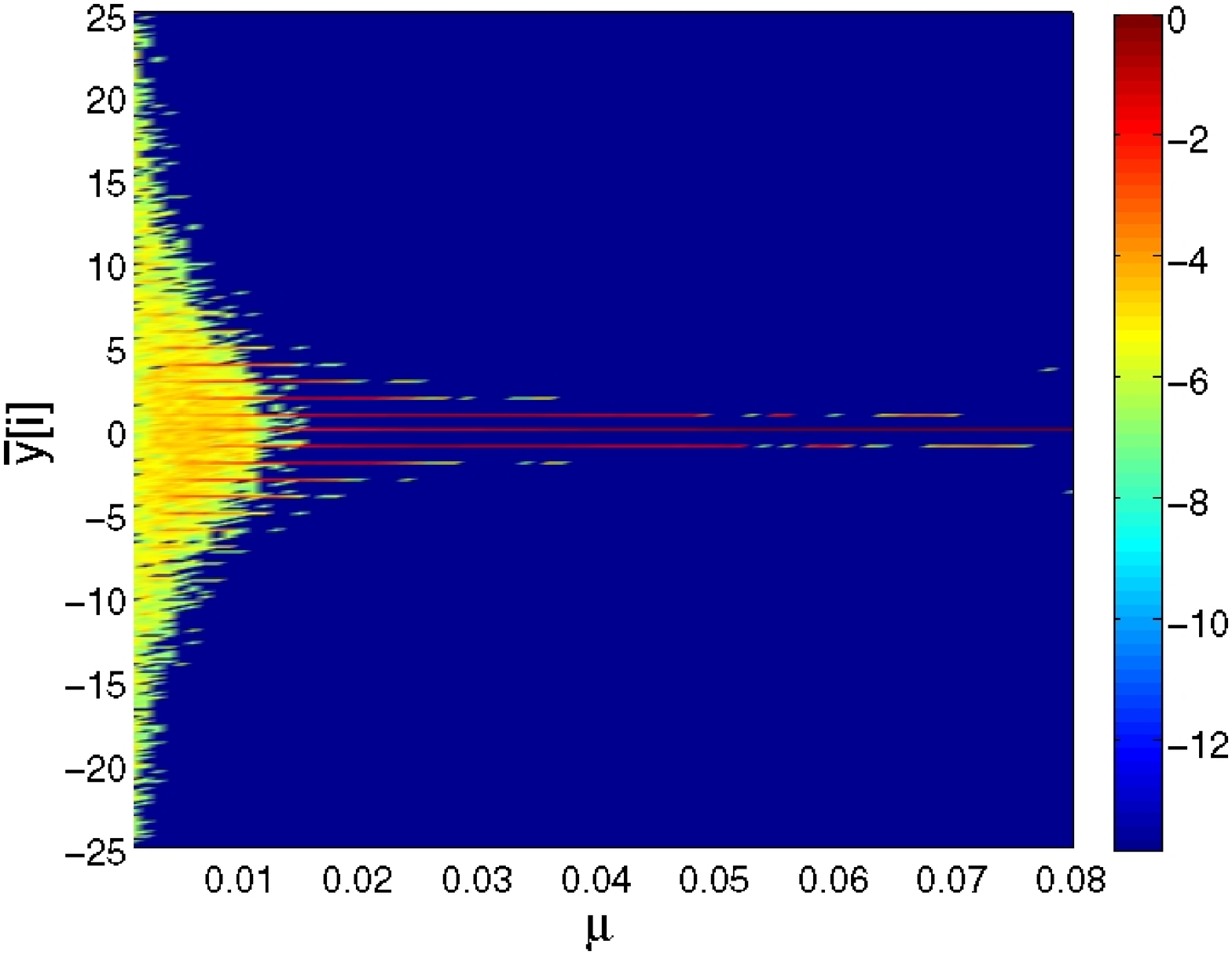}\\
\mbox{(a)} & \mbox{(b)} 
\end{array}$  
\caption[2D color histogram showing the $\bar{y}$-values in function of the coupling $\mu$ for non-delayed system]{2D color histogram showing the $\bar{y}[i]$-values in function of the coupling $\mu$ for non-delayed CCM Eq.(\ref{nodelayCCM}), done equivalently to Fig.\,\ref{fig-ybar-comparisons}. The 4-star's branch node over many initial conditions in (a), and a single initial condition 
for all nodes of the tree in (b).}  \label{fig-nodelay-ybar}
\end{center}
\end{figure}

In Fig.\,\ref{fig-nodelay-nonperiodic} we compare the fractions of non-periodic orbits for delayed vs. non-delayed case, by examining scalefree tree and 4-star. Non-delayed case exhibits similar de-stabilization of emergent dynamics, with apparent dynamical regions. 
Although the profiles are similar in terms of their general shapes, the delayed case still presents somewhat wider dynamical regions in the sense of the corresponding coupling strength intervals. We furthermore investigate the statistics of node's time averages $\bar{y}[i]$ in Fig.\,\ref{fig-nodelay-ybar} for non-delayed CCM of the 4-star and  tree. 
Note that clustering regime is again present, but the persistence and the constancy of clusters is weaker, specially for larger $\mu$-values in the attractor dynamical region. This indicates the non-delayed case exhibits less self-organizational properties than its time-delayed version (cf. Fig\,\ref{fig-ybar-comparisons}), particularly in the attractor region which is most interesting in terms of collective effects. 

The motion of the non-delayed CCM does achieve regularity and periodic orbits organized similarly into dynamical regions, but does not posses the dynamical diversity of the time-delayed case. This suggests further explorations of time-delayed systems is needed, as they might present more realistic collective dynamical effects, which is expected given the universality of the time delay in nature \cite{atay}.\\[0.1cm]

\textbf{The CCM with Different $\e$-values.} Throughout our investigations we keep the standard map's chaotic parameter fixed to $\e=0.9$ in order to be able to comparatively study different topologies and coupling strengths. It is necessary to consider a large $\e$-value as we are interested in the interaction of strongly chaotic maps, examining the ability of various topologies to create regular motion through the inter-node interactions (standard map's chaotic transition occurs for $\e \gtrsim 1.55$ \cite{ll,greene}). 

\begin{figure}[!hbt]
\begin{center}
\includegraphics[height=2.7in,width=4.3in]{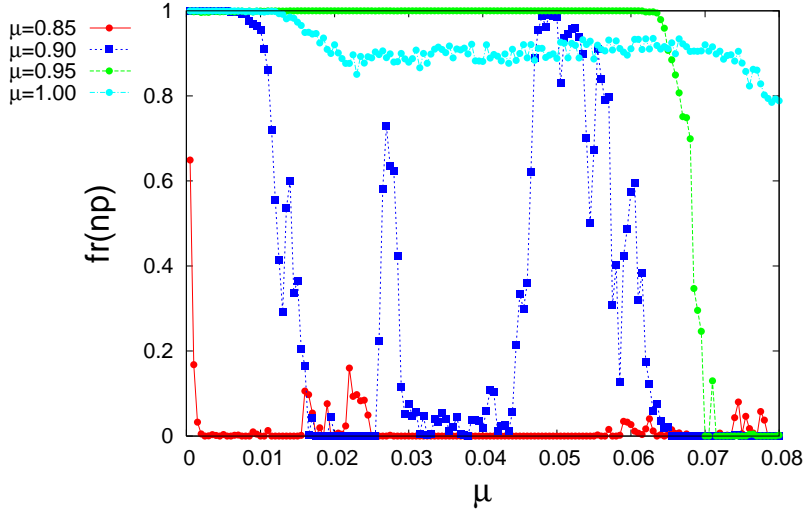} 
\caption[Fraction of non-periodic orbits in function of the coupling strength for systems with various $\e$-values]{Fraction of non-periodic orbits in function of $\mu$-value, averaged over initial conditions. CCM on tree is considered with various $\e$-values, as indicated in the picture.}  \label{fig-different-e}
\end{center}
\end{figure}

Here we investigate the system of CCM given by Eq.(\ref{main-equation}), taking different $\e$-values on the scalefree tree, by considering the profiles for fractions of non-periodic orbits as done previously. The results are shown in Fig.\,\ref{fig-different-e} where we compare the systems with $\e=0.85, 0.95$ and $1.0$ with already examined case of $\e=0.9$. While the network of CCM with $\e=0.85$ stabilizes very quickly, the system with $\e=0.95$  stabilizes only for a very large coupling strength ($\mu \cong 0.07$), and the CCM system with $\e=1.0$ never stabilizes in the studied coupling range (although, it is interesting to note the $\e=1.0$ case exhibits regularity in a part of the tree for $\mu \ge 0.02$). 
The case of $\e=0.85$ as the only one that achieves regularity, shows very small portions of initial conditions leading to non-periodic behavior for larger $\mu$-values (and therefore less potentially interesting collective effects). This case essentially shows no dynamical regions structure, as opposed to a vast spectrum of dynamical behaviors displayed by $\e=0.9$ case. These results suggest that the $\e$-values smaller than 0.85 and bigger than 1.0 are far less interesting (at least in the range of coupling strength investigated here).

Visibly, the case of $\e=0.9$ indeed presents the most interesting example of 2D chaotic standard maps network of CCM given by Eq.(\ref{main-equation}), both as self-organized complex system and as interesting non-symplectic dynamical system.\\[0.1cm]

\textbf{The CCM with Symplectic Coupling.} The coupling form leading to  Eq.(\ref{main-equation}) is non-symplectic by construction, as we favored biological motivation for designing our system over preserving symplectic (area-preserving) character of our model for the isolated unit, namely the chaotic standard map. Typically, systems of CCM are constructed with non-symplectic coupling even for the cases of symplectic units/maps, like logistic map or standard map. However, it is to be noted that standard maps can be coupled with preserving their symplectic character, and yield a symplectic CCM system of arbitrary dimensionality. Examples include already mentioned case studied in \cite{altmann}:
\begin{equation} 
\left(\begin{array}{l}
x[i]_{t+1} \\
y[i]_{t+1}
\end{array}\right)
= 
\left(\begin{array}{l}
x[i]_t' \\
y[i]_t'
\end{array}\right)
+
\frac{\mu}{k_i}
\left(\begin{array}{c}
0 \\
\sum_{j \in {\mathcal K_i}} 2\pi (x[i]_t' - x[j]_t') 
\end{array}\right),
\label{altmann-version} \end{equation}
and the case of metastable states in weakly chaotic coupled standard maps studied in \cite{moyano}: 
\begin{equation} 
\left(\begin{array}{l}
x[i]_{t+1} \\
y[i]_{t+1}
\end{array}\right)
= 
\left(\begin{array}{l}
x[i]_t' \\
y[i]_t'
\end{array}\right)
+
\frac{\mu}{k_i}
\left(\begin{array}{c}
\sum_{j \in {\mathcal K_i}} ( y[j]_t + \e \sin (2\pi x[j]_t) ) \\
0
\end{array}\right),
\label{moyano-version} \end{equation}
and many other. However, these symplectic version of coupling among standard maps generally refer only to the cases of clique and chain network structures, without being applicable to general network topologies (e.g. scalefree tree). Also, these CCM systems do not exhibit regular behavior in the sense of our study here, the phase space dynamics remains chaotic as in Fig.\,\ref{fig-oursm}b (although they exhibit other collective effects related to measure-preserving nature of CCM system, like the anomalous diffusion \cite{altmann} or metastable states \cite{moyano}). 

In order to study the collective dynamics with possible implications for the real complex systems, one needs to go beyond the restrictive symplectic coupling of 2D units. For these reasons we have not devoted a special attention to the symplectic system of standard maps, and we shall continue our examination for the context of previously defined non-symplectic system of CCM.


\chapter{Dynamic Stability of CCM on Networks} \label{Stability of Network Dynamics}

\begin{flushright}
\begin{minipage}{4.6in}
    We study the stability of the emergent motion of our network of CCM using three different techniques. We characterize the dynamical 
    regions further, by observing sharp distinctions between them in terms of stability patterns. A variety of dynamical phenomena related to   
    non-symplectic coupling are investigated, including different types of strange (nonchaotic) attractors and 
    quasi-periodic orbits. Edges of dynamical regions are explained through parametric instability.\\[0.1cm]
\end{minipage}
\end{flushright}

We investigate the stability of the emergent dynamics of CCM Eq.(\ref{main-equation}) on non-directed 4-star and tree, considering standard approaches used in nonlinear dynamics \cite{wiggy}. We are dealing with chaotic maps that due to mutual interactions achieve regular behavior in a wide range of coupling strengths, as shown in detail in the previous Chapter. However, it is of interest to see whether this regular (as opposed to chaotic) motion is also \textit{stable} with respect to small perturbations. By characterizing the stability of our system, we establish a further distinction with the case of the uncoupled maps which is known to be strongly chaotic. Quantitative results indicating emergent motion of nodes to be stable in the sense of nonlinear dynamics, would prove the ability of the considered networks to stabilize the dynamics of very chaotic isolated units. It is important to note that biological complex systems are generally stable and robust to the environmental perturbations, which is necessary for their successful functioning. This implies any complex dynamical system modeling a real biological system ought to exhibit a stable and coherent behavior. 

We will examine the stability of our CCM system mainly focusing on 4-star which, as demonstrated already, contains the core of scalefree tree's dynamical properties. In particular, we will employ the following methods:
\begin{enumerate}
 \item \textit{Finite-time Maximal Lyapunov Exponents (FTMLE)}, used for general characterization of dynamics for all the network structures, based on
 direct computation of divergence of nearby trajectories for a single node
 \item \textit{Standard Maximal Lyapunov Exponents (SMLE)}, computed by the usual technique of Jacobian matrix time-evolution, used for 
precise measurement of Maximal Lyapunov Exponents for 4-star considered as 8-dimensional dynamical system
 \item \textit{The Parametric Instability}, explored using eigenvalues of the system's Jacobian matrix at the thresholds separating different types 
 of motion, used to investigate generation of instabilities through appearance of various types of bifurcations 
\end{enumerate}
By studying the stability of our system's collective dynamics we shall further characterize the dynamical regions already defined in previous Chapter, and explain the differences between them. Study of stability is related to concrete dynamical phenomena exhibited by the system at a single final state after the transient, in contrast to the global statistical properties examined so far. The investigations to follow will complement our previous results and complete the picture of our system's collective dynamics.

\section{Finite-time Maximal Lyapunov Exponents}

Lyapunov Exponent is generally understood as a measure of divergence of nearby trajectories \cite{wiggy,ll,GH}, computed by systematically recording the distance between two initially very close trajectories as they evolve. For a specific phase space point and its trajectory related to a given dynamical system, the \textit{Maximal Lyapunov Exponent} (MLE) is defined to be the fastest divergence rate from this trajectory found in its neighborhood. MLE tells how fast are nearby trajectories diverging from each other or converging towards each other. Divergence of nearby trajectories (which translates into positivity of MLE) is usually related to chaotic dynamics, as opposed to regular dynamics which exhibits only non-positive MLE (implying nearby trajectories are not diverging from each other).

\begin{figure}[!hbt]
\begin{center}
\includegraphics[height=2.8in,width=3.8in]{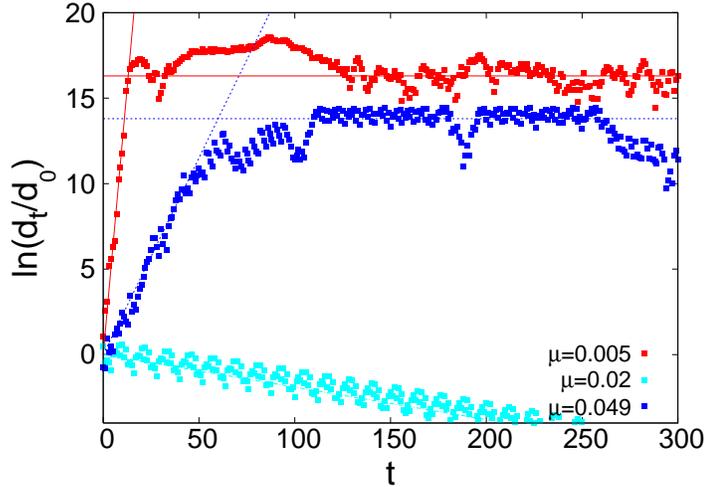} 
  \caption[Time-evolution of distance ratio $\ln (\frac{d_t}{d_0})$ for three characteristic points of the 4-star's branch node]{Time-evolution of
  distance ratio $\ln (\frac{d_t}{d_0})$ for three characteristic points in three dynamical regions of the 4-star's branch node.} 
\label{fig-lyapunovdivergence}  
\end{center}
\end{figure}

MLE are computed by different techniques depending on the system's properties: in this Section we will focus on \textit{Finite-time Maximal Lyapunov Exponents} (FTMLE) that are defined as the \textit{initial divergence rate} of the nearby trajectories \cite{feudel}. We are seeking a method of quantifying trajectory divergence that can be applied for any of the 4-star's or tree's nodes, as this will allow a comparative study of the stability of these structures. We will consider the usual 4-star's branch node in relation to an outer tree node as done previously. For this purpose we examine in Fig.\,\ref{fig-lyapunovdivergence} the divergence between three pairs of nearby trajectories exhibited by 4-star's branch node, picked from three different dynamical regions (similar curves are found for the case of tree nodes). Each curve reports the time-evolution of the distance between two trajectories $d_t$, divided by their initial distance $d_0$. 
For the periodic orbit, the distance ratio shrinks with time, while for the chaotic and the attractor orbit it initially diverges, but eventually settles around a certain value. 

For Finite-time Maximal Lyapunov Exponent related to some phase space point belonging to the node $[i]$ we shall understand the initial slope of the divergence curve, as pictured in Fig.\,\ref{fig-lyapunovdivergence}, and following \cite{ja-jsm}. More precisely, FTMLE $\Lambda_{max}^t (\x_0)$ associated with the point $\x_0 \equiv (x_0,y_0)[i]$ is defined by:
\begin{equation}
\Lambda_{max}^t (\x_0) = \max_{\x \in {\mathcal N}} \; \left\lbrace \mbox{initial slope}  \left[ \frac{1}{t}\ln 
\frac{d(U_t \x,U_t \x_0)}{d(\x,\x_0)} \right]\right\rbrace,
\label{mledefinition}
\end{equation}
where ${\mathcal N}$ stays for a small neighborhood around the point $\x_0$, $d(.,.)$ denotes the distance, and $U_t$ stands for the discrete time-evolution dynamics given by the system of CCM. Note however that the length of the slope should be appropriately selected for each trajectory separately, depending on the dynamical region in question. We shall use FTMLE to characterize the stability of motion of selected nodes separately, comparing 4-star's branch to tree's outer node. FTMLE represent a stability measure related to the initial divergence of trajectories, and are complementary to the Standard Maximal Lyapunov Exponents (SMLE), examined in the next Section. FTMLE depend on the node, as opposed to SMLE which consider the entire system. They are suitable for the present study as they allow a parallel examination of both structures, referring to the orbits exhibited by similarly located nodes (e.g. branch node vs. outer node).

A single $\Lambda_{max}^t (\x_0)$ does not fully describe the dynamics related to the point $\x_0$ as it also depends on the test-point $\x \in {\mathcal N}$. We therefore average over many $\x \in {\mathcal N}$ in order to obtain a better approximation, according to the following algorithm:
\begin{enumerate}
\item consider a final steady state of a given network, and focus on the state $(x_0[i],y_0[i])$ for the chosen node $[i]$
\item consider a random point $(x[i],y[i])$ in this point's close neighborhood, with the distance $d$ between them given as 
"Manhattan distance" (for computational simplicity): $d((x,y),(x_0,y_0))=|x-x_0|+|y-y_0| \ll 1$  
\item iterate the dynamics for both points systematically recording the rate that their distance ratio in Eq.\ (\ref{mledefinition}) changes in time. Determine the time interval with the clearest slope (slope-length) and compute the corresponding slope over this interval (cf.  Fig.\,\ref{fig-lyapunovdivergence}) 
\item Repeat the same procedure for few other random points in the same neighborhood, and determine the maximal among the obtained slopes. This is our approximation of $\Lambda_{max}^t (\x_0)$
\end{enumerate}
Furthermore, the exponent  $\Lambda_{max}^t (\x_0)$ computed this way corresponds to only one point $\x_0$ on the examined orbit. Starting with different points on the same orbit, a distribution of $\Lambda_{max}^t$ can be constructed to characterize the entire orbit. The average value of this spectrum will be called $\lambda_{max}^t$ and understood as the characteristic FTMLE for the whole orbit in question:
\begin{equation} 
\lambda_{max}^t = <\Lambda_{max}^t (\x_0) >_{\x_0 \in \mbox{orbit}}   \label{lambda}
\end{equation}
We will refer to this definition when considering $\lambda_{max}^t (\x_0)$ for different nodes/initial conditions, usually obtained by averaging over many initial conditions. The distribution of $\Lambda_{max}^t (\x_0)$ and the value of $\lambda_{max}^t$ give a good estimate of the initial stability of the orbit in question. For simplicity, we will determine the optimal slope-length for each dynamical region separately (the slope-lengths seem to mostly depend on the dynamical regions), and refer to them for all FTMLE computations.

\begin{figure}[!hbt]
\begin{center}
$\begin{array}{cc}
\includegraphics[height=2.5in,width=3.2in]{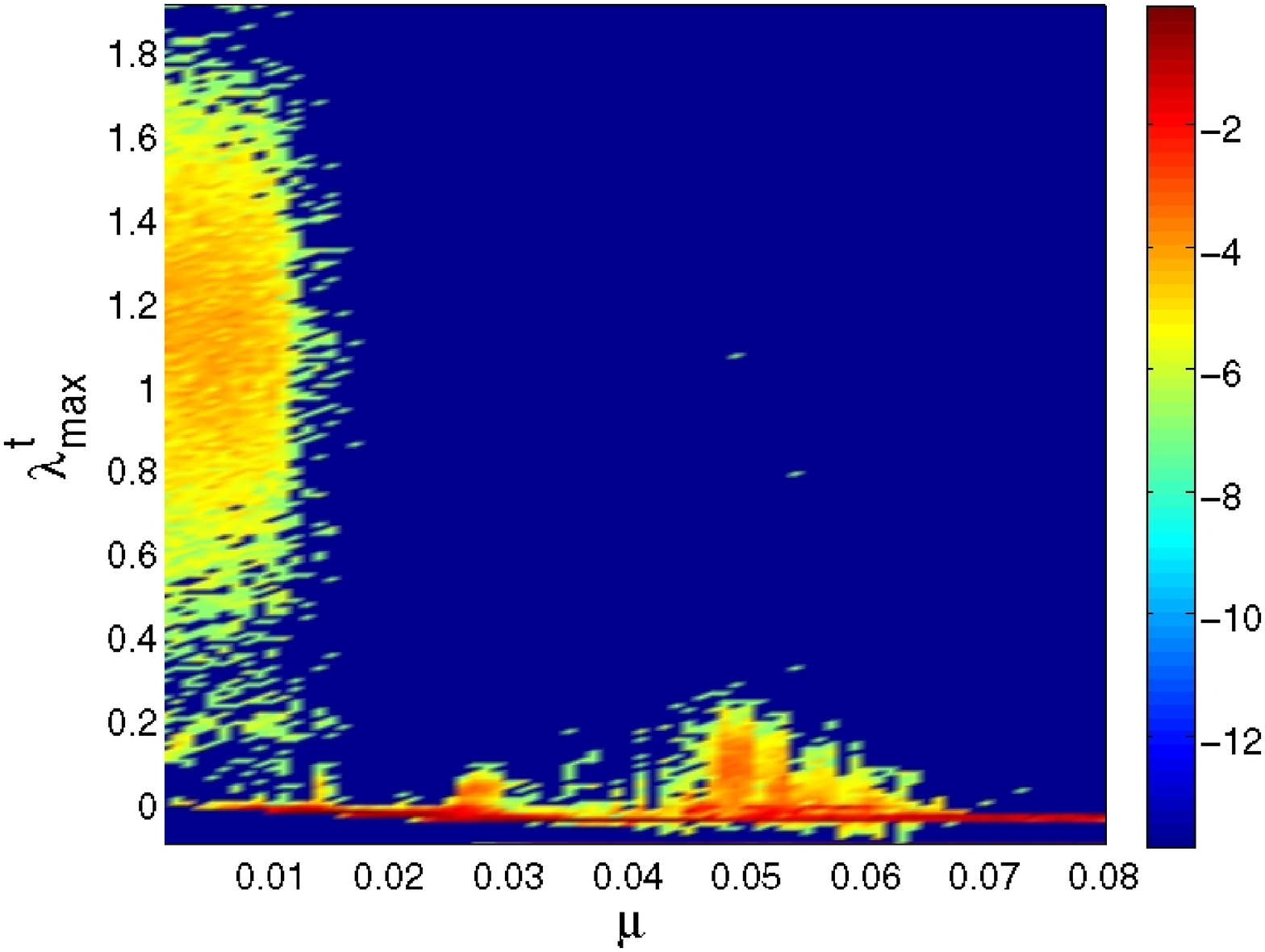} & 
\includegraphics[height=2.5in,width=3.2in]{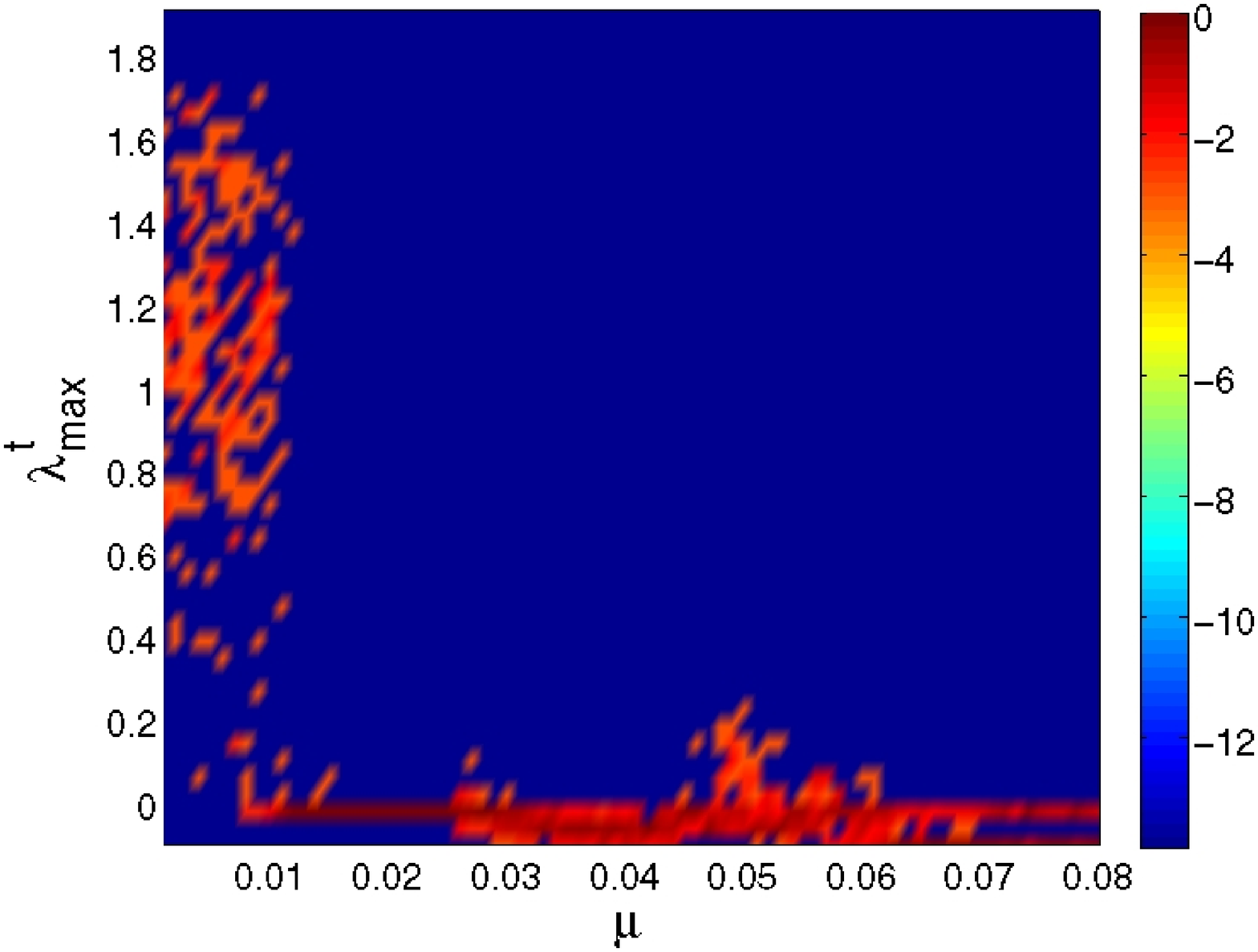} \\
\mbox{(a)} & \mbox{(b)} 
\end{array}$ 
\caption[2D color histograms of $\lambda_{max}^t$ values in function of the coupling for 4-star's branch node and a tree's outer node]{2D color histograms of $\lambda_{max}^t$ values in function of the coupling $\mu$, obtained considering many initial conditions. 4-star's branch node in (a), and a tree's outer node in (b).}
       \label{fig-colorstability}
\end{center}
\end{figure}

In Fig.\,\ref{fig-colorstability}  we report the results of evaluation of $\lambda_{max}^t (\x_0)$ for 4-star's branch node and tree's outer node. For each $\mu$-value we compute the $\lambda_{max}^t (\x_0)$ for many initial conditions, and construct a 2D color histogram as done previously in Fig.\,\ref{fig-ybar-comparisons} (coloration is in log-scale). There is a clear equivalence between profiles for 4-star's branch node and tree's outer node, in terms of similar distribution patterns (the tree node picture includes less initial conditions as it is numerically more demanding). Small $\mu$-values are characterized by large FTMLE $O(1)$ similarly to the uncoupled standard map, whereas the periodic region shows mainly negative FTMLE. In contrast to these two dynamical regions, the attractor region exhibits a larger variety of $\lambda_{max}^t$-values which can be negative and weakly positive $O(10^{-1}$). 
These results indicate a clear analogy between the behavior of 4-star's branch node and tree's outer node in terms of dynamical stability patterns. Given the network location analogy of the two considered nodes, the result also suggests an equivalence between the general stability of the 4-star and the tree. 

\begin{figure}[!hbt]
\begin{center}
\includegraphics[height=2.6in,width=3.5in]{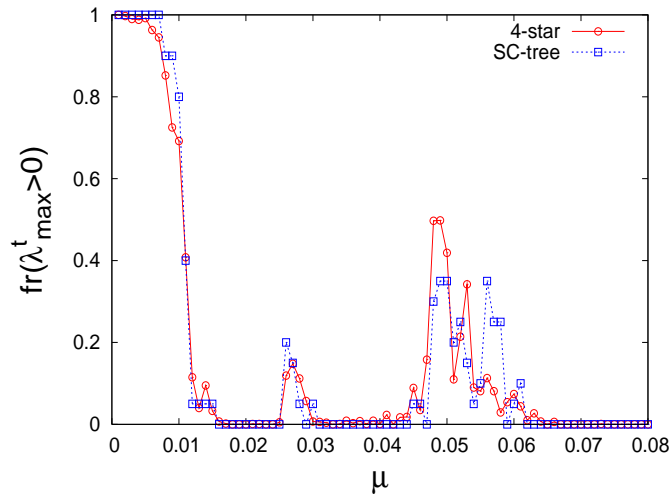}
\caption[Fractions of orbits leading to positive $\lambda_{max}^t$ in function of the coupling $\mu$]{Fraction of orbits leading to positive $\lambda_{max}^t$ in function of the coupling $\mu$ for 4-star's branch node and a tree's outer node, considered over many initial conditions.}    \label{fig-stability-comparison}
\end{center}
\end{figure}

As the unstable motion is related to the positivity of FTMLE, it is illustrative to examine the fraction of initial condition leading to the orbits with positive FTMLE in function of the coupling strength for various nodes/topologies. We construct such plots starting from the data reported in Fig.\,\ref{fig-colorstability} and present the results in Fig.\,\ref{fig-stability-comparison}, showing the fractions of unstable orbits characterized by positive $\lambda_{max}^t$-values in function of $\mu$, for 4-star's branch node and the tree's outer node. 
The profiles show similar patterns, specifically for smaller coupling strengths where the overlap is very precise. Attractor region shows a qualitative overlap in terms of coupling strength interval with a fraction of positive $\lambda_{max}^t$-values. Quasi-periodic regions in both profiles agree as  intervals of coupling strength with some positive FTMLE (quasi-periodic orbits with be studied in detail in the following Sections). \\[1.cm]

\begin{figure}[!hbt]
\begin{center}
$\begin{array}{cc}
\includegraphics[height=2.5in,width=3.12in]{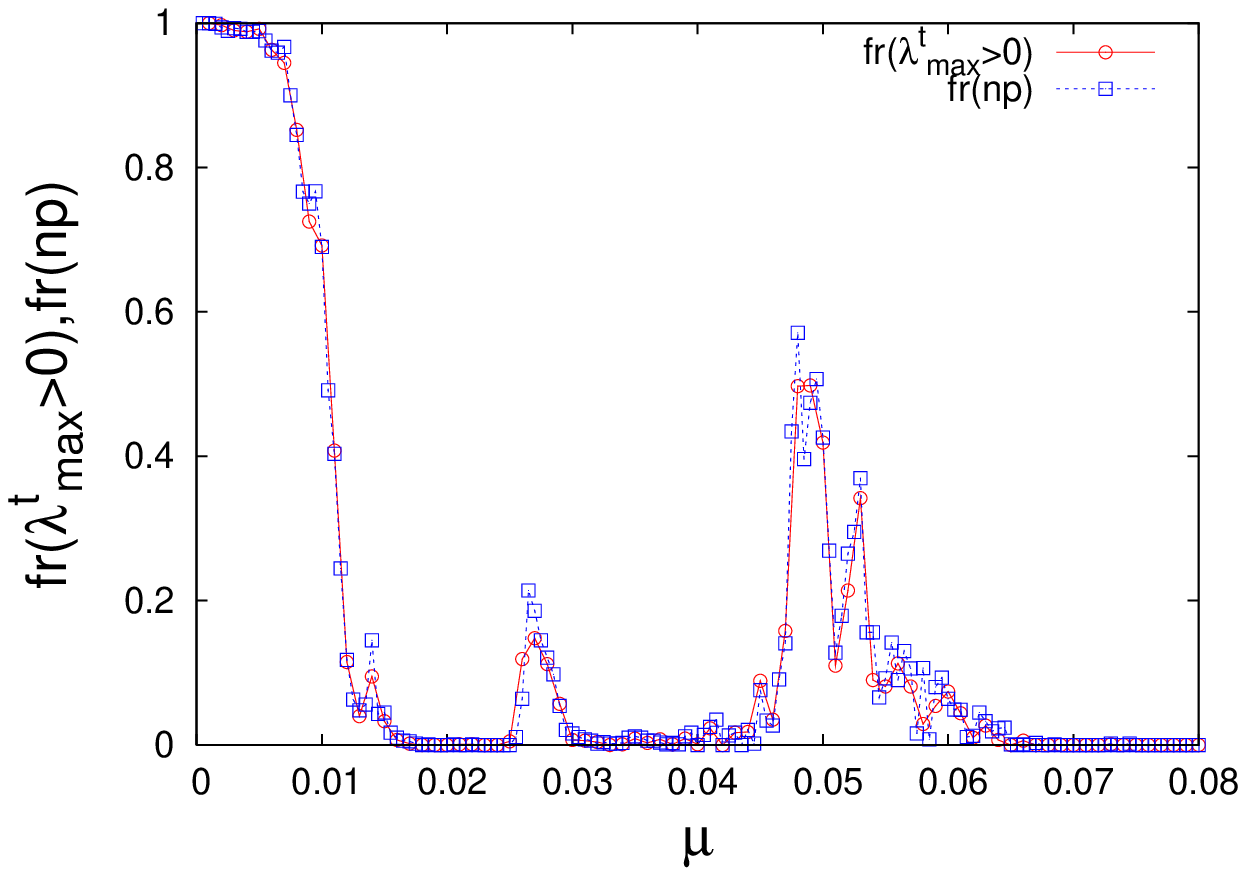} & 
\includegraphics[height=2.5in,width=3.12in]{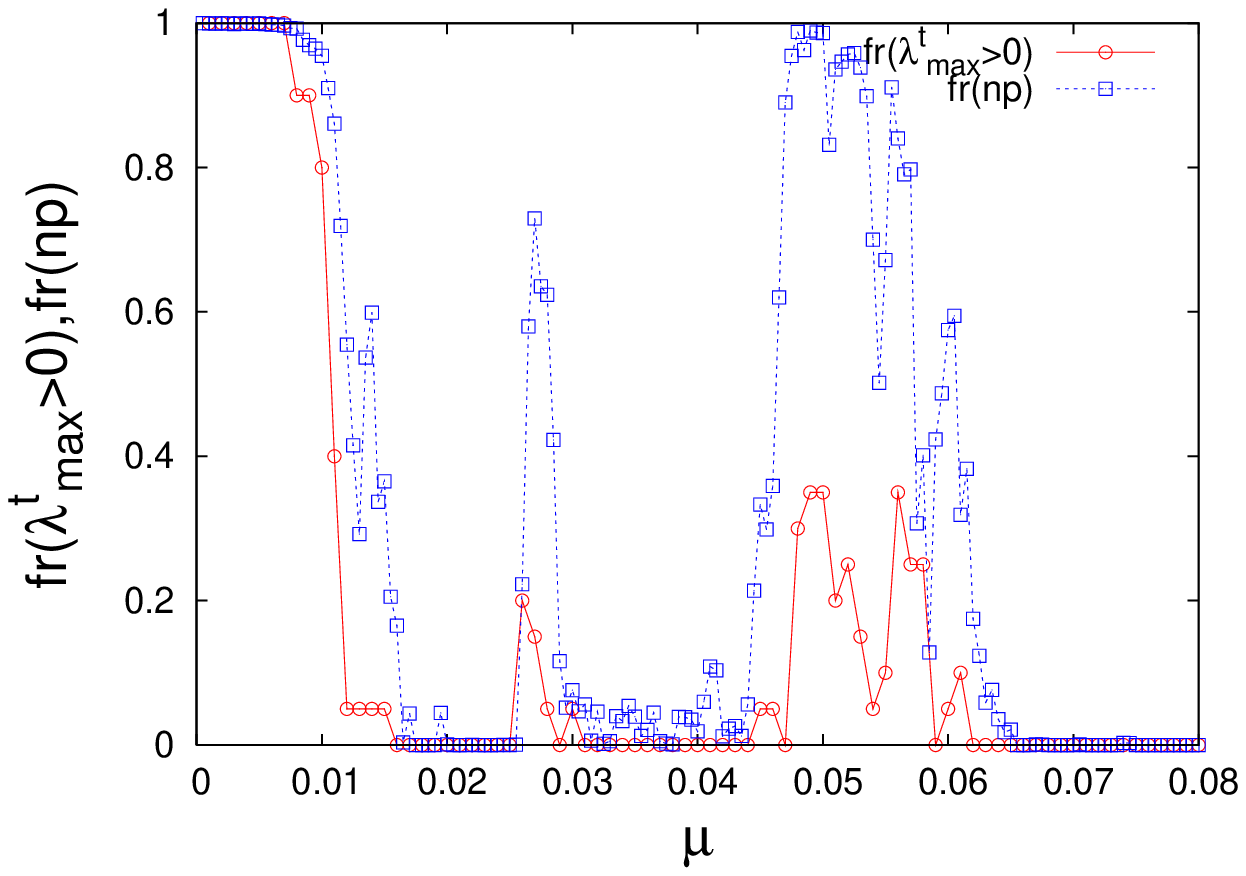} \\ 
\mbox{(a)} & \mbox{(b)} 
\end{array}$ 
\caption[Comparisons of the fractions of orbits having positive $\lambda_{max}^t$ and the fractions of non-periodic orbits]{Comparison of the fraction of orbits having positive $\lambda_{max}^t$ and the fraction of non-periodic orbits, in relation to the coupling strength $\mu$, considered over many initial conditions. 4-star's branch-node in (a), and an outer tree's node in (b).}  \label{fig-stability-nonperiodic}
\end{center}
\end{figure}

\textbf{FTMLE and Dynamical Regions.} The fraction of initial conditions leading to positive FTMLE describes the properties of motion of the specific node in function of the coupling strength, given that the negative FTMLE are basically equivalent to the periodic orbits. We examine the relationship between this fraction and the fraction of non-periodic orbits investigated previously (Figs.\,\ref{fig-nonperiodic}\,\&\,\ref{fig-nonperiodic-comparisons}), and used to define the dynamical regions. Such comparison establishes a quantitative result concerning the relationship between the dynamical regions and the stability of emergent motion within them. The profiles in Fig.\,\ref{fig-stability-nonperiodic} compare the results from Fig.\,\ref{fig-stability-comparison} and the results on non-periodic orbits for cases of 4-star's branch node and tree's outer node. 
Clearly, in case of the 4-star the overlap is almost perfect: essentially all the non-periodic orbits display positive FTMLE, while the periodic ones lead to negative FTMLE, as expected. In the case of tree the overlap is not precise, but the structure of dynamical regions is evident.

It appears the stability of motion of tree's outer node is inherently related to the stability of motion of 4-star's branch node, irrespectively of the coupling strength. Moreover, the characterization of motion of our network of CCM in terms of dynamical regions seems to be confirmed by the stability investigation. Dynamical regions can be equivalently characterized through FTMLE: periodic orbits exhibit negative FTMLE, while non-periodic orbits generally exhibit positive FTMLE. To be emphasized however, is that positive FTMLE in the region of attractors have far smaller values than the ones related to the chaotic orbits for very small $\mu$-values (cf. Fig.\,\ref{fig-colorstability}). As expected, this indicates that although unstable and non-periodic, the type of motion exhibited in the attractor region by 4-star's branch node is intrinsically different from the usual chaotic one. Furthermore, within the attractor region in Fig.\,\ref{fig-stability-nonperiodic}a (specifically, for $\mu \cong 0.048$) there appears to be a small fraction of orbits that are non-periodic, but display negative (or non-positive) FTMLE. This suggest the presence of new dynamical phenomena known as strange nonchaotic attractors, and will be examined in detail in the Sections to follow.

\section{Standard Maximal Lyapunov Exponents}

In this Section we analyze the stability of our system through Standard Maximal Lyapunov Exponents (SMLE) computed using the usual technique involving iterations of linearized system (Jacobian matrix) \cite{wiggy,ll}. Due to computational restrictions, we will examine only the dynamics of 4-star, comparing the results to the ones from the previous Section. As opposed to FTMLE computation model involving only a single node, SMLE is associated with the whole 4-star, considering it as an 8-dimensional dynamical system. SMLE give a precise characterization of coupled system's stability for a given initial condition, whose precision improves with number of iterations considered (in contrast to FTMLE where we had to determine to the optimal slope-length). An 8D dynamical system has 8 independent Lyapunov Exponents; for the purposes of this study we shall limit the discussion to the largest one among them only, called $\zeta$.

The following computational algorithm will be employed in order to evaluate the SMLE $\zeta$ associated with 4-star for a given coupling strength and initial conditions:
\begin{enumerate}
 \item select an initial point $\x_0$ in eight-dimensional phase space of the 4-star, and compute the 4-star's Jacobian at that point (defining 
linearization of the system in the neighborhood of $\x_0$):
    \begin{equation}
             J(\x_0)=  \left. \frac{\partial \mathbf{f} (\x)}{\partial \x} \right|_{\x=\x_0} ,  \label{jacobian}
    \end{equation} 
with $\mathbf{f}$ denoting the full 4-star's map:
    \begin{equation}
         \x_{t+1} = \mathbf{f} (\x_{t}) , \;\; \x \equiv (x[1,\hdots,4],y[1,\hdots,4]). \label{4stareq}
    \end{equation}
 \item chose a random 8D unitary vector $\uu_0$ and compute $\uu_1 = J(\x_0) \cdot \uu_0$. The value 
\[ \zeta_1 = \ln |\uu_1| \] 
represents the lowest order approximation to the SMLE and is recorded. Redefine $\uu_1$ by normalizing it: $\uu_1 \rightarrow \frac{\uu_1}{|\uu_1|}$.
 \item compute the next iteration of 4-star's dynamics $\x_1= \mathbf{f}(\x_0)$, the corresponding $J(\x_1)$, and  $\uu_2 = J(\x_1)\cdot \uu_1$. 
The value 
\[ \zeta_2 = \ln |\uu_2| \]  
is again recoded, and the vector $\uu_2$ is normalized  $\uu_2 \rightarrow \frac{\uu_2}{|\uu_2|}$. 
 \item steps 3 is repeated for the desired $t_{\zeta}$ number of iterations, improving the approximation of the final SMLE. The value
\[ \zeta = \frac{1}{t_{\zeta}} \sum_{k=1}^{k=t_{\zeta}} \zeta_k \]
is defined to be the SMLE for the point $\x_0$ approximated to the $t_{\zeta}$-order of the 4-star's dynamical system given by $\mathbf{f}$. 
\end{enumerate}
The value $\zeta = \zeta (\x_0)$ is related solely to the initial conditions $\x_0$ and independent from the choice of $\uu_0$, at the limit of $t_\zeta \longrightarrow \infty$. At this limit the value of $\zeta$ becomes the usual Infinite-time MLE characterizing a nonlinear dynamical system \cite{wiggy}. 

\begin{figure}[!hbt]
\begin{center}
$\begin{array}{cc}
\includegraphics[height=2.5in,width=3.25in]{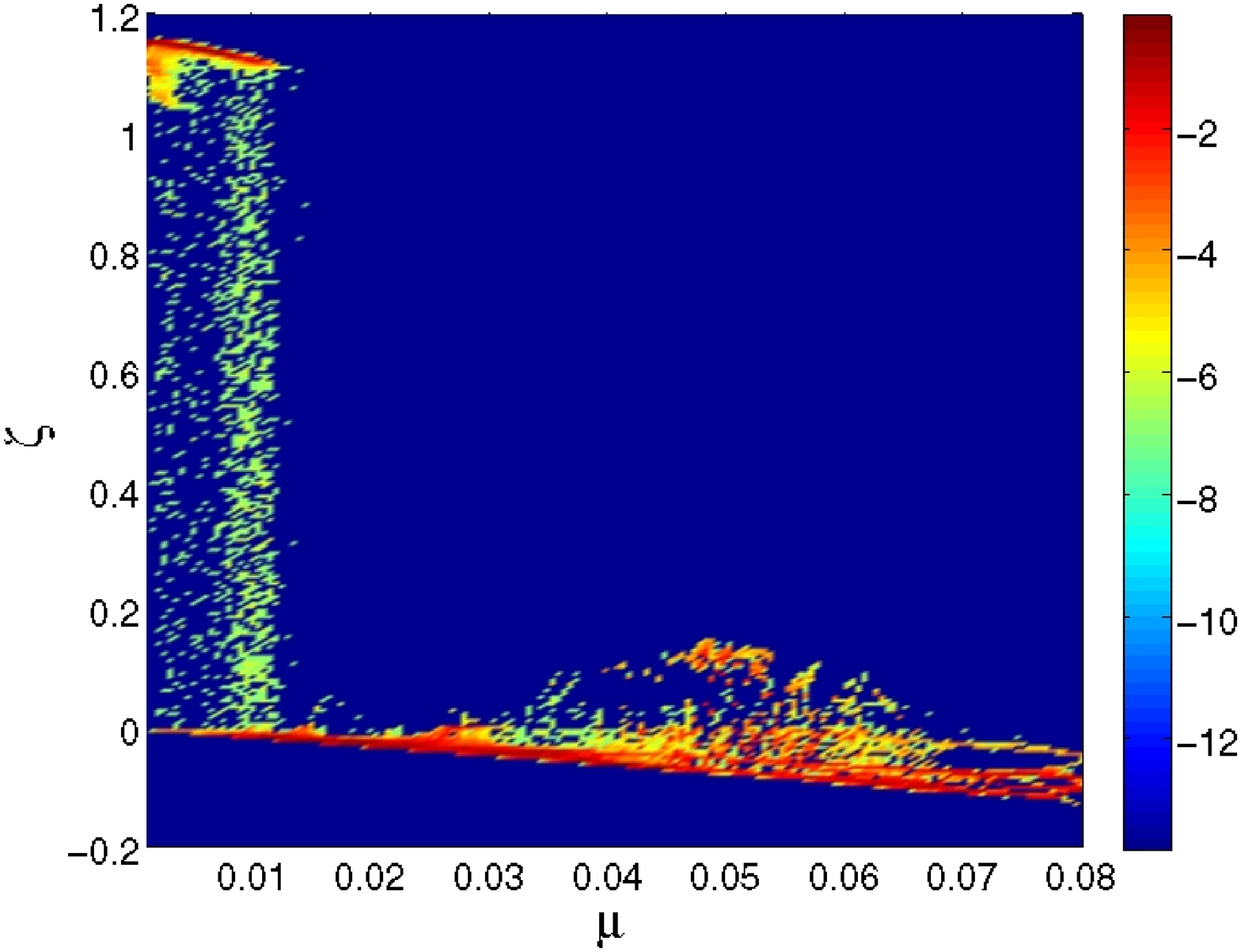} & 
\includegraphics[height=2.5in,width=3.25in]{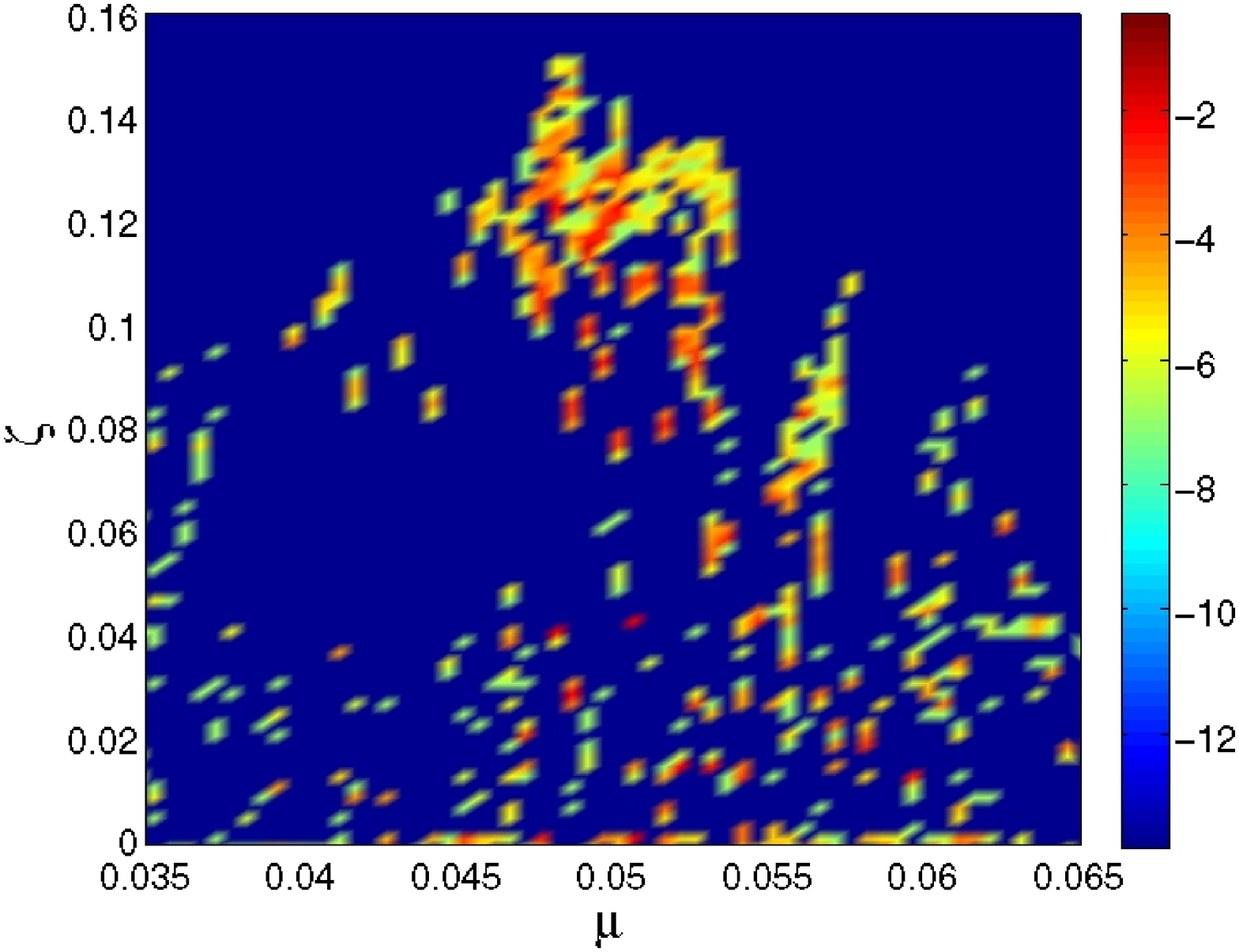} \\ 
\mbox{(a)} & \mbox{(b)} 
\end{array}$ 
\caption[2D color histogram of SMLE $\zeta$ in function of the coupling $\mu$ for the 4-star]{2D color histogram of SMLE $\zeta$ in function of the coupling $\mu$, over many initial conditions for all nodes of the 4-star in (a), and a zoom at the attractor region of positive $\zeta$ for the profile in (a), shown in (b).}  \label{fig-MLE-jacobian}
\end{center}
\end{figure}

SMLE will be hence used to fully characterize the dynamics of 4-star in relation to the coupling strength. In Fig.\,\ref{fig-MLE-jacobian}a we show the 2D color histogram of SMLE values in function of $\mu$ obtained over many initial conditions. 
The profile is complementary to the Fig.\,\ref{fig-colorstability}a providing additional stability analysis. The structure of three dynamical regions is visible, with values of SMLE being similar to the corresponding FTMLE-values. Also, the similarity between named profiles confirms FTMLE to be a valid approximation to the real infinite-time SMLE, in the context of single-node dynamics. We also report the zoom to the attractor region in Fig.\,\ref{fig-MLE-jacobian}b, showing a vast spectrum of dynamical behaviors occurring within this coupling interval. For each $\mu$-value there are certain possible motions (depending on the initial conditions), which does not seem to be changing continuously with $\mu$. This indicates 4-star to be extremely sensitive to both coupling strength and the initial conditions, exhibiting rich dynamical behavior in relation to both. In the next Section, we shall examine the motion for a specific coupling strength value. However, in order to reveal the full spectrum of motions, a much finer discretization of $\mu$-values would be required for the attractor region.

\begin{figure}[!hbt]
\begin{center}
\includegraphics[height=2.6in,width=3.5in]{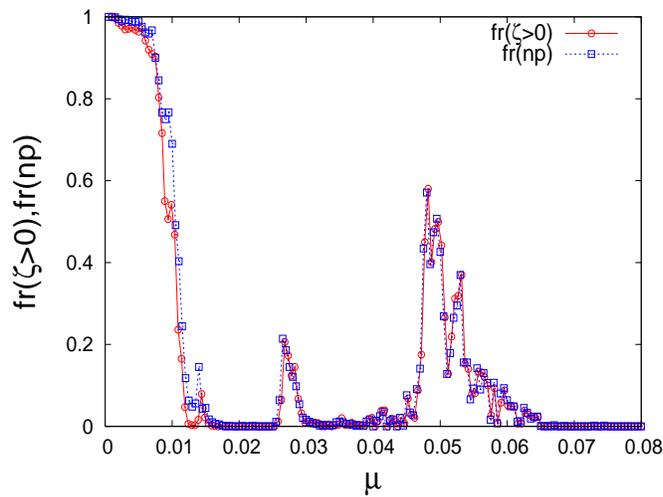}
\caption[Fraction of orbits leading to positive $\zeta$ vs. fraction of non-periodic orbits for 4-star]{Fraction of orbits leading to positive Standard Maximal Lyapunov Exponents $\zeta$ for all nodes of the 4-star vs. fraction of non-periodic orbits for 4-star, averaged over many initial conditions.}  \label{fig-jacobian-nonperiodic}
\end{center}
\end{figure}

In Fig.\,\ref{fig-jacobian-nonperiodic} we report the comparison of non-periodic orbits and positive SMLE for the 4-star, averaged over many initial  conditions. The profile is almost identical to the Fig.\,\ref{fig-stability-nonperiodic}a, re-confirming the general equivalence between non-periodic and unstable orbits for the 4-star. This further confirms the relationship between $\lambda^{t}_{max}$ and $\zeta$ in terms of their correspondence with unstable dynamics. 
The dynamical regions as intervals of coupling strength with distinctive motions are also recognized by SMLE. The positive values of $\zeta$-exponents in the attractor region are far smaller than the ones related to small $\mu$-values (similarly to the case of FTMLE). The study of 4-star via SMLE gave a precise characterization of its dynamics for the considered coupling range, which will be used further to investigate specific dynamical phenomena.

\section{Emergence of Weakly Chaotic Collective Dynamics}

In this Section we present a detailed analysis of the attractor dynamical region, examining dynamical patterns exhibited by the 4-star which display weak chaos. We will investigate the properties of strange attractors (already introduced in Fig.\,\ref{fig-rtstar}c\,\&\,d) appearing in this region, from the nonlinear dynamics viewpoint. For simplicity, we shall focus of a fixed coupling strength of $\mu=0.048$, which is rather typical for attractor dynamical region. We will also quantitatively characterize the quasi-periodic orbits present throughout all dynamical regions.\\[1.cm]

\begin{figure}[!hbt]
\begin{center}
$\begin{array}{cc}
\includegraphics[height=2.5in,width=3.15in]{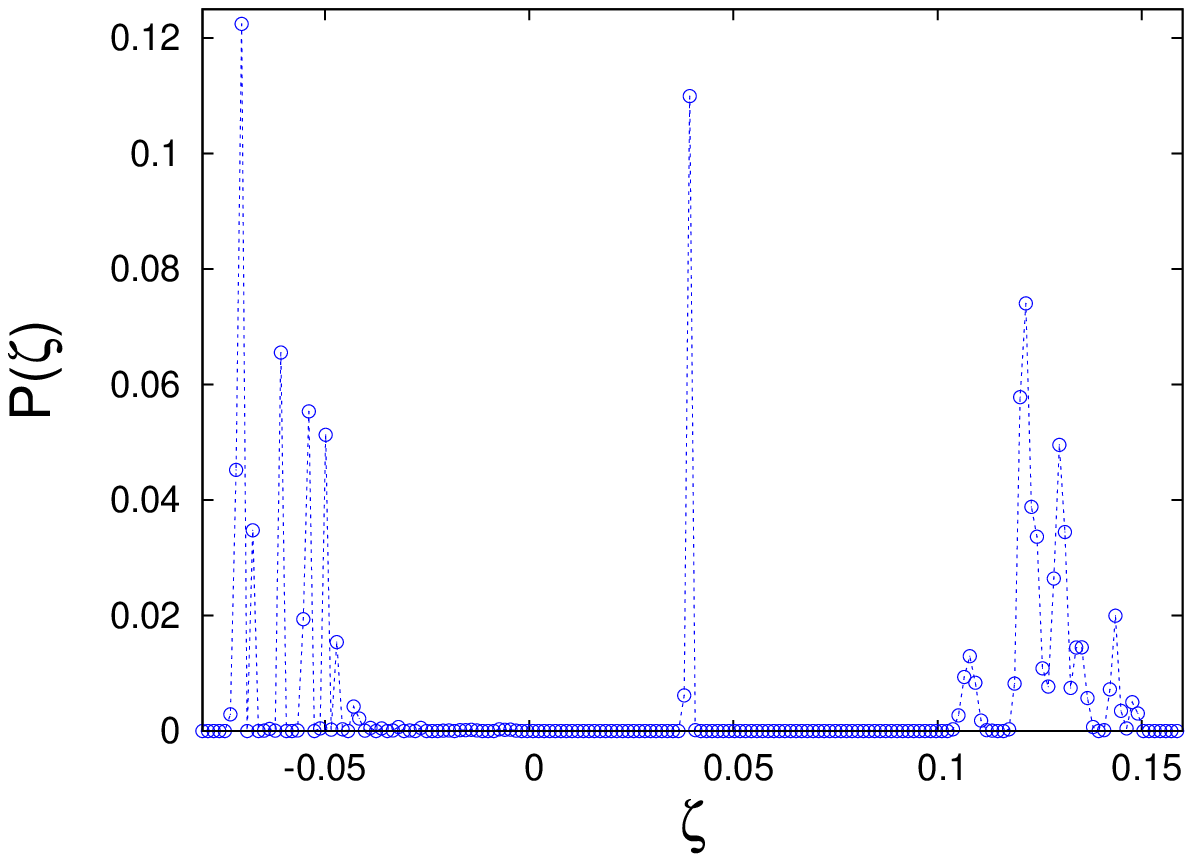} & 
\includegraphics[height=2.5in,width=3.15in]{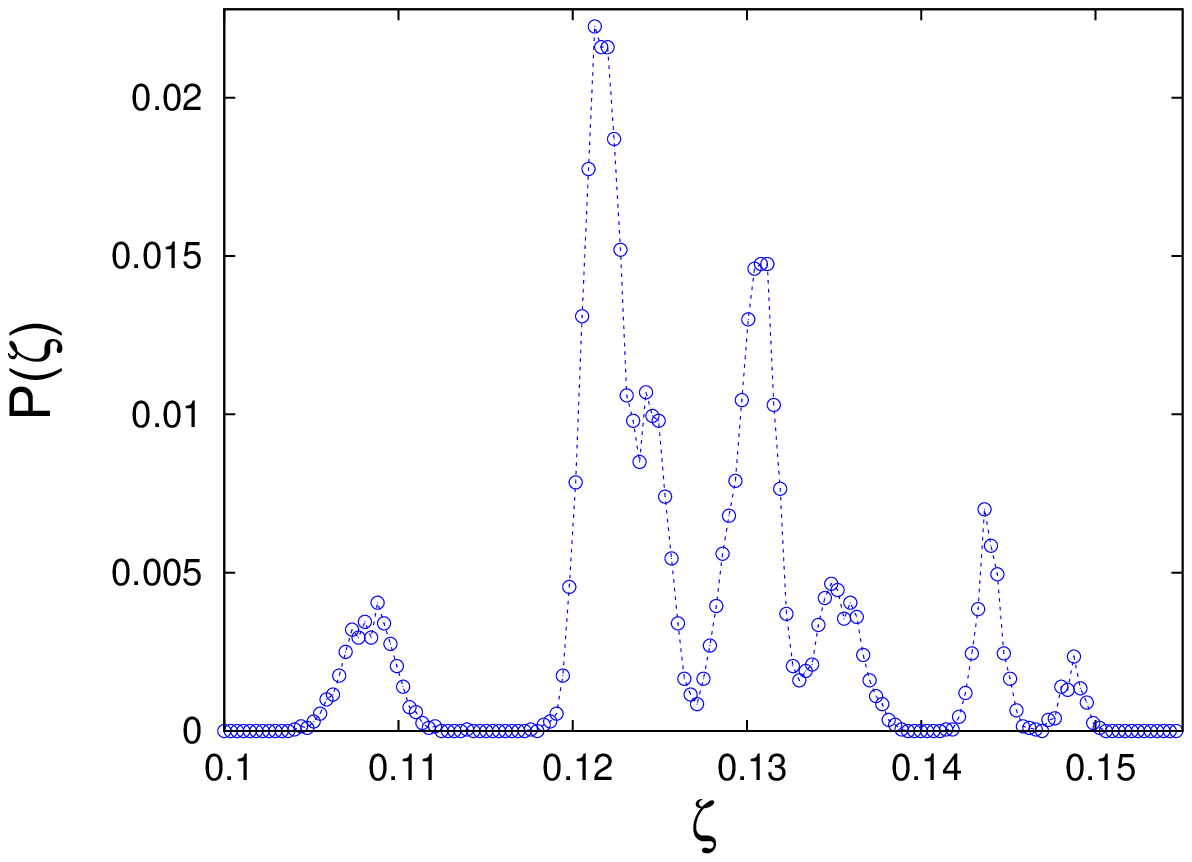} \\ 
\mbox{(a)} & \mbox{(b)} 
\end{array}$ 
\caption[Distribution of SMLE $\zeta$ for 4-star with $\mu=0.048$, with a zoom to the $\zeta \gtrsim 0.1$ part]{Distribution of SMLE $\zeta$ after $10^5$ iterations for many initial conditions for all nodes on the 4-star with $\mu=0.048$ in (a), and the zoom to the  $\zeta \gtrsim 0.1$ part of this distribution in (b).}  \label{fig-ly-048}
\end{center}
\end{figure}

\textbf{Dynamical Patterns of 4-star with $\mu=0.048$.} In Fig.\,\ref{fig-ly-048}a we show the distribution of SMLE $\zeta$ over many initial conditions for the 4-star with fixed coupling strength of $\mu=0.048$. Three main types of motion can be immediately recognized: periodic orbits having negative $\zeta$, strange attractors displaying positive $\zeta \gtrsim 0.1$, and another group of orbits with $\zeta \cong 0.038$. In Fig.\,\ref{fig-ly-048}b we show a zoom to the $\zeta \gtrsim 0.1$ part of the distribution clarifying between different attractor types having similar $\zeta$-exponents. 
Each peak in the distribution of SMLE $\zeta$ represents a group of initial conditions which lead to a certain dynamical pattern on the 4-star. Other $\mu$-values in the attractor region show similar profiles.

\begin{figure}[!hbt]
\begin{center}
$\begin{array}{ccc}
\includegraphics[height=1.85in,width=2.05in]{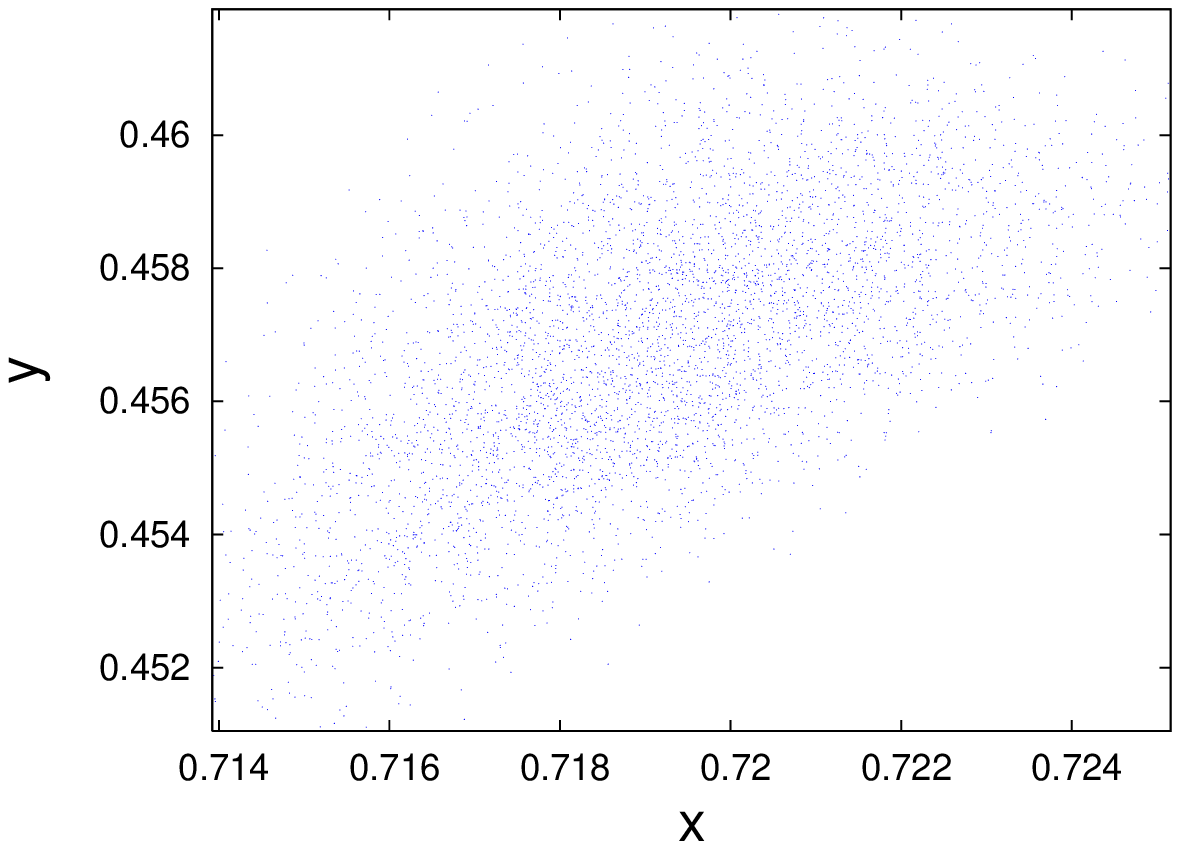} & 
\includegraphics[height=1.85in,width=2.05in]{attractor-starnode1-09-048.eps} & 
\includegraphics[height=1.85in,width=2.05in]{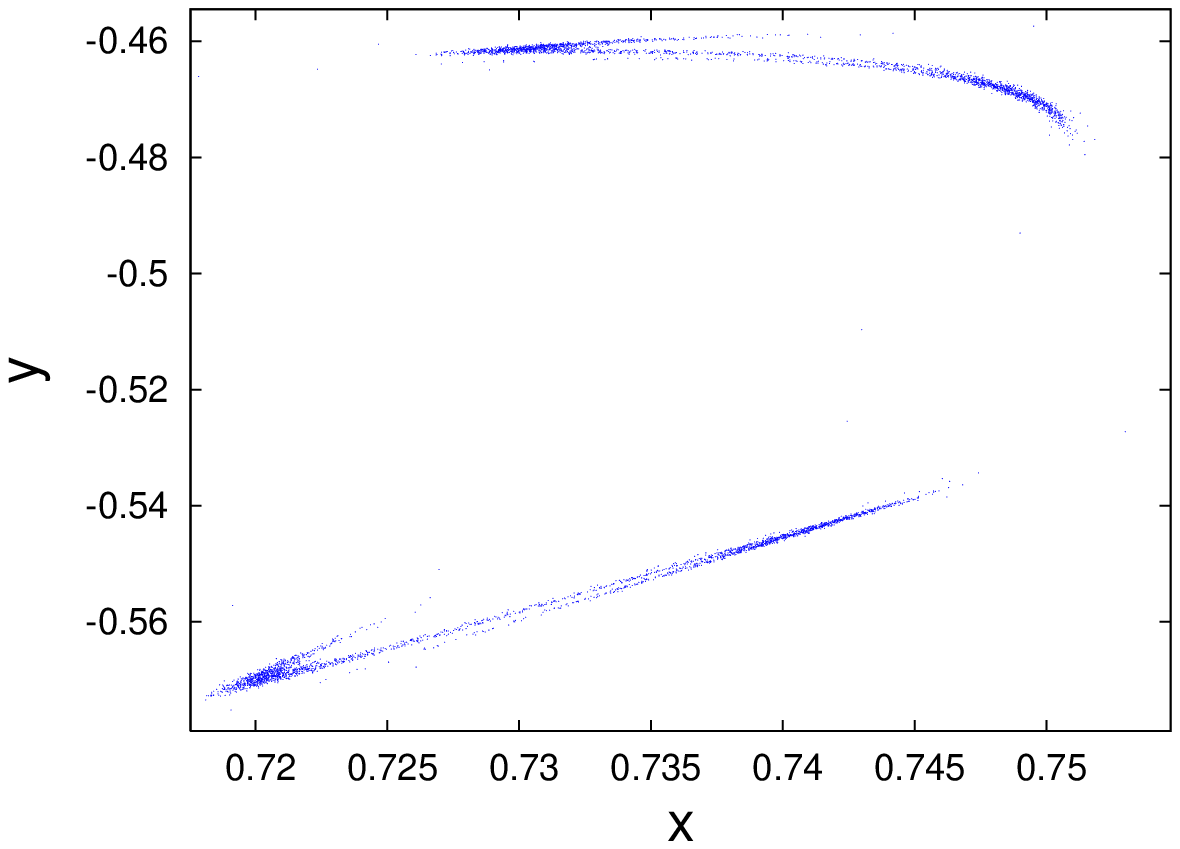} \\
\mbox{(a)} & \mbox{(b)} & \mbox{(c)}
\end{array}$ 
\caption[A dynamical pattern exhibited by 4-star's nodes with $\zeta \cong 0.144$]{Attractors exhibited by 4-star's nodes with $\zeta \cong 0.144$. Hub node in (a), and two branch nodes in (b) and (c). The fourth node's orbit is identical to either (b) or (c), depending on the initial conditions.}  \label{fig-attr-048}
\end{center}
\end{figure}

An attractor corresponding to one of the peaks in Fig.\,\ref{fig-ly-048}b ($\zeta \cong 0.144$) is shown in Fig.\,\ref{fig-attr-048}: the orbits exhibited by the hub and two branch nodes are shown, while the orbit of the third branch node overlaps with one of the first two branch nodes (depending on the initial conditions). Again, as the orbits have two parts, we are showing only a half of each orbit for clarity (see Fig.\,\ref{fig-orbitsexamples}c for illustration). Other peaks in Fig.\,\ref{fig-ly-048}b correspond to the similar situations that include all three branch nodes having attractors as either Fig.\,\ref{fig-attr-048}b or Fig.\,\ref{fig-attr-048}c, while the orbit of the hub node always resembles the one shown in Fig.\,\ref{fig-attr-048}a. 
The strange attractor in Fig.\,\ref{fig-attr-048}b is a proper fractal with fractal dimension $d_f \cong 1.4$, whereas orbits of other nodes (specially the hub) do not display this kind of phase space organization. This is a clear example of strange chaotic attractor as it possesses the usual properties that define it as such \cite{wiggy}.

\begin{figure}[!hbt]
\begin{center}
$\begin{array}{ccc}
\includegraphics[height=1.85in,width=2.05in]{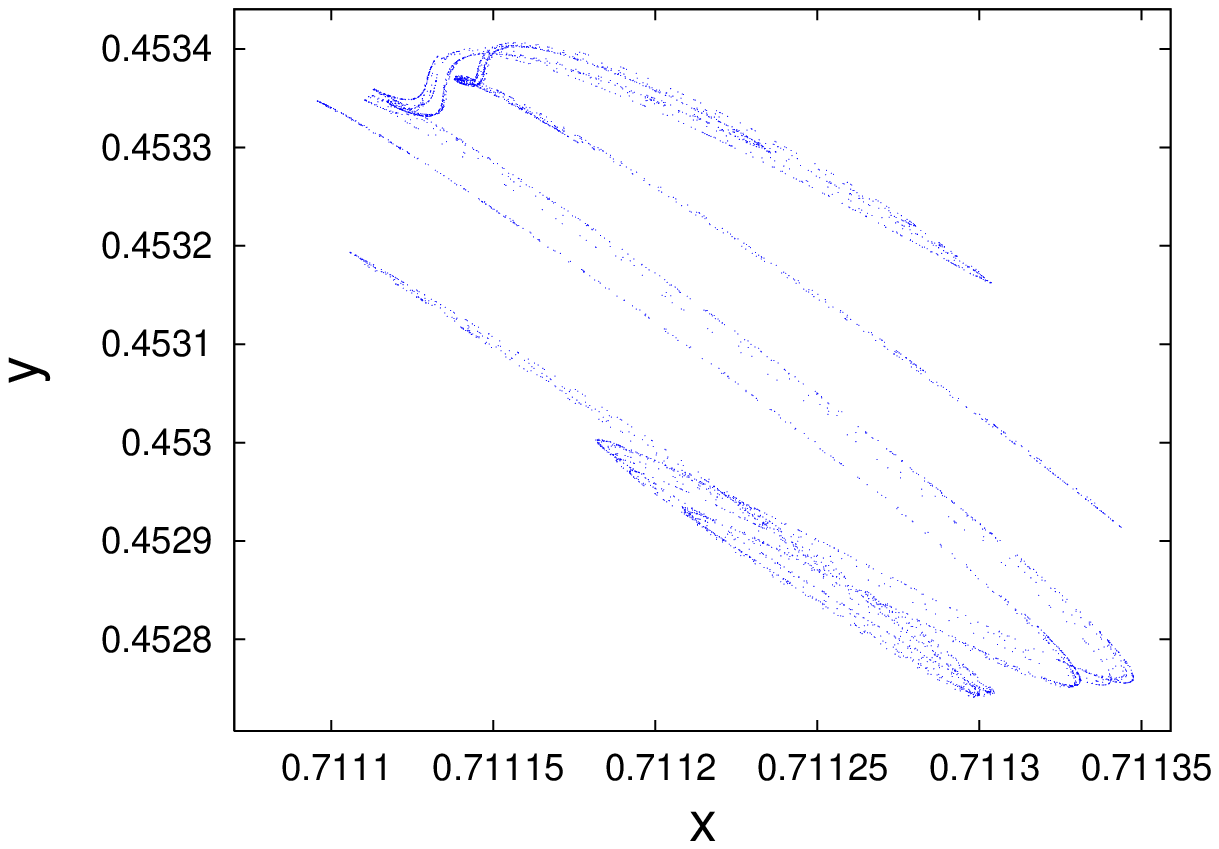} & 
\includegraphics[height=1.85in,width=2.05in]{SNA-starnode1-09-048.eps} & 
\includegraphics[height=1.85in,width=2.05in]{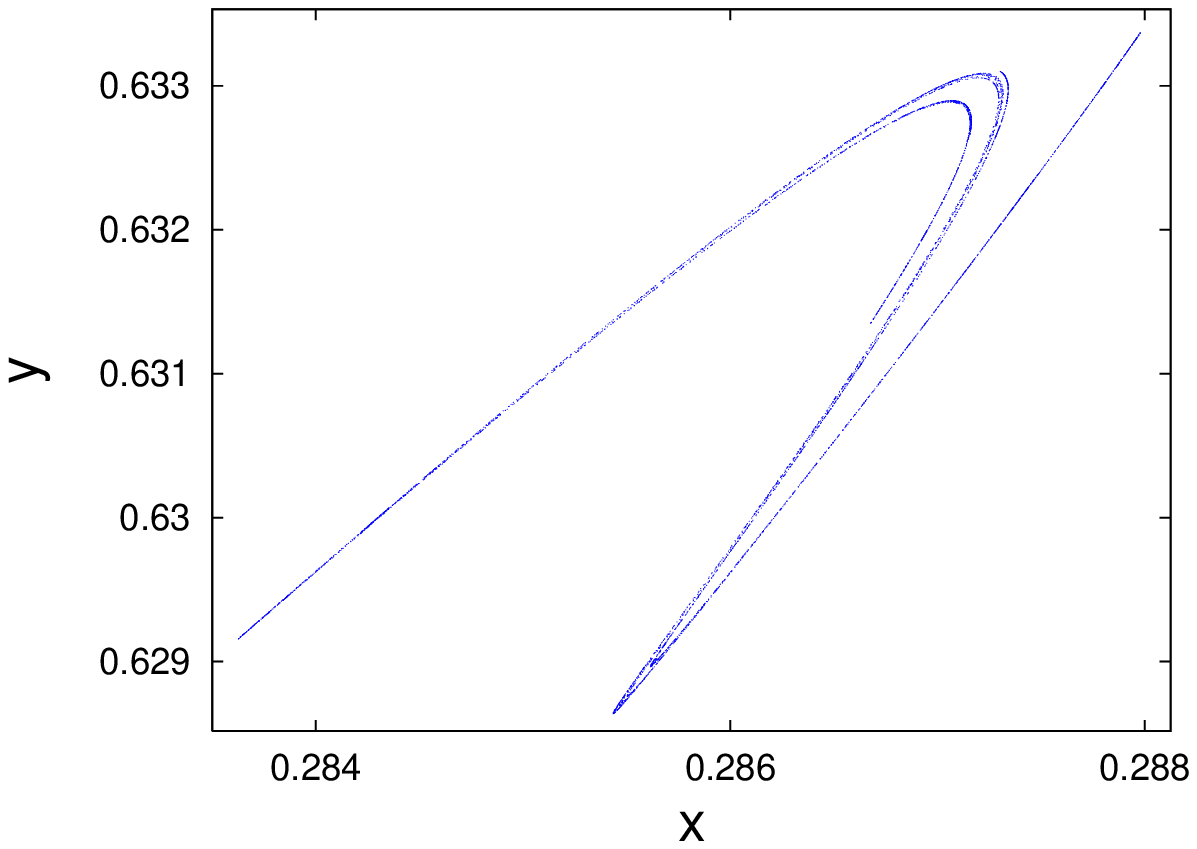} \\
\mbox{(a)} & \mbox{(b)} & \mbox{(c)}
\end{array}$ 
\caption[A dynamical pattern exhibited by 4-star's nodes with $\zeta \cong 0.038$]{Attractors exhibited by 4-star's nodes with $\zeta \cong 0.038$. Hub node in (a), and two branch nodes in (b) and (c). The fourth node's orbits is identical to either (b) or (c), depending on the initial conditions.}  \label{fig-SNA}
\end{center}
\end{figure}

The dynamical situations having their $\zeta$-exponents in the range shown in Fig.\,\ref{fig-ly-048}b are in opposition with the dynamical pattern occurring for $\zeta \cong 0.038$ (central peak in Fig.\,\ref{fig-ly-048}a). Here, all four nodes exhibit strange attractors (with similar fractal dimensions of $d_f \cong 1.4$) that are typically structured as shown in Fig.\,\ref{fig-SNA}. 
Depending on initial conditions leading to this dynamical pattern, two branch node's orbits overlap as either Fig.\,\ref{fig-SNA}b or Fig.\,\ref{fig-SNA}c; regardless of this, the $\zeta$-exponent always converges to $\zeta \cong 0.038$ as $t_\zeta \longrightarrow \infty$. This situation displays much more dynamical organization in terms of inter-node correlation of motion, and it is endemic to the 4-star (as opposed to previously examined strange attractor in Fig.\,\ref{fig-attr-048}b which appears on the outer nodes of the tree as well). Although all nodes are exhibiting fractal attractors, the corresponding SMLE is still relatively small (cf. Fig.\,\ref{fig-ly-048}a). Similar occurrences of all nodes displaying strange attractors are found throughout attractor dynamical region for other $\mu$-values as well, with $\zeta$-exponent always remaining $\zeta \lesssim 0.04$.

\begin{figure}[!hbt]
\begin{center}
$\begin{array}{cc}
\includegraphics[height=2.5in,width=3.15in]{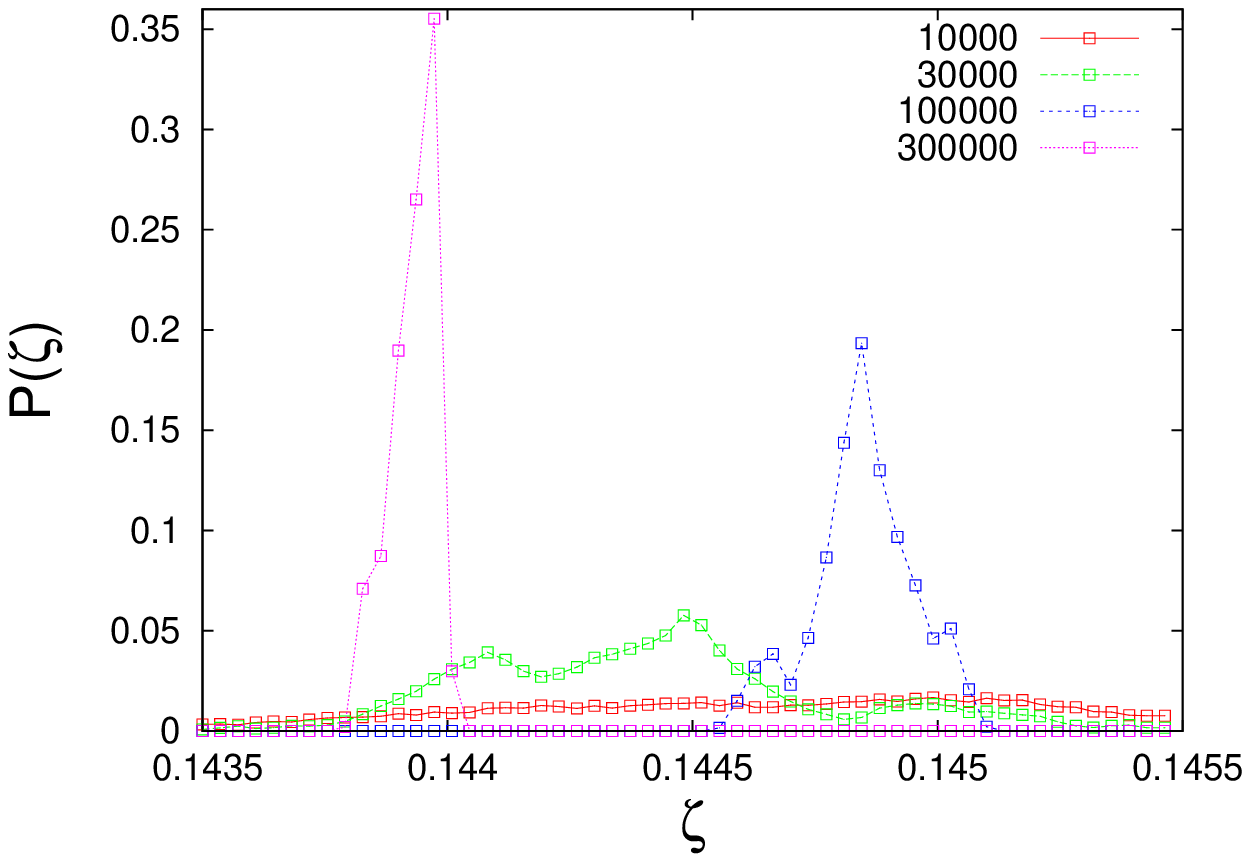} &
\includegraphics[height=2.5in,width=3.15in]{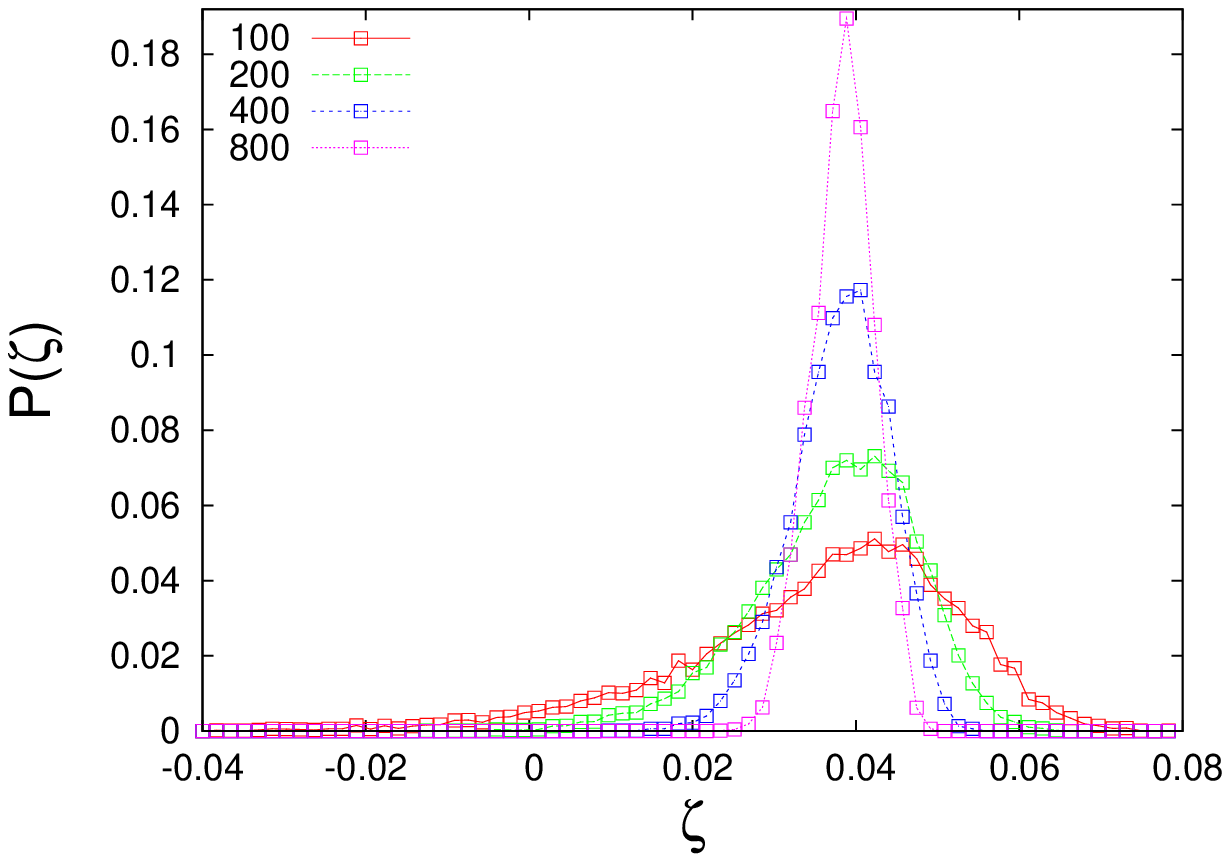} \\
\mbox{(a)} & \mbox{(b)} 
\end{array}$
\caption[Distributions of $\zeta$-values for the 4-star and for different evaluation times $t_\zeta$ for two dynamical situations]{Distribution of  Standard Maximal Lyapunov Exponents $\zeta$ for all 4-star's nodes and  different evaluation times $t_\zeta$, starting from various points on the attractor. Dynamical situation from Fig.\,\ref{fig-attr-048} in (a), and from Fig.\,\ref{fig-SNA} in (b).}  \label{fig-sna-ly-hists}
\end{center}
\end{figure}

These two dynamical patterns were already investigated in terms of their return times distributions, confirming them to be examples of dynamical self-organization (see Fig.\,\ref{fig-rtstar}b). Here we add a convergence study of the SMLE related to the dynamical patterns from Figs.\,\ref{fig-attr-048}\,\&\,\ref{fig-SNA}. In Fig.\,\ref{fig-sna-ly-hists} we show the distribution of $\zeta$-exponents in function of the evaluation time $t_\zeta$.
As expected, in the case of attractor Fig.\,\ref{fig-attr-048} the convergence pattern is quite irregular with final $\zeta$-value clearing only after $t_\zeta \sim O(10^5)$ iterations, whereas for the attractor Fig.\,\ref{fig-SNA} the convergence is much more smooth and fast. Although unstable, the dynamical pattern described in Fig.\,\ref{fig-SNA} is far more dynamically structured, and merits further investigation as a clear example of self-organization in 4-star.\\[0.1cm]

\begin{figure}[!hbt]
\begin{center}
$\begin{array}{cc}
\includegraphics[height=2.5in,width=3.15in]{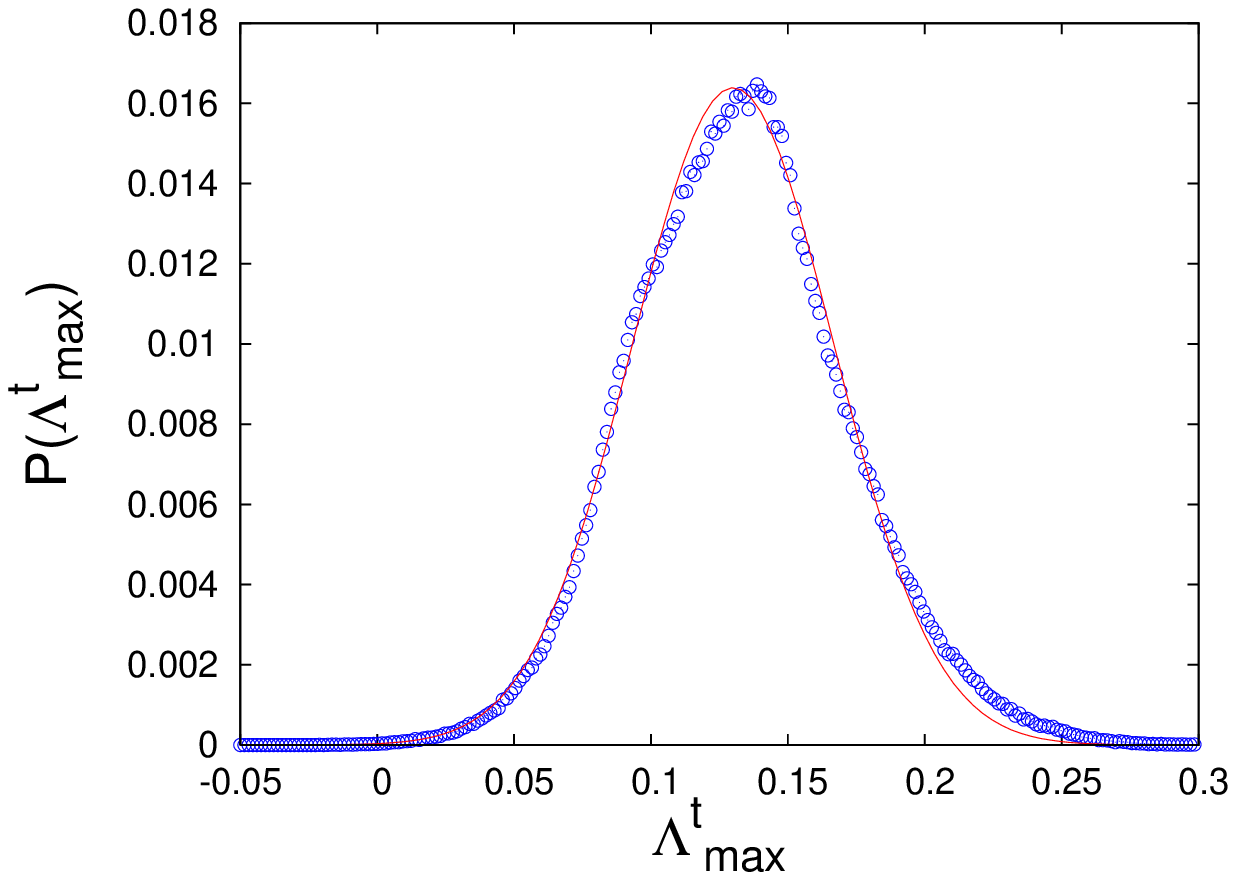} & 
\includegraphics[height=2.5in,width=3.15in]{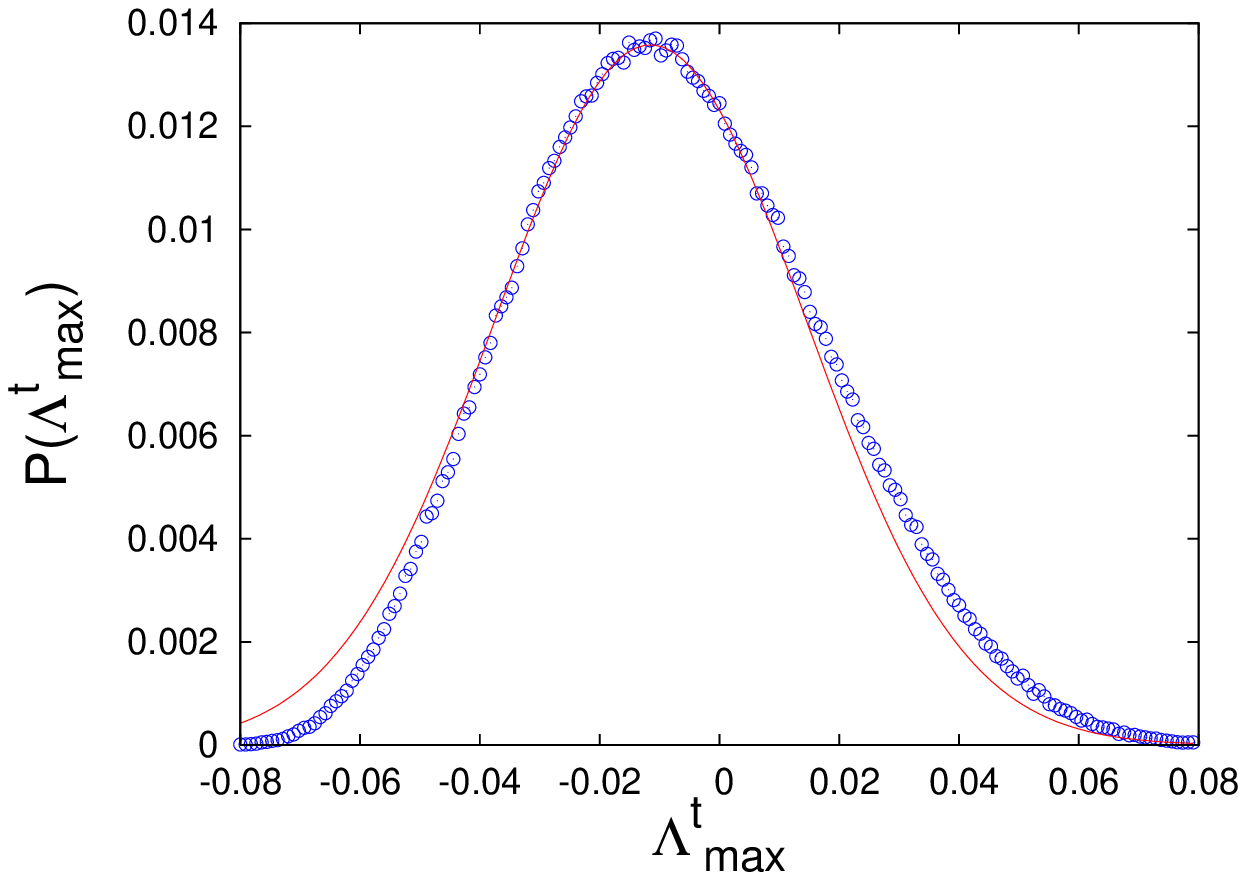} \\
 \mbox{(a)} & \mbox{(b)} 
\end{array}$ 
\caption[Distributions of FTMLE $\Lambda_{max}^t$ two strange attractors from Figs.\,\ref{fig-attr-048}b\,\&\,\ref{fig-SNA}b]{Distributions of FTMLE $\Lambda_{max}^t$ for the strange attractors from Figs.\,\ref{fig-attr-048}b\,\&\,\ref{fig-SNA}b referring to branch-nodes of 4-star, considered over many initial conditions in (a) and (b), respectively. The distributions mean values $\lambda_{max}^t= < \Lambda_{max}^t >_{\x}$ are $\lambda_{max}^t=0.14$ and $\lambda_{max}^t=-0.01$ for (a) and (b), respectively.}   \label{fig-ftmle-distributions}
\end{center}
\end{figure}

\textbf{Additional Analysis of Dynamical Pattern from Fig.\,\ref{fig-SNA}.}  In addition to return times study in Fig.\,\ref{fig-rtstar}b, we consider again the single node strange attractors shown in Figs.\ref{fig-attr-048}b\,\&\,\ref{fig-SNA}b that are exhibited by a 4-star's branch node at the coupling strength of $\mu=0.048$. The corresponding SMLE are computed above and amount to $\zeta=0.144$ and $\zeta=0.038$, respectively. In Fig.\,\ref{fig-ftmle-distributions} we report the distribution of FTMLE $\Lambda_{max}^t$ for the mentioned attractors. The precise values (for infinite computation time $t_\zeta$) of SMLE $\zeta$ lie within the distribution ranges of FTMLE and the distribution for Fig.\,\ref{fig-SNA}b is much more narrow (similarly to Fig.\,\ref{fig-sna-ly-hists}). The centers of distributions from Fig.\,\ref{fig-ftmle-distributions} are respectively equal to $\lambda_{max}^t=0.118$ and $\lambda_{max}^t=-0.005$ for both attractors.
Despite FTMLE being less precisely measured (due to need of slope-length optimization), they still have a precise dynamical meaning: estimate the initial divergence of orbits related to a specific node/trajectory, as opposed to SMLE which gives an overall stability approximation for the entire system. In particular, while the FTMLE distribution in Fig.\,\ref{fig-ftmle-distributions}a confirms the unstable character of the strange attractor from Fig.\ref{fig-attr-048}b, the distribution in Fig.\,\ref{fig-ftmle-distributions}b points to a possibility of strange attractor from Fig.\,\ref{fig-SNA}b exhibiting some form of stable behavior. This is to say that stable behavior might exist in a 2D subspace of 4-star's 8D phase space. The MLE distribution having both positive and negative values indicates presence of stable attractor points together with unstable attractor points.

The strange attractors having similar kinds of MLE properties are referred to as \textit{strange nonchaotic attractors} (SNA) \cite{prasadnegirama,feudel}, and are characterized by having non-positive MLE while still being geometrically strange (e.g. having fractal phase space organization). Their properties and routes to their generation have been extensively studied over the last decade \cite{rama,rama99}, revealing them as a peculiar dynamical phenomena. However, their appearance have so far been related exclusively to the dynamical systems with external driving. The attractor in Fig.\,\ref{fig-SNA}b does possess fractal geometrical properties and shows a SMLE of $\zeta \cong 0.038$, but on a short time-scale seems to exhibit a mixed unstable/stable nature with respect to the small perturbations. As this indicates a possibility of attractor from Fig.\,\ref{fig-SNA}b having SNA properties, we will employ some known methods used to test and describe SNA in order to obtain a better characterization, following \cite{ja-jsm}. 

\begin{figure}[!hbt]
\begin{center}
$\begin{array}{cc}
\includegraphics[height=2.5in,width=3.1in]{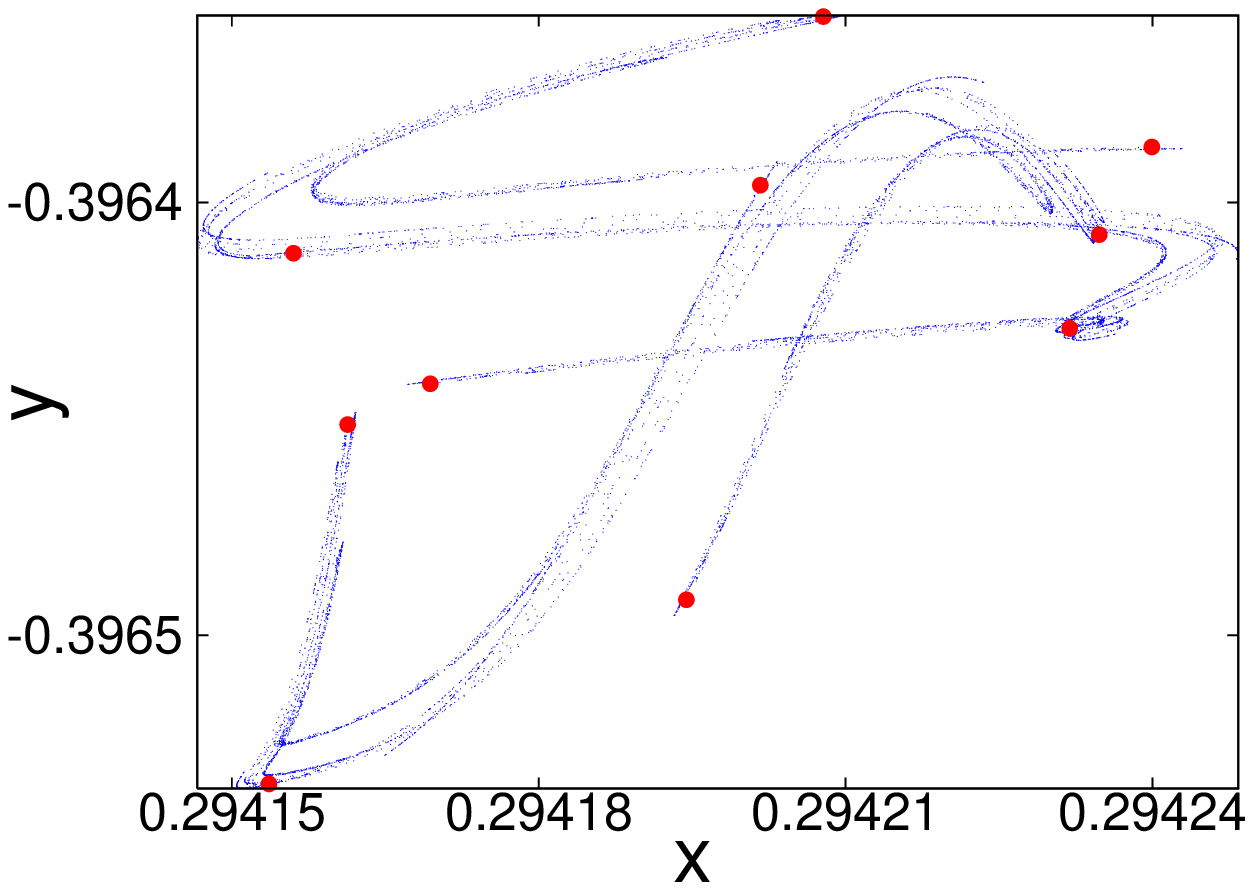} &
\includegraphics[height=2.5in,width=3.1in]{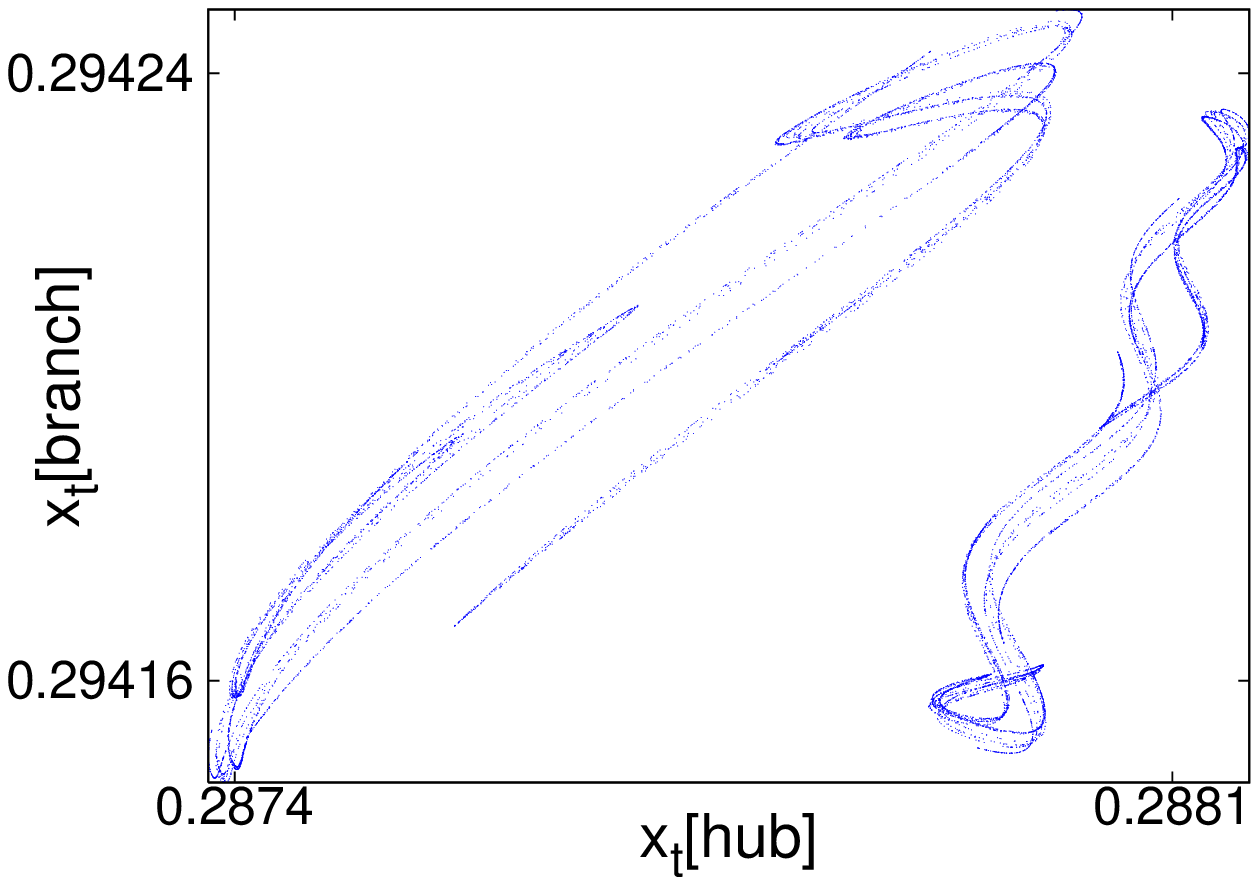} \\
\mbox{(a)} & \mbox{(b)}  \\
\includegraphics[height=2.5in,width=3.1in]{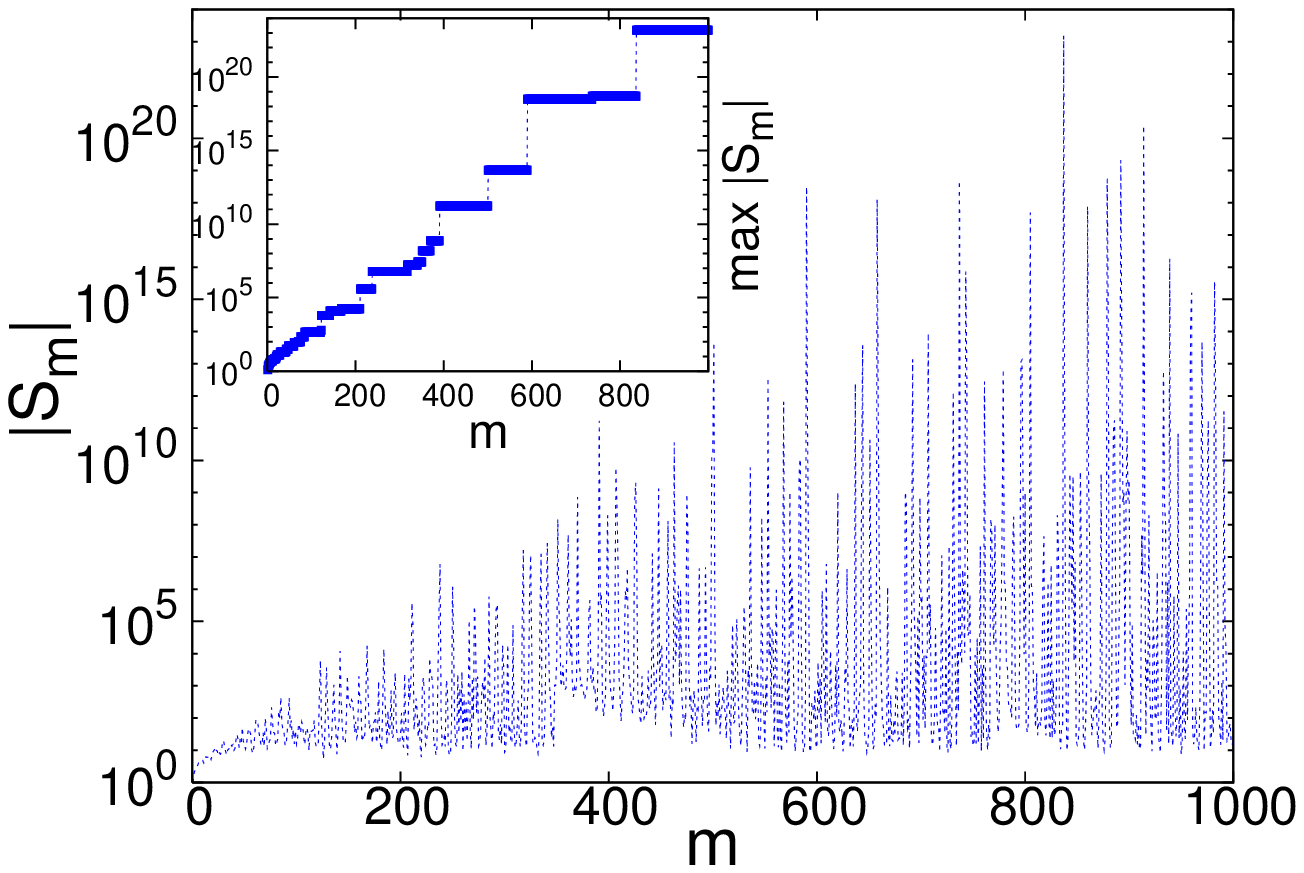} & 
\includegraphics[height=2.5in,width=3.1in]{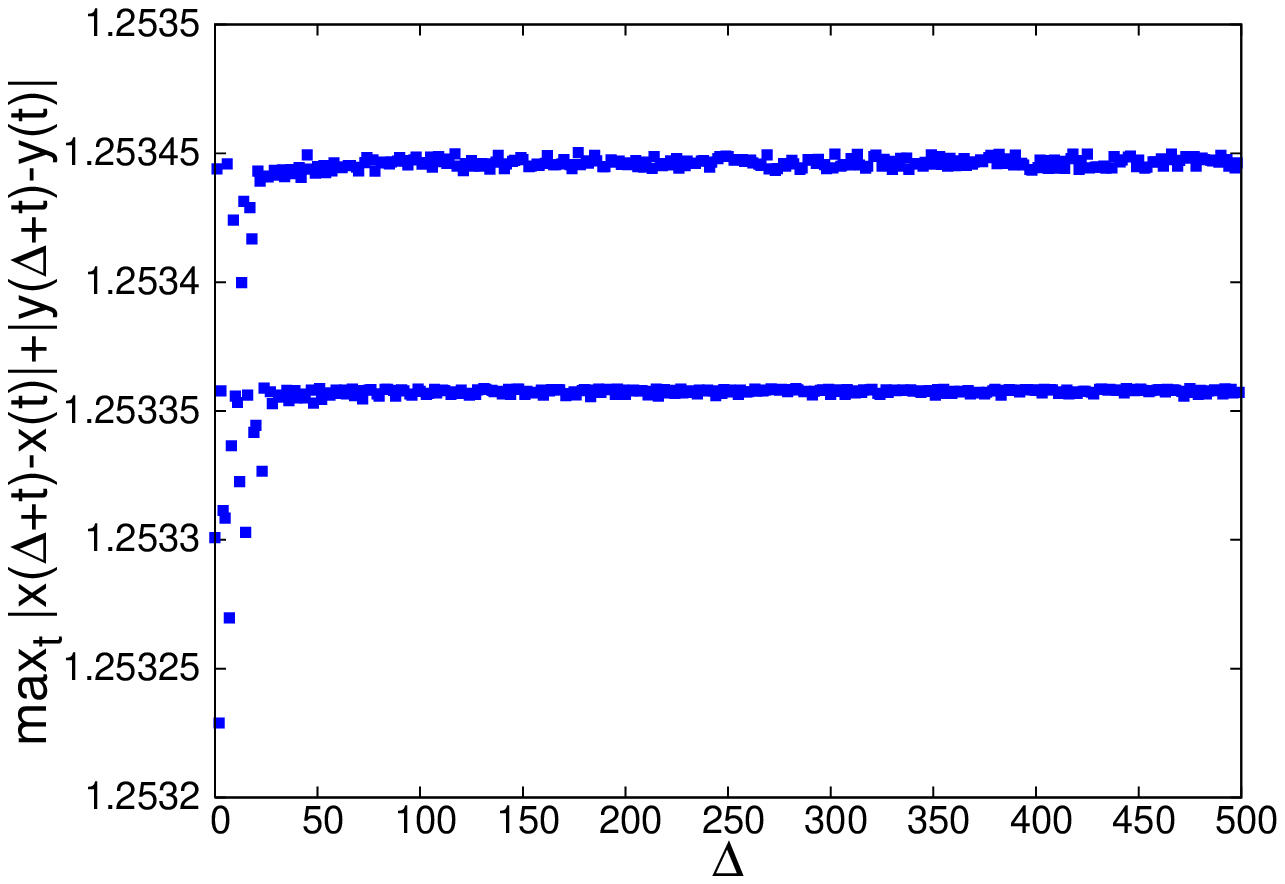} \\
\mbox{(c)} & \mbox{(d)} 
\end{array}$ 
\caption[Detailed characterization of the attractor from Fig.\,\ref{fig-SNA}b]{Properties of the attractor from Fig.\,\ref{fig-SNA}b: a scatter of 10 successive points in (a), phase of the branch node $x_t[branch]$ plotted against the phase of coupled hub node  $x_t[hub]$ in (b), partial sums $\vert S_m\vert$ and their maxima $max\vert S_m\vert$ (inset) in function of the number of terms $m$ in (c), the maximal distance on the attractor against the distance between the iterations $\Delta$.}
       \label{fig-SNA-details}
\end{center}
\end{figure}

We firsty observe that this attractor, in accordance with its fractal structure, does not present phase space continuity, as we demonstrate in Fig.\,\ref{fig-SNA-details}a by overposing 10 successive trajectory iterations to the attractor picture. The apparently uncorrelated successive iterations of the map still create a highly structured attractor. 4-star is an endogenously driven system: hub node and branch node drive each other, with a time-delayed interaction, as pictured in Fig.\,\ref{fig-4star}. The attractor from Fig.\,\ref{fig-SNA}b appears on the 4-star's branch node which is driven by the hub node that exhibits attractor shown in Fig.\,\ref{fig-SNA}a. In Fig.\,\ref{fig-SNA-details}b we report the mutual phase-dependence between the two attractors, showing its structure that clearly indicates the absence of a smooth curve, which is a typical feature of SNA \cite{pikovskyfeudel}. 
Following SNA characterization methods exposed in \cite{pikovskyfeudel}, we test the non-differentiability of the attractor from Fig.\,\ref{fig-SNA}b by considering the properties of the time series $F_h^{t} \equiv \partial x_t[branch]/\partial x_t[hub]$. In particular, one can define the sum \cite{feudel}:
\begin{equation}
S_m = \sum _{k=1}^{k=m} F_h^{k-1} \times \exp\{\Lambda^t_{max} (m-k)\}  \label{pikovskysum}
\end{equation} 
which examines the named time series taking into account the relationship between positive and negative tails in the MLE distribution, as the one shown in Fig.\,\ref{fig-ftmle-distributions}b. We numerically compute the $F_h^{t}$ and show the partial sum series $S_m$ in Fig.\,\ref{fig-SNA-details}c, along with the maximal $S_m$ values reported in the inset. Both profiles seem to indicate the partial sum series $S_m$ grows unbounded with the number of terms $m$, which points to a non-smooth non-differentiable attractor, not excluding SNA. Finally, an additional feature that attractor from Fig.\,\ref{fig-SNA}b shares with the known SNA in non-periodically driven maps \cite{pikovskyfeudel} is the saturation of the maximal distance between the attractor points, independently of the time gap between them. In Fig.\,\ref{fig-SNA-details}d we show the maximal distance between two points on the attractor in function of the time gap $\Delta$ separating them in terms of iterations. For each time gap $\Delta$ we are looking for the maximal distance that two points separated by $\Delta$ can attain. Besides depending on the part of the attractor (recall that Fig.\,\ref{fig-SNA}b is only a part of the total attractor of this node), the profiles are constant, implying that points with arbitrary time gap can be found at the opposite sides of the attractor, in agreement with SNA test in \cite{pikovskyfeudel}.

The SNA are so far exclusively found in externally driven dynamical systems (maps), and are understood as a consequence of non-periodic driving \cite{feudel,prasadnegirama,negirama}. Our 4-star system is however not externally driven, but it includes endogenous driving in the form of time-delayed interaction between the branch-node and the hub (i.e. between the branch node and the rest of the 4-star). Nevertheless, as shown above, the dynamical pattern from Fig.\,\ref{fig-SNA} and specifically the strange attractor shown in Fig.\,\ref{fig-SNA}b, display clear SNA properties in accordance with the usual tests \cite{feudel,pikovskyfeudel}. In particular, these properties are:
\begin{itemize}
\item non-integer fractal dimension $d_f \cong 1.4$ 
\item positive and negative tails in the distribution of FTMLE $\Lambda_{max}^t$ shown in Fig.\,\ref{fig-ftmle-distributions}b 
\item non-differentiability demonstrated through profile of partial sum series $S_m$ from Eq.(\ref{pikovskysum}) shown in Fig.\,\ref{fig-SNA-details}c
\item absence of phase space continuity of the attractor shown in Fig.\,\ref{fig-SNA-details}a
\item saturation of the maximal distance of the points belonging to the attractor shown in Fig.\,\ref{fig-SNA-details}d
\end{itemize}
Our study of SMLE which considers 4-star as 8D dynamical system, showed small positive $\zeta$-values for this dynamical pattern, which suggests the global dynamics of the 4-star to be weakly chaotic. However, all the evidence presented (Fig.\,\ref{fig-SNA-details}) for the case of particular branch-node exhibiting strange attractor shown in Fig.\,\ref{fig-SNA}b, indicates that this might represent an example of SNA in an endogenously driven extended dynamical system.\\[0.1cm]

\begin{figure}[!hbt]
\begin{center}
$\begin{array}{ccc}
\includegraphics[height=1.85in,width=2.05in]{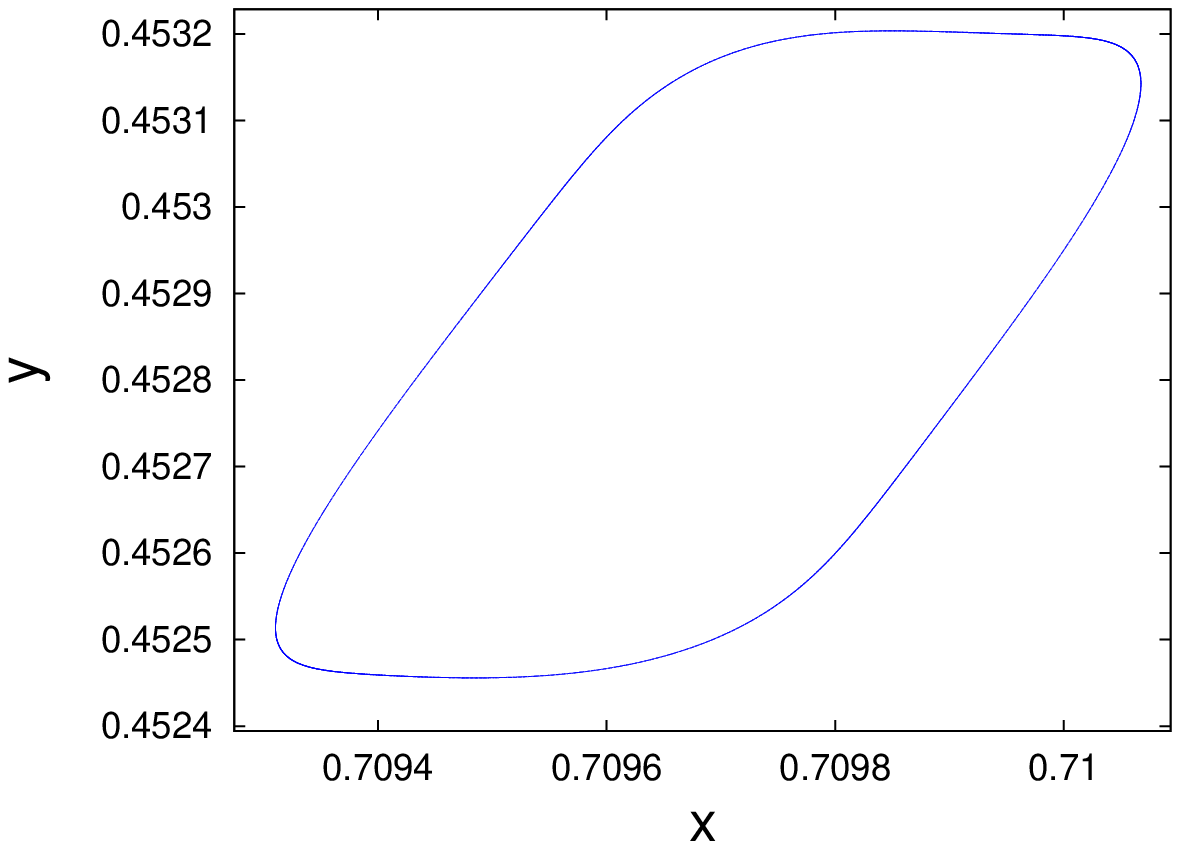} & 
\includegraphics[height=1.85in,width=2.05in]{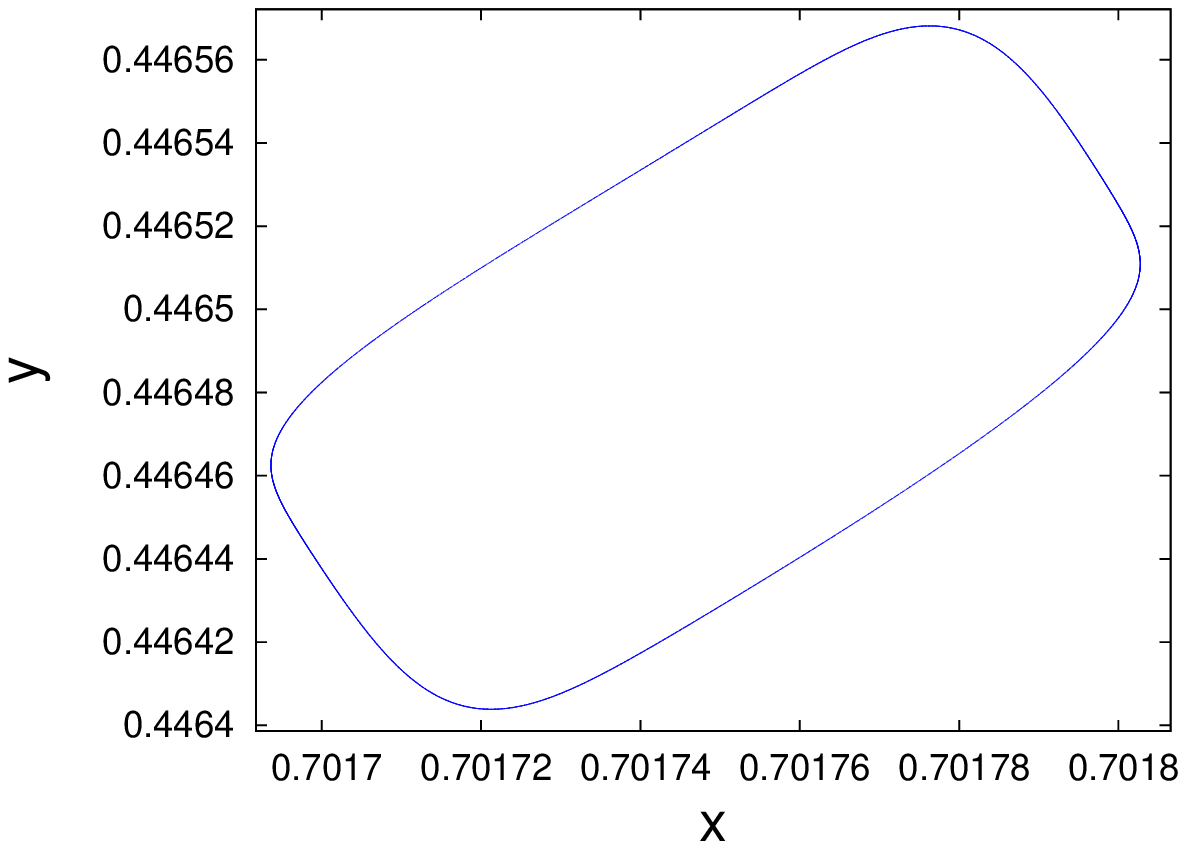} & 
\includegraphics[height=1.85in,width=2.05in]{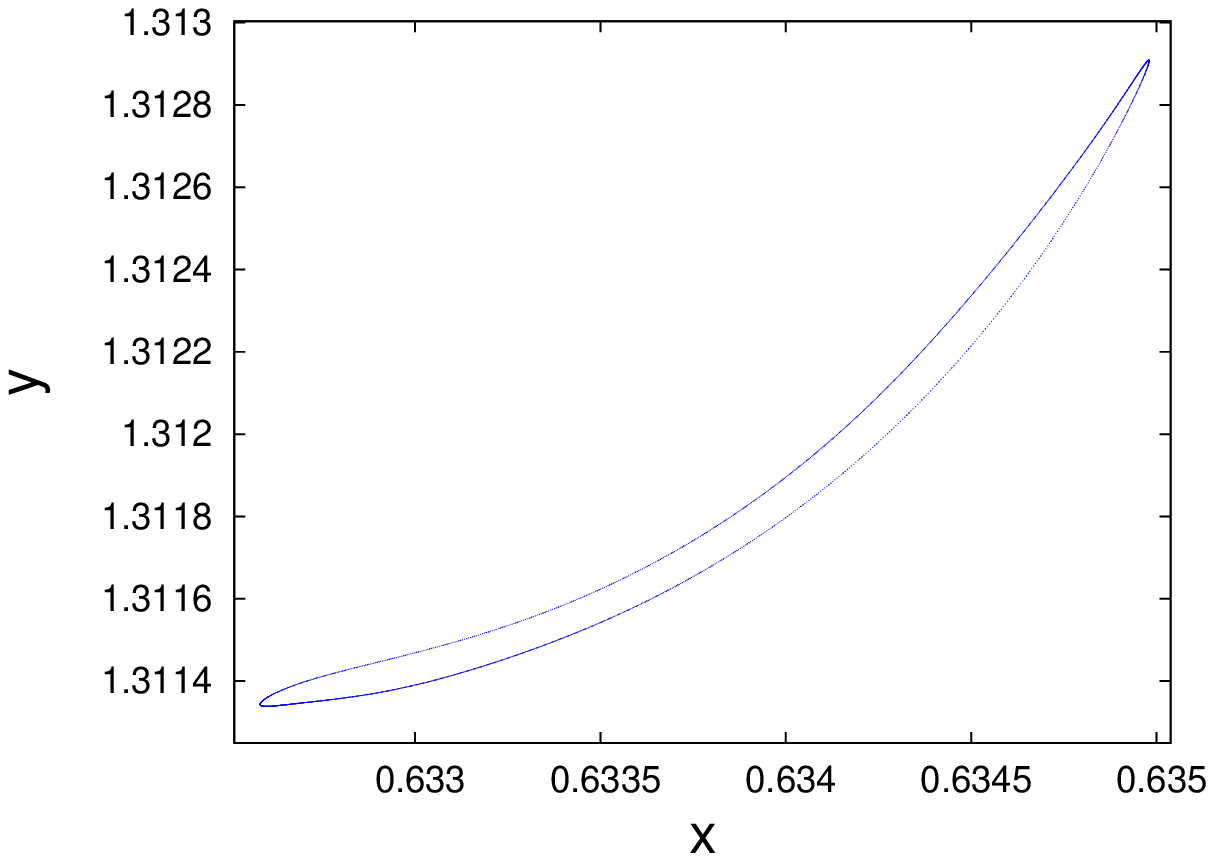} \\
\mbox{(a)} & \mbox{(b)} & \mbox{(c)}
\end{array}$ 
\caption[An example of quasi-periodic orbit exhibited by 4-star at $\mu=0.054$]{Quasi-periodic orbit exhibited by 4-star at $\mu=0.054$. Hub node in (a), and two branch nodes in (b) and (c). The third branch node's orbit overlaps with orbit of either one of branch nodes, depending on the initial conditions.}  \label{fig-qp-orbits}
\end{center}
\end{figure}

\textbf{Quasi-periodic Orbits.} As suggested in the definition of dynamical regions in Fig.\,\ref{fig-nonperiodic}, the system of CCM Eq.(\ref{main-equation}) also exhibits quasi-periodic orbits for various ranges of coupling strength (see Figs.\,\ref{fig-additionalorbits}a\,\&\,b for examples). Quasi-periodic motion is characterized by the presence of two or more incommensurable frequencies, implying it is not periodic (a single frequency present), and not chaotic either (continuous spectrum of frequencies) \cite{wiggy}. A quasi-periodic orbit densely fills a closed curve in phase space, as shown in the additional example in Fig.\,\ref{fig-qp-orbits}.
Our system exhibits quasi-periodic orbits for both 4-star and tree in given coupling ranges, with each node's orbit structured similarly to Figs.\,\ref{fig-qp-orbits}a\,\&\,b\,\&\,c (we are again showing only a half of the whole orbit, which as usually has two parts).

\begin{figure}[!hbt]
\begin{center}
$\begin{array}{cc}
\includegraphics[height=2.55in,width=3.15in]{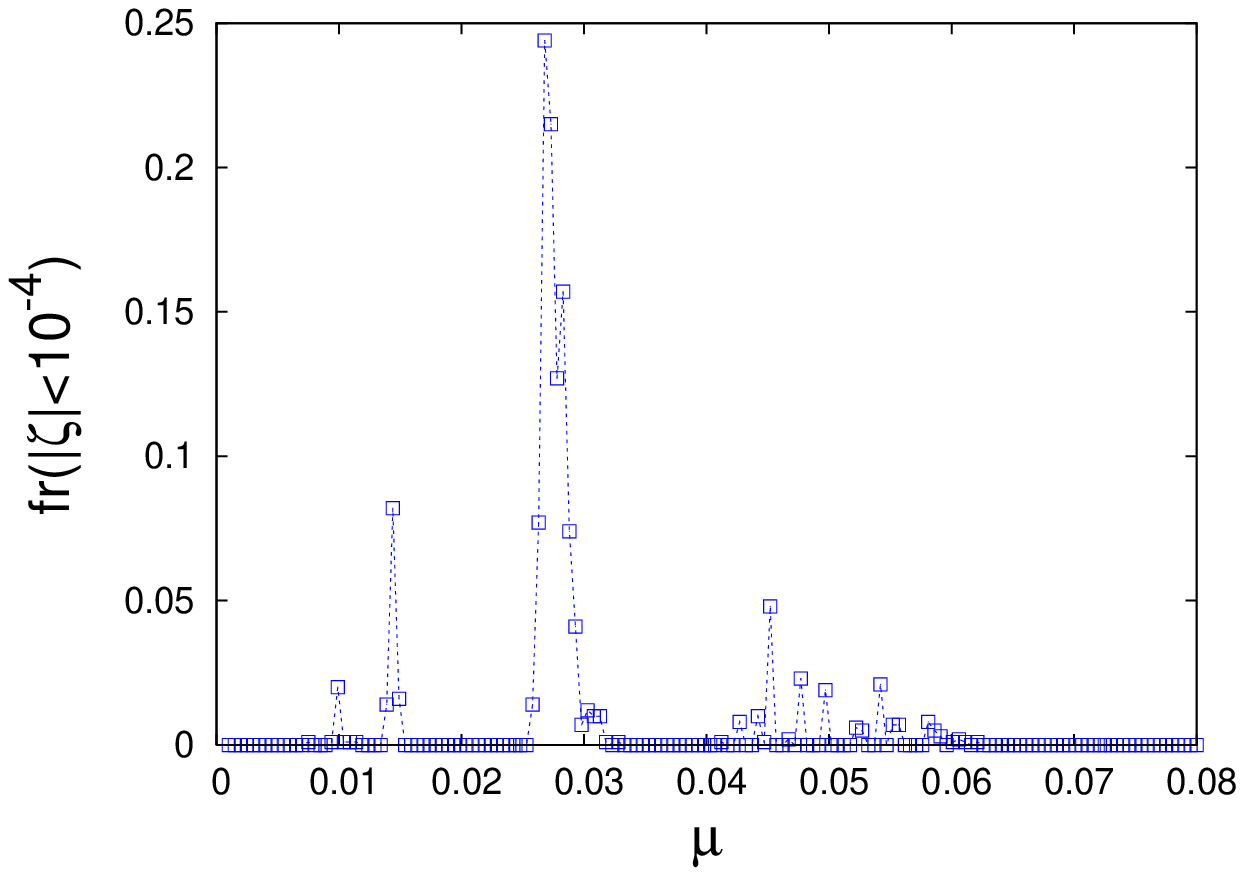} & 
\includegraphics[height=2.55in,width=3.15in]{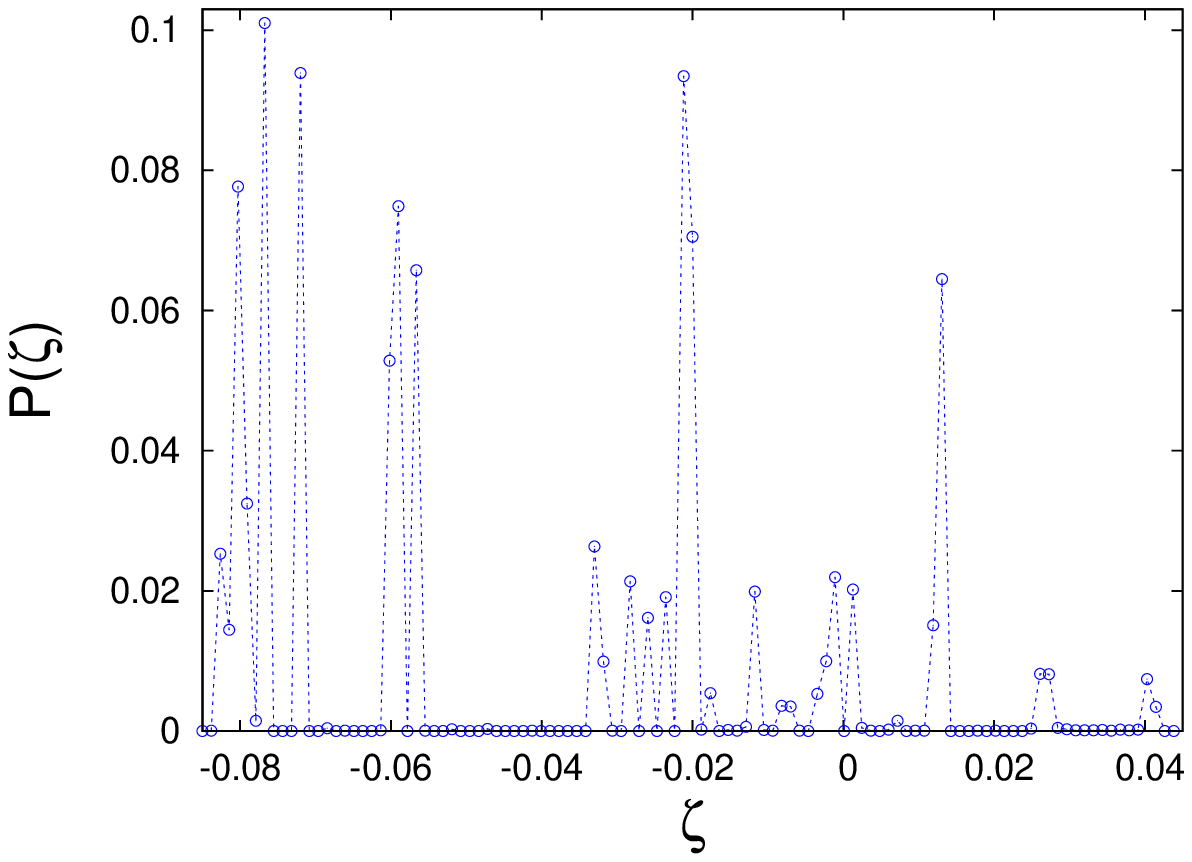} \\ 
\mbox{(a)} & \mbox{(b)} 
\end{array}$ 
\caption[Fraction of quasi-periodic orbits exhibited by 4-star in function of the coupling strength]{Fraction of quasi-periodic orbits exhibited by 4-star (defined as $|\zeta|<10^{-4}$), in function of coupling strength in (a), and distribution of $\zeta$-exponents over many initial conditions for 4-star at fixed $\mu=0.054$ in (b).}  \label{fig-qp-zeta}
\end{center}
\end{figure}

A peculiar characteristic of quasi-periodic motions is given by its SMLE $\zeta$ which converges to zero with the computation time $t_\zeta$. We shall use this to study systematically the appearance and properties of quasi-periodic motion in our CCM on 4-star. In Fig.\,\ref{fig-qp-zeta}a we report the fraction of initial conditions leading to $|\zeta|<10^{-4}$ over the full range of coupling strengths, indicating presence of quasi-periodic orbits in various coupling regions, predominantly around $\mu \sim 0.027$ (cf. Fig.\,\ref{fig-nonperiodic}). Quasi-periodic motion also arises throughout attractor region, although less often. In Fig.\,\ref{fig-qp-zeta}b we show the histogram of $\zeta$-exponents for a fixed coupling strength $\mu=0.054$. As in the previously shown histogram for $\mu=0.048$ (Fig.\,\ref{fig-ly-048}a), there is a clear separation between periodic/stable and attractor/unstable orbits, with both of them showing many peaks corresponding to various types of orbits. 
However, in Fig.\,\ref{fig-qp-zeta}b there is an additional peak around  $\zeta=0$, corresponding to the initial conditions leading to the orbits such as the one shown in Fig.\,\ref{fig-qp-orbits}.

\begin{figure}[!hbt]
\begin{center}
$\begin{array}{cc}
\includegraphics[height=2.55in,width=3.15in]{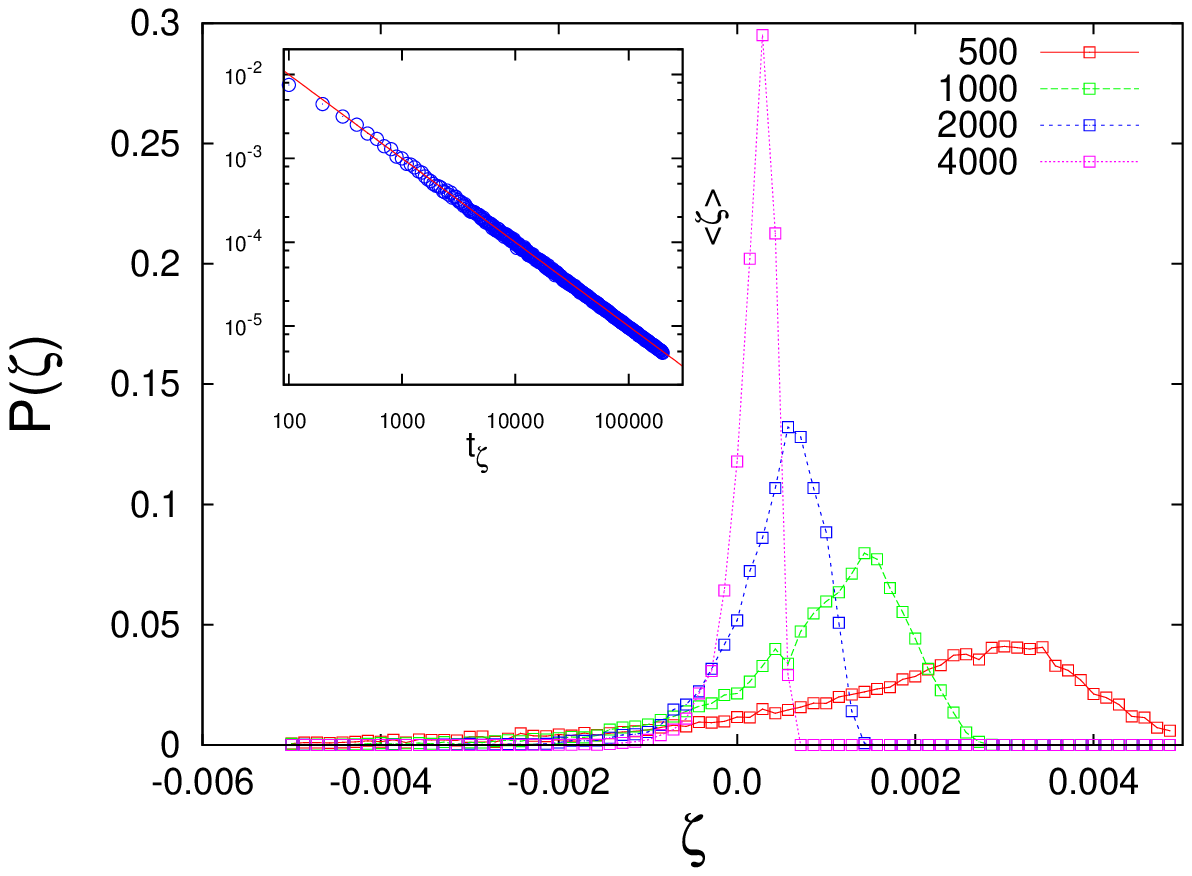} & 
\includegraphics[height=2.55in,width=3.15in]{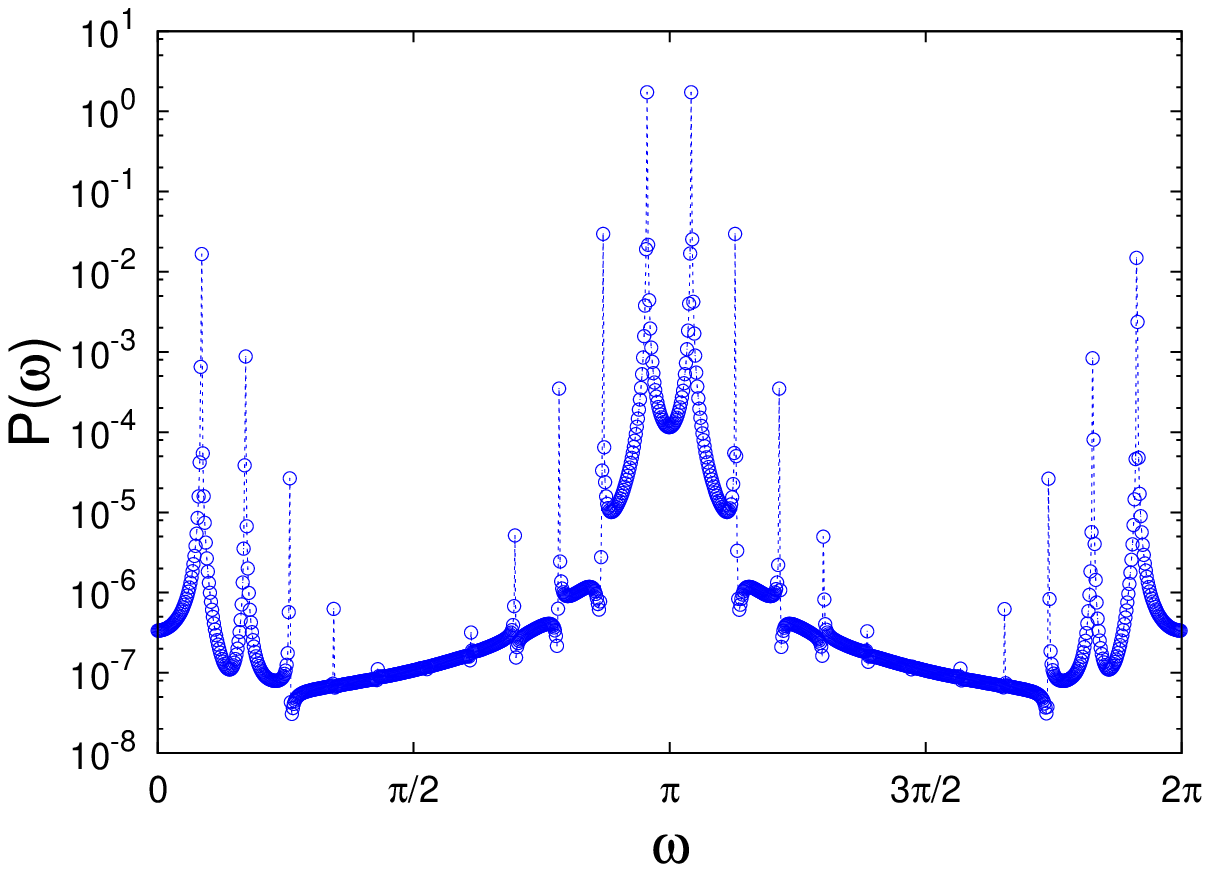} \\ 
\mbox{(a)} & \mbox{(b)} 
\end{array}$ 
\caption[Distributions of $\zeta$-values for various $t_\zeta$ and the convergence of $<\zeta>$ along with a power-spectrum for a quasi-periodic orbit]{Distributions of $\zeta$-values over quasi-periodic orbit points for various $t_\zeta$ in (a); inset: average $\zeta$-value $<\zeta>$ against $t_\zeta$. The frequency (power) spectrum of a quasi-periodic orbit of a single node in (b).}  \label{fig-qp-properties}
\end{center}
\end{figure}

In order to prove quasi-periodic character of examined motion, we investigate in Fig.\,\ref{fig-qp-properties}a the convergence of $\zeta$-exponent to zero with the number of SMLE computation time $t_\zeta$. Distributions of $\zeta$-values for different orbit points quickly peak around zero; 
furthermore, the  $\zeta$-exponent averaged over many orbit points $<\zeta>$ shows power-law convergence to zero in function of computation time $t_\zeta$ (shown in inset in Fig.\,\ref{fig-qp-properties}a), with a slope of 1. This clearly points to $\zeta=0$ at the limit $t_\zeta \rightarrow \infty$. Also, in Fig.\,\ref{fig-qp-properties}b we show a frequency decomposition (power spectrum) of a quasi-periodic orbit, with visible peaks around many $\omega$-values confirming multi-frequency character of these orbits. 

Similarly to periodic orbits, quasi-periodic orbits are rare in the dynamics on uncoupled standard map. Their presence of in the emergent motion of our CCM system is another clear sign of cooperative effects in its dynamics, existing at various network scales and coupling strengths. As mentioned, the dynamics of CCM on the tree also exhibits quasi-periodic orbits on all nodes simultaneously, in the similar range of coupling strengths.

\section{The Parametric Instability between Dynamical Regions}

In this Section we investigate the generation of instabilities at the edges between dynamical regions, explaining the passage from periodic/stable region to non-periodic/unstable region with the change of coupling strength. We consider orbits of the 4-star's branch node, and examine the changes they undergo in relation to the small changes of $\mu$-value, looking for the bifurcations that transform one orbit type into another. This way we study the presence of different types of orbits for the same coupling strength corresponding to different fractions of initial conditions, and the transformations between their ratios with the change of $\mu$-value. We characterize the bifurcations through the eigenvalues of the 4-star's Jacobian matrix Eq.(\ref{jacobian}) corresponding to the phase space point in question. As 4-star is 8D dynamical system Eq.(\ref{4stareq}), we obtain 4 complex conjugate eigenvalue pairs which determine the type of bifurcation \cite{wiggy}.

\begin{figure}[!hbt]
\begin{center}
$\begin{array}{cc}
\includegraphics[height=2.7in,width=2.9in]{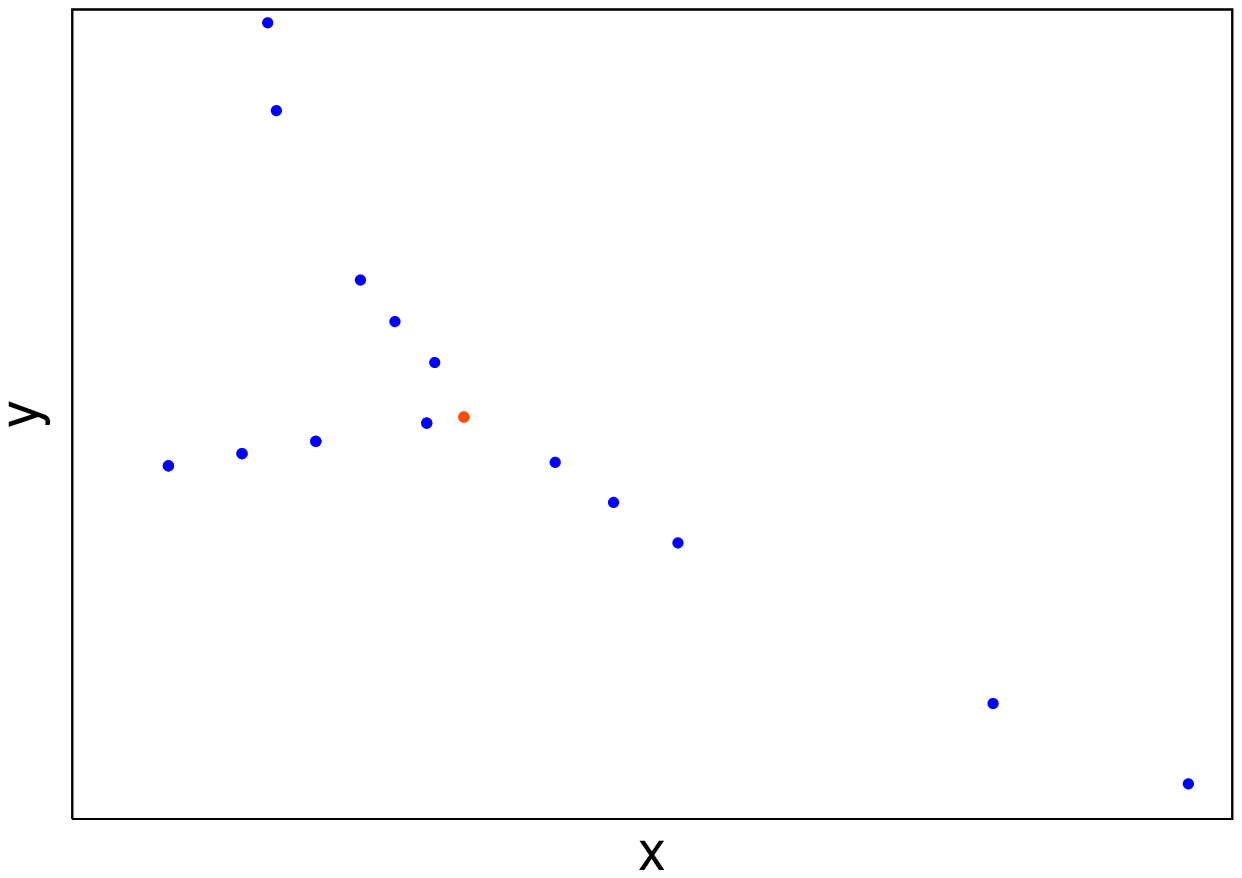} & 
\includegraphics[height=2.7in,width=3.in]{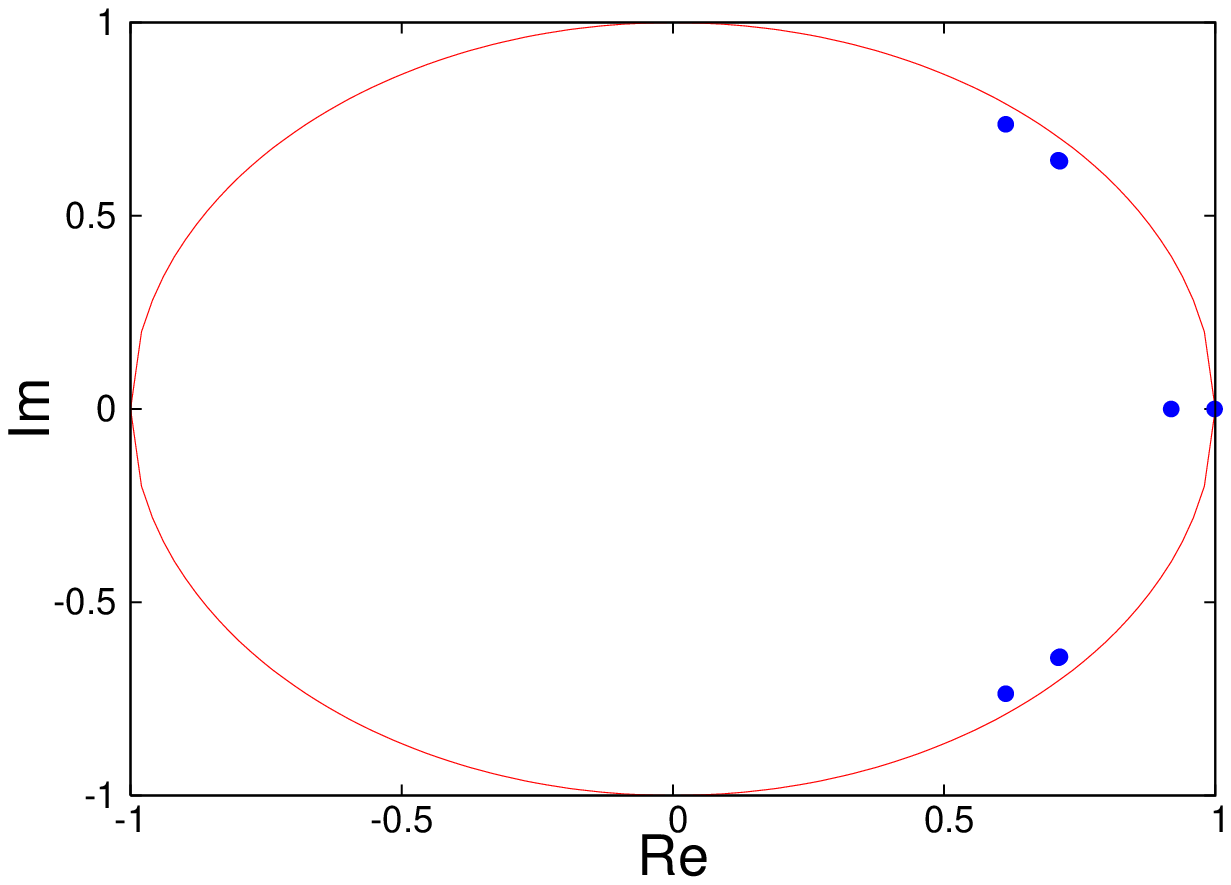} \\ 
\mbox{(a)} & \mbox{(b)} 
\end{array}$ 
\caption[A typical period doubling bifurcation for 4-star's orbit at $\mu \cong 0.015$]{Sequence of orbits of 4-star's branch node (only half is shown) undergoing period doubling bifurcation at $\mu \cong 0.015$ in (a), and the Jacobian matrix eigenvalues computed at the bifurcation point (marked in red) for all 4-star nodes in (b).}  \label{fig-pd}
\end{center}
\end{figure}

We start with periodic region at $\mu \cong 0.015$ and report in Fig.\,\ref{fig-pd}a the evolution of period-2 orbit into a period-4 orbit as $\mu$ is slightly changed (each point represents only a half of one orbit). The corresponding eigenvalues are computed at the marked point in Fig.\,\ref{fig-pd}a and shown in Fig.\,\ref{fig-pd}b: the right-most eigenvalue crosses the unitary circle, confirming this to be a \textit{period-doubling bifurcation}. Other nodes of the 4-star simultaneously undergo the same bifurcation.
The variations of $\mu$-value among the orbits shown in Fig.\,\ref{fig-pd}a are $O(10^{-5})$ -- as already mentioned, our system of CCM is extremely sensitive to both initial conditions and $\mu$-values. We suggest this to be the general mechanism of generation of orbits of different periodicities in all dynamical regions, which is confirmed by the presence of large period values on 4-star and tree (cf. Fig.\,\ref{fig-periodranking}). More complex dynamical patterns might be similarly generated as a fast sequence of period-doubling bifurcations.

\begin{figure}[!hbt]
\begin{center}
$\begin{array}{cc}
\includegraphics[height=2.7in,width=2.9in]{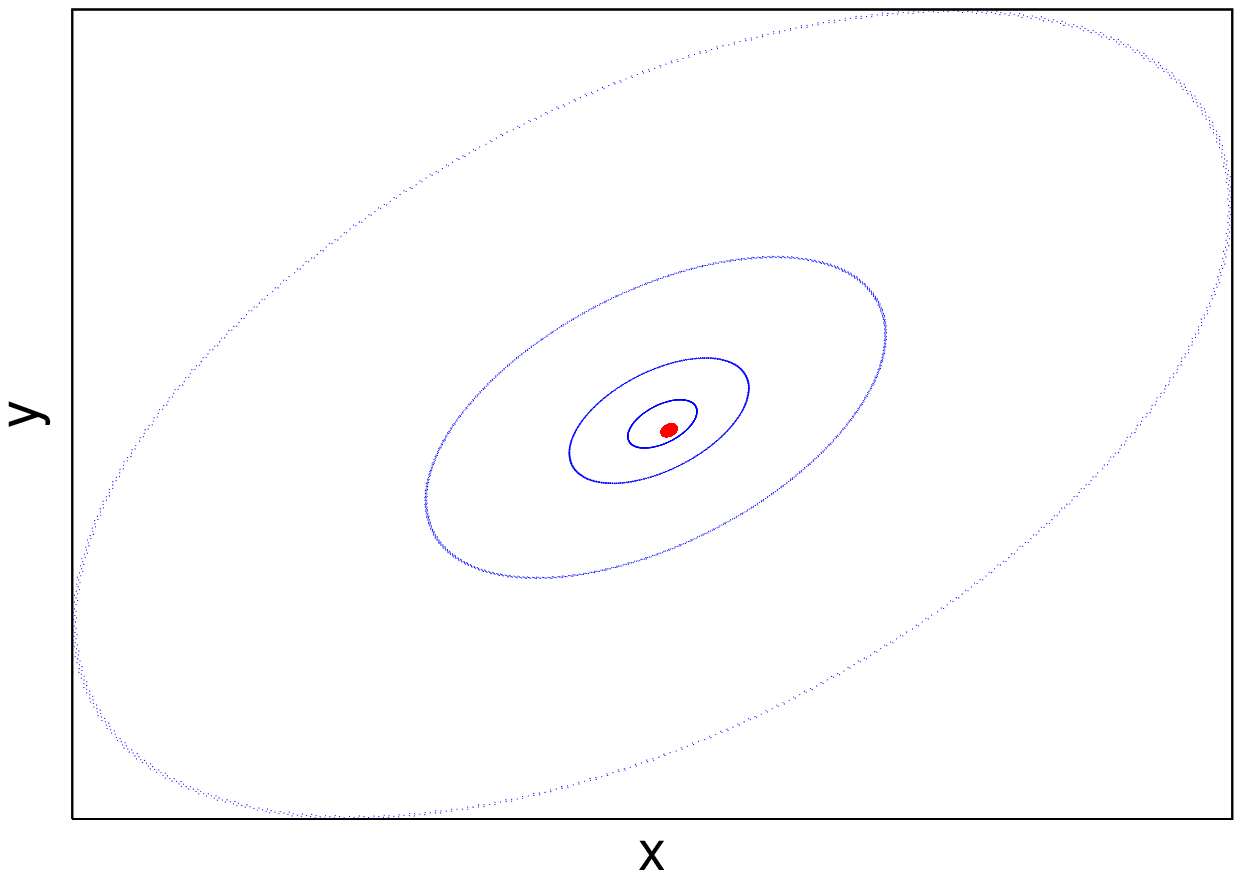} & 
\includegraphics[height=2.7in,width=3.in]{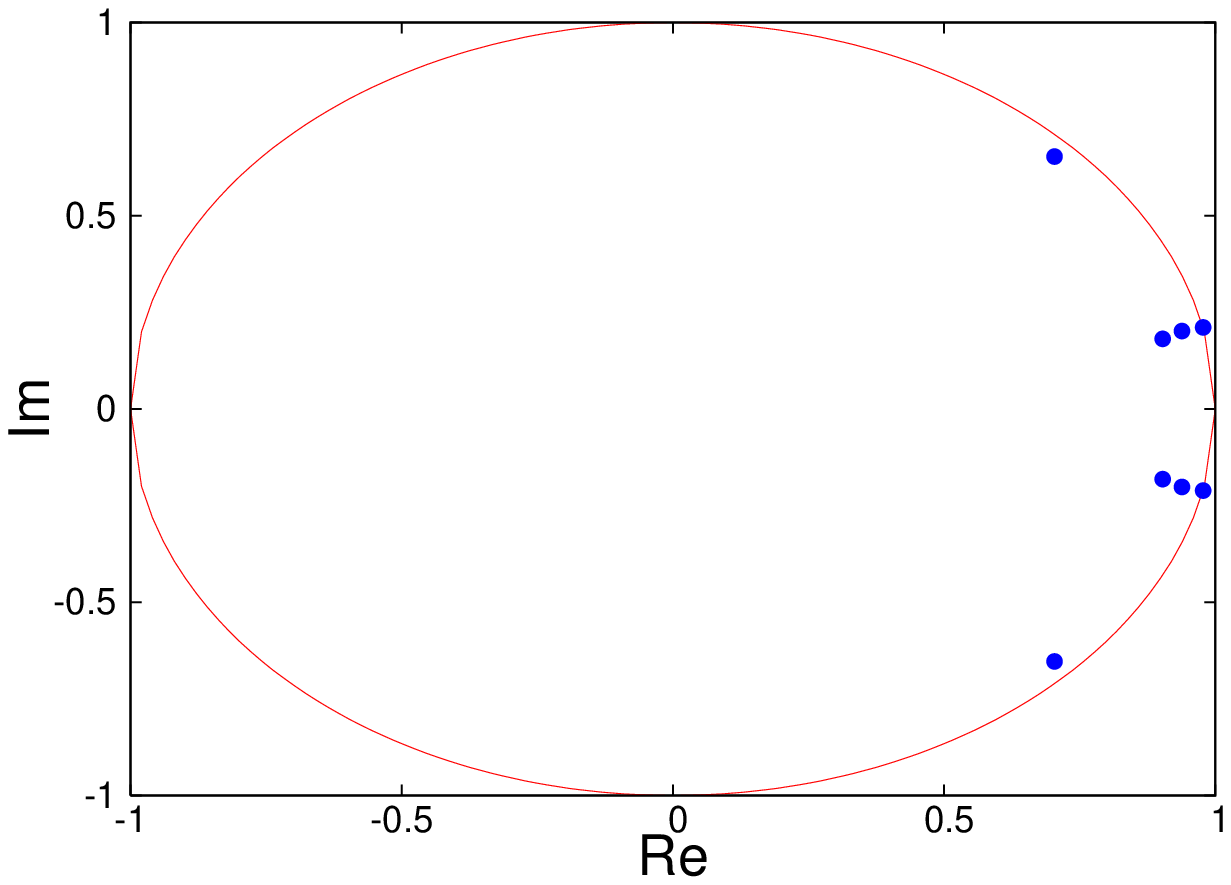} \\ 
\mbox{(a)} & \mbox{(b)} 
\end{array}$ 
\caption[A typical Hopf doubling bifurcation for 4-star's orbit at $\mu \cong 0.014$]{Sequence of quasi-periodic orbits developing after Hopf bifurcation of a stable point at 4-star's branch node with $\mu \cong 0.014$ in (a), the Jacobian matrix eigenvalues computed at the bifurcation point (marked in red) in (b).} \label{fig-hopf}
\end{center}
\end{figure}

In Fig.\,\ref{fig-hopf}a we investigate the generation of a quasi-periodic orbit at $\mu \cong 0.014$ developing from a period-2 orbit (again, we monitor only a half of orbit, seeing quasi-periodic orbit developing from a fixed point). Accordingly, the eigenvalues of the Jacobian matrix at the bifurcation threshold are shown in Fig.\,\ref{fig-hopf}b: right-most complex pair of eigenvalues crosses the unitary circle, indicating this to be a \textit{Hopf bifurcation} \cite{wiggy}. The variations of $\mu$-value in Fig.\,\ref{fig-hopf}a are again $O(10^{-5})$, showing the systems to be equally sensitive here as well. 
Hopf bifurcations are known to often exist in dynamical systems of this sort, and they are likely to be the general mechanism behind the creation of quasi-periodic orbits at all coupling strengths.

Finally, we examine the creation of a strange attractor at $\mu=0.048$ corresponding as the one shown in Fig.\,\ref{fig-SNA}b, arising from a periodic orbit by fast increase of orbit's periodicity. 
\begin{figure}[!hbt]
\begin{center}
$\begin{array}{ccc}
\includegraphics[height=1.85in,width=2.05in]{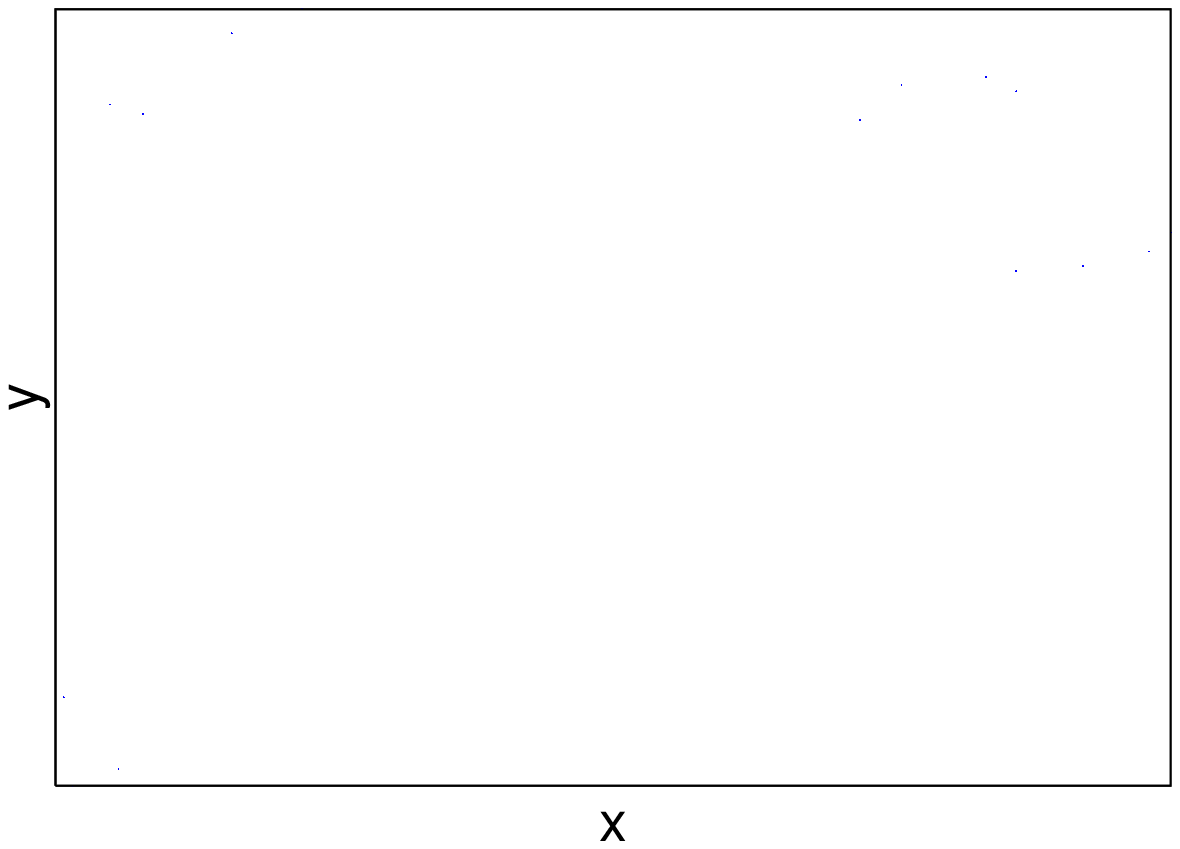} & 
\includegraphics[height=1.85in,width=2.05in]{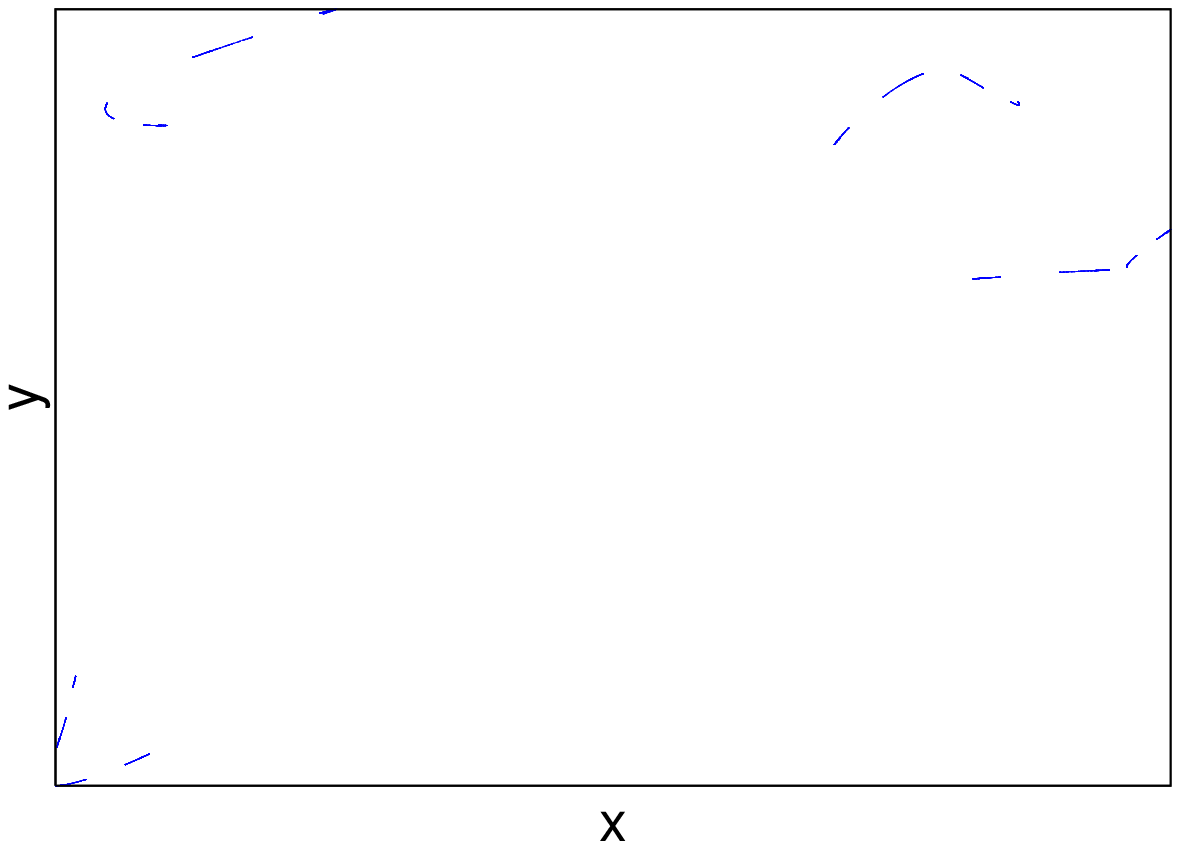} & 
\includegraphics[height=1.85in,width=2.05in]{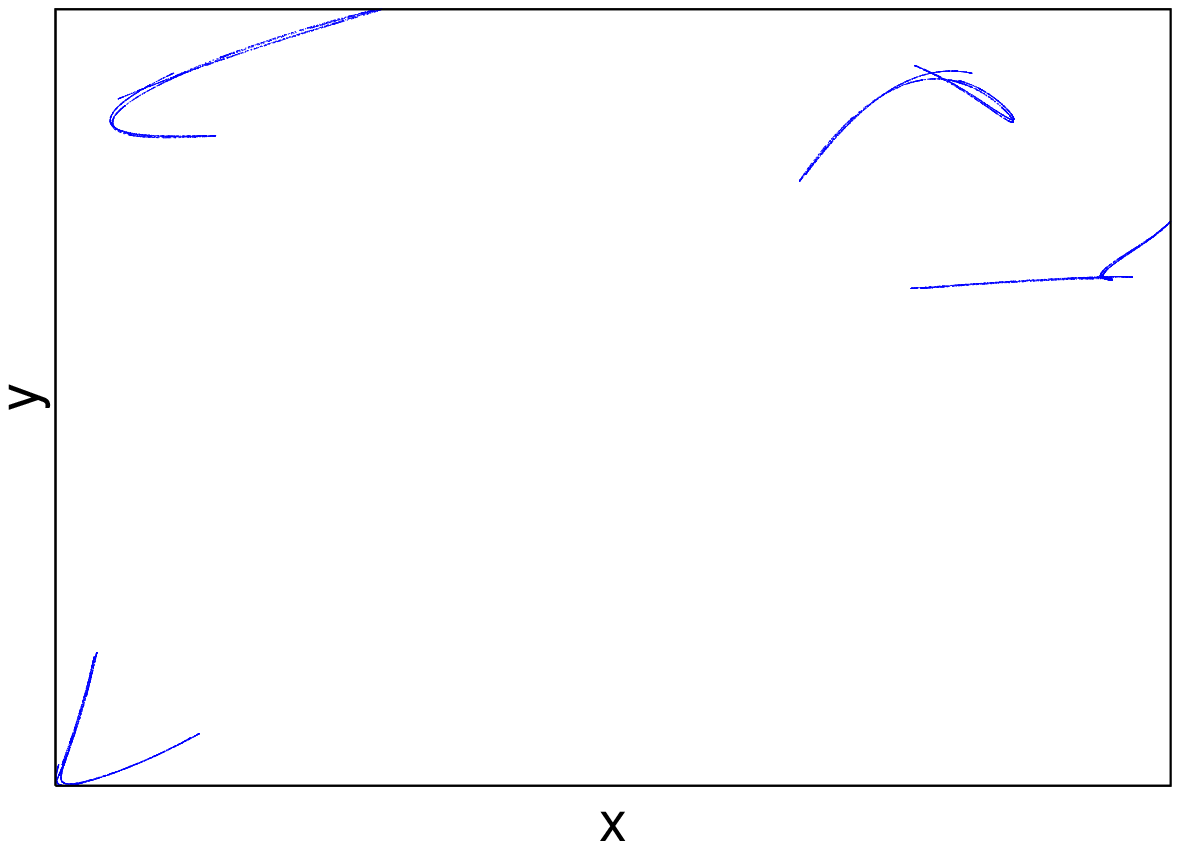} \\
\includegraphics[height=1.85in,width=2.05in]{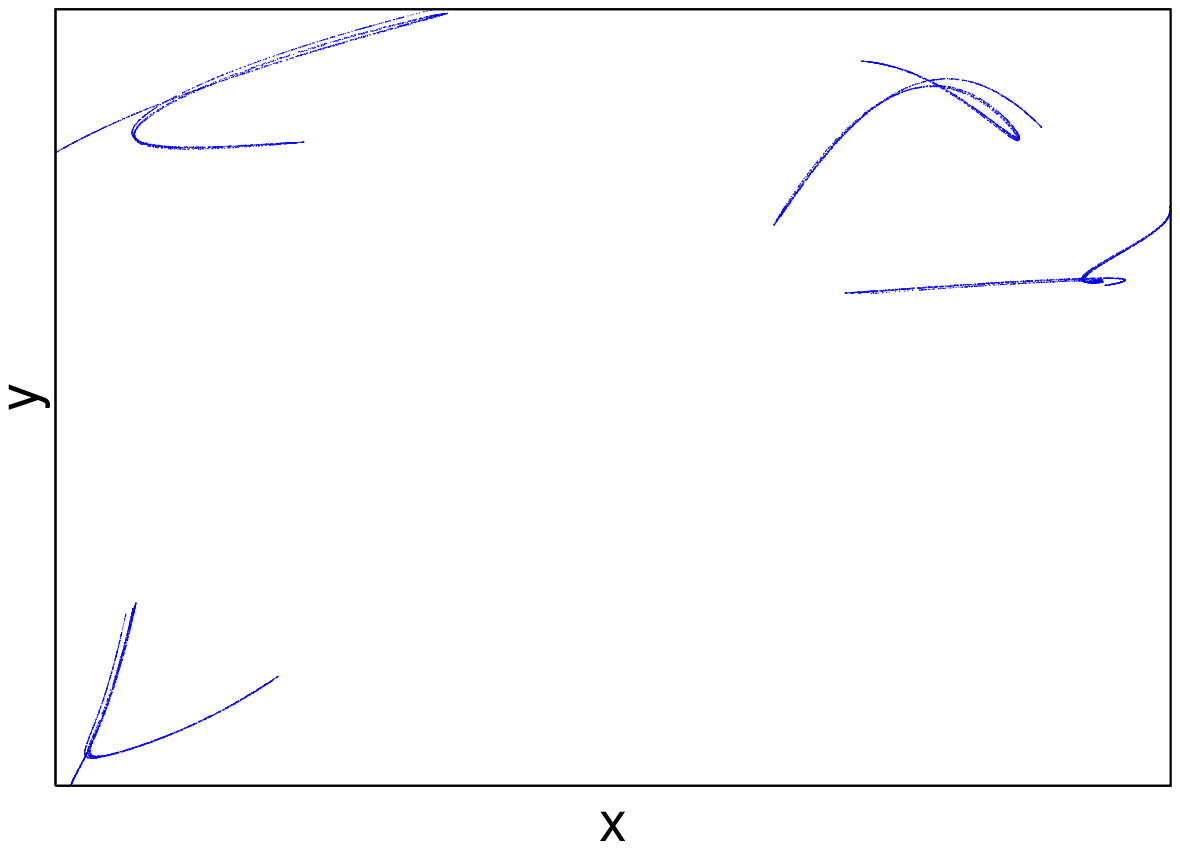} & 
\includegraphics[height=1.85in,width=2.05in]{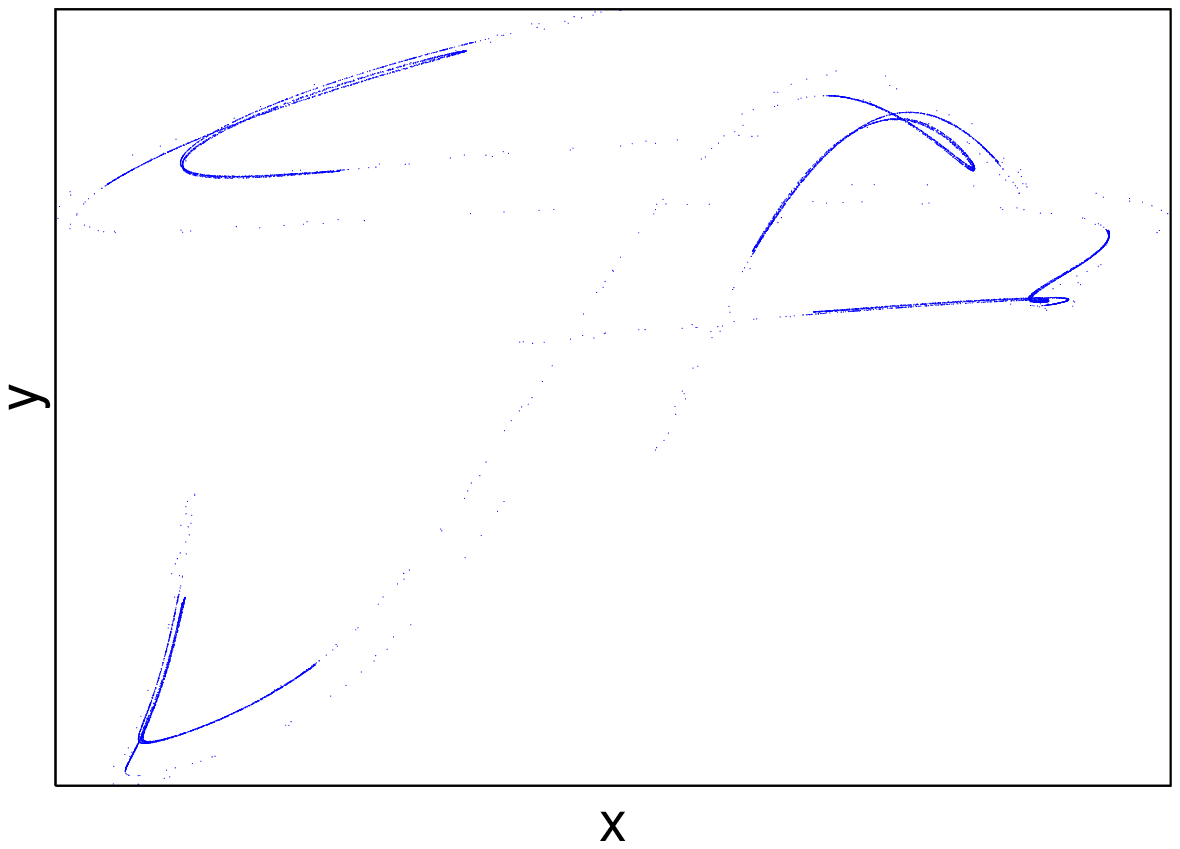} & 
\includegraphics[height=1.85in,width=2.05in]{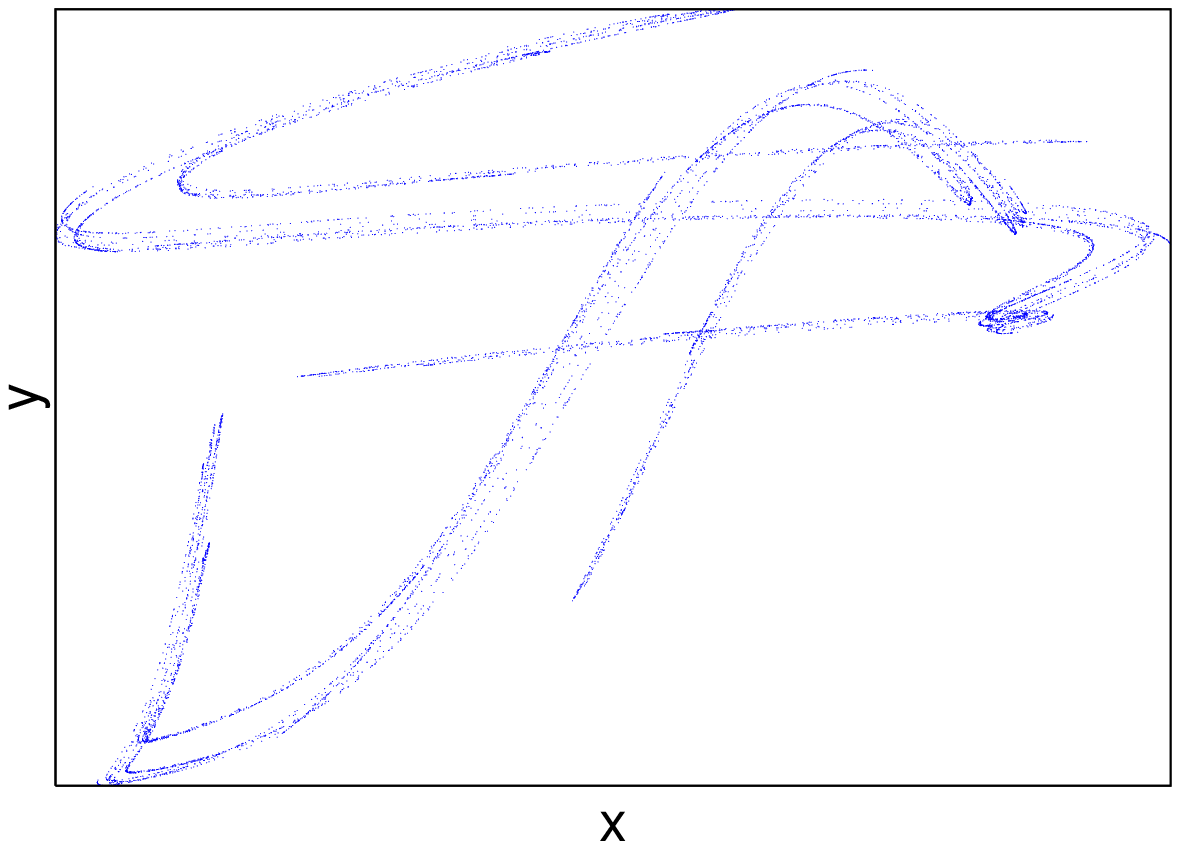} 
\end{array}$ 
\caption[Formation of the attractor from Fig.\,\ref{fig-SNA}b at $\mu \cong 0.048$]{Formation of the attractor from Fig.\,\ref{fig-SNA}b through increase of periodicity of a periodic orbit at $\mu \cong 0.048$. First stage corresponds to $\mu=0.048-10^{4}$ and the last to $\mu=0.048$, while other stages are between these two $\mu$-values.}  \label{fig-attr-formation}
\end{center}
\end{figure}
We show the stages of attractor formation in Fig.\,\ref{fig-attr-formation}, with each stage obtained for the same initial conditions, but with $\mu$-value that vary from $\mu=0.048-10^{-4}$ (first stage) to $\mu=0.048$ (last stage). The process of attractor formation starts with a periodic orbit and displays a sequence of several attractors before it is fully formed. Some intermediate attractors (two middle pictures) have fractal dimensions smaller than 1; attractor reaches its proper fractal dimension of $d_f \cong 1.4$ only in the last stage. 

\begin{figure}[!hbt]
\begin{center}
$\begin{array}{ccc}
\includegraphics[height=2.65in,width=2.85in]{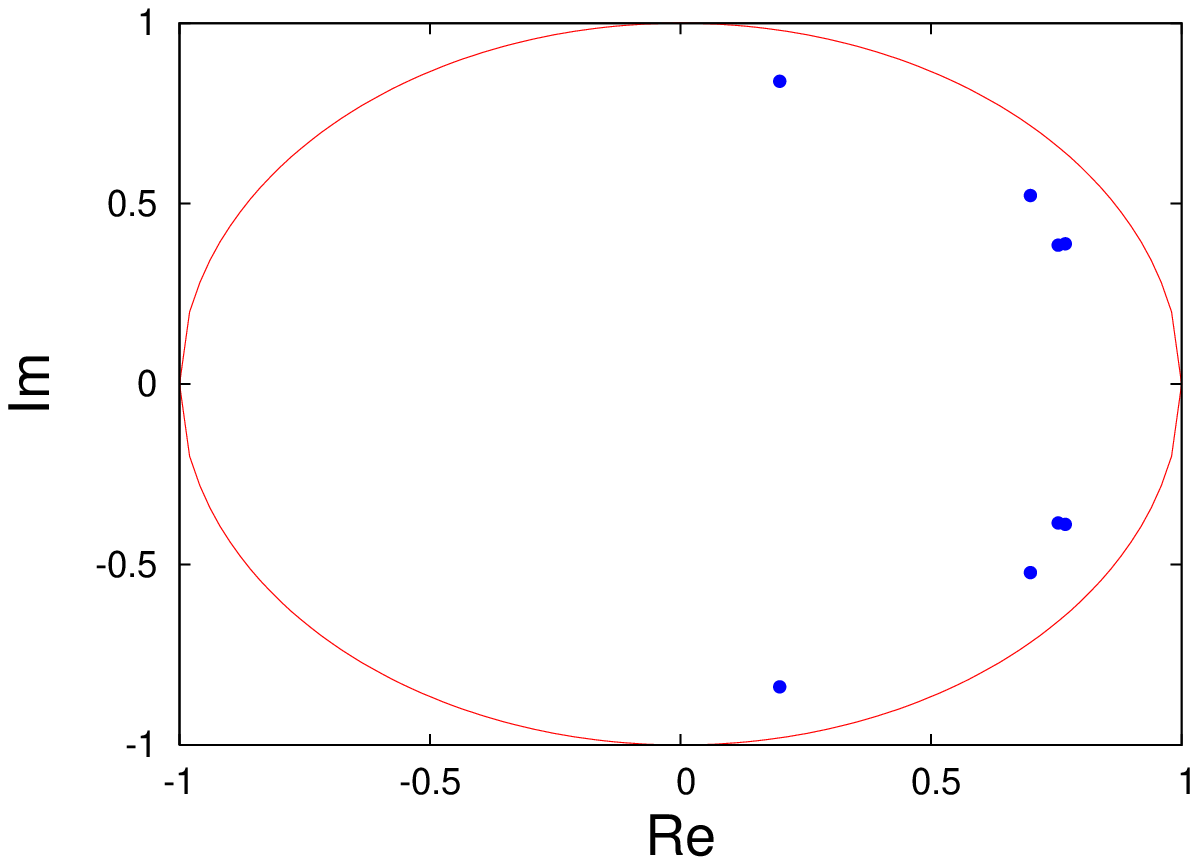} & 
\includegraphics[height=2.65in,width=3.45in]{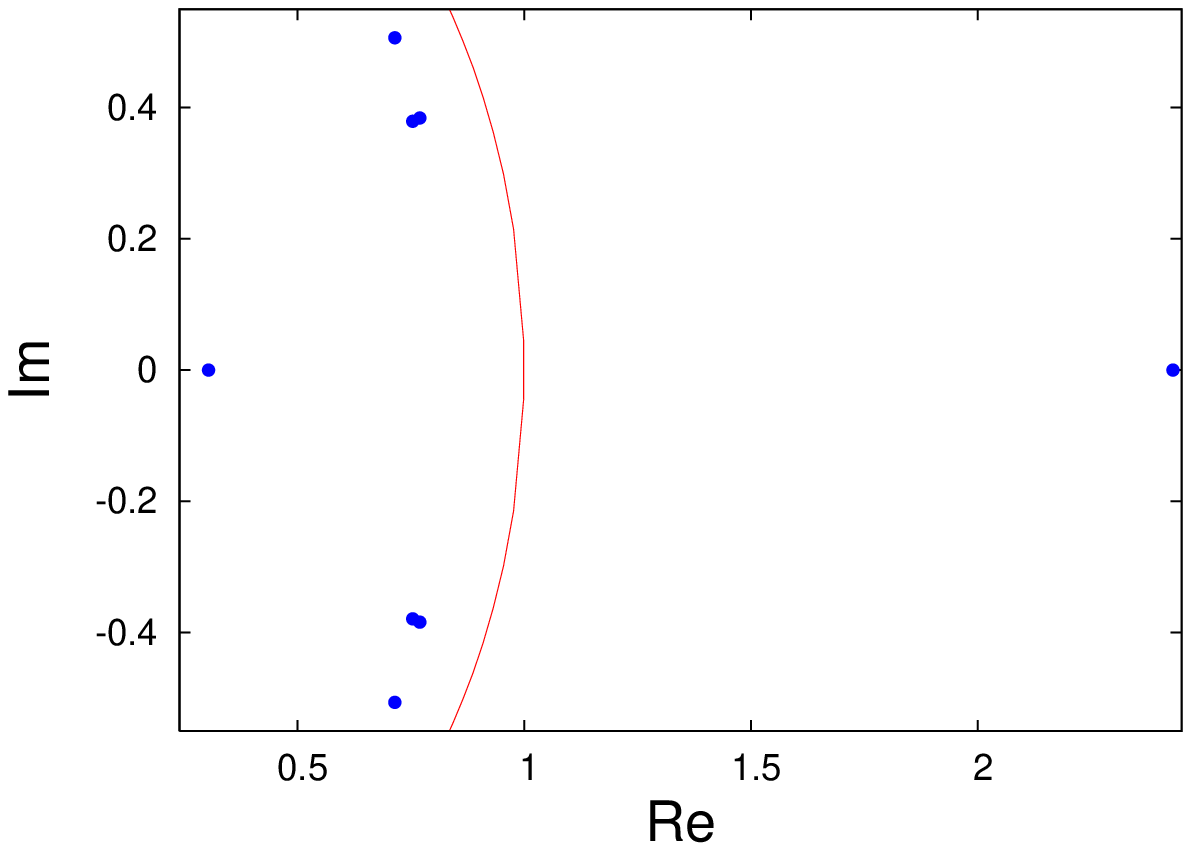} \\
\mbox{(a)} & \mbox{(b)} 
\end{array}$ 
\caption[Eigenvalues of the Jacobian matrix for attractor points at the final stage from Fig.\,\ref{fig-attr-formation}]{Eigenvalues of the Jacobian matrix for some attractor points at the final stage of the attractor formation shown in Fig.\,\ref{fig-attr-formation}. A typical stable situation in (a), and a typical unstable situation in (b), which are interchanging from iteration to iteration. Jacobian matrix eigenvalue with the largest absolute value represents the first-order approximation of SMLE ($t_\zeta=1$).}  \label{fig-attr-bifurcation}
\end{center}
\end{figure}

The Jacobian matrix corresponding to the final stage of attractor formation from Fig.\,\ref{fig-attr-formation} is examined, and the results are shown in  Fig.\,\ref{fig-attr-bifurcation}. The eigenvalues for various attractor points are interchangingly displaying a stable character (all eigenvalues inside the unit circle as in Fig.\,\ref{fig-attr-formation}a) and an unstable character (one eigenvalues outside the unit circle as in Fig.\,\ref{fig-attr-formation}b), without any apparent correlation. The same pattern also occurs if the Jacobian matrix's eigenvalues are computed at any of the previous stages (except for the first one, which is always stable). As opposed to the previous cases of bifurcations, in Fig.\,\ref{fig-attr-formation} one is unable to identify a precise $\mu$-value when the attractor is formed; accordingly, the stable/unstable pattern of Jacobian matrix eigenvalues persists at smaller $\mu$-values as well. Moreover, the stable combinations of eigenvalues (Fig.\,\ref{fig-attr-formation}a) appear far more frequently than the unstable ones, which is even more prominent at the earlier stages of attractor formation, which are closer to the periodic orbit.
This as an additional argument towards the interesting nature of dynamical phenomena from Fig.\,\ref{fig-SNA}; the structure of eigenvalues reports about the stability of linearized system, and in particular, eigenvalue with the largest absolute value corresponds to the one-iteration approximation of SMLE $\zeta$ ($t_\zeta=1$). This indicates the nature of linearized version of our system for the dynamical situation from Fig.\,\ref{fig-SNA} to indeed show both stable and unstable nature, as suggested by the FTMLE distribution in Fig.\,\ref{fig-ftmle-distributions}b. 

The examined bifurcations refer to the orbits exhibited by the 4-star's branch node; however, the transformations of orbit structure occur simultaneously on all nodes. The analysis presented in this Section accounts for formation of different dynamical phenomena by all nodes and at all coupling strengths (with possibly other types of bifurcations also playing a role), hence explaining the evolution of dynamical regions that network of CCM undergoes with variation of coupling. Due to the already established relationship between tree's and 4-star's dynamics, we suggest the same mechanism to be behind the transformations of collective motion between equivalent dynamical regions for the tree, as the bifurcations seem to be unrelated to the topology details.


\chapter{Dynamics on Directed E.Coli Gene Regulatory Network}  \label{Dynamics on E.Coli Network}

\begin{flushright}
\begin{minipage}{4.6in}
    Additional study of our system of CCM is done using the directed gene regulatory network of Escherichia Coli (E.Coli), introduced in  
    Chapter \ref{Coupled Maps System on Networks with Time delay}. Statistical examinations are performed, revealing differences 
    and analogies with non-directed topology. Same network is also employed for investigation of 
    two-dimensional Hill model of E.Coli's gene regulatory network. We particularly examine the flexibility and stability of the 
    emergent behavior, along with its robustness to fluctuations of the environmental inputs.\\[0.1cm]
\end{minipage}
\end{flushright}

In this Chapter we will investigate the collective dynamics on the largest connected component of the directed gene regulatory network of E.Coli shown in Fig.\,\ref{fig-connectedcomponent} \cite{orr,mangan}, with $N=328$ nodes. Two dynamical models will be used: system of coupled chaotic standard maps Eq.(\ref{main-equation}) examined so far, and the continuous-time 2D Hill model \cite{schuster} of gene interaction introduced in Eqs.(\ref{schuster})\&(\ref{hill}). We will examine the dynamics of both models mainly employing the techniques used above. We are comparing two two-dimensional models with very different dynamical natures: while the system Eq.(\ref{main-equation}) involves strongly chaotic maps, the system Eqs.(\ref{schuster})\&(\ref{hill}) displays only regular dynamics regardless of coupling. As opposed to investigations done in previous Chapters, E.Coli network is directed, meaning that interactions among the nodes are generally not mutual, which will affect the properties of the collective dynamics. This network topology also includes topological self-loops, allowing the a node/gene to influence itself. It is also to be noted that  E.Coli network is an example of a real biological network which is empirically found rather than computer generated. This implies each node represents a specific gene of E.Coli's gene regulatory network, and therefore has a given biological purpose which is encoded in its network location.

\section{CCM on Directed Network of E.Coli}

In this Section we consider the dynamical model of CCM given by Eq.(\ref{main-equation}) on the directed network of E.Coli. We adjust the CCM model to the directed nature of the underlying network by excluding the degree-normalization factor in coupling form (as some nodes on directed network may not have incoming links, their in-degree is zero). The modified model reads:
\begin{equation} 
\left(\begin{array}{l}
x[i]_{t+1} \\
y[i]_{t+1}
\end{array}\right)
=(1- \mu) 
\left(\begin{array}{l}
x[i]_t' \\
y[i]_t'
\end{array}\right)
+
\mu
\left(\begin{array}{c}
\sum_{j \in {\mathcal K_i}} (x[j]_t - x[i]_t') \\ 
0
\end{array}\right),
\label{directed-equation} \end{equation}
and contains all the key elements of the CCM Eq.(\ref{main-equation}) (phase coupling, time delay etc.). The prime $(')$ refers to the update of the uncoupled standard map Eq.(\ref{oursm}), and the network neighborhood ${\mathcal K_i}$ includes the nodes $[j]=1,\hdots,N$ having directed links towards the node $[i]$ (which may include the node $[i]$ itself). We will investigate the dynamics of Eq.(\ref{directed-equation}) using the same numerical implementation procedure and study approaches from the previous Chapters. Non-periodic orbits, time-averaged orbits, node by node FTMLE $\lambda_{max}^{t}$ and return times will be examined. The studied coupling range will include $[0,0.08]$ as before. Initial conditions are selected randomly from $(x,y) \in [0,1] \times [-1,1]$, and the averaging over them as previously includes 1000 random initial conditions. We shall often examine color plots showing certain features for each node; in such cases the nodes will be enumerated from 0 to 327.

\begin{figure}[!hbt]
\begin{center}
\includegraphics[height=2.7in,width=4.1in]{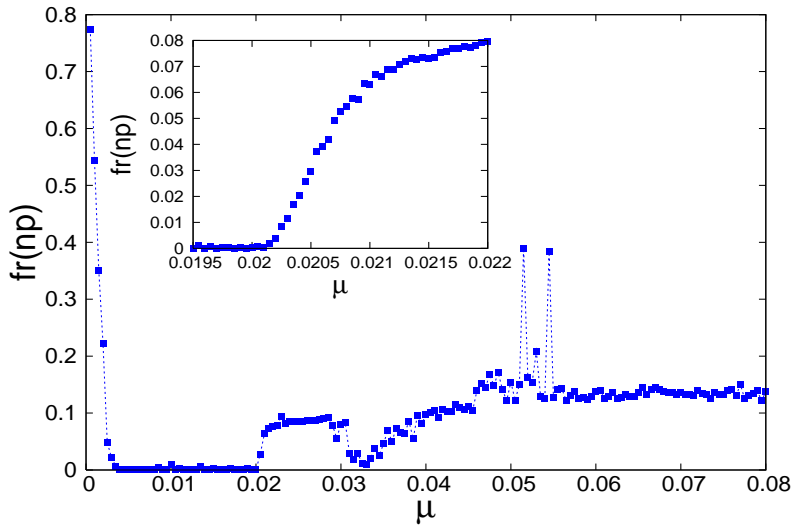}
\caption[Fraction of non-periodic orbits in function of the coupling strength $\mu$ for the directed E.Coli network, with zoom to the transition region]{Fraction of non-periodic nodes/orbits after a transient of $10^{6}$ iterations in function of the coupling strength $\mu$, averaged over many initial conditions. Inset: zoom to the dynamical transition region around $\mu \sim 0.022$.}  \label{fig-directed-nonperiodic}
\end{center}
\end{figure}

We start by reporting the fraction of non-periodic orbits in function of coupling strength $\mu$ averaged over initial conditions in Fig.\,\ref{fig-directed-nonperiodic}. The profile structure at small $\mu$-values resembles the profile for non-directed tree: after a quick initial transient the orbits of all nodes for all initial conditions stabilize into periodic orbits. 
However, with increase of coupling strength, a certain fraction of the network nodes destabilizes again into non-periodic orbits (inset in Fig.\,\ref{fig-directed-nonperiodic} provides detailed view of this dynamical region with $\mu \sim 0.022$). As the $\mu$-value is increased further, the network of CCM re-stabilizes in a short region around $\mu \sim 0.033$, after which remains with a relatively constant number of non-periodic nodes for the remaining part of studied coupling range. The concrete properties of specific node's orbits will be studied in more detail later.

\begin{figure}[!hbt]
\begin{center}
$\begin{array}{cc}
\includegraphics[height=2.45in,width=3.15in]{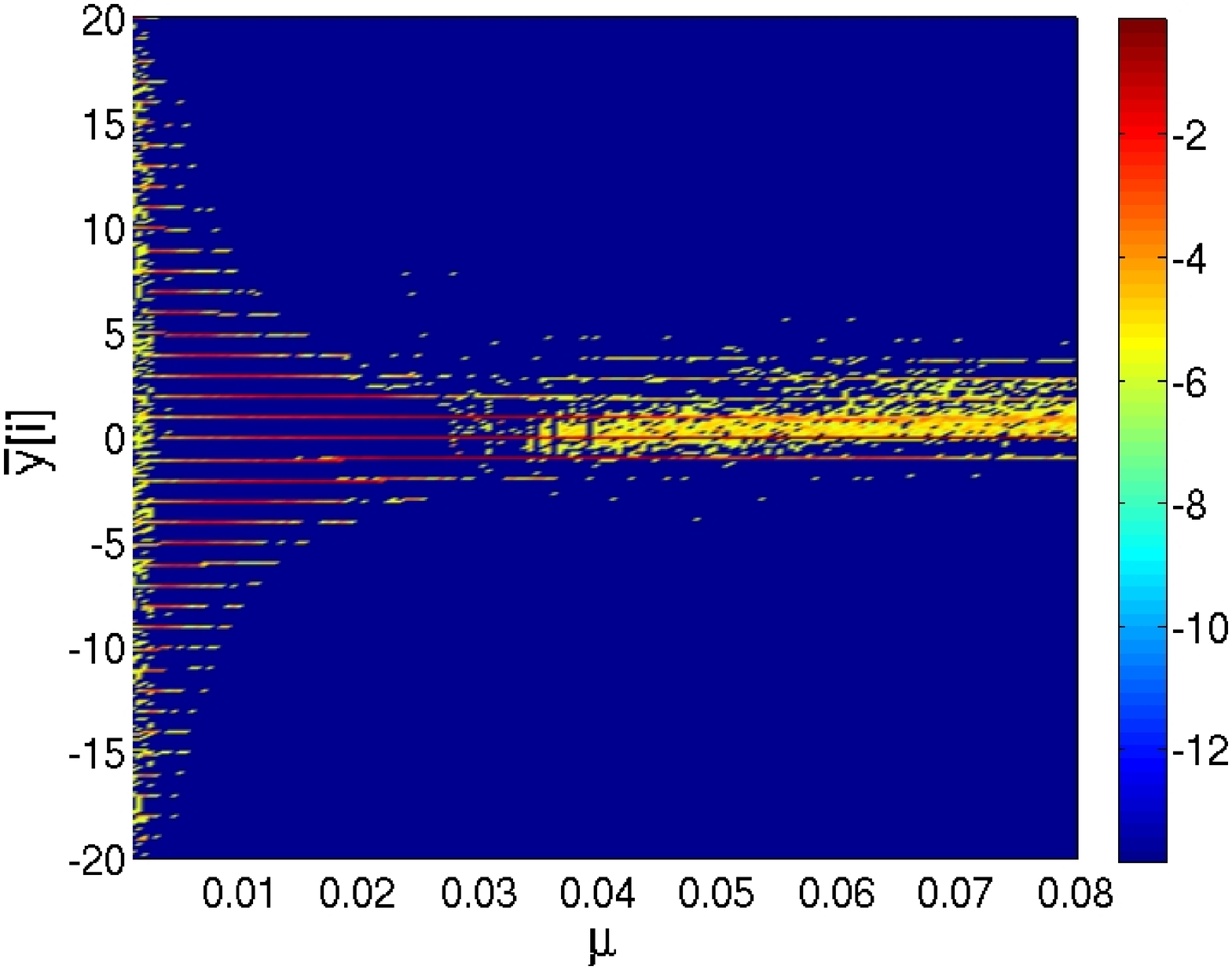} & 
\includegraphics[height=2.45in,width=3.15in]{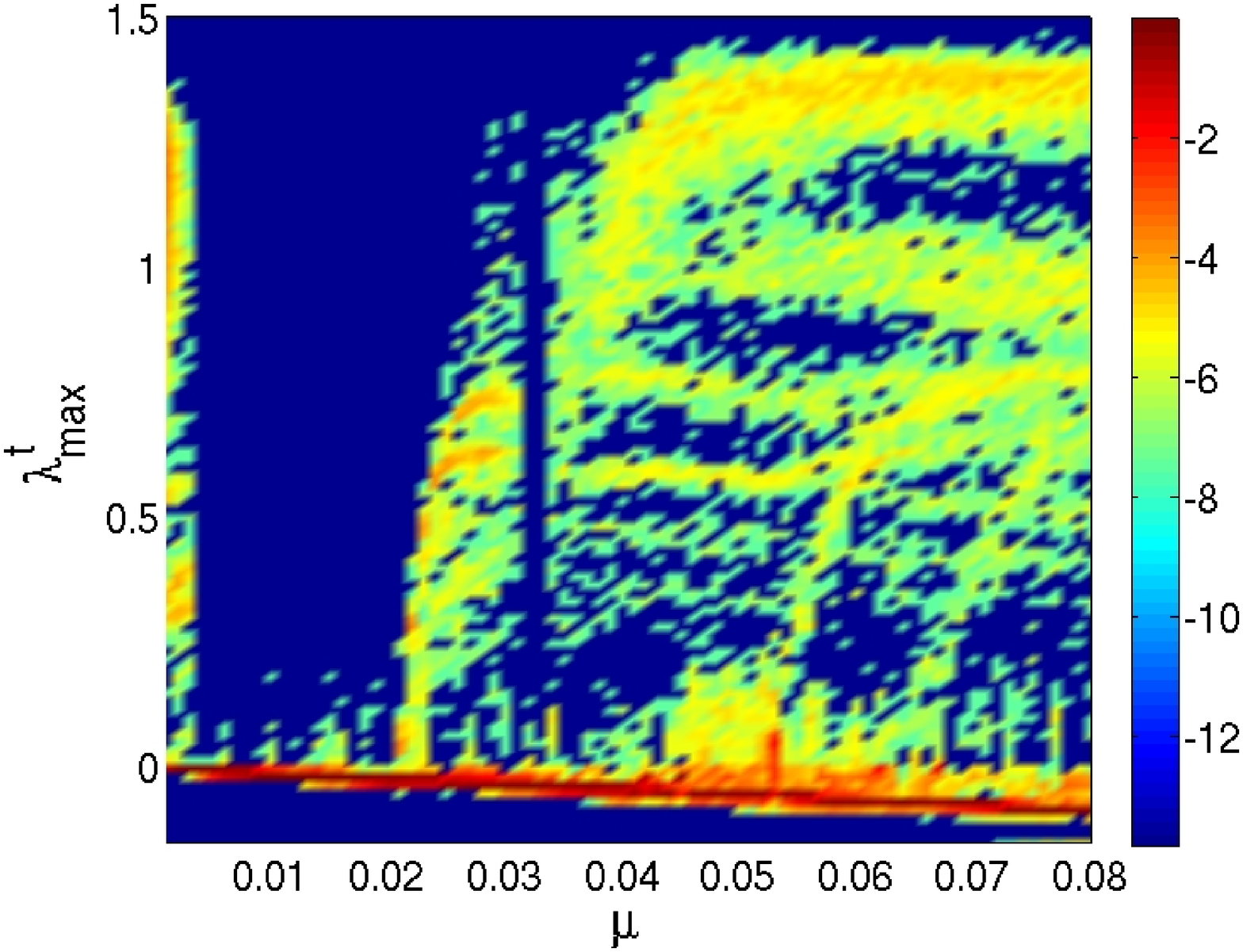} \\ 
\mbox{(a)} & \mbox{(b)} 
\end{array}$ 
\caption[2D color histogram of $\bar{y}$-values and $\lambda_{max}^{t}$-values in function of the coupling strength]{2D color histogram of $\bar{y}[i]$-values in function of coupling strength $\mu$ for a single initial condition in (a), 2D color histogram  of $\lambda_{max}^{t}$-values in function of $\mu$, averaged over many initial conditions in (b).} \label{fig-directed-00do08}
\end{center}
\end{figure}

In Fig.\,\ref{fig-directed-00do08}a we show the 2D color histogram of $\bar{y}[i]$-values in function of the coupling strength, equivalent to Fig.\,\ref{fig-ybar-comparisons}c, which confirms the dynamical region structure from Fig.\,\ref{fig-directed-nonperiodic}. For small $\mu$-values the system presents fully clustered organization of periodic orbits in analogy with non-directed case, while for larger $\mu$-values a destabilization is visible. Note that certain clusters are re-appearing in destabilization region. We also examine FTMLE defined as in Eqs.(\ref{mledefinition})\&(\ref{lambda}) by computing $\lambda_{max}^{t}$-values node by node, following the same procedure explained in Chapter \ref{Stability of Network Dynamics}. The 2D color histogram equivalent to Fig.\,\ref{fig-colorstability} is reported in Fig.\,\ref{fig-directed-00do08}b: the same dynamical region structure is again visible in relation to the coupling strength. 
However, as opposed to previously considered non-directed case, after destabilization some nodes show $\lambda_{max}^{t} > 1$ indicating presence of strongly chaotic behavior similar to the uncoupled case. Nevertheless, majority of the nodes have $\lambda_{max}^{t} < 0$ pointing to periodic orbits, and some nodes display $\lambda_{max}^{t}$-values in a wide range between 0 and 1.5, allowing the possibility of various dynamical behaviors of nodes.

In the reminder of this Section we perform a detailed analysis of cooperative dynamical patterns at two particular coupling strength values: at the onset of destabilization at $\mu=0.022$, and the irregular region at $\mu=0.05$ characterized by $\lambda_{max}^{t} > 1$.\\[0.1cm]

\begin{figure}[!hbt]
\begin{center}
\includegraphics[height=2.7in,width=4.6in]{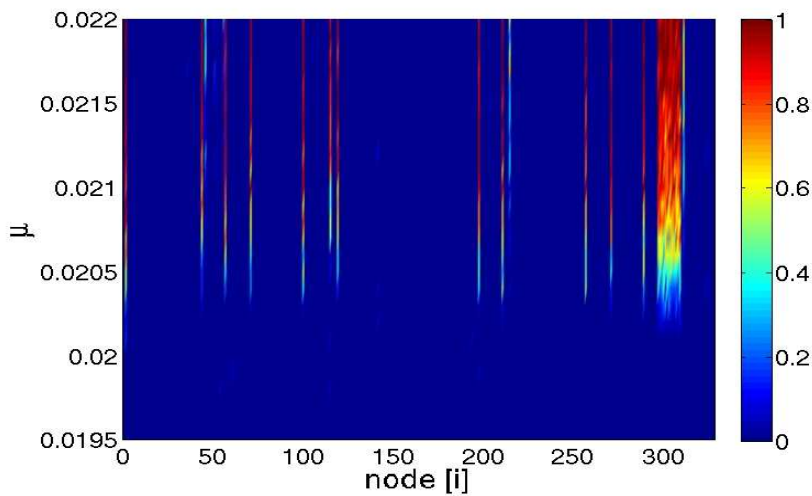}  
\caption[Node be node color representation of the instability transition near $\mu=0.022$]{2D plot of the instability transition near $\mu=0.022$: fraction of non-periodic orbits for each node (enumerated from 0 to 327) as function of coupling strength $\mu$, averaged over many initial conditions.} \label{fig-directed-0022transition}
\end{center} 
\end{figure}

\textbf{Onset of Destabilization at $\mu=0.022$.} We consider the directed dynamics of CCM Eq.(\ref{directed-equation}) on the E.Coli network Fig.\,\ref{fig-connectedcomponent} near $\mu=0.022$. In Fig.\,\ref{fig-directed-0022transition} we examine the profile of the transition considering the fraction of initial conditions leading to non-periodic orbits for each node separately. As the coupling strength increases, the instability develops on a specific group of nodes rather than over the whole network. Even after the transition, the instability does not spread to the rest of the network, but remains confined to the initially destabilized sub-network.
This is in sharp contradiction with what observed in the case of non-directed dynamics on tree, where the dynamical regions are characterized by all the nodes sharing similar dynamical behaviors. 

\begin{figure}[!hbt]
\begin{center}
$\begin{array}{cc}
\includegraphics[height=2.85in,width=3.1in]{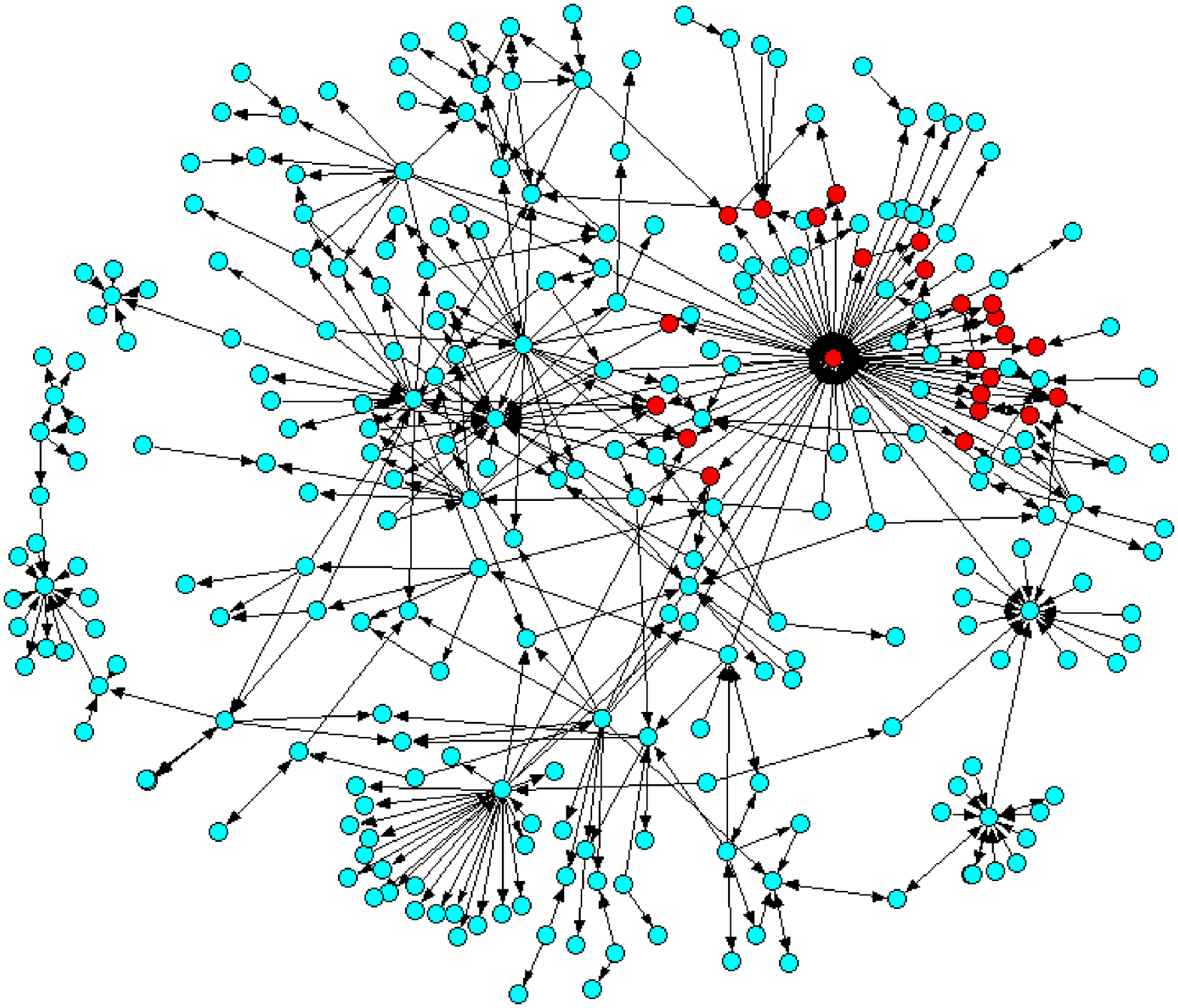} & 
\includegraphics[height=2.85in,width=3.1in]{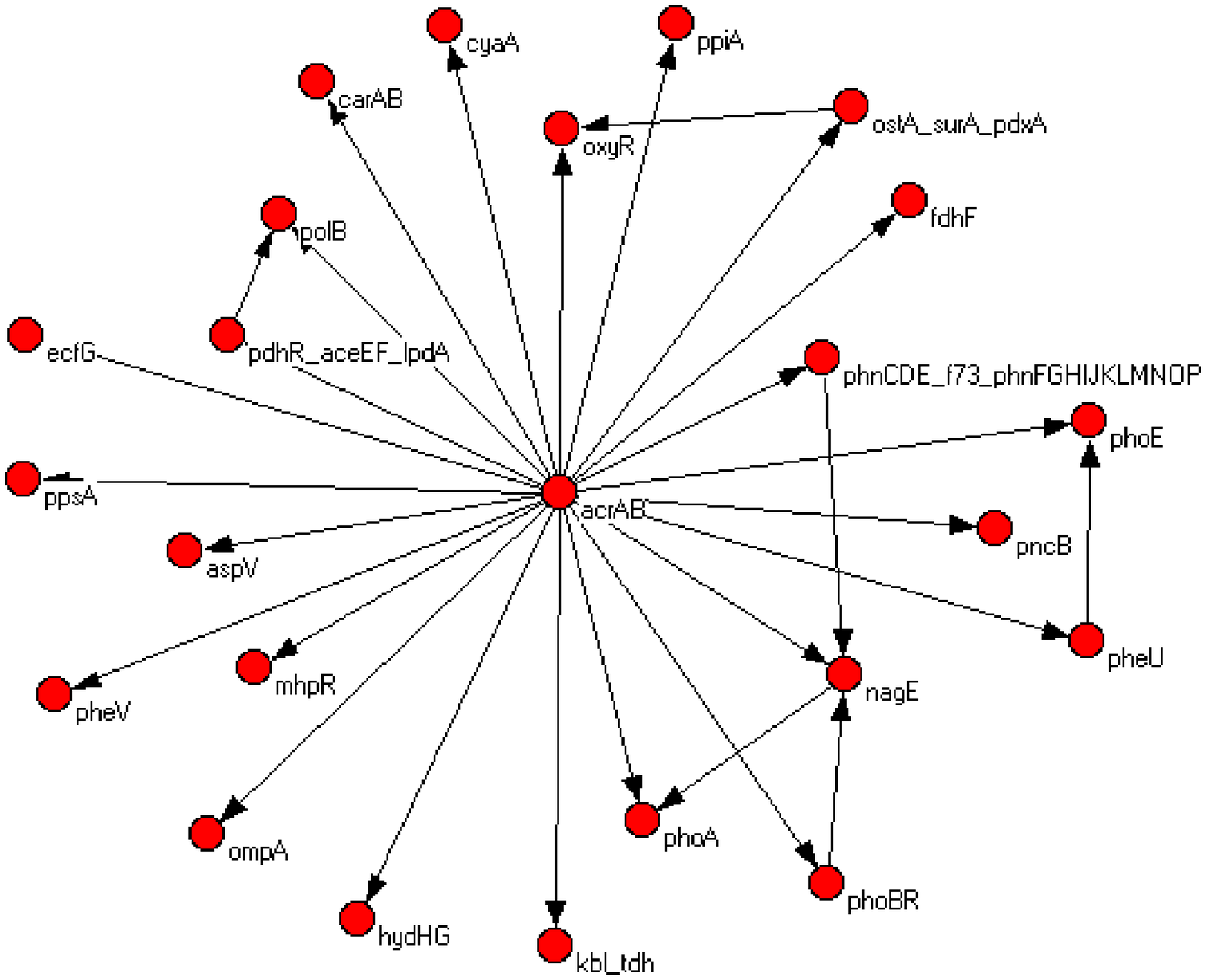} \\ 
\mbox{(a)} & \mbox{(b)} 
\end{array}$ 
\caption[Graphical representations of the sub-network undergoing destabilization at $\mu=0.022$]{Instability transition at $\mu=0.022$. E.Coli's largest connected component with destabilized nodes (marked in red) in (a), the sub-network undergoing destabilization with nodes/genes biological names in (b).} \label{fig-directed-0022pajek}
\end{center}
\end{figure}

We investigate the structure of the unstable sub-network by showing a graphical representation of the E.Coli network (from Fig.\,\ref{fig-connectedcomponent}) with different coloration for the unstable nodes in Fig.\,\ref{fig-directed-0022pajek}a. The unstable sub-network includes the hub node (having biggest number of in-links and out-links) and other surrounding nodes which are integrated with the rest of the network. In 
Fig.\,\ref{fig-directed-0022pajek}b we show the destabilized sub-network and denote each node/gene with its biological name.
The unstable sub-network has both incoming and outgoing links with the rest of the network, suggesting this to be a non-trivial dynamical phenomenon related to this particular CCM system and the architecture of the directed E.Coli network.

\begin{figure}[!hbt]
\begin{center}
\includegraphics[height=2.2in,width=2.05in]{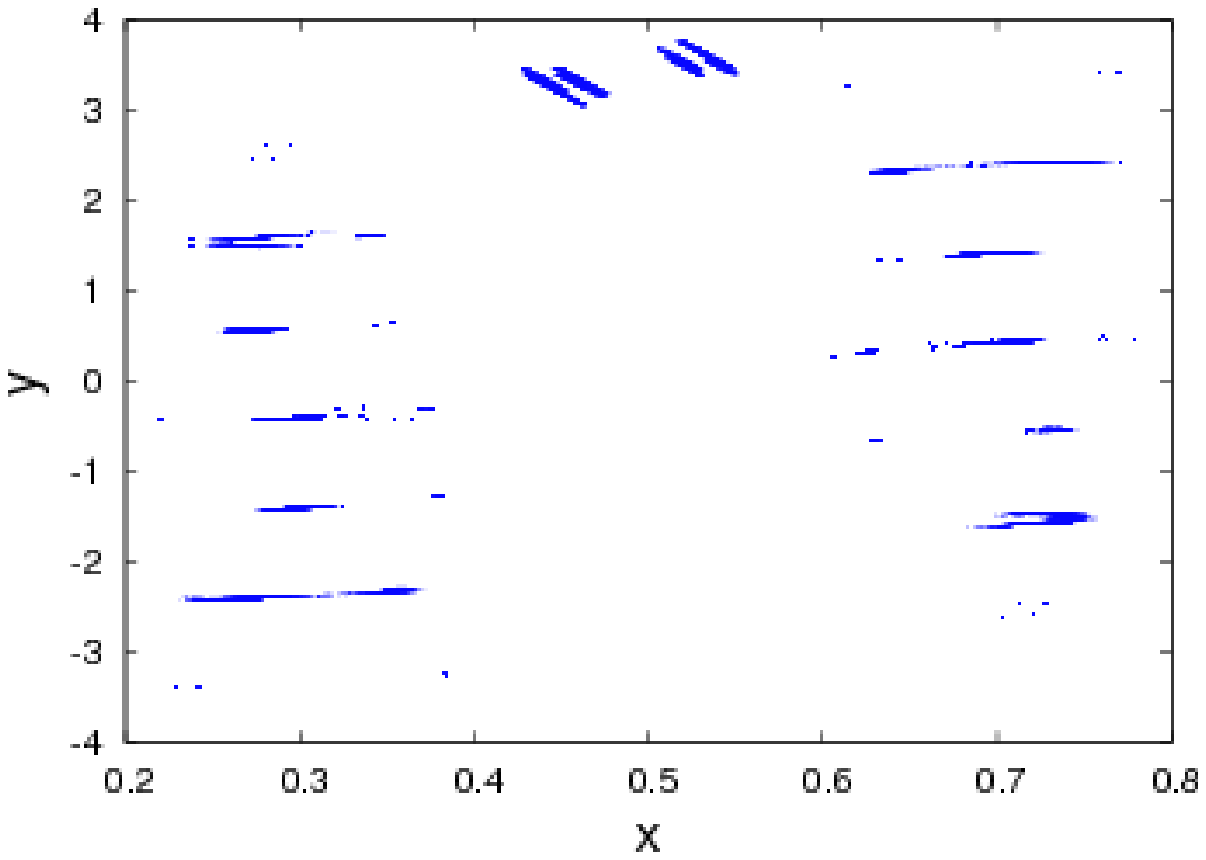}
\includegraphics[height=2.2in,width=2.05in]{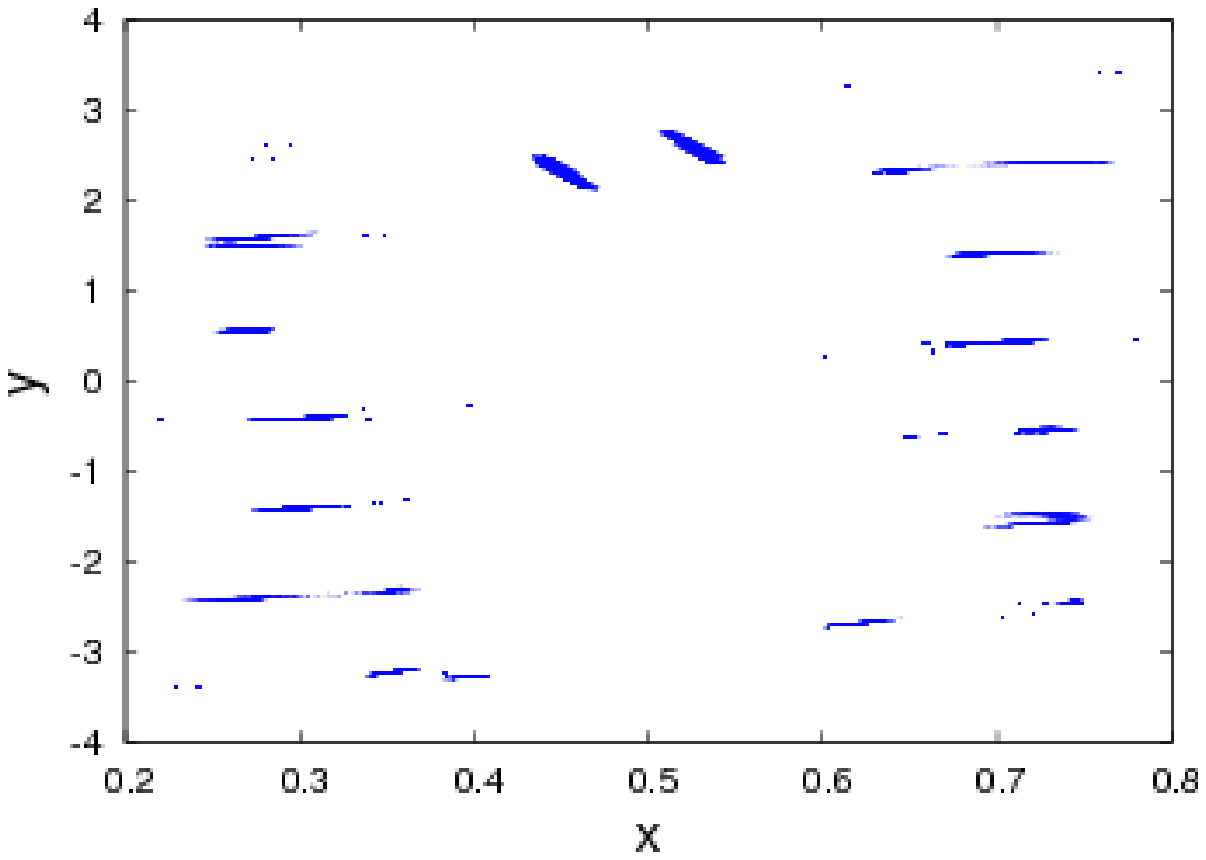}
\includegraphics[height=2.2in,width=2.05in]{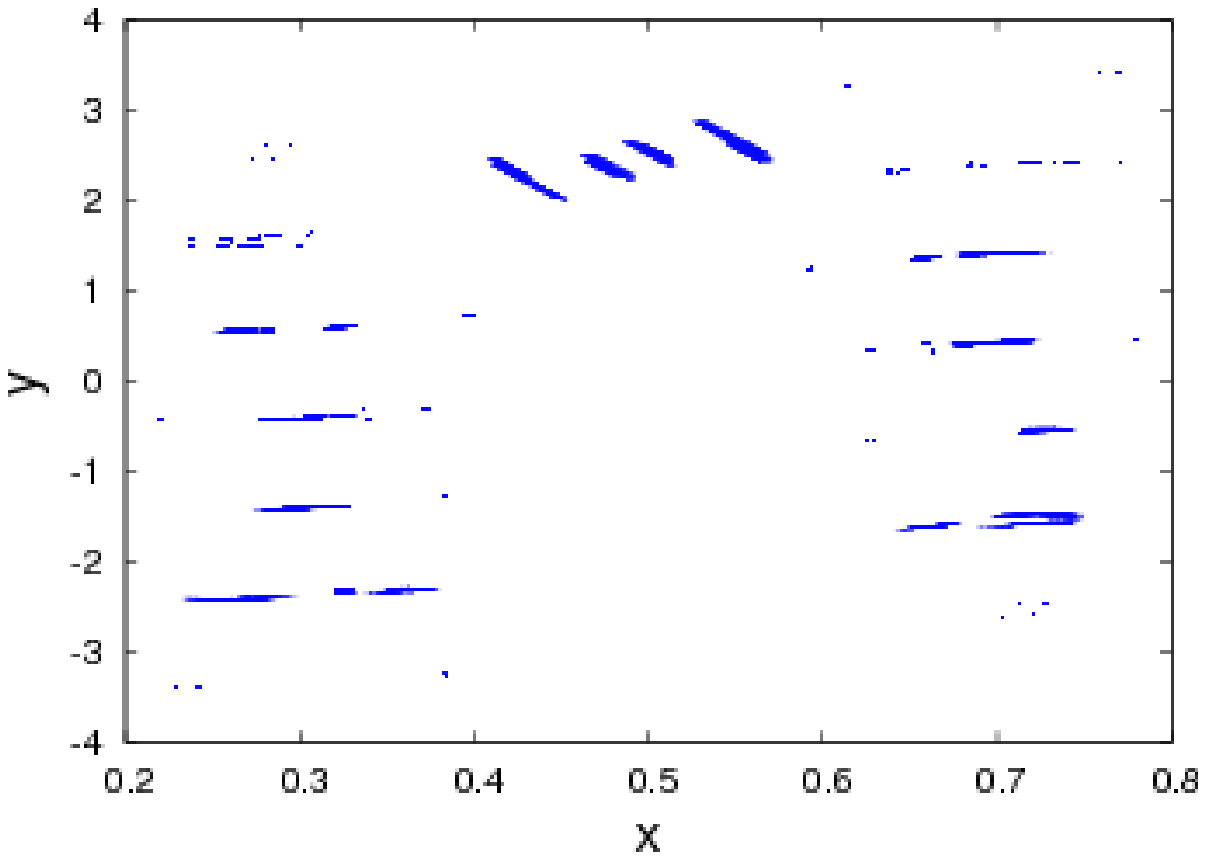}
\caption[Examples of the final dynamical steady states of directed CCM with $\mu=0.022$]{Three examples of final dynamical steady states of all nodes of  directed CCM Eq.(\ref{directed-equation}), each corresponding to a single initial condition at $\mu=0.022$. Shown are 10000 iterations of each node on the same plot.} \label{fig-directed-0022steady}
\end{center} 
\end{figure}

\begin{figure}[!hbt]
\begin{center}
$\begin{array}{ccc}
\includegraphics[height=1.85in,width=2.02in]{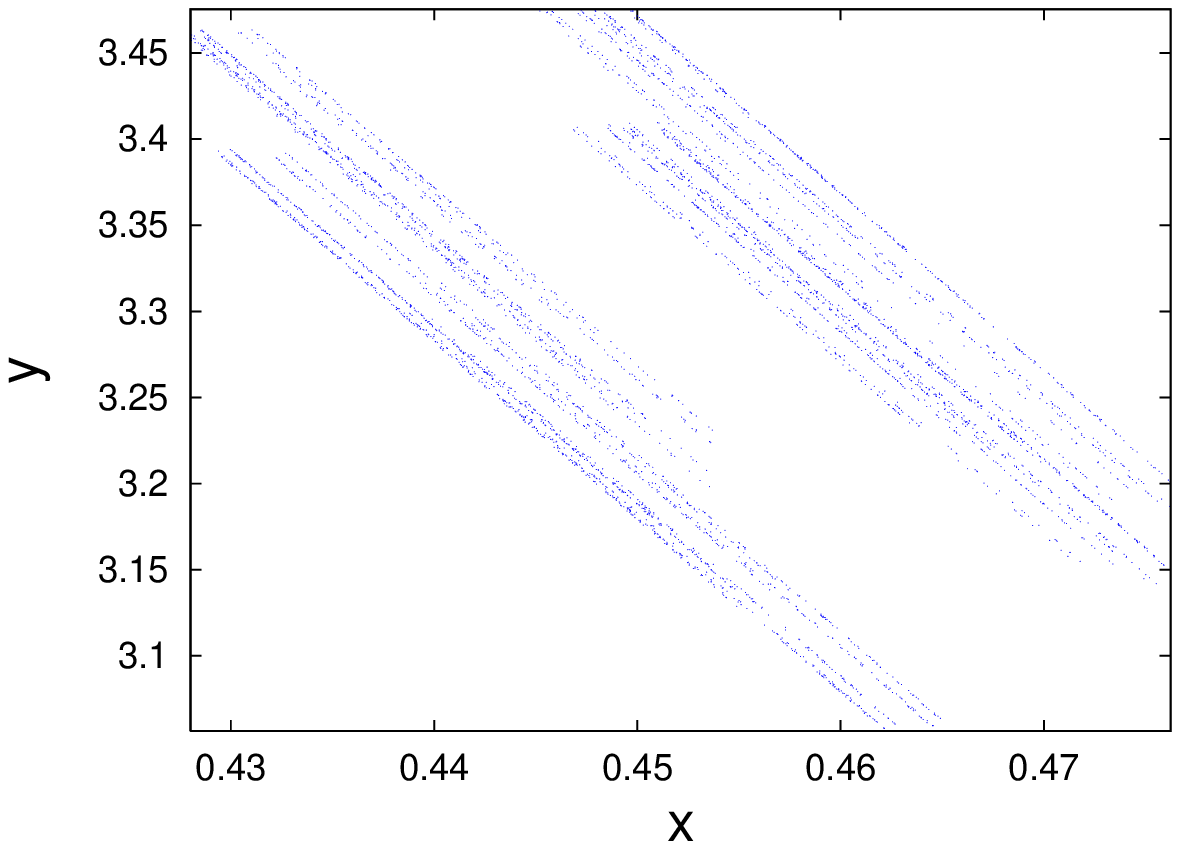} &
\includegraphics[height=1.85in,width=2.02in]{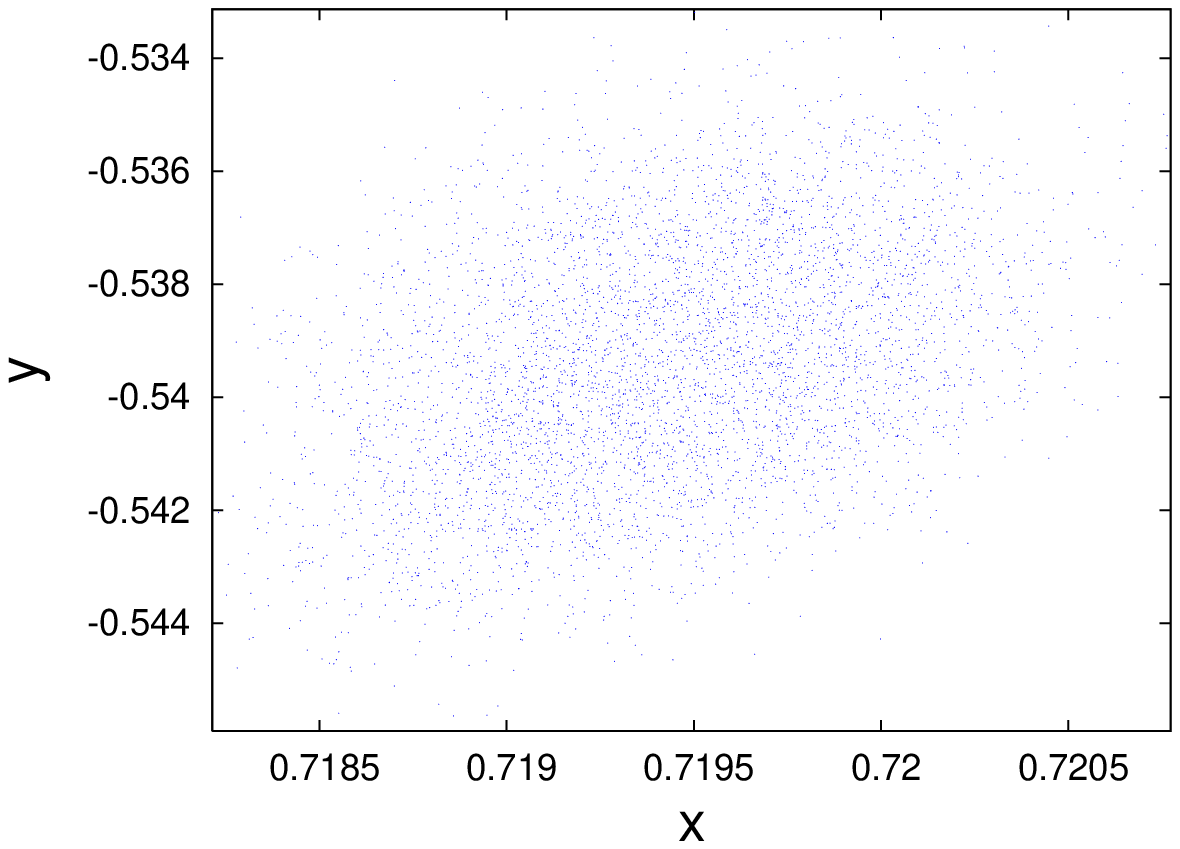} &
\includegraphics[height=1.85in,width=2.02in]{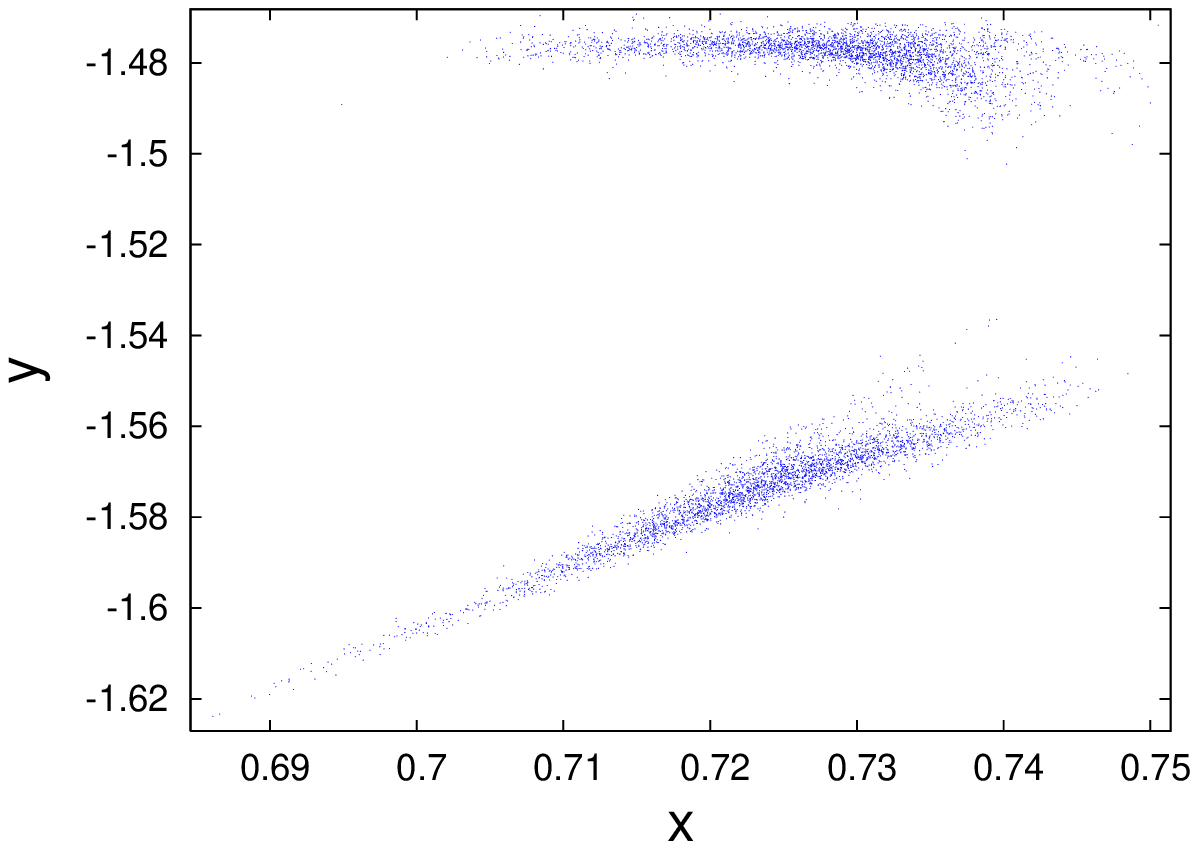} \\
\mbox{(a)} & \mbox{(b)} & \mbox{(c)}
\end{array}$
\caption[Examples of single-node orbits taken from data in Fig.\,\ref{fig-directed-0022steady}]{Examples of single-node orbits taken from data in Fig.\,\ref{fig-directed-0022steady}. The strange attractor of node 2 (hub) in (a), and typical orbits of node 214 in (b) and node 46 (c) (as previously, we do not show the entire orbit but only its representative part).} \label{fig-directed-0022orbits}
\end{center} 
\end{figure}

In Fig.\,\ref{fig-directed-0022steady} we report three examples of final dynamical steady states at $\mu=0.022$ by plotting 10000 iterations of each node on the same phase space (as done in Fig.\,\ref{fig-clustering}). Besides usual cluster-organization involving large majority of nodes, the system also exhibits other kinds of attractors. Specifically, the attractor visible in the upper-most region of the phase space corresponds to the hub node, and its position shifts depending on the initial conditions, although it qualitatively remains the same. 
To investigate this further, we show in Fig.\,\ref{fig-directed-0022orbits} three typical attractors/orbits appearing on various network nodes at this coupling strength. In Fig.\,\ref{fig-directed-0022orbits}a we see the fractal structure of the hub node's attractor, indicating this to be a strange attractor. At this coupling strength the hub node always settles in a strange attractor of this type regardless of the initial conditions, while no other node ever develops the same attractor. 
Other network nodes either display usual cluster-oscillations identical to the non-directed case (cf. Fig.\,\ref{fig-orbitsexamples}b), or develop weakly chaotic orbits as the one in Figs.\,\ref{fig-directed-0022orbits}b\,\&\,c, also present in the tree's non-directed dynamics (but with larger coupling strength).

\begin{figure}[!hbt]
\begin{center}
\includegraphics[height=2.6in,width=3.4in]{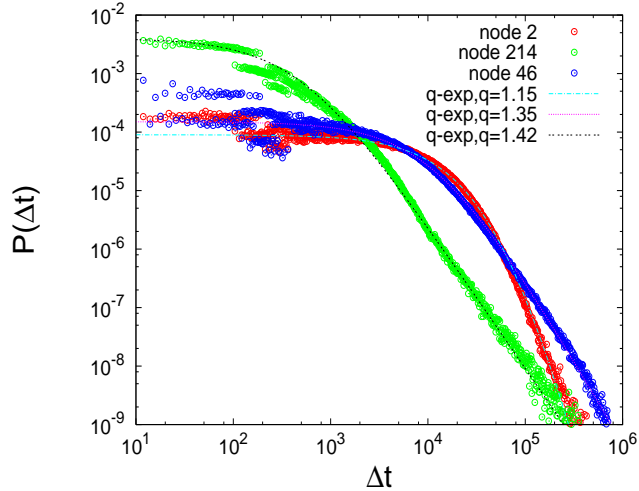}  
\caption[Distributions of return times to phase space partition in $x$-coordinate for single-node orbits with $\mu=0.022$]{Distributions of return times to phase space partition (100000 cells) in $x$-coordinate for single-node orbits from  Fig.\,\ref{fig-directed-0022orbits} with $\mu=0.022$, fitted with q-exponential distributions.} \label{fig-directed-0022rt}
\end{center} 
\end{figure}

We examine the statistical properties of various single-node orbits by computing their return time distributions with respect to phase space partition in $x$-coordinate. The results are shown in Fig.\,\ref{fig-directed-0022rt} where we compare the distributions of return times for the orbits from  Fig.\,\ref{fig-directed-0022orbits}. 
All the distributions indicate presence of long-term correlations in their motion, and can be fitted with the q-exponential function Eq.(\ref{qexp}) with various values of $q$.

\begin{figure}[!hbt]
\begin{center}
$\begin{array}{cc}
\includegraphics[height=2.45in,width=3.15in]{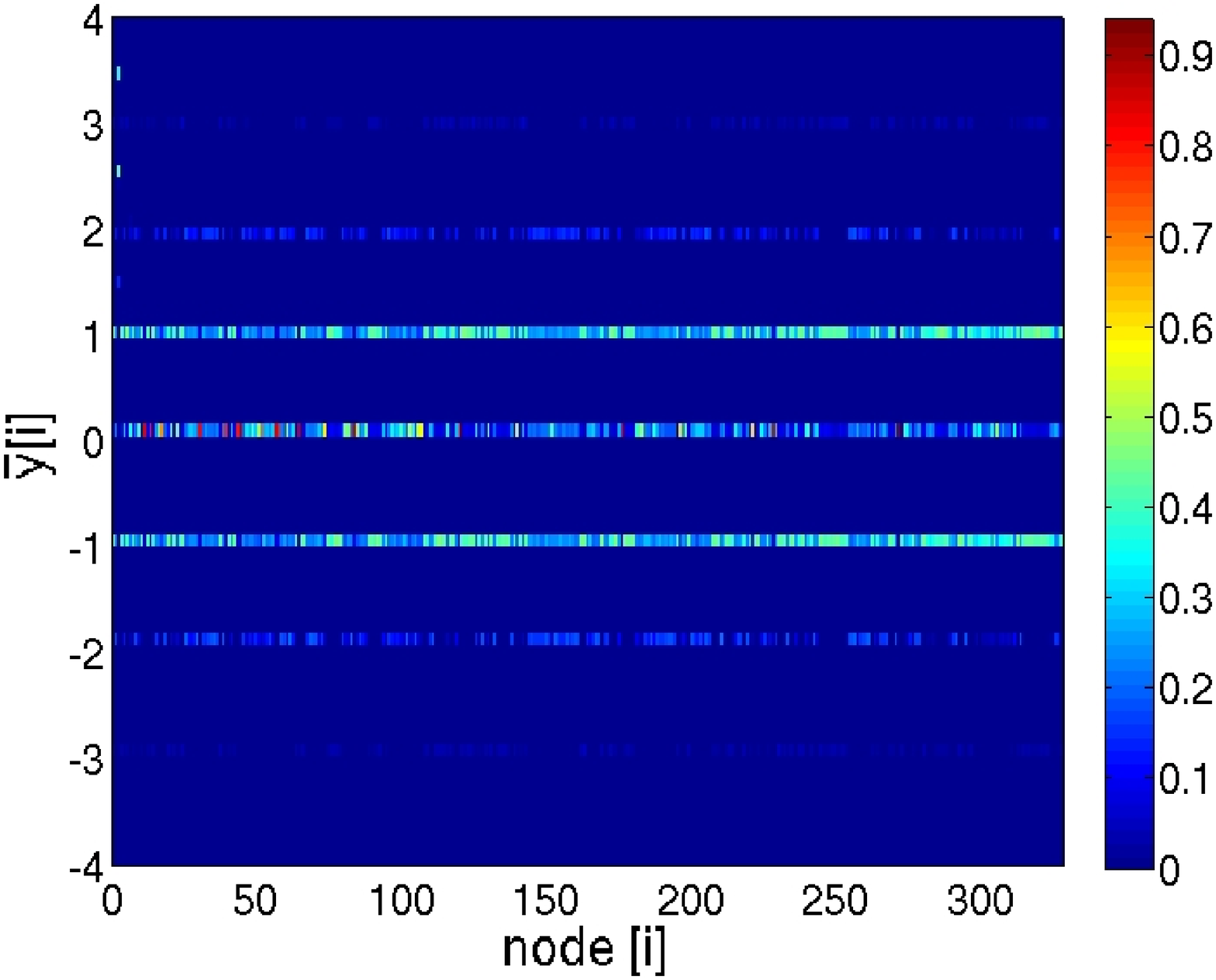} & 
\includegraphics[height=2.45in,width=3.15in]{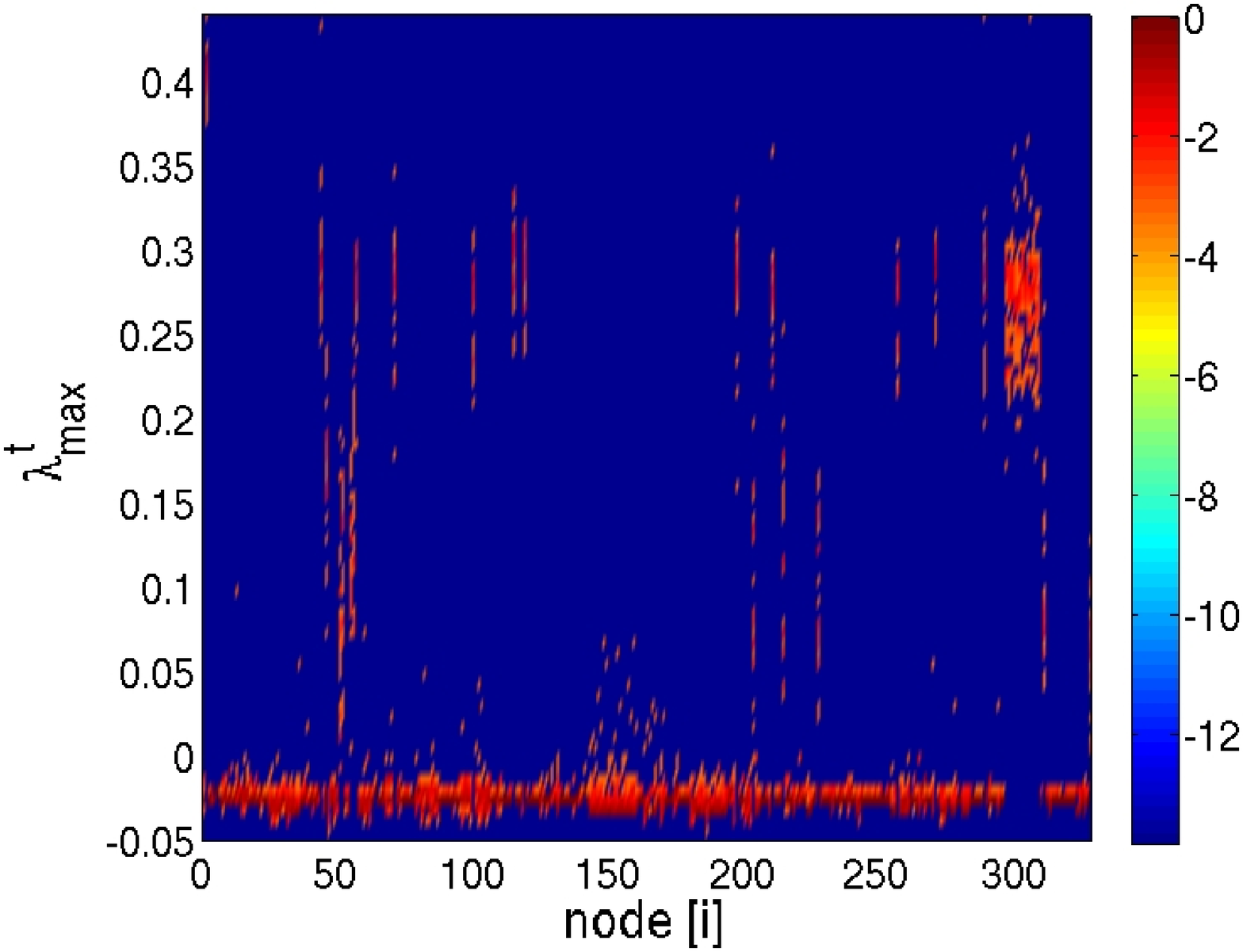} \\ 
\mbox{(a)} & \mbox{(b)} 
\end{array}$ 
\caption[2D color histograms showing node by node distributions of $\bar{y}$-values and $\lambda_{max}^t$-values at $\mu=0.022$]{2D color histograms showing node by node distributions of $\bar{y}[i]$-values in (a) and FTMLE $\lambda_{max}^t$ in (b), averaged over many initial conditions at $\mu=0.022$.} \label{fig-directed-0022stat}
\end{center}
\end{figure}

Furthermore, we examine the time-averaged orbits $\bar{y}[i]$ in Fig.\,\ref{fig-directed-0022stat}a by showing the histogram of $\bar{y}[i]$-values for each node separately. Note that hub node (node 2) has $\bar{y}[i]$-value out of usual cluster $\bar{y}[i]$-values (which are almost integer values), whereas other nodes mostly belong to three central clusters with $\bar{y}\cong -1$, $\bar{y}\cong 0$ and $\bar{y}\cong 1$. 
FTMLE $\lambda_{max}^t$ defined as in non-directed case are studied for this coupling strength node by node. In Fig.\,\ref{fig-directed-0022stat}b we report 2D color histogram showing distribution of $\lambda_{max}^t$-values averaged over many initial conditions, for each node. Clearly, the destabilized nodes exhibit positive FTMLE (cf. Fig.\,\ref{fig-directed-0022transition}), while other nodes mainly show $\lambda_{max}^t \sim -0.02$. Also, unstable nodes show $\lambda_{max}^t \sim 0.3$ which is much less than the uncoupled case with $\lambda_{max}^t \sim 1.5$. Moreover, each unstable node shows a specific range of $\lambda_{max}^t$-values, indicating a presence of self-organized network behavior at this coupling strength displayed as a clear departure from a very chaotic nature of the uncoupled standard map.

\begin{figure}[!hbt]
\begin{center}
\includegraphics[height=4.in,width=4.55in]{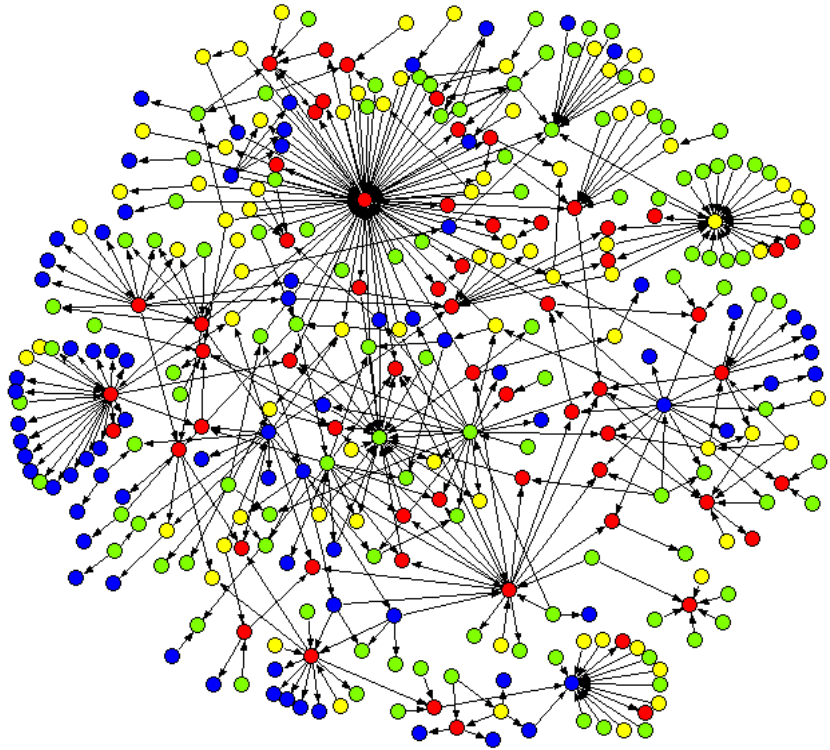}
\caption[Graphical representation of the standard deviations $\sigma(\bar{y})$ for all network nodes for $\mu=0.022$]{Standard deviation $\sigma[i](\bar{y})$ for each node averaged over many initial conditions for $\mu=0.022$. Red: $\sigma[i](\bar{y})>1.26$, yellow $1.15<\sigma[i](\bar{y})<1.26$, green $1.02<\sigma[i](\bar{y})<1.15$, blue $\sigma[i](\bar{y})<1.02$.} \label{fig-sigmaybar-0022}
\end{center}
\end{figure}

Finally, we study the flexibility or emergent motion by considering the distribution of $\bar{y}[i]$-values node by node for many initial conditions, and computing the standard deviation $\sigma[i](\bar{y})$ for each node. Small $\sigma[i](\bar{y})$ indicates node prefers to settle in a specific cluster only, while large $\sigma[i](\bar{y})$ points to a certain flexibility of node's final motion. In Fig.\,\ref{fig-sigmaybar-0022} we show the E.Coli's network, indicating the flexibility of nodes.
While centrally located nodes with many ingoing and outgoing connections are typically having small $\sigma[i](\bar{y})$ (blue), outer nodes are generally more flexible (red) with bigger $\sigma[i](\bar{y})$, i.e. can be found in various motion types depending on the initial conditions. This suggests that the network exhibits a self-organization in terms of each node showing a certain range of possible behaviors that depend on initial conditions.\\[0.1cm]

\textbf{E.Coli Network Dynamics at $\mu=0.05$.}  In addition to the instability transition around $\mu=0.022$ we study the instability region around $\mu=0.05$ where some nodes remain clustered with periodic orbits, while some other nodes develop fully chaotic motion equivalent to the uncoupled dynamics. However, except for full chaos, at this coupling strength the network displays various types of non-periodic motion which are specifically dependent on the node in question.

\begin{figure}[!hbt]
\begin{center}
$\begin{array}{cc}
\includegraphics[height=2.85in,width=3.1in]{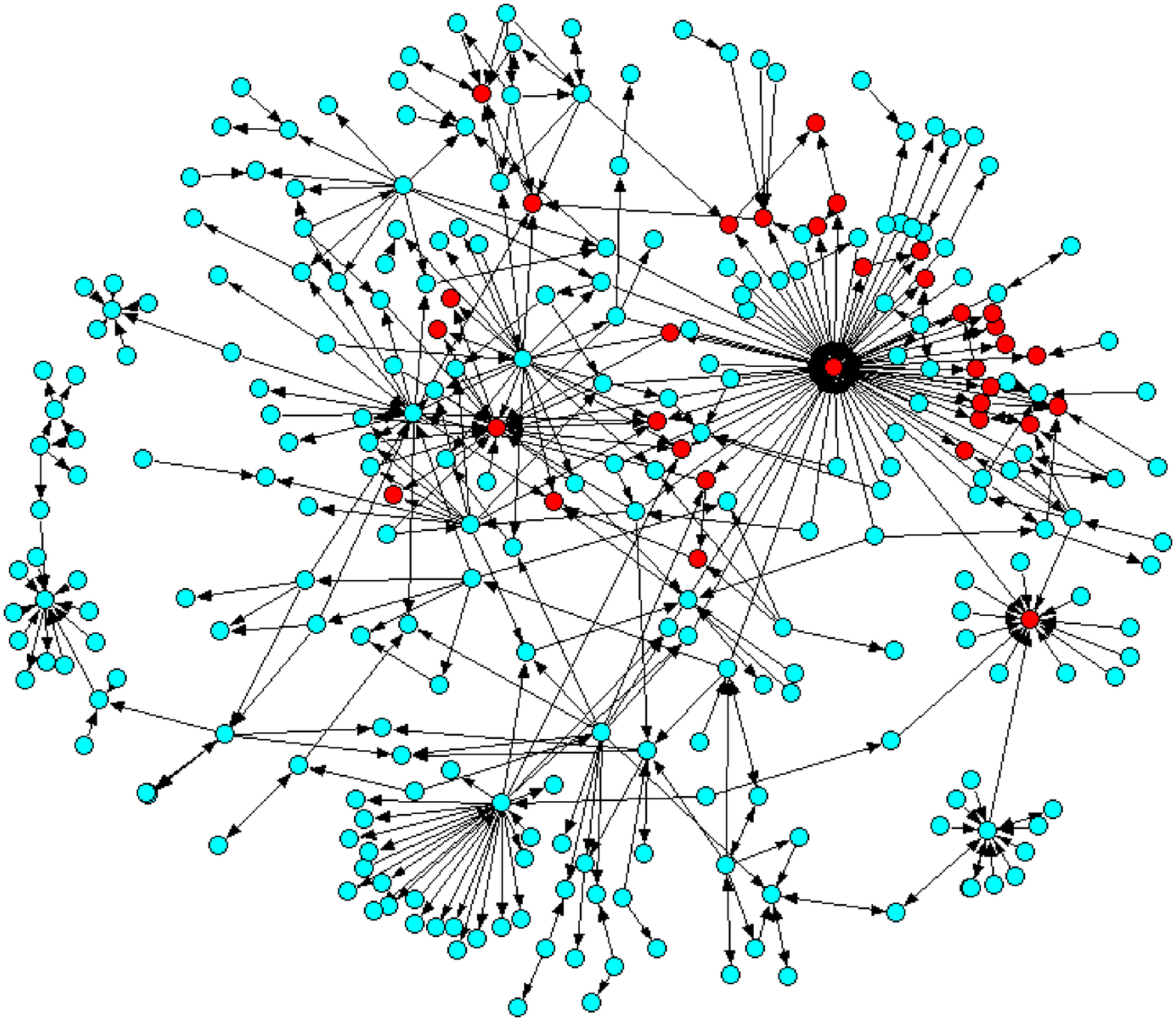} & 
\includegraphics[height=2.85in,width=3.1in]{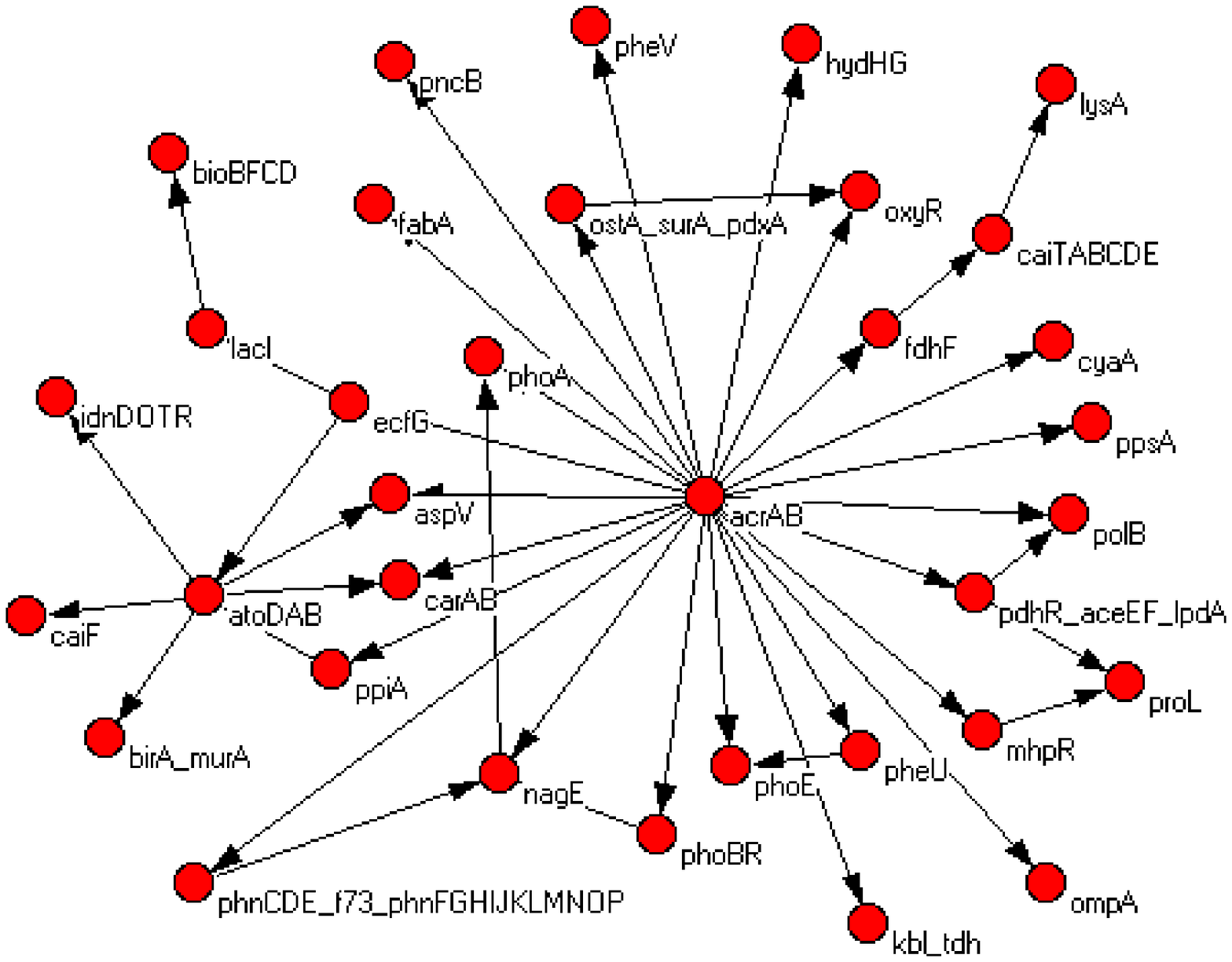} \\ 
\mbox{(a)} & \mbox{(b)} 
\end{array}$ 
\caption[Graphical representations of the sub-network undergoing destabilization at $\mu=0.05$]{Instability region at $\mu=0.05$. E.Coli's largest connected component with destabilized nodes (marked in red) in (a), the sub-network undergoing destabilization with nodes/genes biological names in (b).} \label{fig-directed-005pajek}
\end{center}
\end{figure}

In Fig.\,\ref{fig-directed-005pajek}a we show the E.Coli network's largest connected component with red color for the non-periodic nodes at $\mu=0.05$: again, the irregular behavior is localized to a sub-network, although this sub-network is fully connected to the rest of the network. In Fig.\,\ref{fig-directed-005pajek}b we report the graphical representation of the destabilized sub-network, with respective biological names for the nodes/genes.
In addition to other nodes, this unstable sub-network includes the unstable sub-network for $\mu=0.022$ from Fig.\,\ref{fig-directed-0022pajek}b.

\begin{figure}[!hbt]
\begin{center}
$\begin{array}{ccc}
\includegraphics[height=1.85in,width=2.02in]{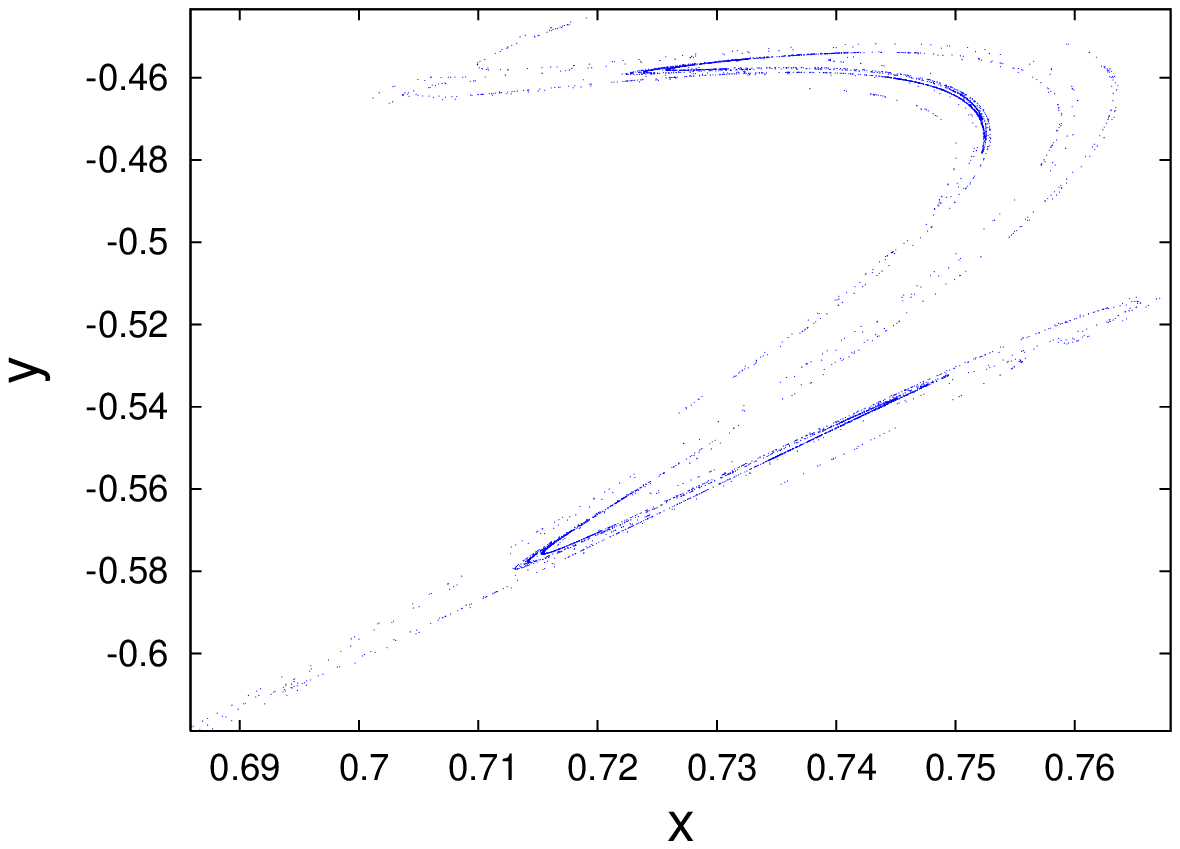} &
\includegraphics[height=1.85in,width=2.02in]{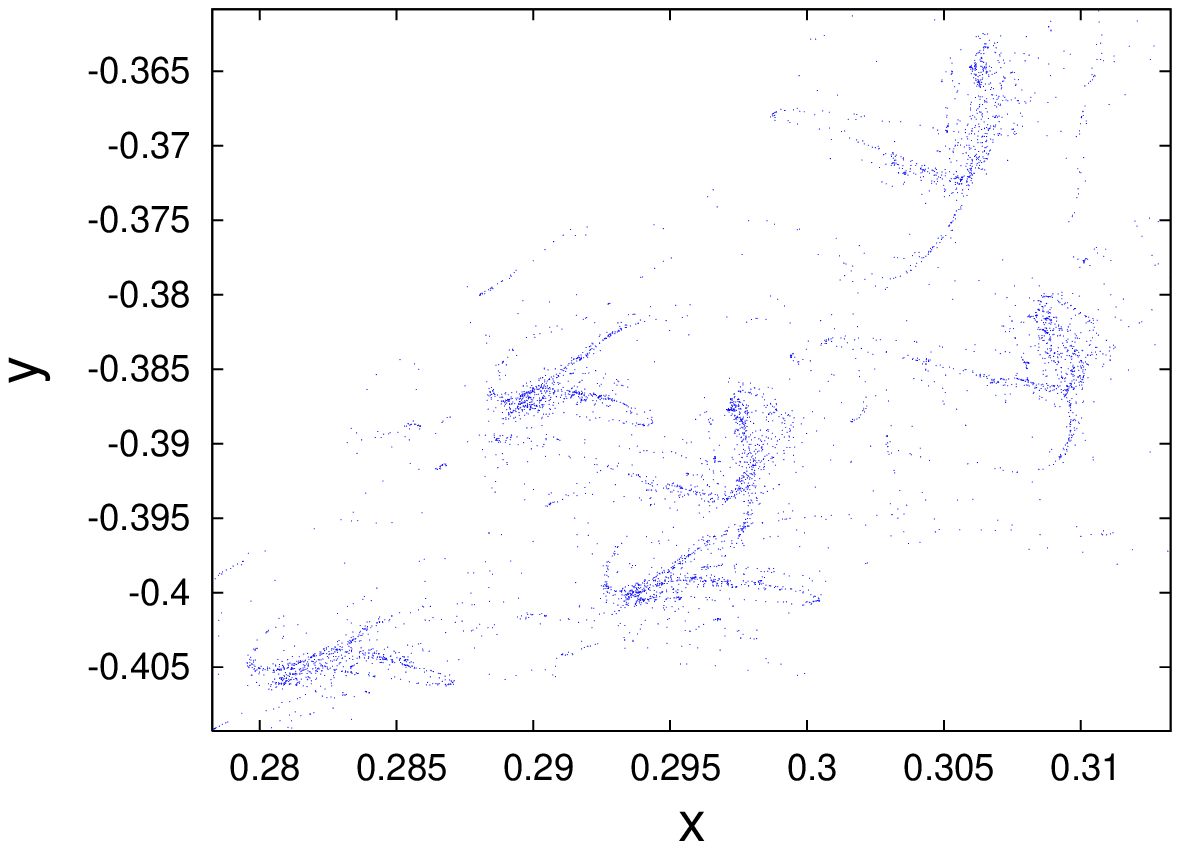} &
\includegraphics[height=1.85in,width=2.02in]{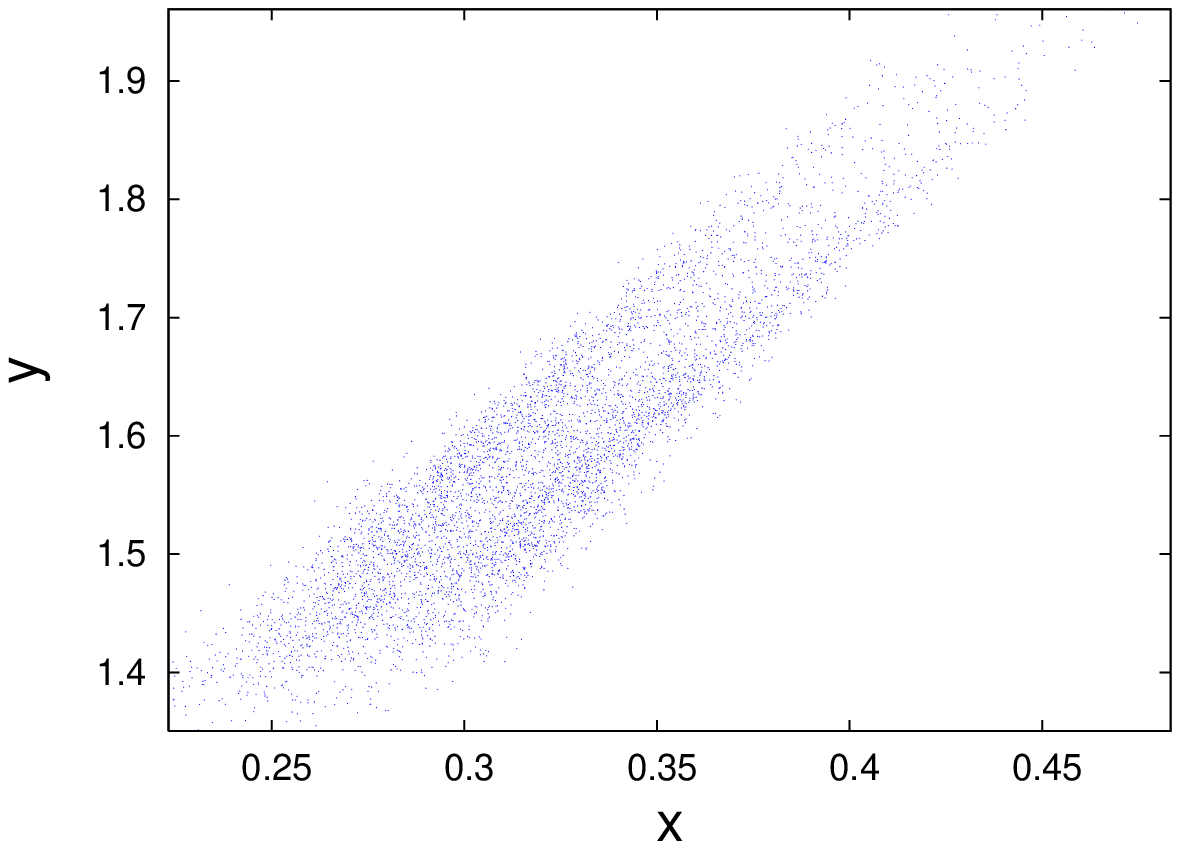} \\
\mbox{(a)} & \mbox{(b)} & \mbox{(c)}
\end{array}$
\caption[Examples of single-node orbits with $\mu=0.05$]{Examples of single-node orbits of CCM on E.Coli's directed gene regulatory network with $\mu=0.05$. Node 18 in (a), node 86 in (b) and node 119 in (c) (as usually, we do not show the entire orbits but only their representative parts).} \label{fig-directed-005orbits}
\end{center} 
\end{figure}

\begin{figure}[!hbt]
\begin{center}
\includegraphics[height=2.6in,width=3.4in]{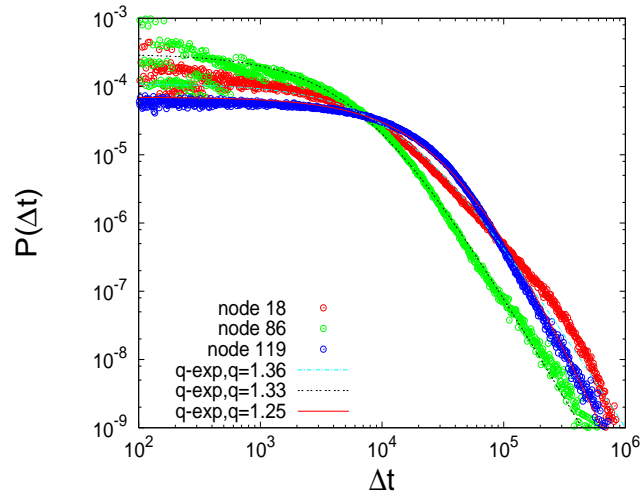}  
\caption[Distributions of return times to phase space partition in $x$-coordinate for single-node orbits with $\mu=0.05$]{Distributions of return times to phase space partition in $x$-coordinate (100000 cells) for single-node orbits from  Fig.\,\ref{fig-directed-005orbits} with $\mu=0.05$, fitted with q-exponential distributions.} \label{fig-directed-005rt}
\end{center} 
\end{figure}

Contrary to what was observed in the case of $\mu=0.022$, at this coupling strength the hub node does not display any specific attractor, but it exhibits non-localized strongly chaotic motion regardless of the initial conditions. However, the behavior of other non-periodic nodes shows a variety of structural patterns as illustrated in Fig.\,\ref{fig-directed-005orbits} where we show three typical node orbits. 
Interestingly, the orbit in Fig.\,\ref{fig-directed-005orbits}a resembles the chaotic attractor found in the non-directed 4-star case from Fig.\,\ref{fig-attr-048}b. This attractor  appears to be inherently related to this particular coupling form among the standard maps, rather than to the directedness of graph. The other 4-star attractor pattern Fig.\,\ref{fig-SNA} was not found in the directed case. We consider the distributions of return times to phase space partition in $x$-coordinate for the single node orbits from Fig.\,\ref{fig-directed-005orbits} and show the results in Fig.\,\ref{fig-directed-005rt}. As previously, the distributions are fitted with q-exponential forms, indicating presence of similar collective effects at this coupling strength. 
Note that respective values of $q$ in both cases are in the same range, although they always depend on the node.

\begin{figure}[!hbt]
\begin{center}
$\begin{array}{cc}
\includegraphics[height=2.45in,width=3.15in]{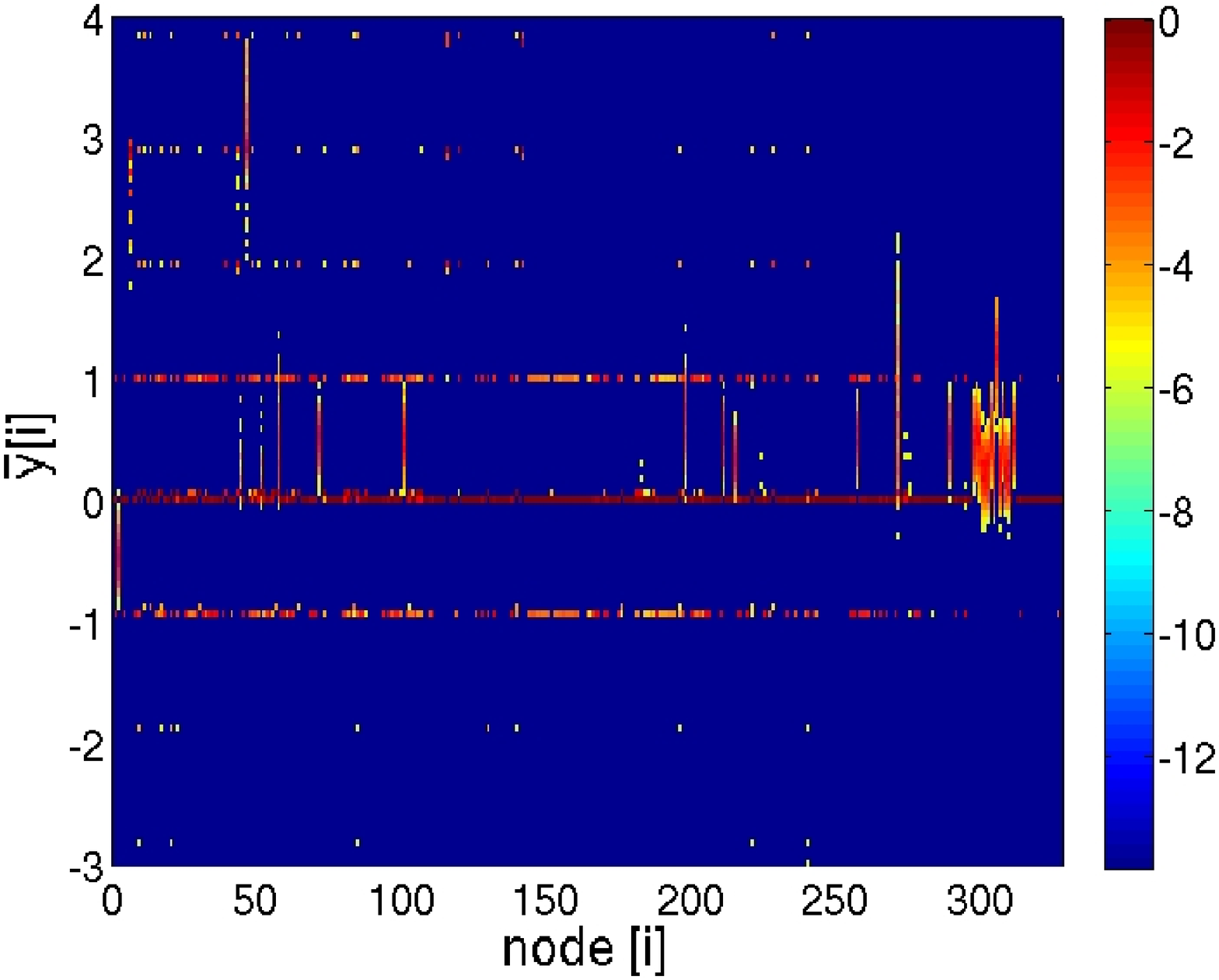} & 
\includegraphics[height=2.45in,width=3.15in]{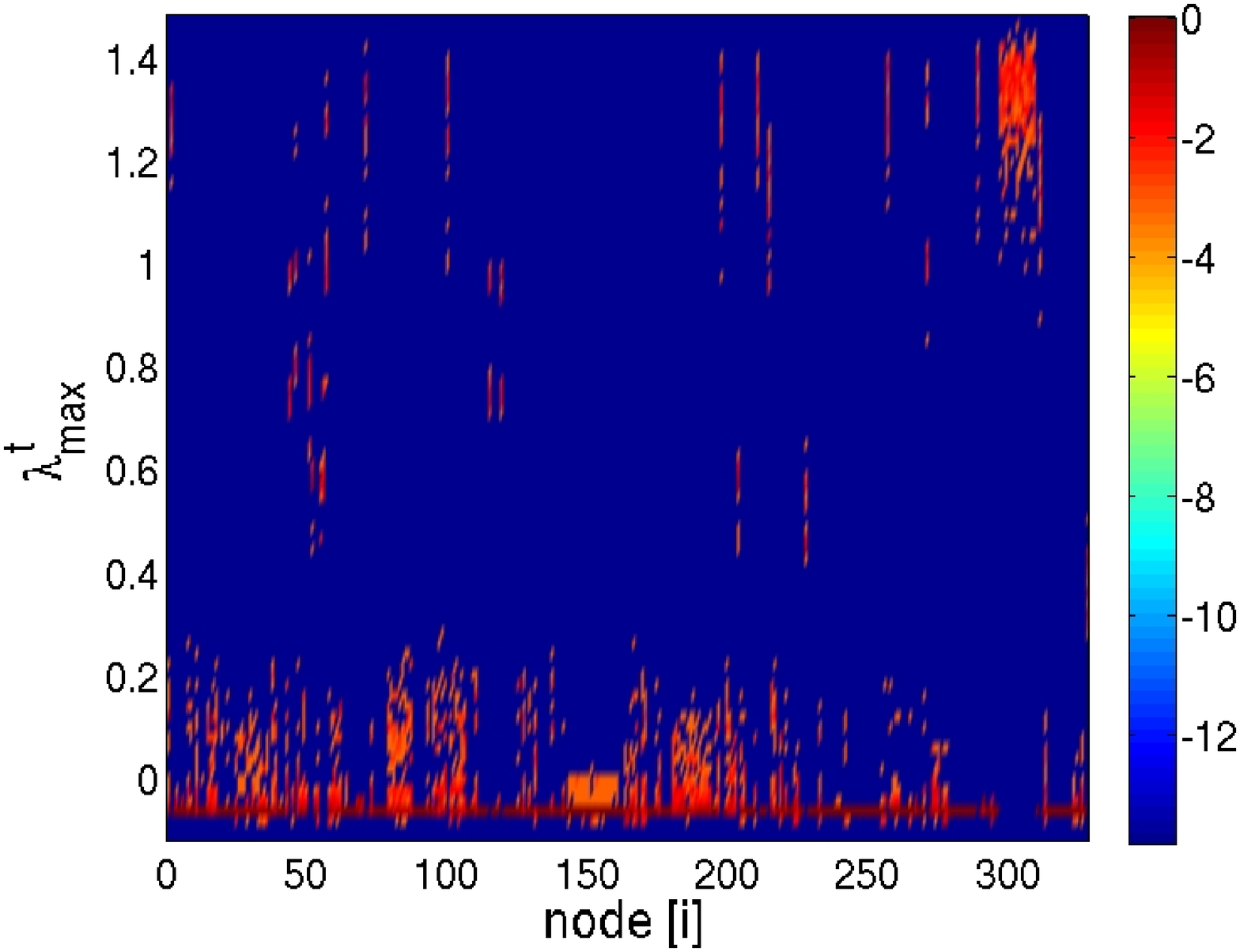} \\ 
\mbox{(a)} & \mbox{(b)} 
\end{array}$ 
\caption[2D color histograms showing node by node distributions of $\bar{y}$-values and $\lambda_{max}^t$-values for $\mu=0.05$]{2D color histograms showing node by node distributions of $\bar{y}[i]$-values in (a) and $\lambda_{max}^t$ in (b), averaged over many initial conditions at $\mu=0.05$.} \label{fig-directed-005stat}
\end{center}
\end{figure}

\begin{figure}[!hbt]
\begin{center}
\includegraphics[height=4.in,width=4.55in]{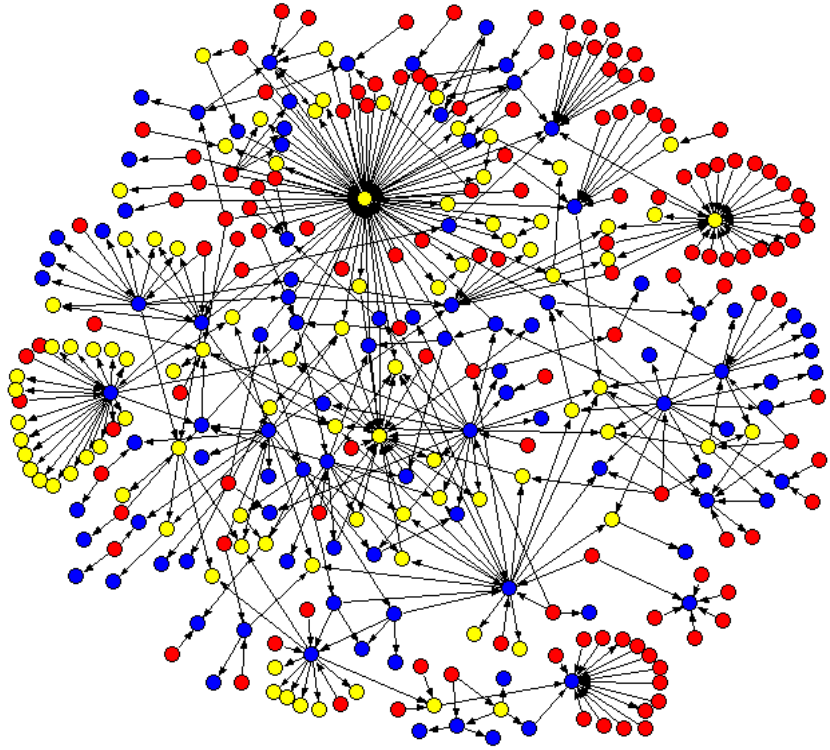}
\caption[Graphical representation of the standard deviations $\sigma(\bar{y})$ for all network nodes for $\mu=0.05$]{Standard deviation $\sigma[i](\bar{y})$ for each node averaged over many initial conditions for $\mu=0.05$. Red: $\sigma[i](\bar{y})>0.33$, yellow $0.03<\sigma[i](\bar{y})<0.33$, blue $\sigma[i](\bar{y})<0.03$.} \label{fig-sigmaybar-005}
\end{center}
\end{figure}

The global statistical properties of the directed CCM Eq.(\ref{directed-equation}) at $\mu=0.05$ are shown in Fig.\,\ref{fig-directed-005stat} where we show 2D color histograms of $\bar{y}[i]$-values and FTMLE $\lambda_{max}^t$-values. As opposed to $\mu=0.022$ case, the statistics of $\bar{y}[i]$-values depends on the node in question, with different nodes exhibiting very different $\bar{y}$ distributions. It seems each node has a certain "role" in the network' collective motion, given by its spectrum of possible behaviors in function of the initial conditions.
The distribution of $\lambda_{max}^t$-values is also node dependent; while most of the nodes are stable with $\lambda_{max}^t<0$, some nodes display a double nature by showing (depending on initial conditions) either positive or negative FTMLE. In particular, the node 18, similarly to 4-star's branch node, can display an attractor (cf. Fig.\,\ref{fig-directed-005orbits}a) as well as a periodic orbit. Also, while some unstable nodes show large FTMLE $\lambda_{max}^t \sim 1$ similarly to the uncoupled case, some other nodes show moderate FTMLE values of $\lambda_{max}^t \sim 0.5$ relating them to the $\mu=0.022$ case. Still, the majority of nodes show a stable behavior with negative FTMLE. We also examine the network distribution of standard deviations $\sigma[i](\bar{y})$ with data from Fig.\,\ref{fig-directed-005stat}a, describing the flexibility of each node. The results are reported in Fig.\,\ref{fig-sigmaybar-005} where, as previously, we attach a color to each node expressing its $\sigma[i](\bar{y})$-value.
In contrast to $\mu=0.022$ case, many outer less connected nodes are now fixed with very small $\sigma[i](\bar{y})$-values, while inner more connected nodes display flexibility of their final dynamical states in relation to the initial conditions.

\begin{figure}[!hbt]
\begin{center}
$\begin{array}{ccc}
\includegraphics[height=1.85in,width=2.02in]{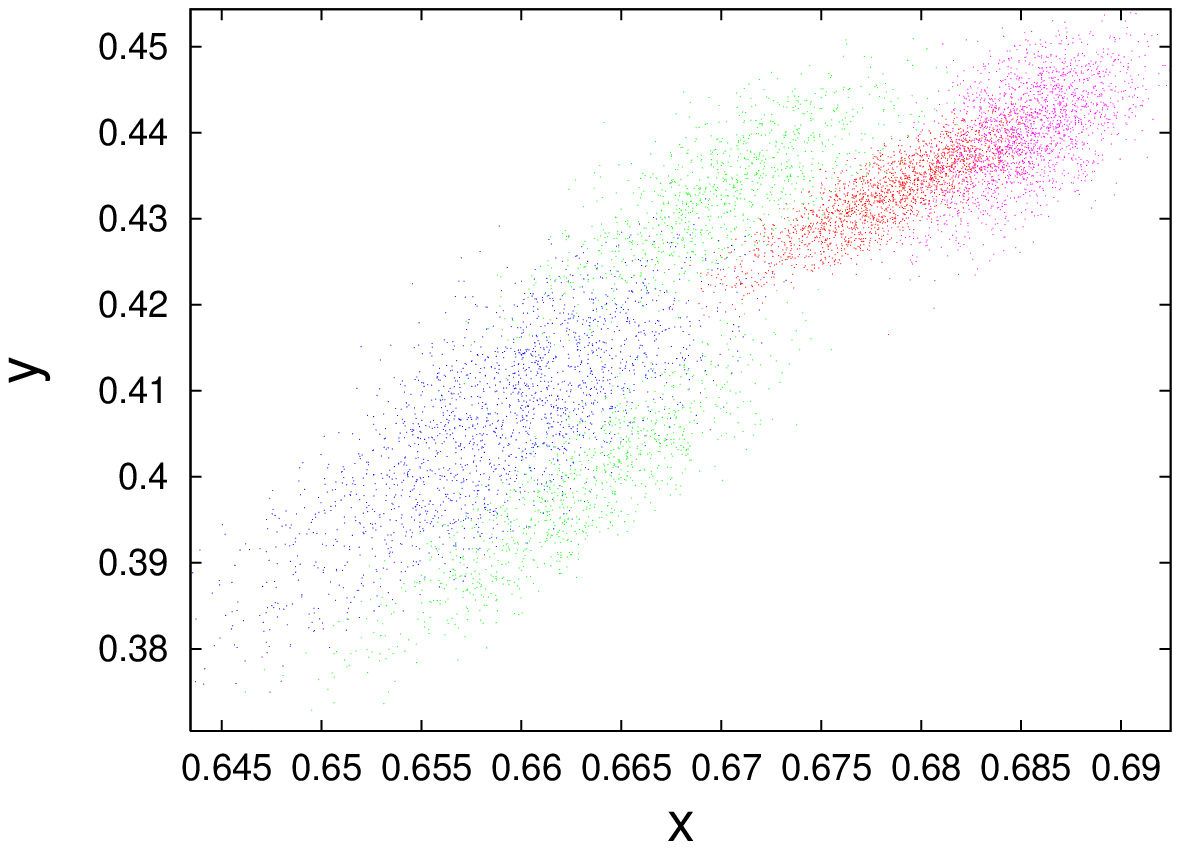} &
\includegraphics[height=1.85in,width=2.02in]{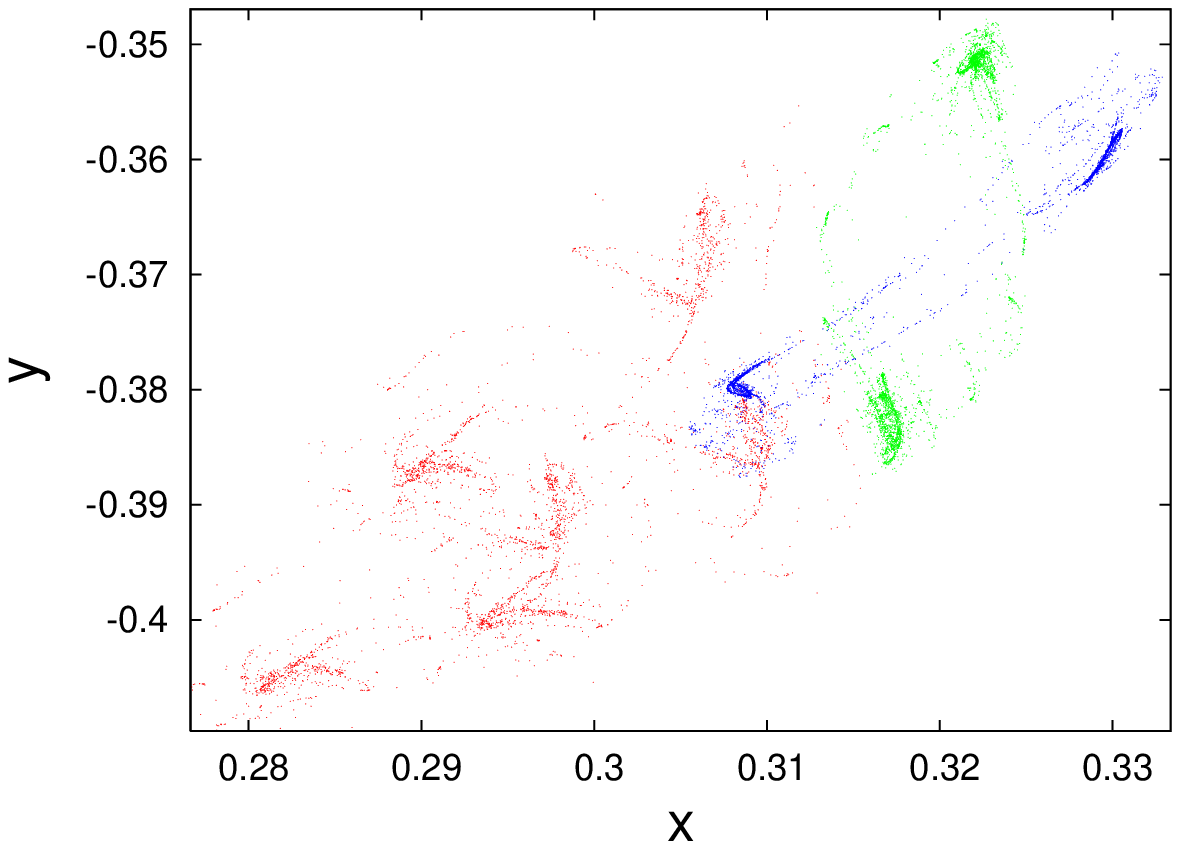} &
\includegraphics[height=1.85in,width=2.02in]{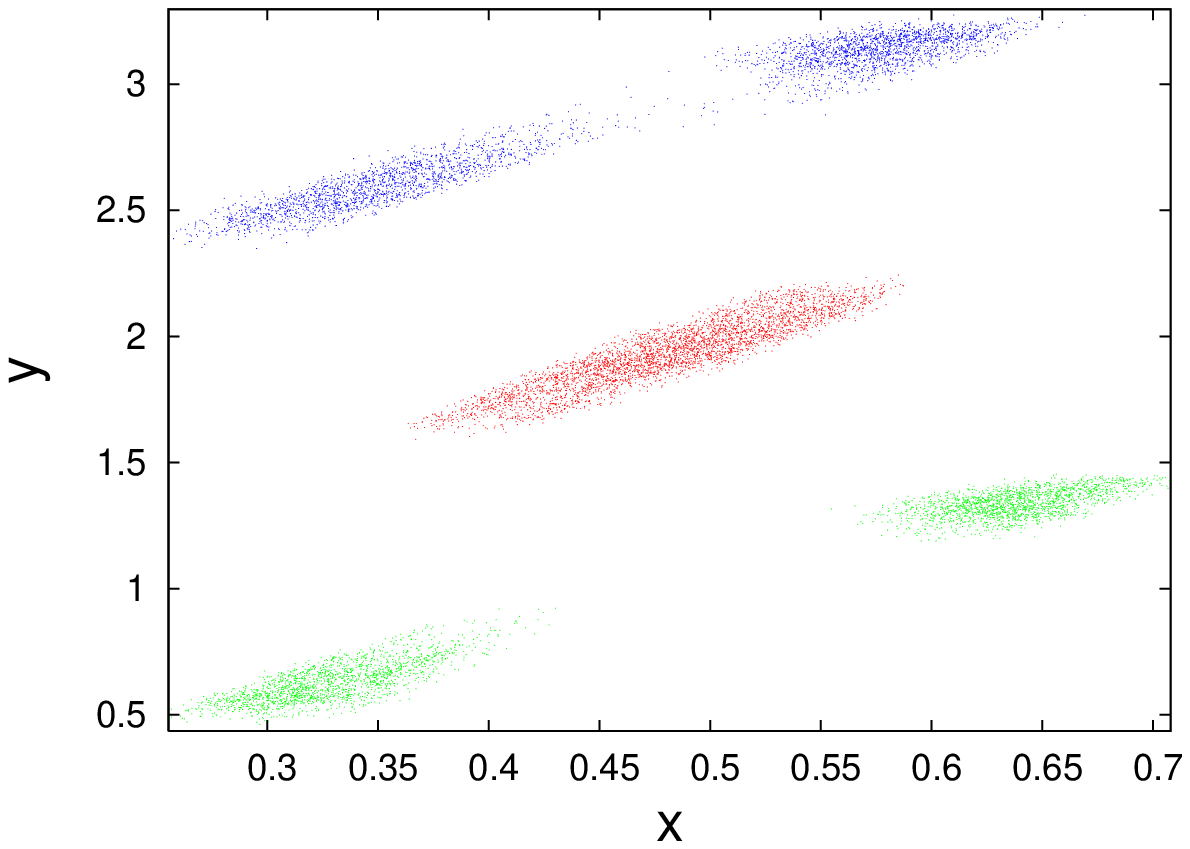} \\
\mbox{(a)} & \mbox{(b)} & \mbox{(c)}
\end{array}$
\caption[Examples of final single-node orbits in relation to different initial conditions at $\mu=0.05$, for some nodes]{Examples of final single-node orbits (different colors) in relation to different initial conditions at $\mu=0.05$. Node 56 in (a), node 86 in (b) and node 119 in (c).} \label{fig-directed-005nodestates}
\end{center} 
\end{figure}

This indicates the unstable motion of specific nodes for the coupling strength of $\mu=0.05$ can exhibit different patterns that however do share common statistical properties. We illustrate this further in Fig.\,\ref{fig-directed-005nodestates} where we show different orbits  corresponding to various initial conditions for three network nodes.
The dynamics of a given non-periodic node does maintain certain motion properties: note the similarities among the orbit structures and locations in all three cases. Further examination of the same nodes reveals constant presence of the given type of motion, which is qualitatively independent of initial conditions. Some other nodes show this behavior displaying different motion patterns, although most of the nodes are still periodic (cf. Fig.\,\ref{fig-directed-005stat}b). This kind of motion flexibility is characteristic for biological collective dynamics: e.g., activity of a gene is expected to be adaptive to the cellular needs while maintaining the key behavioral properties. We furthermore examine the distributions of return times with respect to a phase space partition in $x$-coordinate. The shown profiles refer to the node orbits emerging for different initial conditions for node 86 (cf. Fig.\,\ref{fig-directed-005nodestates}b) and node 119 (cf. Fig.\,\ref{fig-directed-005nodestates}c). 
\begin{figure}[!hbt]
\begin{center}
$\begin{array}{cc}
\includegraphics[height=2.45in,width=3.1in]{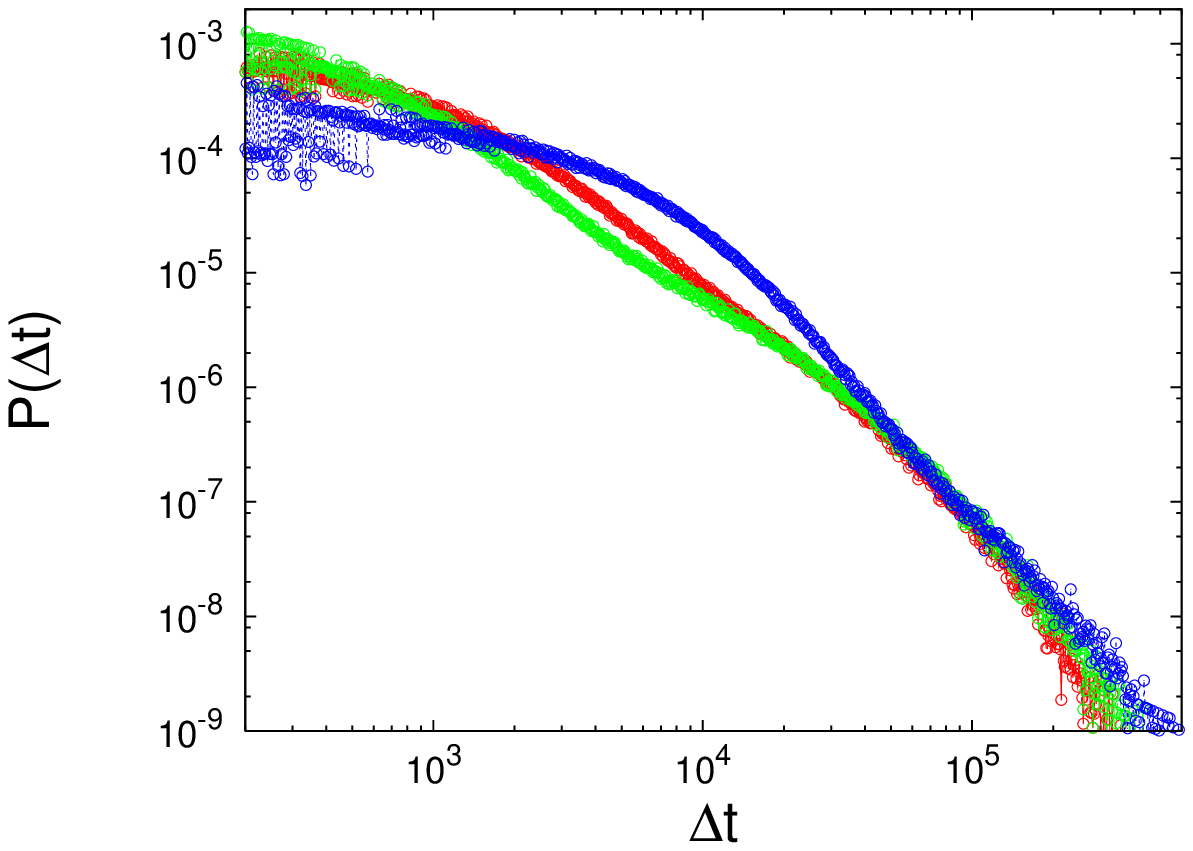} & 
\includegraphics[height=2.45in,width=3.1in]{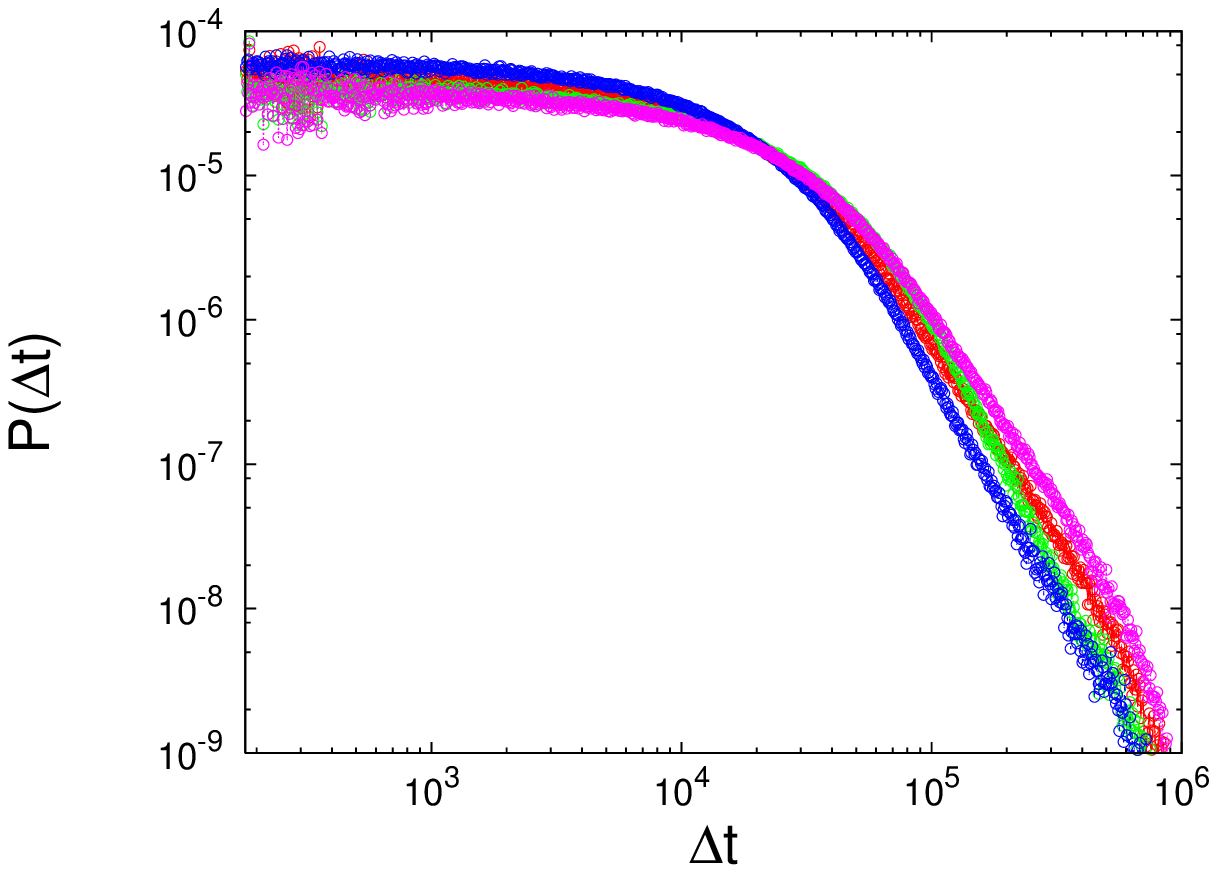} \\ 
\mbox{(a)} & \mbox{(b)} 
\end{array}$ 
\caption[Distributions of return times with respect to a phase space partition in $x$-coordinate for various orbits with $\mu=0.05$]{Distributions of return times with respect to a phase space partition in $x$-coordinate (100000 cells) for orbits corresponding to various initial conditions with $\mu=0.05$. Node 86 in (a) and node 119 in (b).} \label{fig-directed-rt005nodes}
\end{center}
\end{figure}
The distribution profiles show minor variations, while still maintaining common structural properties regardless of initial conditions. 

This suggests the emergent dynamics of given unstable nodes is not fully irregular, but seems to posses certain flexibility with respect to node/initial conditions. As observed already, a feature of this type might be related to biological origin of the studied network, in terms of node's/gene's "adaptability". As we shall describe in the reminder of this Chapter, the flexibility of emergent dynamics with respect to initial/boundary conditions is the crucial characteristic of the continuous-time Hill model of gene interactions.

\section{Hill Dynamics of E.Coli Gene Regulatory Network}

\begin{figure}[!hbt]
\begin{center}
\includegraphics[height=4.in,width=4.7in]{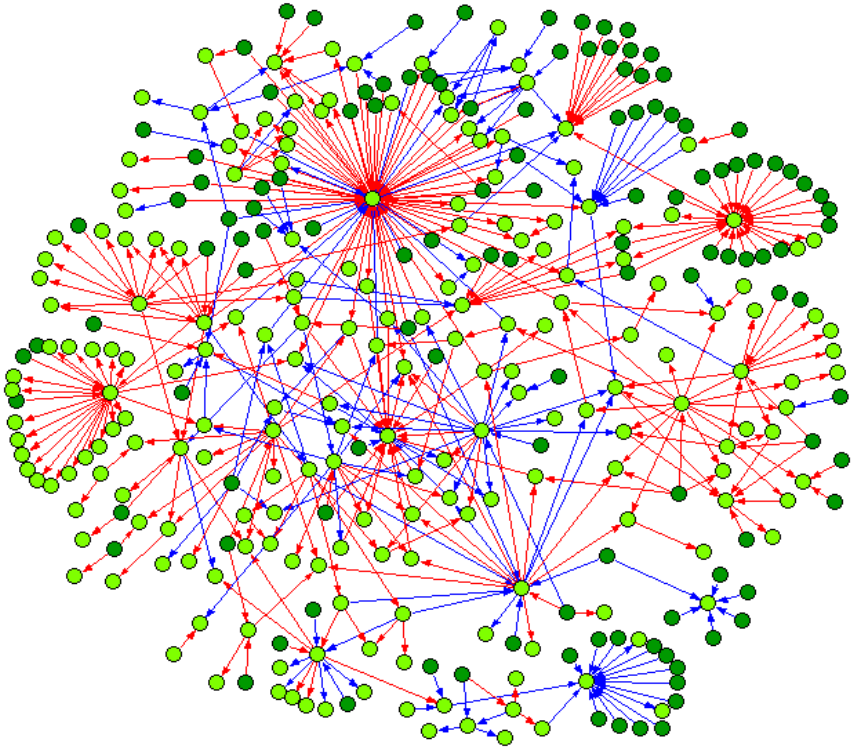}
\caption[Largest connected component of the E.Coli gene regulatory network with the interaction types pictured as activatory or repressory]{Largest connected component of the E.Coli gene regulatory network (cf. Fig.\,\ref{fig-connectedcomponent}) with the interaction types pictured as red (activatory) or blue (repressory). The nodes without incoming links (expect for self-loops) are in dark green, and other nodes are in light green.} \label{fig-ecoli-colorpajek}
\end{center}
\end{figure}

In this Section we study the two-dimensional Hill model of gene regulation dynamics on the largest connected component of E.Coli directed network from   Fig.\,\ref{fig-connectedcomponent}. The behavior of a single gene (node) will be described as a continuous-time 2D dynamical system defined in Eqs.(\ref{schuster})\&(\ref{hill}). Besides the directed nature of network, the interactions among the genes (nodes) are given as activation or repression, which we show in a new graphical representation of E.Coli network that includes interaction types as colors of links Fig.\,\ref{fig-ecoli-colorpajek} (following \cite{orr,mangan}). 

Two versions of the system are considered, differing by the logic gate model used for multiply regulated genes: the \textit{SUM model} and  \textit{AND model} will have all their logic gates described by the SUM-gates and AND-gates, respectively. The dynamical state of a single node/gene $[i]$ within a network with $N$ nodes is given by the pairs of variables $(q[i],p[i])$, and evolves with time according to:
\begin{equation}
\begin{array}{ccl}
 \dfrac{dq[i]}{dt}  & = & {\mathcal G}_i \{ F_{1i} (p[j]),\hdots , F_{Ni} (p[j]) \} - \gamma^{Q}_{i} q[i]  \\ 
   & & \\
 \dfrac{dp[i]}{dt}  & = & \alpha^{P}_{i} q[i] - \gamma^{P}_{i} p[i] 
\end{array}  \label{ecoli-model}
\end{equation}  
where in the case of SUM-gates the operator ${\mathcal G}$ is given as the average value of all the inputs:
\begin{equation}
 {\mathcal G}_i^S = \frac{1}{k_i^{in}} \sum_{j=1}^{j=N} F_{ji} (p[j])
\end{equation}
$k_i^{in}$ being in-degree of the node $[i]$, while in the case of AND-gates ${\mathcal G}$ is defined to be the minimal of all the inputs \cite{uribook}:
\begin{equation}
  {\mathcal G}_i^A = \min_{j=1,\hdots,N} F_{ji} (p[j]).
\end{equation}
The function $F_{ji}$ describes the way node $[j]$ influences node $[i]$ in case a link between the two nodes exists, and can be either activatory or repressory:
\begin{equation}
\begin{array}{ccc}
 F^{+}_{ji} (p[j]) & = \xi_{ji} + \dfrac{\beta_{ji} p[j]^{n_{ji}}}{p[j]^{n_{ji}} + {T_{ji}}^{n_{ji}}} , & \;\;\; \mbox{activation}  \\
   & & \\
 F^{-}_{ji} (p[j]) & = \xi_{ji} + \dfrac{\beta_{ji} {T_{ji}}^{n_{ji}}}{p[j]^{n_{ji}} + {T_{ji}}^{n_{ji}}} , & \;\;\; \mbox{repression}
\end{array}  \label{ecoli-dynamicalsystem}
\end{equation}
where $T$ denotes the interaction threshold, $n$ is the Hill exponent (\cite{schuster}), $\xi$ allows leaky transcription and $\beta$ defines the maximum expression level (cf. Eq.(\ref{hill})).

A SUM-gate corresponds to the usual paradigm of coupled maps systems as it involves averaging all the transcription inputs, whereas an AND-gate models the fact that a gene's expression rate generally depends on the "worst" transcription input (minimal activation or maximal repression). As already mentioned, real logic gates involved in E.Coli network are typically complicated and depend on the gene \cite{uribook}, although can sometimes be seen as a combination of SUM and AND gates. For the purposes of this work, we will assume all network gates to belong to only one of these categories. 

Despite being non-linear, the dynamical system given by Eq.(\ref{ecoli-dynamicalsystem}) is fully regular for all the parameter value options. We are therefore expecting the emergent dynamics of the coupled system Eq.(\ref{ecoli-model}) to be regular as well, which is required for description of a biological system.  For this reason we will include the arguments like network's flexibility and robustness to noise in the reminder of this Section.\\[0.1cm]

\textbf{The Numerical Set-up of the Model.}  We consider the largest connected component of E.Coli directed interaction network with $N=328$ nodes (Fig.\,\ref{fig-ecoli-colorpajek}) with the state of each gene given by $(q[i](t),p[i](t))$ at time $t$. The system's time-evolution is governed by the Eq.(\ref{ecoli-model}) under the network and parameter constraints. In order to model the cellular input to which the gene regulatory network responds, we add a regulatory interaction by an external transcription factor (ETF) to every gene that does not have any incoming links (cf. Fig.\,\ref{fig-widderscheme}), or has only one coming from a self-loop (dark green nodes in Fig.\,\ref{fig-ecoli-colorpajek}). 
An ETF for a given gene consists of a fixed value of $p$ which is influencing the corresponding gene through an interaction of a fixed type. This way we are able to monitor the response/adjustment of the network in relation to the environmental inputs defined by the sequence of ETFs, regulating those genes that do not have incoming links from the rest of the network.

The equations are integrated using classical $4^{th}$-order Runge-Kutta integration method, in its time dependent version. Given the non-stiff nature of our equations, this explicit method provides sufficient precision of the results. For simplicity, we fix the values $\alpha=\gamma^{P}=\gamma^{Q}=1$ for all genes, and quench the parameters describing the interaction properties by selecting randomly for all the links (both gene$\rightarrow$gene and ETF$\rightarrow$gene):
\begin{itemize}
 \item $\beta_{ji}$-values from log-uniform distribution on the interval $[1,10]$
 \item $T_{ji}$-values from log-uniform distribution on  the interval $[1,10]$
 \item $n_{ji}$-values between $\{2,3,4\}$ (only cooperative binding is considered)
 \item $\xi_{ji}$-values from log-uniform distribution on  the interval $[0.1,1]$
\end{itemize}
We employ log-uniform scale (uniform distribution on the log-scale) in order to emphasize the small values of the variables, i.e. the probability for bigger values exponentially decreases. The initial states of genes and the values of ETFs are the only quantities that vary from simulation to simulation. The computation algorithm used to run a simulation is the following:
\begin{enumerate}
 \item upload the adjacency matrix describing the network and quench the (pre-selected) values of the parameters as described above
 \item set the values of ETFs by choosing them randomly from log-uniform distribution on $[1,10]$ for each gene without other external input
 \item set the initial conditions on all the genes $(q[i](t=0),p[i](t=0))$ by choosing randomly from log-uniform distribution on $[1,10]$ for both $q$ 
 and $p$ coordinates separately 
 \item run the network dynamics as described above until the final state at time $t_f$ is reached. The considered Runge-Kutta time-step is $h=0.01$ 
(selected as optimal), and a simulation involves $\frac{t_f}{h}$ integration steps. Typically, a simulation is run until $t_f=100$ 
 \item consider the final network state $(q[i](t_f),p[i](t_f))$ (generally corresponding to the system's attractor) and perform the desired data analysis 
\end{enumerate}
In the reminder of this Section we will be employing the algorithm described above for gaining insights into the dynamics behind the collective behavior of E.Coli gene regulatory network. For simplicity, we represent the results only using the coordinate $p$ (concentration of protein) which is biologically more relevant. The dynamics in $q$-coordinate is qualitatively similar.

Note that similarly to the system of CCM, here we study two-dimensional units on each node, running the coupled dynamics in $2 \times 328$-dimensional phase space. Moreover, each node's local dynamics is given by two equations, with equation in $q$ influencing the equation in $p$ (dynamical self-loop). This is in analogy with the case of CCM, where two-dimensional standard map also involved a dynamical self-loop.\\[0.1cm]

\begin{figure}[!hbt]
\begin{center}
\includegraphics[height=2.6in,width=4.1in]{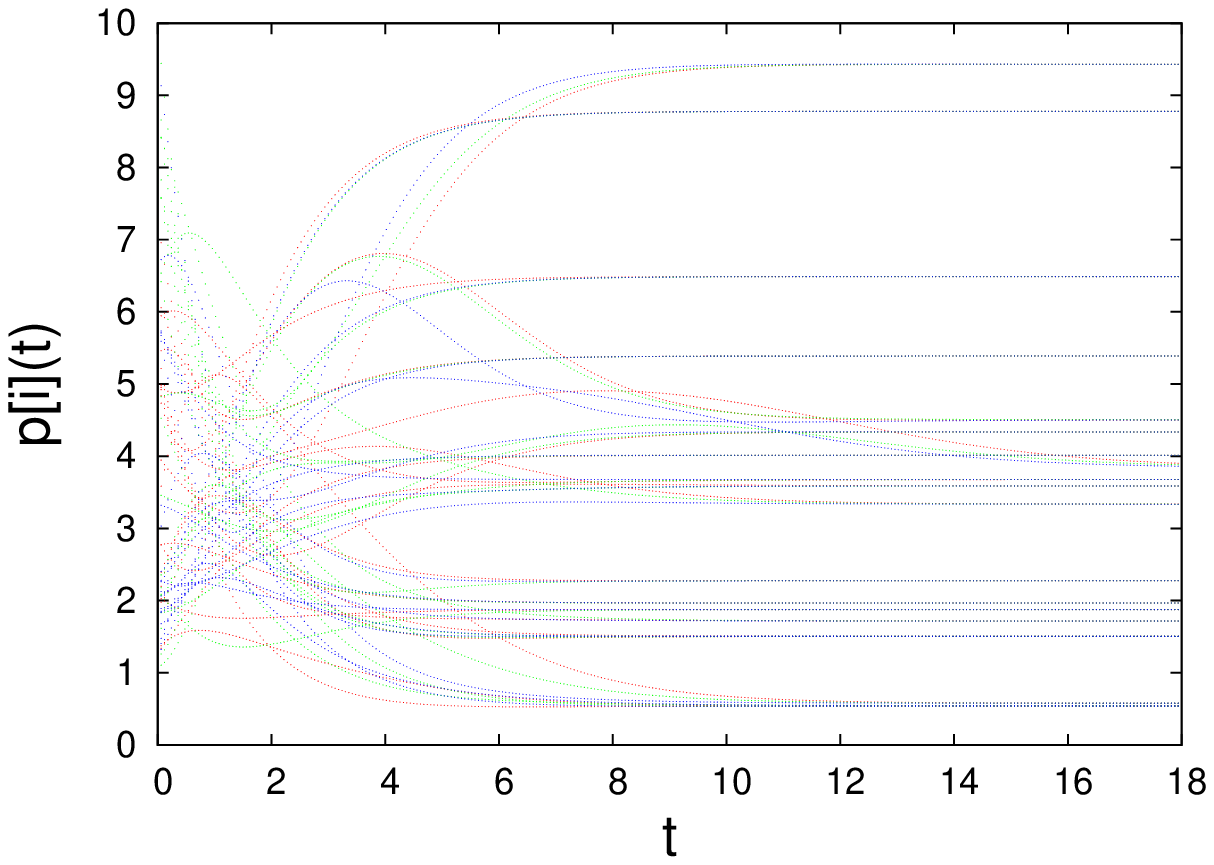}
\caption[Trajectories of some nodes of E.Coli network for three different initial conditions, with the same ETF-values]{Three trajectories of 20 representative nodes of E.Coli network for three different initial conditions, with the same ETF-values for the SUM model.} \label{fig-trj-fixtf}
\end{center}
\end{figure}

\textbf{The Homeostasis and Flexibility of Network's Response.}  In Fig.\,\ref{fig-trj-fixtf} we show three trajectories of 20 representative nodes with fixed ETF-values, but starting from three different initial condition on the nodes (values of $p[i]$ and $q[i]$). After a short transient time, all the trajectories settle into a set of constant values which are independent from the initial conditions, and related solely to the nodes. We examine only the SUM model here, as the trajectories for AND model are qualitatively similar. Repeating this simulation for many different initial conditions, we concluded that once the ETF-values are fixed, trajectory of each network node eventually settles into a given constant values of $p$ regardless of initial conditions. The set of final values of $p$ define the system's attractor.
In the context of network-averaged orbit used in previous Chapters (cf. Eq.(\ref{naeo})), the final system's attractor is therefore a single point, whose domain of attraction is the entire $2 \times 328$-dimensional phase space of all the nodes. We note this to be relevant for the biological interpretation: regardless of initial concentrations of proteins and mRNA, all the gene expression rates quickly reach their values that correspond to the environmental inputs, modeled through EFT-values. The system's attractor is determined exclusively by the sequence of ETF-values, and represents system's response to it.

\begin{figure}[!hbt]
\begin{center}
\includegraphics[height=2.4in,width=6.35in]{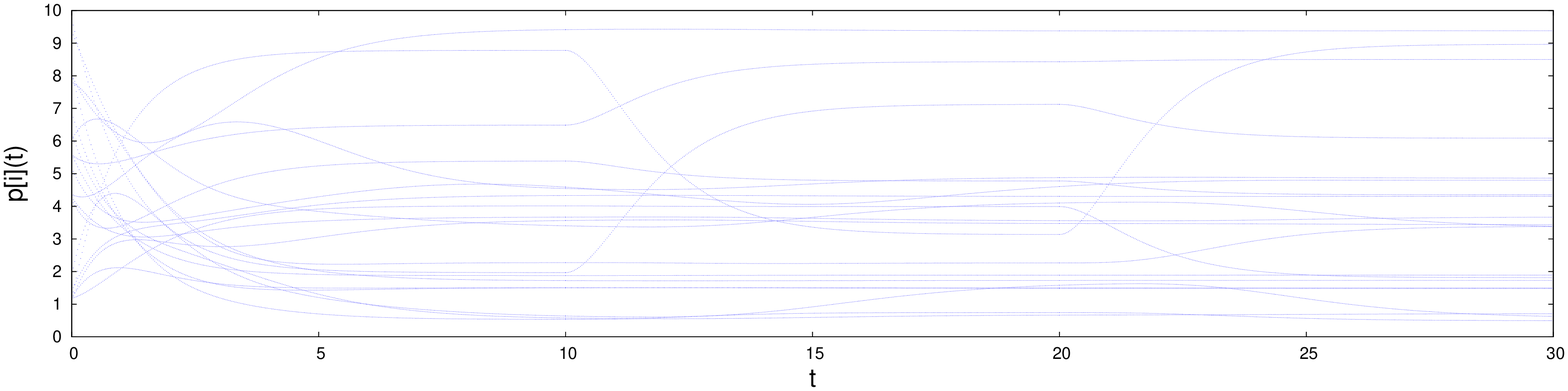}
\caption[Trajectories of some E.Coli nodes undergoing two changes of EFT-values]{Trajectories of 20 representative nodes undergoing two changes of EFT-values for SUM model. ETFs are changed instantaneously to a new random set of values at $t_1=10$ and then again at $t_2=20$.} \label{fig-trj-3tf}
\end{center}
\end{figure}

We furthermore consider the response of the system to a sudden change of ETF-values. In Fig.\,\ref{fig-trj-3tf} we consider the evolution of trajectories of 20 representative nodes, with network's ETF-values changed after it settles in the respective attractor (at $t_1=10$), and then changed again after the dynamics settles in a new attractor ($t_2=20$). The system responds quickly, with all the values $p[i]$ reaching new attractor after a short transient time.
This is another consequence of the biological motivation behind the problem: change of environmental conditions induces a response from the network, which results in a fast adjustment to the new input. Again, the new attractor is given as a constant value of $p$ for each node that is determined solely by ETF-values.

\begin{figure}[!hbt]
\begin{center}
$\begin{array}{cc}
\includegraphics[height=2.45in,width=3.2in]{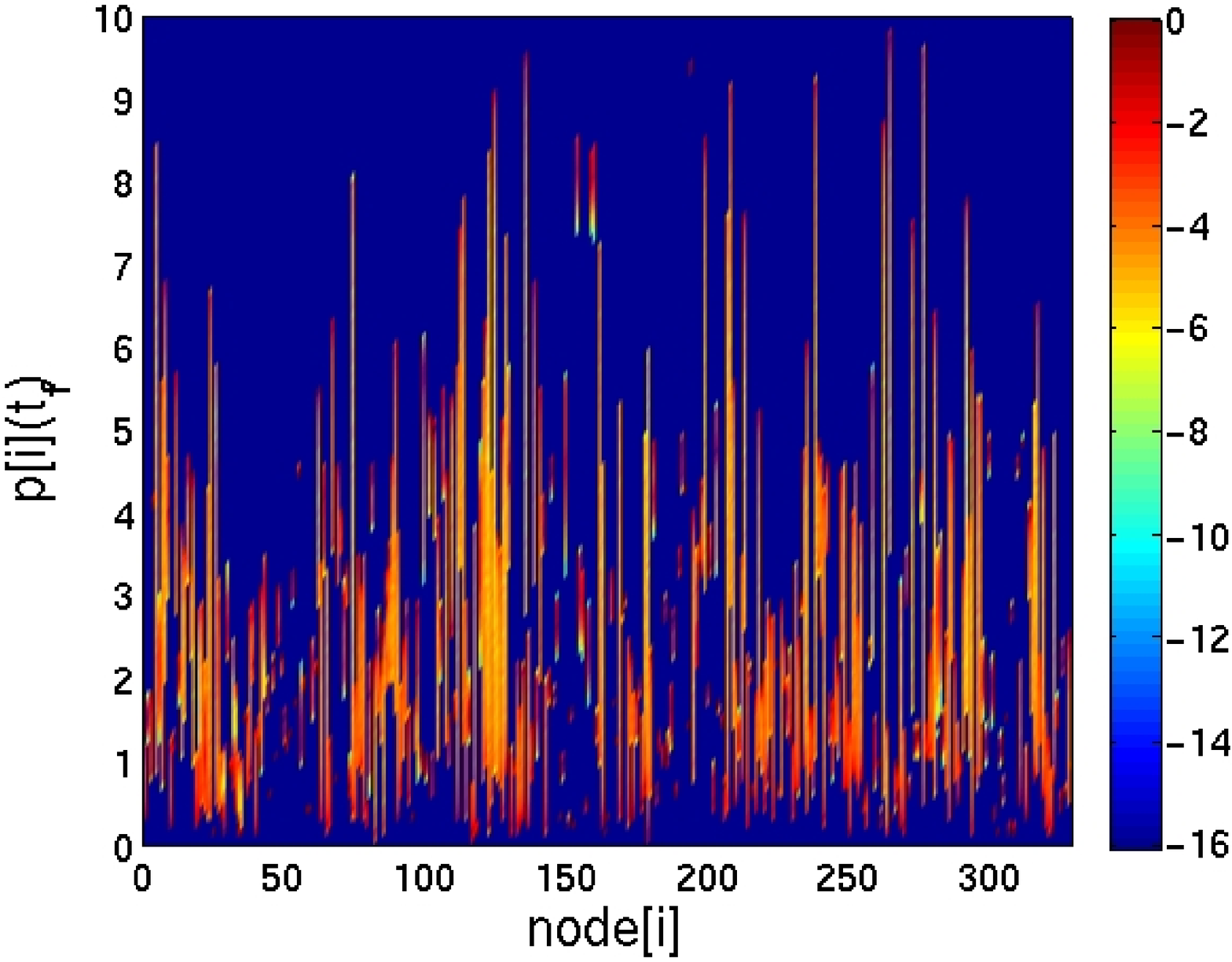} & 
\includegraphics[height=2.45in,width=3.2in]{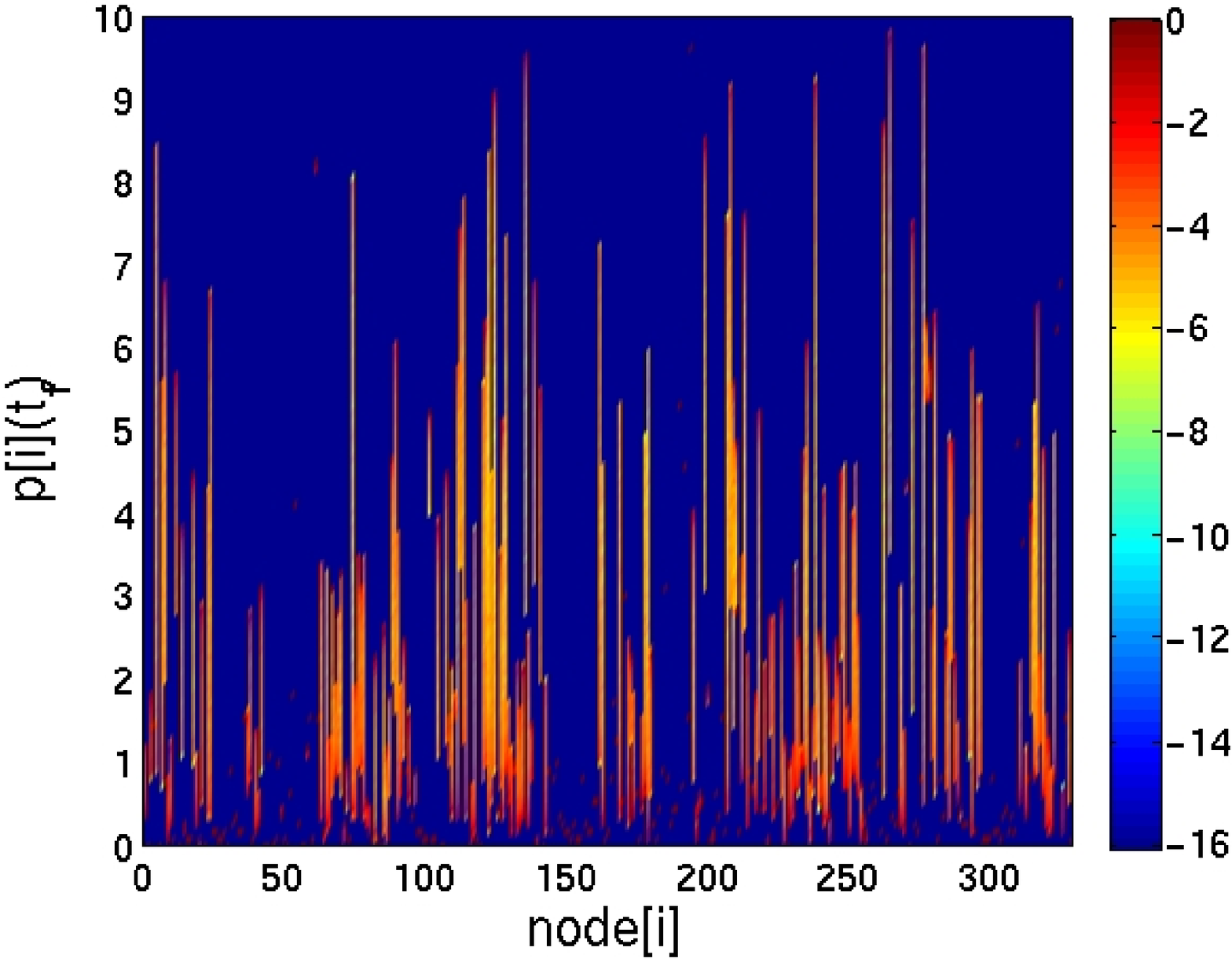} \\ 
\mbox{(a)} & \mbox{(b)} 
\end{array}$ 
\caption[2D color histograms of attractor values of $p$ for SUM and AND model]{2D color histograms of attractor values of $p$ for SUM model in (a) and AND model in (b), obtained by considering attracting values of $p[i]$ for all the nodes in relation to many ETF combinations.}
\label{fig-ecoli-flex}
\end{center}
\end{figure}

\begin{figure}[!hbt]
\begin{center}
$\begin{array}{cc}
\includegraphics[height=2.45in,width=3.15in]{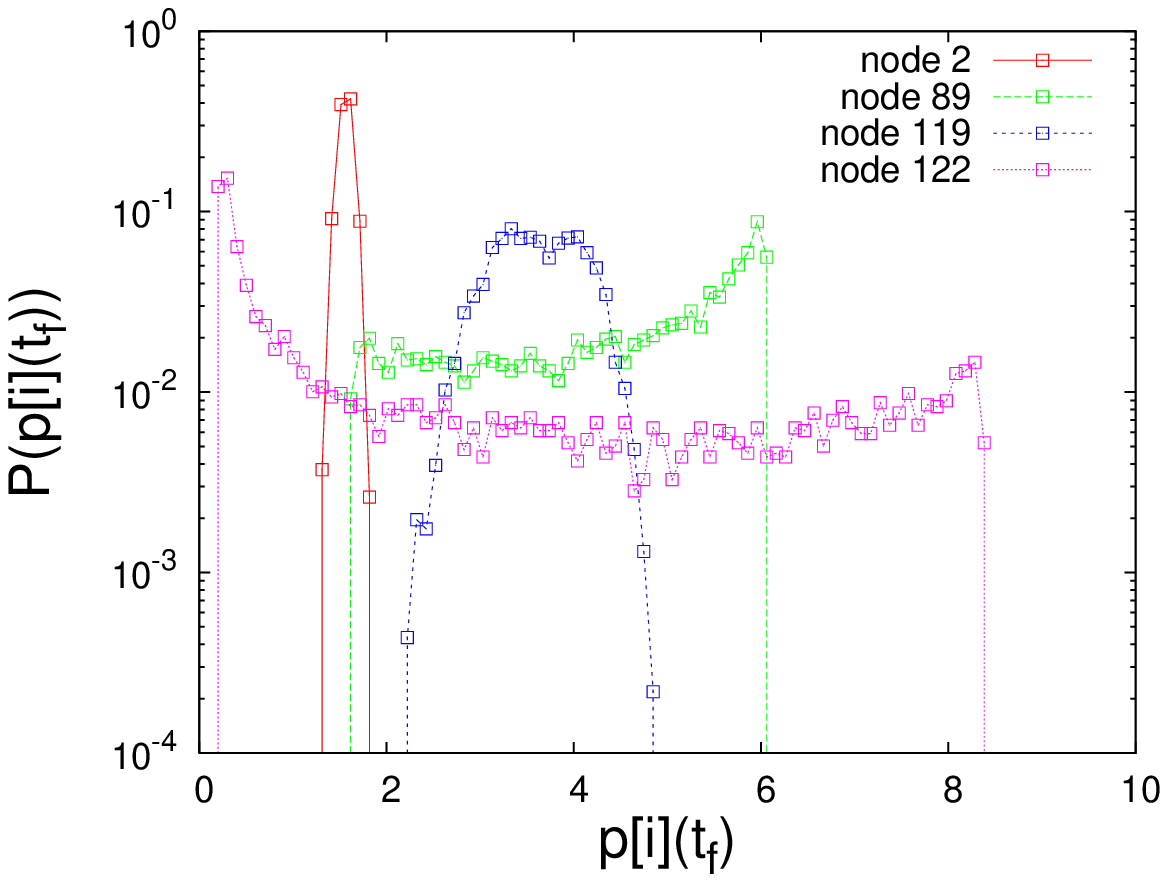} & 
\includegraphics[height=2.45in,width=3.15in]{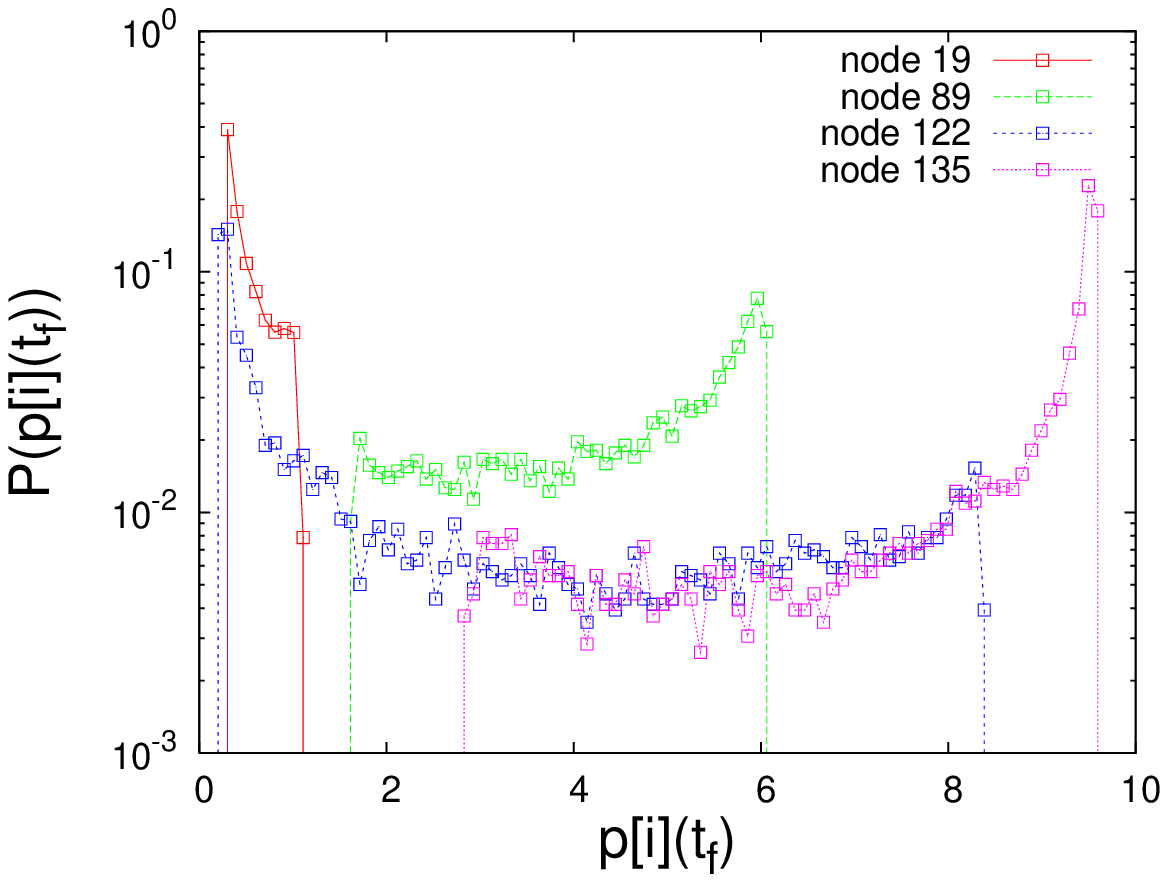} \\ 
\mbox{(a)} & \mbox{(b)} 
\end{array}$ 
\caption[Distributions of attracting values $p$ in relation to ETFs for some nodes, and for SUM model and AND model]{Distributions of attracting values $p[i]$ (same data from Fig.\,\ref{fig-ecoli-flex}) in relation to ETFs for four nodes displaying typical profiles. SUM model in (a) and AND model in (b).} \label{fig-ecoli-flex-nodes}
\end{center}
\end{figure}

A question of biological importance regards the network's adaptability to environmental changes. In the context of our model, that refers to the range of possible attractors that system can display in relation to various environmental inputs, i.e. ETF-values. To this end we examine the distributions of attractor values of $p$-coordinate for all the nodes for many random sequences of ETFs. Results are reported in Fig.\,\ref{fig-ecoli-flex} where we show the 2D color histogram of final values of $p$ for each network node, considered over many combinations of ETFs, for SUM model and AND model separately. 
There is a considerable difference among various nodes in both models: while some nodes are relatively fixed to a specific value of $p$-coordinate, other nodes are more flexible and exhibit a vast range of possible attracting values of $p$. The SUM model is visibly more adaptable than AND model, as it shows far less nodes with inflexible behavior. This is expected, as a SUM-gate includes all the node's inputs, while an AND-gate considers the smallest one only, which easily leads to final $p \cong 0$ situation. For illustration of nodes' responses to various ETFs, we show in Fig.\,\ref{fig-ecoli-flex-nodes} the distributions of attracting values $p[i]$ for four typical nodes in SUM model and AND model. Distributions mostly have sharp edges and prominent peaks, but always show at lest some flexibility to ETF-values, which in some cases extends to a vast range of values of $p$.
Two profiles for a same node in two models have generally nothing in common, with some exceptions like the node 89 in Fig.\,\ref{fig-ecoli-flex-nodes}.

\begin{figure}[!hbt]
\begin{center}
$\begin{array}{cc}
\includegraphics[height=2.45in,width=3.15in]{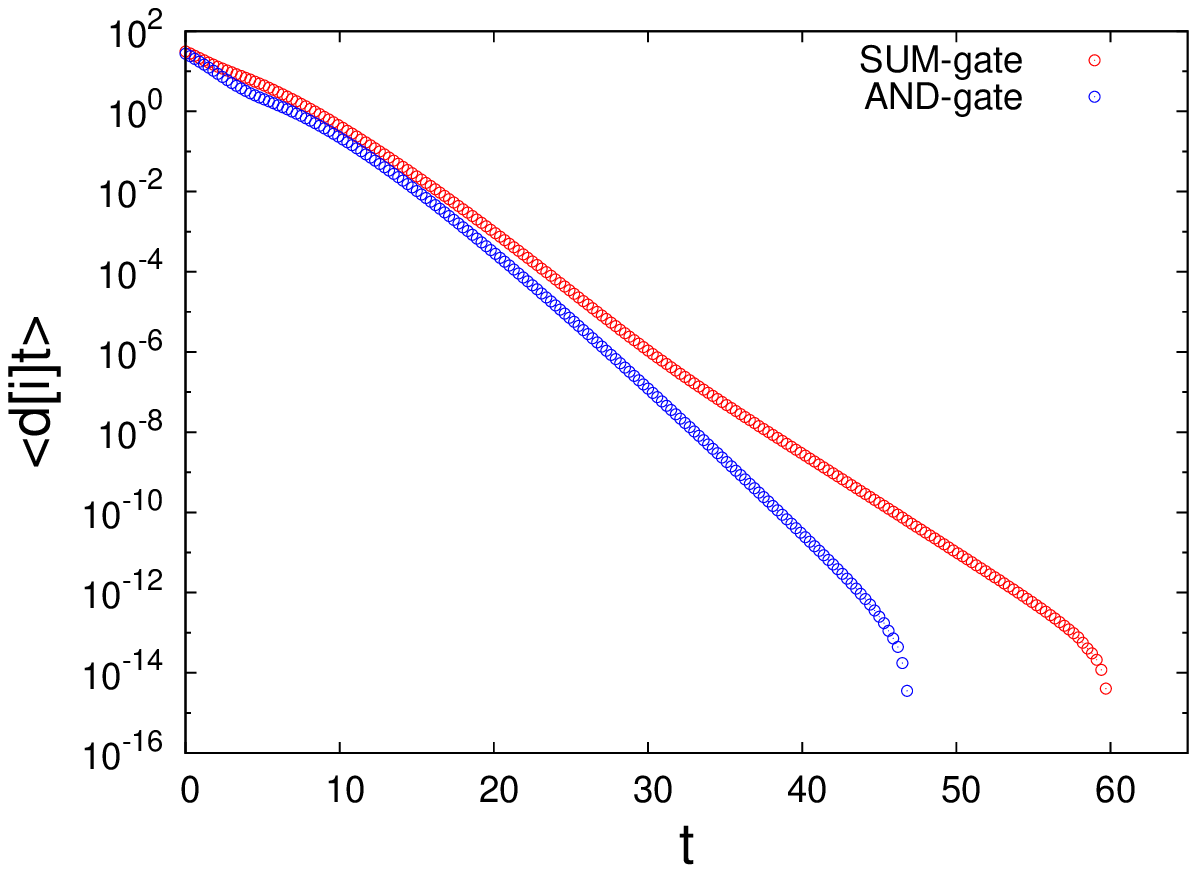} & 
\includegraphics[height=2.45in,width=3.15in]{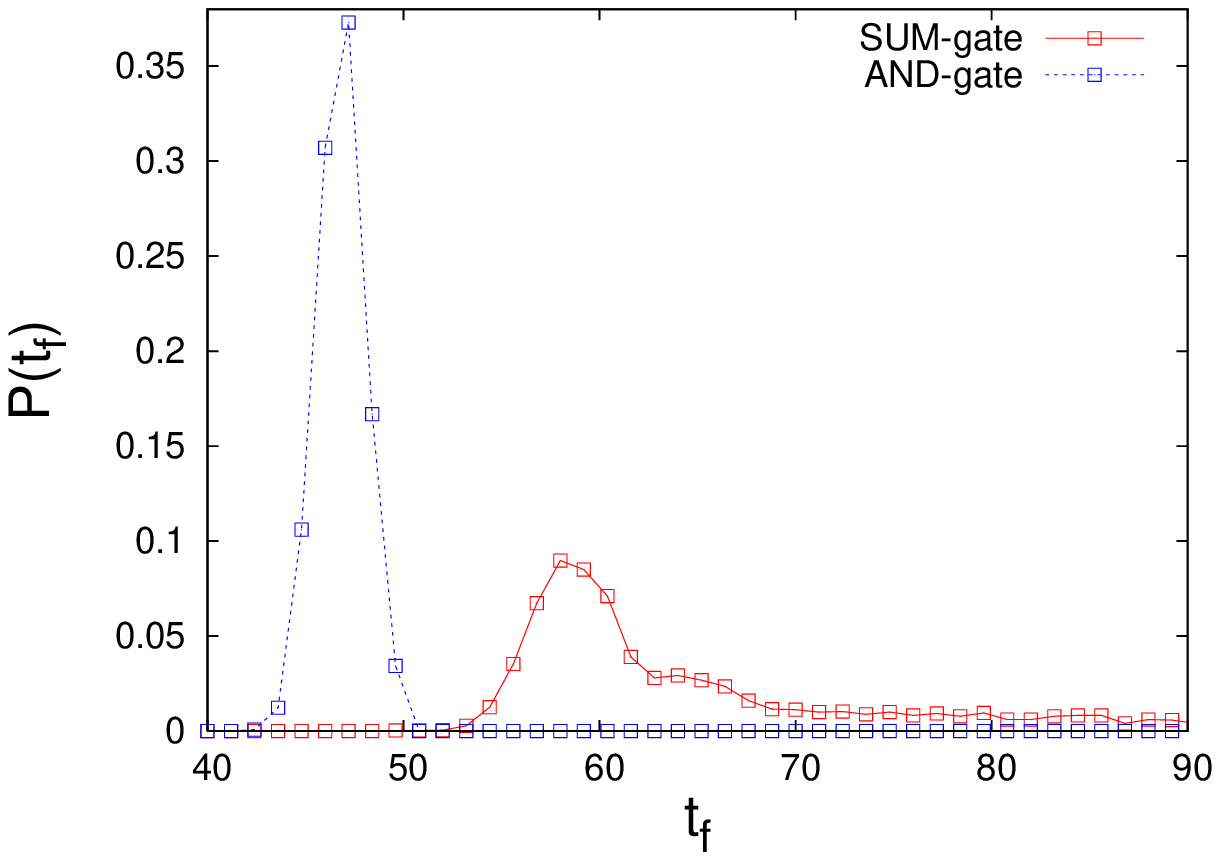} \\ 
\mbox{(a)} & \mbox{(b)} 
\end{array}$ 
\caption[Time-evolution of node-averaged distance to attractor $<d(t)>$ for both models, and the respective distributions of transient times]{The time-evolution of node-averaged distance to attractor $<d[i](t)>$ for both models under the same ETFs and initial conditions in (a). Distributions of $t_f$-values with $<d[i](t_f)> \; < \; 10^{-15}$ for both models taken over many ETFs and initial conditions in (b).} \label{fig-ecoli-tt}
\end{center}
\end{figure}

To complete the investigation of network's flexibility, we study the transient times needed for all trajectories to settle in their attractors, once a new sequence of ETFs is applied. We run the dynamics for a very long time $t_f$, record the final node values $p[i](t_f)$, and consider the time-evolution of the distance:
\begin{equation}
   d[i](t) = | p[i](t) - p[i](t_f) |
\end{equation} 
which measures the speed of $[i]$-node's trajectory approaching its final value $p[i](t_f)$. In order to illustrate the global network's approach to attractor, we consider the evolution averaged over all the network's nodes $<d[i](t)>$. In Fig.\,\ref{fig-ecoli-tt}a we report an example of $<d[i](t)>$ evolution for a SUM model trajectory and an AND model trajectory using the same ETFs and initial conditions. Visibly, the AND model's trajectory is faster to arrive to the respective attractor. 
To gain quantitative insight into the distribution of transient times, we re-define the attractor time to be $t_f$ such that  $<d[i](t_f)> \; < \; 10^{-15}$. In Fig.\,\ref{fig-ecoli-tt}b the distributions of transient times $t_f$ is shown for both models, considered over many ETF-values. The AND model clearly shows a much faster approach to the attracting values $p[i]$ regardless of the environmental conditions. 

It is interesting to note that each model has one of the biologically needed advantages: while the SUM model provides more flexibility for network adjustment to changes in the environmental inputs, the AND model is much faster to adopt to the changes. This suggests the real biological logic gates in the E.Coli gene regulatory network might have characteristics of both of the exposed models.\\[0.1cm]

\textbf{Stability of the Systems' Attractors.}  In order to quantitatively characterize the stability of an attractor corresponding to a sequence of ETF-values, we examine the divergence of the nearby trajectories once the system relaxes in the attracting constant values of $p$. We compute the Finite-time Maximal Lyapunov exponents FTMLE $\Lambda^t_{max}$ (as done previously with network of CCM Eq.(\ref{main-equation}) and Eq.(\ref{directed-equation})) for each systems orbit, obtaining a measure of systems' attractor stability. The procedure is equivalent to the one exposed previously (cf. Chapter \ref{Stability of Network Dynamics}.) and involves time-evolution of the distance between the attractor point $p[i](t)$ and a point in its close neighborhood $\tilde{p}[i](t)$:
\begin{equation}
  \frac{d_t[i]}{d_0[i]} = \frac{d(p[i](t>t_f),\tilde{p}[i](t>t_f))}{d(p[i](t_f),\tilde{p}[i](t_f))} 
\end{equation} 
thus measuring the trajectory divergence for the node $[i]$. The FTMLE $\Lambda^t_{max}$ is then defined as the initial slope of the divergence curve (on the log scale). The values $\Lambda^t_{max}$ are here equivalent to previously considered $\lambda^t_{max}$. The averaging over the orbit is redundant, as the orbit (trajectory) is composed of a single point (attracting value of $p[i]$).

\begin{figure}[!hbt] 
\begin{center}
\includegraphics[height=2.6in,width=4.3in]{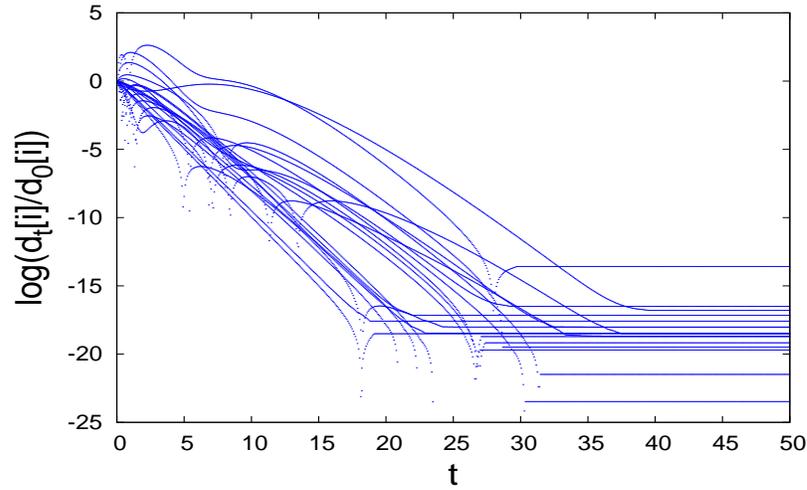}
\caption[Divergence of nearby trajectories at the attracting state for some E.Coli's nodes]{Divergence of nearby trajectories at the attracting state for a few representative nodes in the E.Coli SUM model (AND model shows a similar behavior).} \label{fig-ecoli-slopes}
\end{center}
\end{figure}

\begin{figure}[!hbt]
\begin{center}
$\begin{array}{cc}
\includegraphics[height=2.45in,width=3.15in]{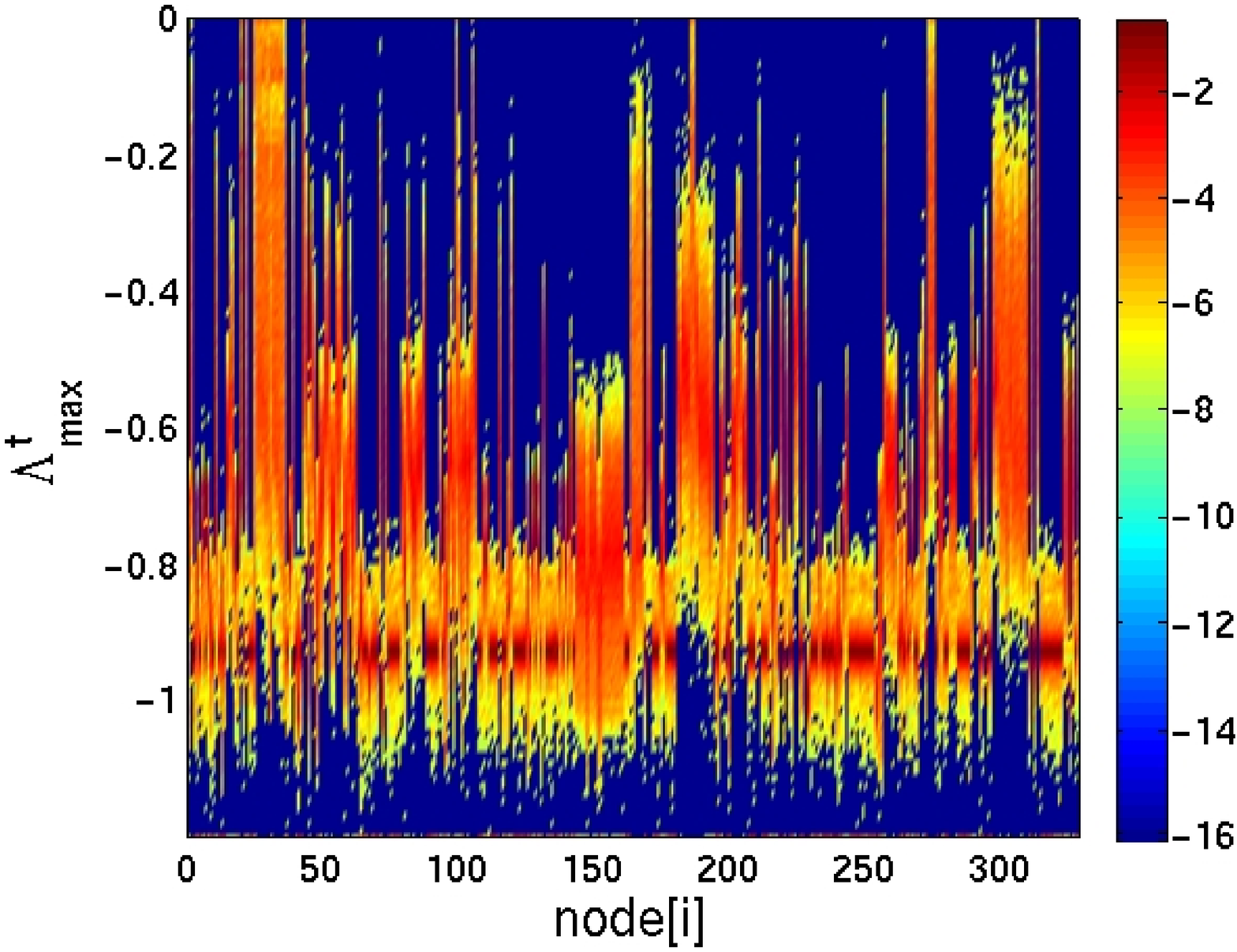} & 
\includegraphics[height=2.45in,width=3.15in]{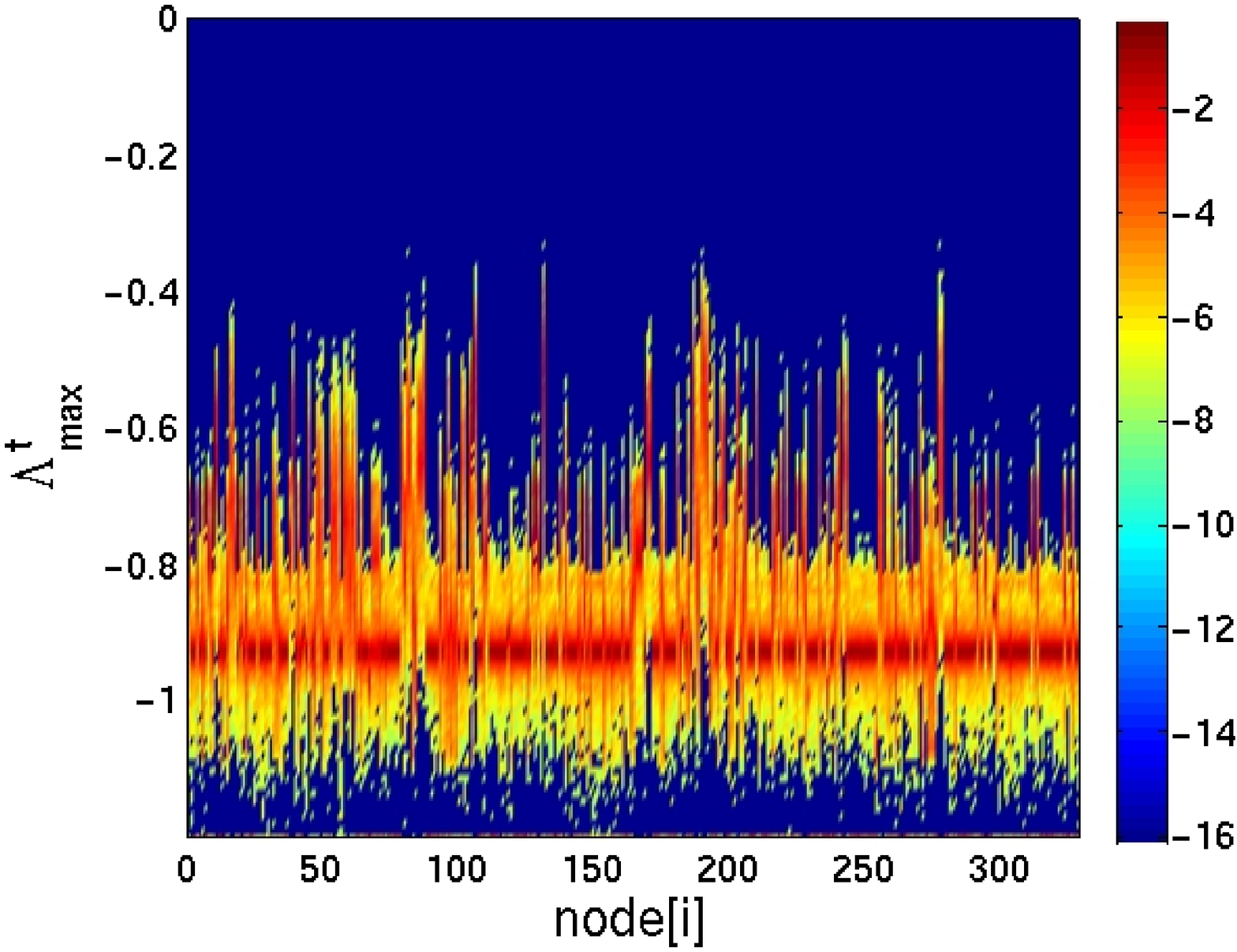} \\ 
\mbox{(a)} & \mbox{(b)} 
\end{array}$ 
\caption[2D color histogram showing the distributions $\Lambda_t^{max}$ for both models]{2D color histogram showing the distributions of FTMLE $\Lambda_t^{max}$ over many combinations of ETFs for each node. SUM model in (a) and AND model in (b).}  \label{fig-ecoli-lyapunov-nodes}
\end{center}
\end{figure}

In Fig.\,\ref{fig-ecoli-slopes} we show the evolution of $\frac{d_t[i]}{d_0[i]}$ for a few representative nodes for E.Coli's network in their attractor states. All the nodes initially exhibit exponential convergence of nearby trajectories, after which the distance settles to a fixed  value which is smaller than $O(10^{-15})$ (and therefore small enough to be confused with zero in the context of finite numerical precision).
The initial slope of the divergence curve is defined as the FTMLE $\Lambda^t_{max}$, which are measured accordingly for each node separately. As the attractor properties are solely given by the ETF-values (and the SUM/AND model in question), the respective values of $\Lambda^t_{max}$ node by node are also defined by them. We therefore compute the distributions of  $\Lambda^t_{max}$-values for each node, considered over many combinations of ETFs, for both models. Results are reported in Fig.\,\ref{fig-ecoli-lyapunov-nodes} in form of 2D color histograms. Clearly, both models exhibit non-positive FTMLE for all nodes and for essentially all ETFs.
This implies the finial attractor system's state to be stable regardless of ETF-values. This is of large importance in the context of biology: for each functional network mode (i.e., for all possible combinations of ETF-values), the network's final steady state is stable to small perturbations. This means the network's function is robust to small changes of protein concentrations that may occur due to e.g. thermal noise within the cell. Notably, AND model shows somewhat more uniform stability pattern which is less variable from node to node, which is a further advantage of the AND model.\\[0.1cm]

\textbf{Models' Robustness to the Noise.} We extend our model by assuming a weak time-dependence of ETF-values. We modify ETFs by adding a gaussian noise to each of them centered at pre-selected ETF-value and having intensity (standard deviation) $\eta$. The value of $\eta$ is equal for all the ETFs, and is much smaller than the values of ETFs themselves (typically $O(10^{-2})$, while ETF-values are $O(10^{1})$). In this way we are closer to the model of the living cell, where the presence of external transcription factors giving input to the gene regulatory network always fluctuates within some standard deviation.

\begin{figure}[!hbt]
\begin{center}
$\begin{array}{cc}
\includegraphics[height=2.55in,width=3.15in]{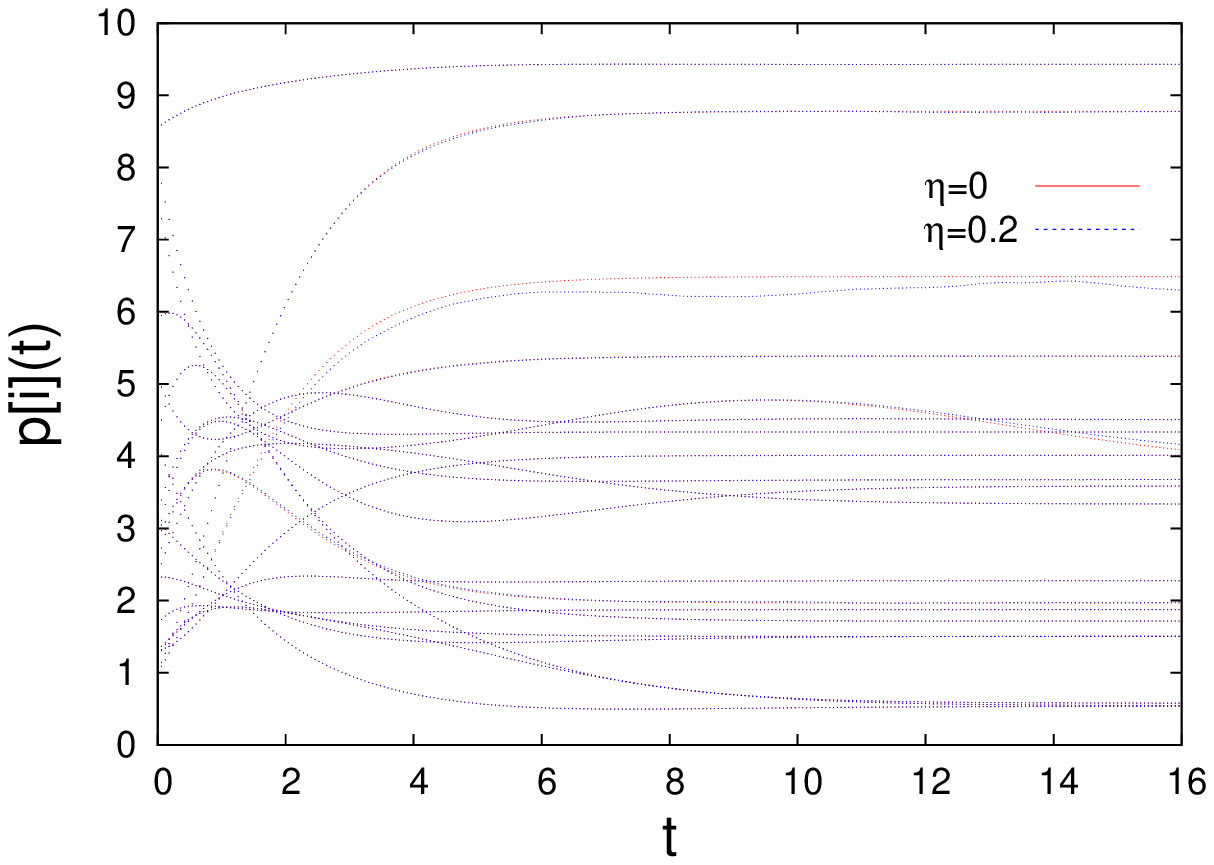} & 
\includegraphics[height=2.55in,width=3.15in]{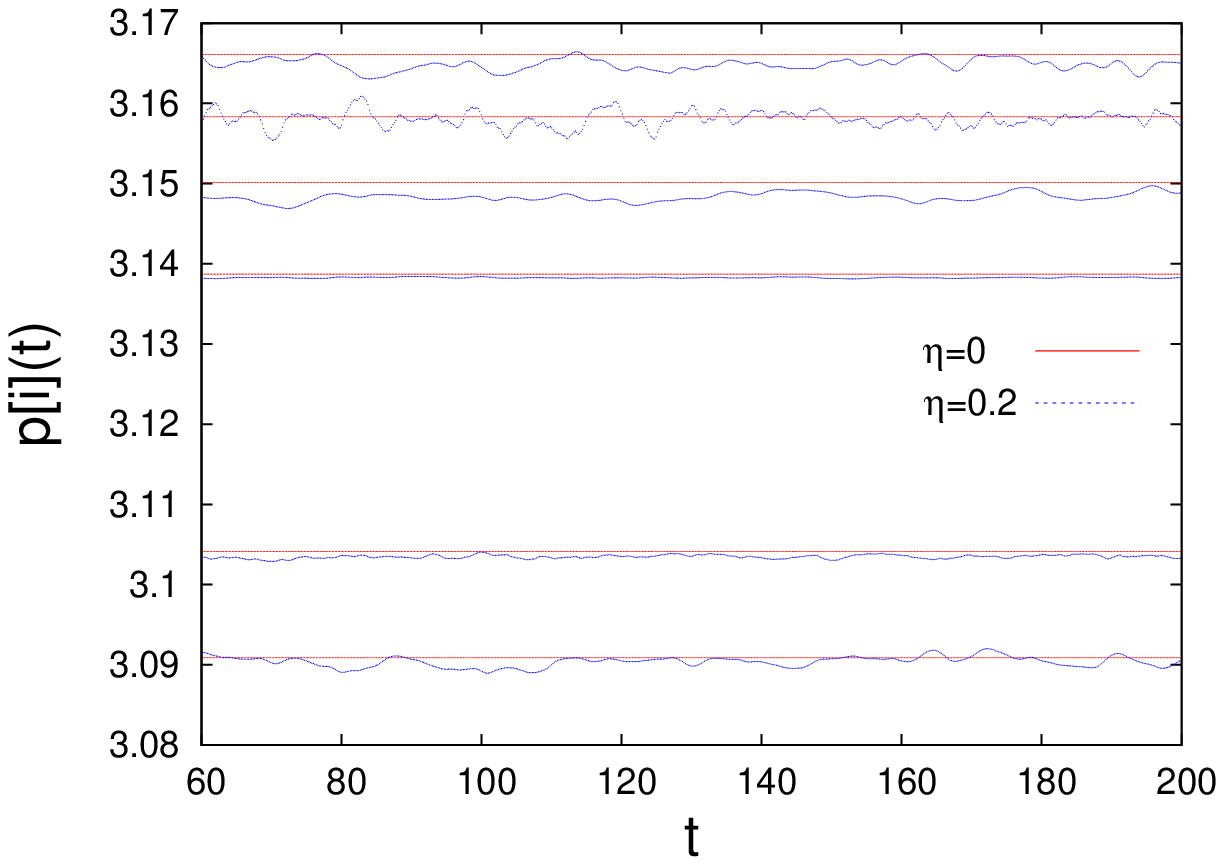} \\ 
\mbox{(a)} & \mbox{(b)} 
\end{array}$ 
\caption[Fluctuating trajectories of with noise $\eta=0.2$, in comparison with the corresponding non-fluctuating trajectories]{Fluctuating trajectories of few representative nodes with noise $\eta=0.2$, in comparison with the corresponding non-fluctuating trajectories for SUM model. Time-evolution of trajectories towards the attracting values in (a), and the fluctuations of trajectories after transients in (b).} \label{fig-ecoli-trj-noise}
\end{center}
\end{figure}

\begin{figure}[!hbt] 
\begin{center}
\includegraphics[height=2.45in,width=6.35in]{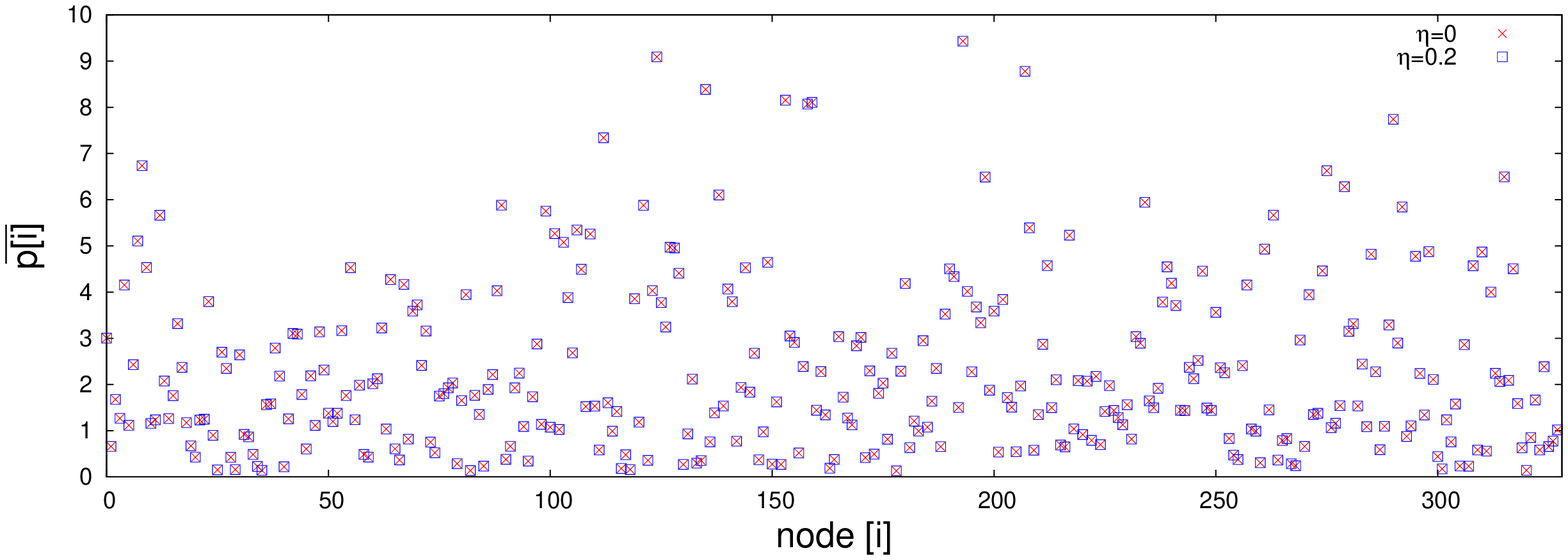}
\caption[The attracting values of $p$ of a non-fluctuating system, and the means of $p$ for the same system with ETFs fluctuating with $\eta=0.2$]{The attracting values of $p$ of a non-fluctuating system and the means of $p$ (averaged over a long time interval away from transients) for the 
system with the same ETFs fluctuating with $\eta=0.2$.} \label{fig-finalp-noise}
\end{center}
\end{figure}

As as result of model modification, the system never reaches the precise attracting (constant) values of $p$, but the trajectories $p[i](t)$ of all the nodes $[i]$ always fluctuate. However, the time-average of each node-trajectory $\overline{p[i]}$ considered over a long time interval (and after transient time) coincides with the non-fluctuating attracting values of $p$ for the same sequence of ETFs. This indicates the network to be robust in the response to the fluctuations (noise), which is crucial for the functional operation of biological networks. We illustrate this in Fig.\,\ref{fig-ecoli-trj-noise}a showing the trajectories of fluctuating and non-fluctuating system for the same ETFs as they evolve, and in Fig.\,\ref{fig-ecoli-trj-noise}b where we examine the fluctuation properties of trajectories for the same two systems away from the transients.
The noisy system (blue in Fig.\,\ref{fig-ecoli-trj-noise}) settles to fluctuate around the attracting values of the non-fluctuating system with a speed which is in agreement with transient times examined in Fig.\,\ref{fig-ecoli-tt}. To confirm this quantitatively, we show in Fig.\,\ref{fig-finalp-noise} the attracting non-fluctuating values of $p$ and the means of $p[i]$ of the fluctuating trajectories for each node, 
which show a clear overlap for all the network nodes, as suggested. Biologically, this indicates the system to be able to "on average" maintain the protein concentrations at their respective levels defined by the sequence of ETF-values, despite the ETF fluctuations. Robustness to noise of this type is generally present in all biological systems.

\begin{figure}[!hbt]
\begin{center}
$\begin{array}{cc}
\includegraphics[height=2.45in,width=3.15in]{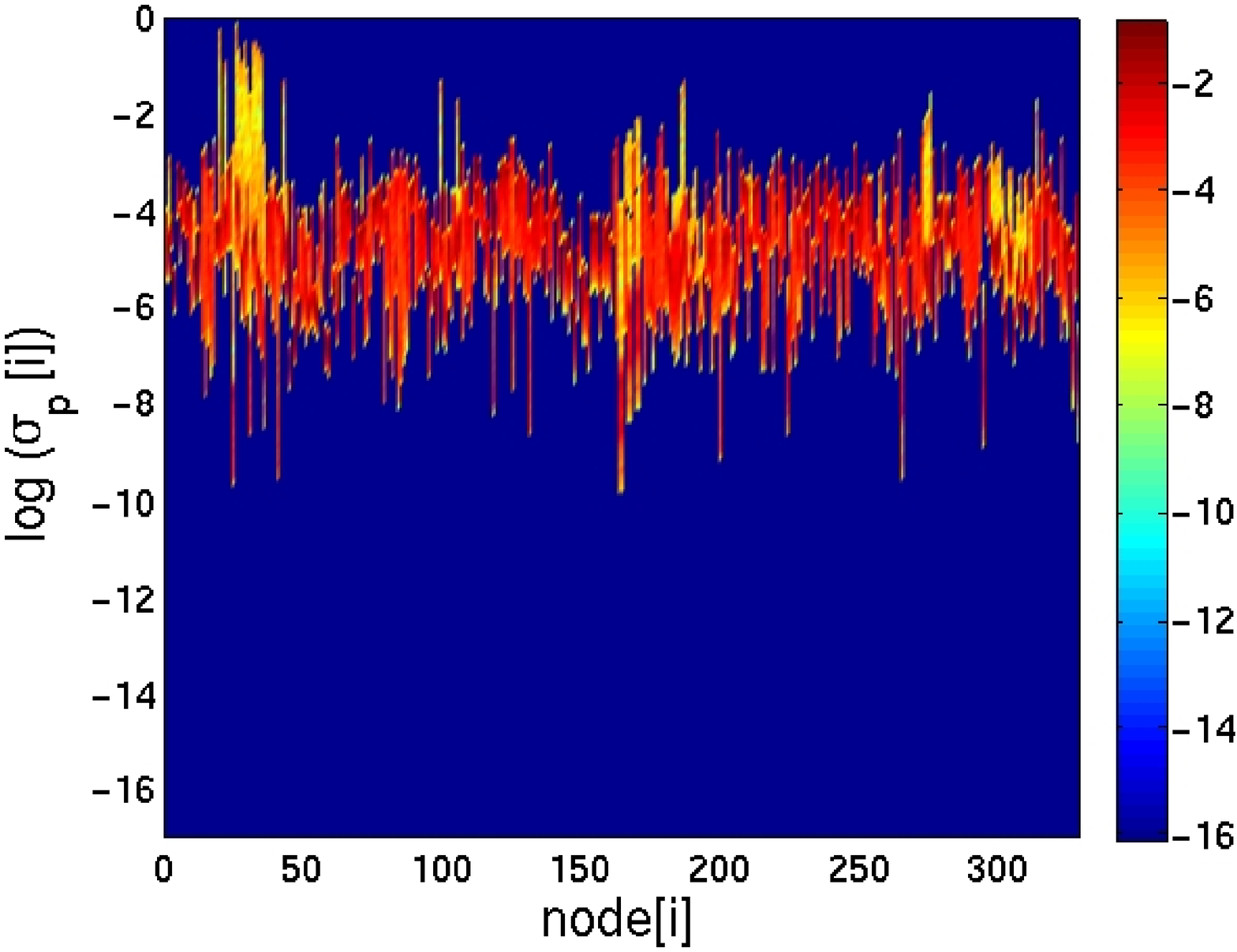} & 
\includegraphics[height=2.45in,width=3.15in]{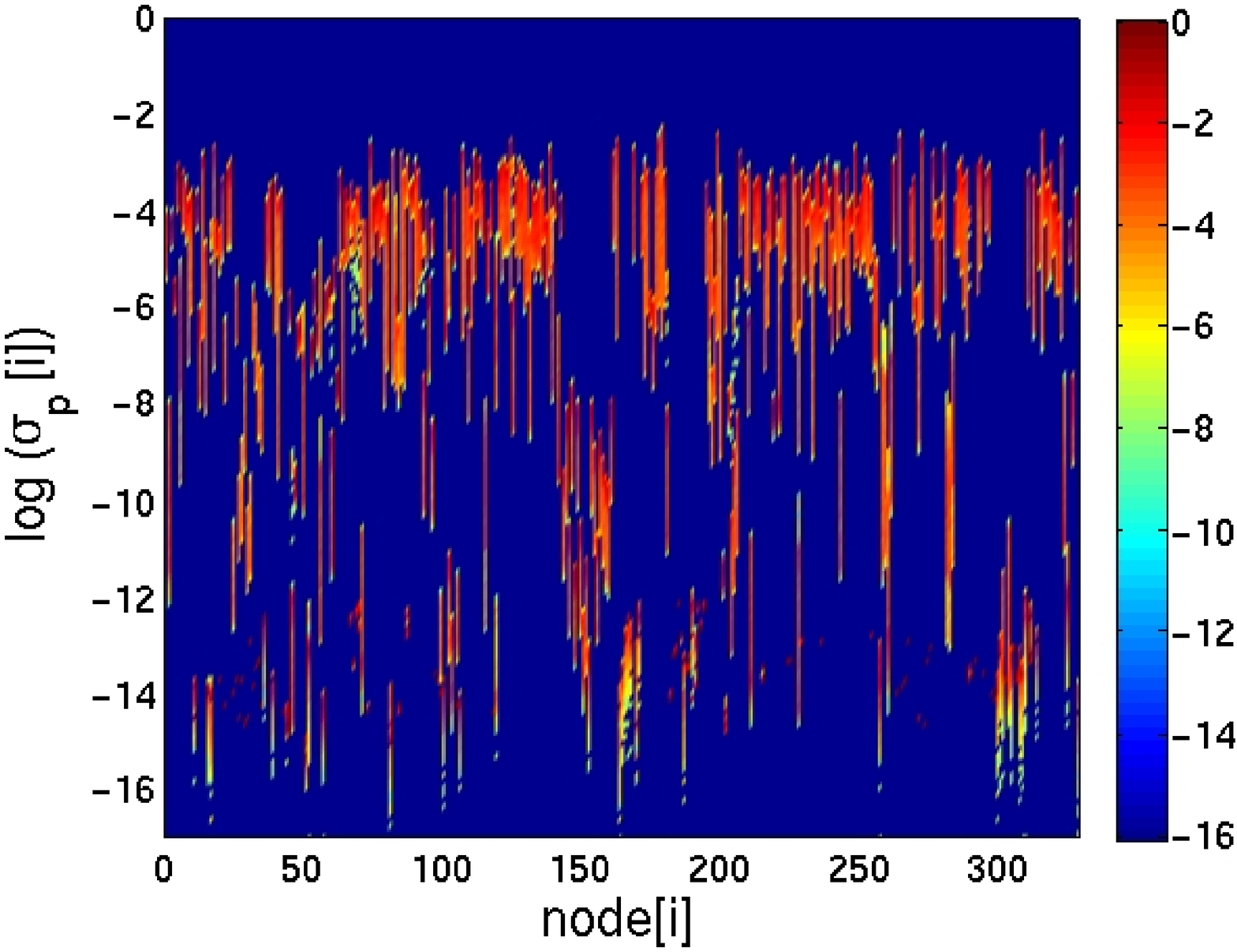} \\ 
\mbox{(a)} & \mbox{(b)} 
\end{array}$ 
\caption[2D color histogram of $\sigma_p$-values for each node with noise $\eta=0.01$ for both models]{2D color histogram of $\sigma_p [i]$-values over many combinations of ETFs for each node with noise $\eta=0.01$. 
SUM model in (a), AND model in (b).}  \label{fig-ecoli-noise-nodes}
\end{center}
\end{figure}

After transients, all the nodes fluctuate around their corresponding non-fluctuating values of $p$, with the properties that depend on the node in question and the noise intensity $\eta$. We quantify the trajectory fluctuations after transients by evaluating their standard deviation $\sigma_p [i]$. The value of $\sigma_p [i]$ (if computed over a sufficiently long time interval) depends on the node, ETFs and the noise intensity $\eta$. In order to systematically study the systems' response to ETF fluctuations, we consider the distribution of $\sigma_p [i]$ for each node over many combinations of ETFs. The results are reported in Fig.\,\ref{fig-ecoli-noise-nodes} for SUM model and AND model separately for the noise of $\eta=0.01$. While the value of $\sigma_p [i]$ depends on both node in question and the ETF-value, it is immediately clear that E.Coli gene regulatory network in both models acts as a noise inhibitor. 
While the input fluctuations of the ETFs are $\eta=10^{-2}$, in both models the fluctuations on the node trajectories are typically smaller by few orders of magnitude. The AND model inhibits the noise better with some nodes output fluctuations being only $O(10^{-15}$ (which is partially due to some of these nodes always displaying zero attracting value of $p[i]$). Both models exhibit a specific range of $\sigma_p [i]$ for each node, which can be attributed to the biological network preferring some nodes more robust than the others. In computational terms, the inhibition of ETF fluctuations is due to the shape of the Hill's functions: the range of fluctuations of ETFs values $p$ is generally larger than the range of fluctuations of $F^{+}(p[ETF])$ and $F^{-}(p[ETF])$. 

\begin{figure}[!hbt] 
\begin{center}
\includegraphics[height=2.5in,width=3.in]{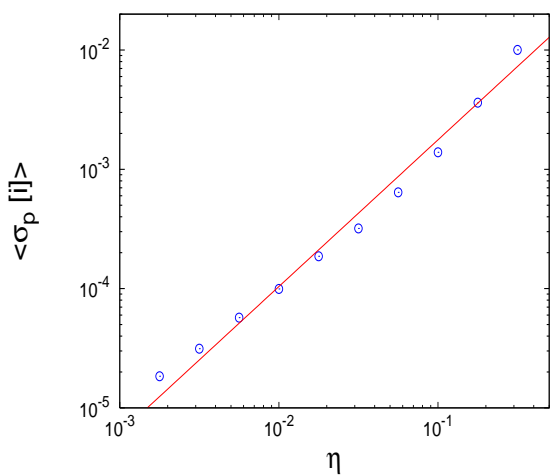}
\caption[The node-averaged standard deviation of all the trajectories fluctuations $<\sigma_p>$ in function of the noise $\eta$]{The node-averaged standard deviation of all the trajectories fluctuations $<\sigma_p [i]>$ in function of the noise $\eta$, fitted with 
slope of 1.23. Consideration is done for the SUM model.} \label{fig-ecoli-noise-sigma}
\end{center}
\end{figure}

To illustrate more directly the inhibition of fluctuations which characterizes our models, we compute $<\sigma_p [i]>$, the average of $\sigma_p [i]$ over all the nodes for a few $\eta$-values, and present the results in Fig.\,\ref{fig-ecoli-noise-sigma}. The value of $<\sigma_p [i]>$ is constantly smaller by about 2 orders of magnitude, as already observed. Moreover, there appears to be a power-law relationship between the values of $<\sigma_p [i]>$ and the corresponding value of noise $\eta$, as suggested by the power-law fit in Fig.\,\ref{fig-ecoli-noise-sigma} with the slope of 1.23.
These results for the SUM model indicate noise inhibition to be even more efficient for the case of AND model. Visibly, the examined models of gene regulation in E.Coli posses a high degree of self-organization and are strongly robust to the fluctuations of input ETFs, maintaining their emergent operation only weakly influenced by the presence of the noise. 

The studied models of gene dynamics of E.Coli gene regulatory network correctly predict the main features that a biological system of this sort is expected to have: homeostasis, flexibility of response to changes in external transcription factors (ETFs), stability of system's final attracting state  regardless of particular combination of ETFs and the robustness to the fluctuations of ETFs and their inhibition. Furthermore, our finding agree with the fact that E.Coli's gene logic gates are a complicated composition of SUM-gates and AND-gates \cite{uribook}, by showing that each of these models possesses the biologically required characteristics to a certain degree.


\chapter{Conclusions} \label{Conclusions}

\begin{flushright}
\begin{minipage}{4.6in}
    We conclude the present Thesis, summarizing and discussing the results reported in the previous Chapters. 
    A list of open questions related to the studied topics is also provided. \\[0.1cm]
\end{minipage}
\end{flushright}

We have exposed the research results according to the study directions established in Chapters \ref{Introduction}. and \ref{Coupled Maps System on Networks with Time delay}. We considered networks of 2D chaotic maps coupled with time delay, with a variety of directed and non-directed topologies. Our finding indicate that the considered networks are able to introduce regularity and stability in the motion of very chaotic maps due to inter-node interactions. Moreover, we found a clear dynamical relationship between a large network structure and its typical dynamical motif, arising as a consequence of topological relationship between the two structures. We recognize the presence of three main types of cooperative dynamics: regular,  weakly chaotic, and self-organized motion with MLE close to zero. Ways of transition from periodic to non-periodic orbits were also investigated. In our 4-star system that can be seen as 8D nonlinear dynamical system we have found evidence of specific dynamical phenomena termed strange nonchaotic attractors.

\section{Outline of Main Results}

In this Section we give a more systematic presentation of the main results discussed in this Thesis. \\[0.1cm]

\n \textbf{The Regularization of Collective Dynamics of CCM on Networks.}  For any non-zero value of coupling strength $\mu$ the dynamics of chaotic two-dimensional standard maps on both directed and non-directed network becomes regular, which is for small values of coupling strength realized in form of periodic orbits with different periodicities. Due to the inter-node interactions each node develops a periodic motion after some transient time, despite very chaotic nature of the isolated standard map. The mechanism behind the regularization process involves the inhibition of chaotic diffusion characteristic for standard map, which with time-evolution increases the correlations among the motion of the nodes. The transient time of the regularization process depends on the size of the network structure and the coupling strength. For large scalefree tree the process typically starts from outer less connected nodes, and moves along the tree branches towards the hub node. The final dynamical steady state is characterized by periodic orbits on all the nodes, which are stable in the sense of having negative FTMLE and have period values that might vary from node to node. Clearly, the scalefree tree topology is able to introduce regularity into the motion of 2D strongly chaotic maps, for even very small values of the coupling strength. At larger coupling strengths, the motion of the nodes is not periodic, but nevertheless displays other self-organizational features.\\[0.1cm]

\n \textbf{Dynamical Regions of CCM in Relation to the Coupling Strength.} With increase of coupling strength, the network of CCM does not simply maintain the structure of periodic orbits, but instead develops a variety of new collective behaviors. In particular, depending on the value of coupling strength within the investigated coupling range $[0,0.08]$, we recognized four dynamical regions: chaotic region ($0 < \mu \lesssim 0.012$), periodic region ($0.012 \lesssim \mu \lesssim 0.026$), quasi-periodic region ($0.026 \lesssim \mu \lesssim 0.04$) and attractor region ($0.04 \lesssim \mu \lesssim 0.065$). The regions are characterized by specific types of collective dynamics that can occur within the respective coupling ranges for given initial conditions \cite{ja-jsm}. The types of motion are also somewhat interwoven, in the sense that periodic orbits occur at all coupling strengths, for the appropriate initial conditions; quasi-periodic orbits also appear at some specific coupling values outside the quasi-periodic region. Stability investigation via Maximal Lyapunov Exponents (MLE) over the whole coupling range confirms the dynamical regions: chaotic orbits are highly unstable with large positive MLE, periodic orbits are always stable with negative MLE, quasi-periodic orbits are neutrally stable with zero MLE, and strange attractors are mostly weakly unstable with small positive MLE (although evidence of strange nonchaotic attractors have been found, with very small MLE). Directed network possesses differently organized dynamical regions, although the periodic region is still present with the same properties in a similar coupling range. The changes between the dynamical regions can be explained through the parametric instability of an emergent orbit. In particular, we have found and investigated the cases of period-doubling and Hopf bifurcations occurring at the edges of dynamical regions and inducing the changes between the types of motion.\\[0.1cm]

\n \textbf{Characterization of Self-organization Properties.} The dynamics of the network of CCM for directed and non-directed structures in all dynamical regions (with exception of chaotic region) is characterized by a variety of self-organization effects, which were studied through their statistical and stability properties. For all structures, the regularization of dynamics is accompanied by the clusterization of periodic (and some non-periodic) orbits. Each node's periodic orbit belongs to one of the clusters defined by a certain $y$-value of the orbits' time-averages $\bar{y}[i]$. The $\bar{y}$-values considered over the entire network make a discrete set, with almost precise integer spacing in $y$-coordinate. The distribution of network distances between the nodes belonging to the same cluster is different from the distribution of topological distances on the network, with prevailing distance of 2 links \cite{ja-lncs-1,arenasguilera}. For non-directed topologies, periodic region shows power-law tails in the period distribution. The attractor region is characterized by q-exponential distributions of return times with respect to the phase space partitioning. Anomalous diffusion in phase space of similar sort was found for all values of the coupling strength in the motion of the tree. Directed topologies show destabilization of dynamics at larger coupling strengths, with motion properties of some nodes being identical to the case of the isolated maps, i.e. chaotic. The destabilized nodes are mutually connected and construct a sub-network which is integrated with the rest of the network. Nevertheless, directed graph exhibit variety of non-periodic weakly unstable orbits, displaying strange attractors and q-exponential distributions of return times to phase space partitions for particular nodes and coupling strengths. \\[0.1cm]

\newpage
\n \textbf{Dynamical Patterns in CCM on 4-star.} Various manifestations of non-symplectic coupling were found in the dynamics of CCM on 4-star motif at particular values of the coupling strength. This include a wide spectrum of strange attractors in the attractor region having positive MLE, but with much smaller values ($O(0.1)$) with respect to the isolated map case ($O(1)$), and quasi-periodic orbits. A dynamical situation was revealed with all four 4-star nodes exhibiting strange attractors, and with a distribution of FTMLE with negative peak value having tails on both negative and positive FTMLE values. SMLE for this case was found to be $\zeta \cong 0.038$. However, considering the attractor on a 4-star's branch node (coupled to the rest of the 4-star), we found all the standard characteristics of a strange nonchaotic attractors \cite{feudel} present. This is the first example that the evidence of SNA was found in an endogenously driven system (while SNA are so far known to arise only in externally driven dynamical systems) \cite{ja-jsm}. Collective effects in the motion of the CCM on 4-star are a further example of self-organization in our system which results in entirely different dynamical patterns than for the case of uncoupled standard maps.\\[0.1cm]

\n \textbf{Dynamical Relationship of Scalefree tree vs. 4-star.} The expected relationship between the collective dynamics on a large-scale (scalefree tree) and a small-scale (the typical dynamical motif capturing the tree topology, 4-star) was found. The dynamical patterns observed on the large scalefree tree seem to be induced by the dynamics exhibited on the 4-star, which suggests the known topological relationship between these two structures to extend to the context of dynamical behavior. The identified dynamical regions for the tree and the 4-star overlap in terms of the coupling strength range, with same motion types being present in all of them (although sometimes for different fractions of initial conditions). Main statistical properties of the emergent motion, along with the stability profile in function of the coupling strength is also shared by these two structures. By comparing the clustering properties of 4-star and 4-clique motif, we found the 4-star to resemble much better the clustering properties observed for the scalefree tree. However, the clustering properties of both 4-star and 4-clique seem to be shared with the modular network, grown from the scalefree tree with the addition of cliques. \\[0.1cm]

\n \textbf{Other Coupling Forms.} CCM with different coupling forms were also studied for comparison. Specifically, the equivalent non-delayed system showed similar general properties, although with less structured collective effects (e.g. less organized cluster structure). This confirmed the advantage of the time delay in the node coupling for the context of two-dimensional chaotic maps. The systems with different standard map's chaotic parameters $\e$ were also examined, showing very little collective effects (larger $\e$) or showing periodic orbits as basically the only collective effect (smaller $\e$). This study confirmed our choice of $\e$-value ($\e=0.9$) as optimal for the purposes of investigation done here, which was primarily oriented towards exploration of the self-organizational effects arising in collective dynamics of coupled 2D chaotic maps.\\[0.1cm]

\n \textbf{Two-dimensional Hill Model of E.Coli Gene Regulatory Network.} The continuous-time model of gene regulatory network of E.Coli based on Hill two-dimensional approach to the gene interactions was implemented on the same directed network. The system exhibits the expected collective behavior in the context of biologically motivated functional network. This includes homeostasis (final attractor state is independent from the initial states of the genes/nodes), flexibility of response to environmental inputs (quick adjustment to new external transcription factors with wide range of respective attractors) and stability of the attractor (negative FTMLE for all nodes regardless of ETF-values). The version of the system with logic-gates modeled as SUM-gates shows wider flexibility of response to ETFs, while the system with AND-gates is faster in responding to ETF changes. Both models are robust to fluctuation in the ETF-values, reducing the standard deviation of fluctuations from ETFs to the node's attractor trajectories by almost two orders of magnitude. Directed E.Coli network does not only produce a largely stable system with very chaotic standard maps, but also possesses the needed functionality in the context of biological systems.

\section{Future Directions}

The results presented in this Thesis are by no means exhaustive in regard of study of two-dimensional dynamical units on complex networks. In this Section we outline few directions that can be taken in future studies, in order to extend/complete the arguments presented here. 
\begin{itemize}
\item CCM with alternative coupling forms (e.g. other versions of diffusive angle-coupling or possibly action-coupling) might be designed appropriately
in the context of specific 2D dynamical system in use. Different chaotic 2D maps can be employed, with varying levels of chaoticity. As the time
delay appears to be relevant for oscillator-motivated dynamical systems, a different structure of time delays might be of interest (for example,
time delays that vary from link to link according to a prescribed rule). Additionally, disordered systems might be of interest, where each node/map has 
a different level of chaoticity (that is for example a system of coupled standard maps with varying $\e$-values).
\item The CCM considered here appears to be relatively robust to the underlying topology: it would be interesting to examine how far the robustness 
persists, i.e. for how diverse spectrum of networks this CCM would still maintain its key collective properties (clustering, periodic orbits, anomalous diffusion, etc.). More generally, investigations of 2D CCM on various networks are still very scarce. An open question revolves around the issue of modularity: how much insight on the behavior of a large structure can be gained from studying the behavior of its building blocks (modules), i.e. by following a "bottom-up" approach?
\item In our study of CCM on 4-star motif we revealed a large spectrum of collective dynamical effects, which include various strange attractors with small Lyapunov Exponents. Some of them have properties that are possessed by a special category of attractors referred to in the literature as strange nonchaotic attractors (SNA). These dynamical phenomena are so far known to arise only in externally driven dynamical systems (whereas our system of 4-star is endogenously driven). Our evidence opens a question regarding possible mechanism that can lead to SNA in dynamical systems without external driving, where further investigation is needed.
\item We have demonstrated some results regarding the collective dynamics of 2D units on a directed topology of a real biological system -- the gene regulatory network of bacterium E.Coli. A wider investigation of dynamics on directed structures is still to be done, specially in the context of 2D CCM. Furthermore, 2D description of genes and their interactions might help design better models of gene regulatory networks, allowing a vast range of studies and applications. Realistic model of gene interactions could also be implemented on computer generated networks, and followed by studies of optimal design of functional gene networks.
\end{itemize}


\newpage
\addcontentsline{toc}{chapter}{Acknowledgments}
\chapter*{Acknowledgments}

\n Many Thanks to my adviser Prof. Bosiljka Tadi\'c for her guidance, support, suggestions, critiques, patience and everything else I need to get where I am now. \\[0.1cm] 

\n I gratefully acknowledge the support by the Program P10044 of the Ministery of Higher Education, Science and Technology of the Republic of Slovenia. \\[0.1cm]

\n I am grateful to Professors Ram Ramaswamy, Toma\v{z} Prosen, Alexander Mikhailov, Albert D\'iaz-Guilera, Ra\v sa Pirc, Geoff Rodgers and Thilo Gross for very useful suggestions and discussions. \\[0.1cm]

\n Very special Thanks to Milovan \v Suvakov for teaching me essentially all the \texttt{C++} programming I know today.\\[0.1cm] 

\n Thanks to Lev Vidmar, Marija Mitrovi\'c and Jelena Gruji\'c for being my colleagues and friends.\\[0.1cm]

\n Thanks to all my other colleagues and co-workers, which range from Ljubljana via Belgrade to New Delhi. \\[0.1cm] 

\n Thanks to all the personnel and staff from the Department of Theoretical Physics, Jo\v zef Stefan Institute, and specifically to Prof. Rajmund Krivec for administrative support of computer facilities that made this work possible. \\[0.1cm]

\n Thanks to Ur\v ska for her love and endless patience, specially during my Thesis work. \\[0.1cm]

\n And of course, I am infinitely grateful to my mother and my family for living with me through all this. \\[0.1cm]

\n Finally, let me not forget my father, my uncle, and my grandma, who would be proud if they were here.


\bibliographystyle{acm}
\newpage\addcontentsline{toc}{chapter}{Bibliography}
\bibliography{references-thesis.bib}

\newpage\addcontentsline{toc}{chapter}{List of Figures}
\listoffigures

\newpage\newpage\chapter*{Appendix} \addcontentsline{toc}{chapter}{Appendix}
We append two papers containing parts of this Thesis, in particular:
\begin{itemize}
\item Z. Levnaji\'c and B. Tadi\'c: "Self-organization in Trees and Motifs of Two-Dimensional Chaotic Maps with Time Delay", Journal of 
Statistical Mechanics: Theory and Experiment, P03003, 2008 (\cite{ja-jsm}) 
\item Z. Levnaji\'c and B. Tadi\'c: "Dynamical Patterns in Scalefree Trees of Coupled 2D Chaotic Maps", ICCS 2007, Lecture Notes on 
Computer Science 4488, p.633-640, 2007 (\cite{ja-lncs-1})
\end{itemize}

\end{document}